\documentclass[english]{fitthesis} 
\usepackage[nopostdot]{glossaries}
\usepackage{glossary-superragged}
\makeglossaries 
\csdoublequotesoff

\usepackage{url}

\usepackage{amsmath}
\usepackage[framemethod=tikz]{mdframed}
\usepackage{tikz}
\usetikzlibrary{shapes,arrows,positioning,calc}
\usetikzlibrary{decorations.pathreplacing}
\usetikzlibrary{shadows.blur}
\usetikzlibrary{shapes.symbols}
\usetikzlibrary{bayesnet}
\usetikzlibrary{babel}

\usepackage{breakcites}

\usepackage{algorithmicx}
\usepackage{algorithm}
\usepackage[noend]{algpseudocode}
\usepackage{siunitx}
\usepackage{pifont}
\usepackage{xcolor}
\usepackage{comment}
\usepackage{multirow,makecell}
\usepackage{enumitem}
\usepackage{caption}
\usepackage{subcaption}
\usepackage{tablefootnote}
\usepackage{graphicx}
\usepackage{multicol}

\ifWis
\ifx\pdfoutput\undefined 
\else
  \usepackage{color}
  \usepackage[unicode,colorlinks,hyperindex,plainpages=false,pdftex]{hyperref}
  \definecolor{hrcolor-ref}{RGB}{223,52,30}
  \definecolor{hrcolor-cite}{HTML}{2F8F00}
  \definecolor{hrcolor-urls}{HTML}{092EAB}
  \hypersetup{
	linkcolor=hrcolor-ref,
	citecolor=hrcolor-cite,
	filecolor=magenta,
	urlcolor=hrcolor-urls
  }
\fi
\else 
\ifx\pdfoutput\undefined 
\else
  \usepackage{color}
  \usepackage[unicode,colorlinks,hyperindex,plainpages=false,pdftex,urlcolor=black,linkcolor=black,citecolor=black]{hyperref}
  \definecolor{links}{rgb}{0,0,0}
  \definecolor{anchors}{rgb}{0,0,0}

\fi
\fi
\usepackage[all]{hypcap}

\newcommand*\XX{\mathbf{X}}
\newcommand*\YY{\mathbf{Y}}
\newcommand*\ZZ{\mathbf{Z}}
\newcommand*\xx{\mathbf{x}}
\newcommand*\yy{\mathbf{y}}
\newcommand*\VV{\mathbf{V}}
\DeclareMathOperator{\tr}{tr}

\newcommand\myatopthree[3]{\left[\substack{#1 \\ #2 \\ #3}\right]} 

\newcommand{\textoverline}[1]{$\overline{\mbox{#1}}$}

\usepackage{tocloft}
\projectinfo{
  project={DR},            
  year={2024},             
  date=\today,             
  title.cs={Od modulárních k celostním systémům diarizace řečníka},  
  title.en={From Modular to End-to-End Speaker Diarization}, 
  title.length={11cm}, 
  author.name={Federico Nicolás},   
  author.surname={Landini},   
  author.title.p={Lic.}, 
  department={UPGM}, 
  supervisor.name={Lukáš},   
  supervisor.surname={Burget},   
  supervisor.title.p={doc. Ing.},   
  supervisor.title.a={Ph.D.},    
  cosupervisor.name={Mireia},   
  cosupervisor.surname={Diez Sánchez},   
  cosupervisor.title.p={},   
  cosupervisor.title.a={M.Sc., Ph.D.},    
  keywords.cs={Diarizace mluvčího, VBx, neuronová diarizace end-to-end, simulované rozhovory, DiaPer.}, 
  keywords.en={Speaker diarization, VBx, end-to-end neural diarization, simulated conversations, DiaPer.}, 
  abstract.cs={
  Diarizace mluvčího se obvykle popisuje jako úloha, která určuje, „kdo kdy mluvil“ v nahrávce. Ještě před několika lety byly všechny konkurenceschopné přístupy modulární, tj. detekce hlasové aktivity, segmentace, extrakce embeddingů, shlukování a detekce a zpracování překrývající se řeči byly řešeny různými subsystémy a aplikovány jeden po druhém. Takto konstruované systémy dosahovaly ve většině scénářů nejlepších výsledků, ale měly velké potíže vypořádat se s překrývající se řečí. V poslední době vzbudil velkou pozornost nástup end-to-end modelů, které jsou schopny řešit všechny aspekty diarizace mluvčího pomocí jediného modelu a mají lepší úspěšnost, pokud jde o překrývající se řeč.
  
  Tato práce vznikla v období koexistence těchto dvou trendů. Popisujeme systém založený na bayesovském skrytém Markovově modelu používaném ke shlukování x-vektorů (embeddingů mluvčích získaných pomocí neuronové sítě), známý jako VBx, který prokázal pozoruhodnou úspěšnost na různých souborech dat a evaluačních v různých kampaních. Popíšeme jeho výhody a omezení a vyhodnotíme výsledky na různých relevantních korpusech. Poté přejdeme k metodám end-to-end neuronové diarizace (EEND). Vzhledem k potřebě velkých trénovacích sad pro trénování těchto modelů a nedostatku ručně anotovaných diarizačních dat v dostatečném množství spočívá kompromisní řešení v umělém generování trénovacích dat. Popíšeme přístup ke generování syntetických dat, která se z hlediska změn a překryvů mluvčích podobají skutečným rozhovorům. Ukážeme, jak tato metoda generování ,,simulovaných konverzací'' umožňuje dosáhnout lepší úspěšnosti než dříve navržená metoda vytváření ,,simulovaných směsí'' při trénování populárního EEND s enkodér-dekodér atraktory (EEND-EDA). Navrhneme také nový model založený na EEND, který nazýváme DiaPer, a ukážeme, že může dosáhnout lepších výsledků než EEND-EDA, zejména při práci s mnoha mluvčími a při zpracování překrývající se řeči.
  Nakonec porovnáváme oba systémy, VBx i DiaPer, na široké škále datových sad a prodiskutejeme výhody každého z nich.
  }, 
  abstract.en={
  Speaker diarization is usually referred to as the task that determines ``who spoke when'' in a recording. Until a few years ago, all competitive approaches were modular, i.e. voice activity detection, segmentation, embedding extraction, clustering and overlapped speech detection and handling were tackled by different sub-systems and applied in a cascaded fashion. Systems based on this framework reached state-of-the-art performance in most scenarios but had major difficulties dealing with overlapped speech. More recently, the advent of end-to-end models, capable of dealing with all aspects of speaker diarization with a single model and better performing regarding overlapped speech, has brought high levels of attention.
  
  This thesis is framed during a period of co-existence of these two trends. We describe a system based on a Bayesian hidden Markov model used to cluster x-vectors (speaker embeddings obtained with a neural network), known as VBx, which has shown remarkable performance on different datasets and challenges. We comment on its advantages and limitations and evaluate results on different relevant corpora. Then, we move towards end-to-end neural diarization (EEND) methods. Due to the need for large training sets for training these models and the lack of manually annotated diarization data in sufficient quantities, the compromise solution consists in generating training data artificially. We describe an approach for generating synthetic data which resembles real conversations in terms of speaker turns and overlaps. We show how this method generating ``simulated conversations'' allows for better performance than using a previously proposed method for creating ``simulated mixtures'' when training the popular EEND with encoder-decoder attractors (EEND-EDA). We also propose a new EEND-based model, which we call DiaPer, and show that it can perform better than EEND-EDA, especially when dealing with many speakers and handling overlapped speech.
  Finally, we compare both VBx-based and DiaPer systems on a wide variety of corpora and comment on the advantages of each technique.
  }, 
  declaration={I hereby declare that this doctoral thesis was prepared as an original work by the author under the supervision of Dr. Lukáš Burget and Dr. Mireia Diez Sánchez.
 I have listed all the literary sources, publications and other sources, which were used during the preparation of this thesis.},
  acknowledgment={Throughout my Ph.D. studies, I had the chance to interact with many people and I would like to thank them all as everyone probably shaped this work in some way or another. However, some deserve special mention.

First of all, my utmost gratitude goes to my advisors, Lukáš and Mireia, Mireia and Lukáš; of whom, I believe, I am their first common ``academic child''. Each of you has played an immense role in the work I have done during the past years. Mireia, always available to discuss ideas and results on a day-to-day basis and provide feedback without delays. Lukáš, always happy to talk about research ideas, even if that meant being late for his next appointment. A huge thank you goes to you two; you certainly know how much of this work would not have happened if it were not for your guidance. Thank you for allowing me to work quite freely and keeping a cheerful spirit despite my bad mood when things did not work as expected.

Next in line, Honza Černocký, who has also served as supervisor by shielding much of the administrative burden. I also thank him for giving me the opportunity to join BUT-Speech@FIT and for running the group as a non-bossy boss. This thesis would definitely not have happened without his support. My appreciation extends to all the staff at BUT who made it possible for me to work here during this time.

It has been a great experience to be in the BUT-Speech@FIT group, discussing work- and non-work-related topics. Newer members who met me towards the end of my Ph.D. might not know it, but I was certainly less bitter before. My advice to my past self would be to worry less and enjoy the process more. In any case, my gratitude goes to Speech@FIT as a whole: past/current permanent/temporary members. Special mentions go to the Speaker ID crew, with whom I interacted more at the beginning of my studies, and to the Diarization subgroup, who had to put up with me during the last time. Most of you have certainly shaped, if not co-authored, part of this work. Special among them is Pavel Matějka who showed me how pleasant running in the forests around Brno can be. Thank you for igniting what has become my passion for ultra-running.

The few internships I was able to do during these years were also very enriching and I thank everyone with whom I had the chance to work. Even if the work from the internships is not directly reflected in the thesis, the stays have certainly given me the chance to learn many things.

I would also like to thank my former advisor, Luciana Ferrer, with whom I took my first steps in the field of speech technologies. She made the contact with the folks in Brno that allowed me to join them for my Ph.D. so this work would not have happened without her help.

Then, I want to thank my parents, Adriana and Fabián, who encouraged the scientific side in me when I was little and who always supported me to keep learning. My gratitude extends to the rest of my family who I only saw a couple of times during these years but who were present all along the way. The same goes for my friends who, in spite of the distance, were always with me, even if I was not with them.

Last but not least, the biggest thanks are to Katka, my wife, for all her support through these years specially the last one. Thank you for putting up with me, be it when I am in a bad mood or with whatever upcoming ultra race I decide to train for.},
  extendedabstract={
  Diarizace mluvčího se obvykle označuje jako úloha, která určuje ``kdo kdy mluvil'' v nahrávce. Ještě před několika lety byly všechny konkurenční přístupy modulární, tj. detekce hlasové aktivity, segmentace, extrakce embeddingu, shlukování a detekce a zpracování překrývající se řeči byly řešeny různými subsystémy a aplikovány kaskádovitě. Tyto systémy dosahovaly ve většině scénářů nejlepších výsledků, ale měly velké potíže s překrývající se řečí. V nedávné době vzbudil velkou pozornost nástup end-to-end modelů, které jsou schopny řešit všechny aspekty diarizace mluvčího pomocí jediného modelu a mají lepší výkonnost, pokud jde o překrývající se řeč. Tato práce je zarámována do období koexistence těchto dvou trendů.

  První část této práce popisuje VBx, systém založený na bayesovském skrytém Markovově modelu (HMM), který se používá ke shlukování x-vektorů (embeddingů mluvčích získaných pomocí neuronové sítě). Tento přístup vychází z vysoké kvality x-vektorů jako reprezentací mluvčích a pravděpodobnostního modelu lineární diskriminační analýzy (PLDA) natrénovaného k porovnávání takových embeddingů, přičemž oba jsou natrénovány na velkém množství dat od tisíců mluvčích. Stavy v HMM, reprezentující každého z mluvčích v konverzaci, mají specifické modely mluvčích odvozené z PLDA, díky čemuž je celý systém velmi robustní vůči různým mluvčím a akustickým podmínkám. Model navíc dokáže velmi přesně zjistit počet mluvčích v rozhovorech, jak ukazují výsledky na několika datových sadách.

  V době zveřejnění dosáhl tento přístup nejlepších výsledků na třech relevantních souborech dat: Callhome, AMI a DIHARD 2. VBx byl základním modulem našeho vítězného systému v rámci druhé výzvy DIHARD a systémem, který se umístil na druhém místě v soutěži VoxSRC 2020. Díky silným výsledkům a veřejným receptům, které jsme zveřejnili, byl VBx použit jako výchozí systém organizátory nedávných výzev, jako jsou M2MeT a Ego4D, a jako výchozí systém v bezpočtu publikací.

  Zatímco kaskádové metody jsou poněkud těžkopádné kvůli nutnosti aplikovat různé subsystémy jeden po druhém, poměrně často jsou výsledky přístupů založených na shlukování, jako je VBx, prezentovány ve zjednodušených nastaveních, jako je použití detekce hlasové aktivity nebo bez zpracování překrývající se řeči. Za účelem poskytnutí spravedlivých referenčních hodnot pro komunitu byl v této práci vyhodnocen celoplošný systém založený na VBx na více než deseti různých souborech dat, přičemž u většiny z nich byly prokázány silné schopnosti.

  Pro odstranění hlavních nevýhod modulárních systémů se zbytek práce zaměřuje na modely neuronové diarizace end-to-end (EEND). Tyto modely představují výhodu v tom, že všechny aspekty úlohy, které v modulárních systémech zpracovávají různé a obvykle nezávislé systémy, řeší jediný modul. End-to-end modely však vyžadují velké množství trénovacích dat a vzhledem k nedostatku ručně anotovaných diarizačních dat v těchto množstvích je třeba vytvářet data syntetická. Představujeme alternativní přístup k původnímu schématu ,,simulovaných směsí'' pro generování ,,simulovaných konverzací''. Simulované směsi byly inspirovány technikami běžně používanými k přípravě trénovacích dat pro separaci řeči, která obsahují vysoké množství překrývající se řeči. To však neodráží vzorce v běžných rozhovorech. Abychom to vyřešili, jsou naše simulované konverzace navrženy s využitím statistik ze skutečných konverzací o délkách pauz a překrývání. Tato strategie přesněji napodobuje skutečné interakce a při použití jako trénovacích dat pro end-to-end modely jim umožňuje dosáhnout lepšího výkonu. Rozsáhlá srovnání našeho přístupu s předchozím ukázala, že použití simulovaných konverzací téměř zcela snižuje potřebu dolaďování modelu pomocí dat v doméně ve scénáři telefonování. Zveřejnili jsme kód pro generování trénovacích dat, jakož i naši implementaci modelu EEND s atraktory kodéru a dekodéru (EEND-EDA) v jazyce PyTorch a modely natrénované na volných veřejných datech. Díky tomu byly simulované rozhovory již převzaty do dalších publikací.

  Dále představujeme nový end-to-end model DiaPer, který představuje zobecněný systém pro neautoregresivní dekódování atraktorů v kontextu EEND. Ukazujeme, že může dosáhnout konkurenceschopných výsledků na telefonních hovorech s menším počtem parametrů než jiné modely EEND. Hlavními výhodami tohoto přístupu je lepší zvládání konverzací s mnoha mluvčími a také lepší schopnost zpracovávat překrývající se řeč ve srovnání se základním modelem EEND-EDA. DiaPer také vyhodnocujeme na několika širokopásmových datových sadách, které poskytují komplexní výchozí údaje pro budoucí publikace a na některých z nich vykazují velmi konkurenceschopný výkon. Kód a modely natrénované na volně přístupných veřejných datech jsou rovněž otevřeny.

  Nakonec porovnáváme systémy založené na VBx a DiaPer na široké škále korpusů a komentujeme výhody každé techniky vzhledem ke zvláštnostem každého souboru dat. Celkově si modulární systémy vedou lépe na konverzacích s vysokým počtem mluvčích. End-to-end systémy mají výhodu při práci s překryvy a nahrávkami s malým počtem mluvčích, avšak s jednostupňovým inferenčním procesem.
  
  },
  extabstract.odd={true}, 
  faculty.cs={Fakulta informačních technologií}, 
  faculty.en={Faculty of Information Technology}, 
  department.cs={Ústav počítačové grafiky a multimédií}, 
  department.en={Department of Computer Graphics and Multimedia} 
}

\newlist{checklist}{itemize}{1}
\setlist[checklist]{label=$\square$}

\begin{document}
  \maketitle
  \setlength{\parskip}{0pt}

  {\hypersetup{hidelinks}\tableofcontents}
  
  %

  \ifenglish
    \renewcommand*\glossaryname{List of acronyms}%
    \renewcommand*\entryname{Acronym}
    \renewcommand*\descriptionname{Meaning}
  \fi
  \newglossaryentry{AHC}{
    name={AHC},
    description={agglomerative hierarchical clustering}}
  \newglossaryentry{ASR}{
    name={ASR},
    description={automatic speech recognition}}
  \newglossaryentry{BCE}{
    name={BCE},
    description={binary cross-entropy}}
  \newglossaryentry{BHMM}{
    name={BHMM},
    description={Bayesian hidden Markov model}}
  \newglossaryentry{BIC}{
    name={BIC},
    description={Bayesian information criterion}}
  \newglossaryentry{BLSTM}{
    name={BLSTM},
    description={bidirectional long short-term memory}}
  \newglossaryentry{CNN}{
    name={CNN},
    description={convolutional neural network}}
  \newglossaryentry{DER}{
    name={DER},
    description={diarization error rate}}
  \newglossaryentry{DOVER}{
    name={DOVER},
    description={diarization output voting error reduction}}
  \newglossaryentry{DNN}{
    name={DNN},
    description={deep neural network}}
  \newglossaryentry{DVBx}{
    name={DVBx},
    description={discriminatively trained Bayesian HMM clustering of x-vector sequences}}
  \newglossaryentry{ECAPA-TDNN}{
    name={ECAPA-TDNN},
    description={emphasized channel attention, propagation and aggregation time-delay neural network}}
  \newglossaryentry{EEND}{
    name={EEND},
    description={end-to-end neural diarization}}
  \newglossaryentry{EEND-EDA}{
    name={EEND-EDA},
    description={end-to-end neural diarization with encoder-decoder attractors}}
  \newglossaryentry{EEND-VC}{
    name={EEND-VC},
    description={end-to-end neural diarization vector clustering}}
  \newglossaryentry{ELBO}{
    name={ELBO},
    description={evidence lower bound objective}}
  \newglossaryentry{FT}{
    name={FT},
    description={fine-tuning}}
  \newglossaryentry{GMM}{
    name={GMM},
    description={Gaussian mixture model}}
  \newglossaryentry{GNN}{
    name={GNN},
    description={graph neural network}}
  \newglossaryentry{GLR}{
    name={GLR},
    description={generalized likelihood ratio}}
  \newglossaryentry{HDP-HMM}{
    name={HDP-HMM},
    description={hierarchical Dirichlet process hidden Markov model}}
  \newglossaryentry{HMM}{
    name={HMM},
    description={hidden Markov model}}
  \newglossaryentry{JER}{
    name={JER},
    description={Jaccard error rate}}
  \newglossaryentry{JFA}{
    name={JFA},
    description={joint factor analysis}}
  \newglossaryentry{LDA}{
    name={LDA},
    description={linear discriminant analysis}}
  \newglossaryentry{LSTM}{
    name={LSTM},
    description={long short-term memory}}
  \newglossaryentry{MFCC}{
    name={MFCC},
    description={Mel frequency cepstral coefficients}}
  \newglossaryentry{MHA}{
    name={MHA},
    description={multi-head attention}}
  \newglossaryentry{MHSA}{
    name={MHSA},
    description={multi-head self-attention}}
  \newglossaryentry{MSCE}{
    name={MSCE},
    description={mean speaker counting error}}
  \newglossaryentry{NIST}{
    name={NIST},
    description={National Institute of Standards and Technology}}
  \newglossaryentry{NN}{
    name={NN},
    description={neural network}}
  \newglossaryentry{OSD}{
    name={OSD},
    description={overlapped speech detection}}
  \newglossaryentry{PIT}{
    name={PIT},
    description={permutation invariant training}}
  \newglossaryentry{PLDA}{
    name={PLDA},
    description={probabilistic linear discriminant analysis}}
  \newglossaryentry{PSE}{
    name={PSE},
    description={power set encoding}}
  \newglossaryentry{RIR}{
    name={RIR},
    description={room impulse response}}
  \newglossaryentry{ROVER}{
    name={ROVER},
    description={recognizer output voting error reduction}}
  \newglossaryentry{RNN}{
    name={RNN},
    description={recurrent neural network}}
  \newglossaryentry{RT}{
    name={RT},
    description={Rich Transcriptions}}
  \newglossaryentry{SAD}{
    name={SAD},
    description={speech activity detection}}
  \newglossaryentry{SA-EEND}{
    name={SA-EEND},
    description={self-attention end-to-end neural diarization}}
  \newglossaryentry{SD}{
    name={SD},
    description={speaker diarization}}
  \newglossaryentry{SC}{
    name={SC},
    description={simulated conversations}}
  \newglossaryentry{SC-EEND}{
    name={SC-EEND},
    description={speaker-wise chain end-to-end neural diarization}}
  \newglossaryentry{SCD}{
    name={SCD},
    description={speaker change point detection}}
  \newglossaryentry{SM}{
    name={SM},
    description={simulated mixtures}}
  \newglossaryentry{SOT}{
    name={SOT},
    description={serialized output training}}
  \newglossaryentry{SNR}{
    name={SNR},
    description={signal-to-noise ratio}}
  \newglossaryentry{SR}{
    name={SR},
    description={speaker recognition}}
  \newglossaryentry{SRE}{
    name={SRE},
    description={speaker recognition evaluation}}
  \newglossaryentry{TDNN}{
    name={TDNN},
    description={time-delay neural network}}
  \newglossaryentry{TSE}{
    name={TSE},
    description={target speaker extraction}}
  \newglossaryentry{TS-VAD}{
    name={TS-VAD},
    description={target speaker voice activity detection}}
  \newglossaryentry{UBM}{
    name={UBM},
    description={universal background model}}
  \newglossaryentry{VAD}{
    name={VAD},
    description={voice activity detection}}
  \newglossaryentry{VB}{
    name={VB},
    description={variational Bayesian}}
  \newglossaryentry{VBx}{
    name={VBx},
    description={Bayesian HMM clustering of x-vector sequences}}
   
  \setglossarystyle{superragged}
  \glsadd{AHC}
  \glsadd{ASR}
  \glsadd{BCE}
  \glsadd{BHMM}
  \glsadd{BIC}
  \glsadd{BLSTM}
  \glsadd{CNN}
  \glsadd{DER}
  \glsadd{DNN}
  \glsadd{DOVER}
  \glsadd{DVBx}
  \glsadd{ECAPA-TDNN}
  \glsadd{EEND}
  \glsadd{EEND-EDA}
  \glsadd{EEND-VC}
  \glsadd{ELBO}
  \glsadd{FT}
  \glsadd{GMM}
  \glsadd{GNN}
  \glsadd{GLR}
  \glsadd{HDP-HMM}
  \glsadd{HMM}
  \glsadd{JER}
  \glsadd{JFA}
  \glsadd{LDA}
  \glsadd{LSTM}
  \glsadd{MFCC}
  \glsadd{MHA}
  \glsadd{MHSA}
  \glsadd{MSCE}
  \glsadd{NIST}
  \glsadd{NN}
  \glsadd{OSD}
  \glsadd{PIT}
  \glsadd{PLDA}
  \glsadd{PSE}
  \glsadd{SAD}
  \glsadd{SA-EEND}
  \glsadd{SC}
  \glsadd{SC-EEND}
  \glsadd{SM}
  \glsadd{SOT}
  \glsadd{RIR}
  \glsadd{ROVER}
  \glsadd{RNN}
  \glsadd{RT}
  \glsadd{SCD}
  \glsadd{SD}
  \glsadd{SNR}
  \glsadd{SR}
  \glsadd{SRE}
  \glsadd{TDNN}
  \glsadd{TSE}
  \glsadd{TS-VAD}
  \glsadd{UBM}
  \glsadd{VAD}
  \glsadd{VB}
  \glsadd{VBx}
  
  \printglossary[type=main,style=long,nonumberlist]

  \ifODSAZ
    \setlength{\parskip}{0.5\bigskipamount}
  \else
    \setlength{\parskip}{0pt}
  \fi

  \iftwoside
    \cleardoublepage
  \fi

  \ifenglish


\newcommand\TODO[1]{\textcolor{red}{#1}}
\newcommand\REWRITE[1]{\textcolor{blue}{#1}}

\chapter{Introduction}
\label{sec:introduction}

Speech processing technologies have advanced considerably during the last decades due to more computing capacity, more available data, and big efforts from the research community to propose new techniques and improve the performance of systems. Many speech-related tasks have progressed remarkably and one of them is speaker diarization, usually referred to as the task of deciding ``who spoke when''. More formally, it is the task of assigning speaker labels to segments of speech in a given recording. This task is of interest per se for allowing segmentation of conversations or search of speech from a speaker of interest, among others. However, it is also useful as an upstream task for other well-studied tasks such as information retrieval, speech recognition, or speaker verification.

Speaker diarization is applied on diverse domains. While originally constrained to telephone conversations, recorded meetings, or broadcast news, in the last few years, large amounts of broadcast data (not only news) have also gained relevance as target data. Even more, the availability of small devices has facilitated recording speech in more varied situations. Common examples are recordings in courtrooms or parliaments as well as conversations in many locations, such as restaurants. Besides, some specific applications are of interest, for example, diarization on clinical interviews to analyze sessions with patients and diarization on conversations in households with children to evaluate language acquisition. Furthermore, the recent global pandemic has increased the number of meetings over web-based platforms and speaker diarization can be useful to enrich transcriptions. Even more, ubiquitous speech-controlled devices often require to know which user is interacting with the device, thus benefiting from automatic diarization. All these applications provide new challenges and such conditions demand robust diarization systems able to cope with different acoustic channels, styles of speech and numbers of speakers.

Since the origins of speaker diarization around 20 years ago, several approaches have been proposed. We describe the techniques used in the most popular past systems as well as the models that attain state-of-the-art performance nowadays. Systems can be framed under two big families, namely modular ones: those formed by different sub-modules such as voice activity detection, embedding extraction, and clustering (among others); and neural network-based end-to-end systems that solve the task with a single module. State-of-the-art systems under these two frameworks are described, evaluated and compared in this work.

\newpage
\section{Contributions}
\label{sec:contributions}

The contributions of our work are the following:

\begin{itemize}
    \item We present the Bayesian HMM clustering of x-vectors (VBx) as a method for clustering of speaker embeddings for diarization. VBx was presented in \cite{landini2022bayesian}, showing state-of-the-art performance (at the time) on three relevant datasets. At the time of that publication, AMI, one of the most standard diarization datasets, was used with different evaluation setups, effectively impeding proper comparison between works of different authors. We pointed this out and showed the performance of VBx in the most widely used setups. With the aim of avoiding future confusion, we released the annotations derived from the official transcriptions following a clear and detailed procedure \url{https://github.com/BUTSpeechFIT/AMI-diarization-setup}
    
    We also released the code and x-vector extractors needed to run the recipes: 
    
    \url{https://github.com/BUTSpeechFIT/VBx}
    
    Due to the continuous maintenance of the repository, ease of use and competitive results, VBx has served as baseline in numerous publications.
    \item VBx was also a core module of the winning system of the Second DIHARD Challenge~\cite{landini2020but} and the second-ranked system in VoxSRC 2020~\cite{landini2021analysis}. We released the code for both systems too.
    
    \url{https://github.com/BUTSpeechFIT/VBx/tree/v1.0_DIHARDII}
    
    \url{https://github.com/BUTSpeechFIT/VBx/tree/v1.1_VoxConverse2020}
    
    Thanks to the strong performance and public recipes, VBx has been used as the baseline in recent challenges such as M2MeT and Ego4D~\cite{yu2022m2met,grauman2022ego4d}. 
    \item This thesis also presents the results obtained with a fully modular system based on the public VBx recipe and public voice activity detection and overlapped speech detection systems. These results are presented on a variety of datasets and serve not only as baseline in the context of this work but for other researchers in the field.
    \item End-to-end neural diarization (EEND) models require large amounts of training data and, due to the lack of manually annotated data in those quantities, synthetic data need to be generated. We introduce an alternative approach to the original ``simulated mixtures'' scheme for generating ``simulated conversations'' (which resemble real conversations regarding the lengths of pauses and overlaps), used as training data for end-to-end models. Extensive comparisons between our approach and the former one in \cite{landini22_interspeech, landini2023multi} showed that using simulated conversations reduces almost completely the need for fine-tuning the model using in-domain data in the telephony scenario. We released the code for generating training data as well as our PyTorch implementation of the EEND with encoder-decoder attractors model and models trained on free public data:
    
    \url{https://github.com/BUTSpeechFIT/EEND_dataprep}
    
    \url{https://github.com/BUTSpeechFIT/EEND}

    Thanks to this, simulated conversations have already been adopted in other publications.
    \item We present the novel end-to-end DiaPer model, which presents a generalized framework for non-autoregressive attractor decoding in the context of EEND. We show that it can reach competitive results on telephone conversations with fewer parameters than other EEND models. We also evaluate DiaPer on several datasets providing a comprehensive baseline. This work~\cite{landini2023diaper} has not yet been published in any venue at the time of writing this thesis. We also share the code and models trained on free public data: \url{https://github.com/BUTSpeechFIT/DiaPer}
\end{itemize}

It should be noted that not all the work done during the PhD is mentioned above, namely the work related to publications~\cite{diez2018but,landini2020but,diez2020optimizing,zmolikovabut,landini2021but,landini2021analysis}. However, those works have shaped what is presented in this document. At the same time, some of the content of the thesis has not been presented in previous publications, especially the comparisons in Chapter~\ref{sec:conclusion}.

\section{Organization of the thesis}
\label{sec:organization}

The rest of the thesis is organized as follows:
\begin{itemize}
    \item Chapter~\ref{sec:preliminaries} formally defines the task of speaker diarization describing how models are evaluated and the connection with other tasks.
    \item Chapter~\ref{sec:cascaded_diarization} presents modular diarization systems and describes in detail VBx, one of the clustering methods considered state-of-the-art nowadays, which is one of the main contributions of this thesis. The performance of VBx is analyzed on different datasets.
    \item Chapter~\ref{sec:e2e_diarization} presents end-to-end diarization systems covering the different works that have been recently published. Given the ``data-hungry'' nature of these models, we describe a new method for generating synthetic training data, coined ``simulated conversations'', and show its advantages. Then, we present DiaPer, a novel method based on the end-to-end framework and analyze its performance on different datasets.
    \item Chapter~\ref{sec:conclusion} concludes the thesis discussing the comparison of the aforementioned modular and end-to-end systems and the future challenges of speaker diarization.
\end{itemize}

\chapter{Preliminaries of speaker diarization}\label{sec:preliminaries}
\section{Task definition}\label{sec:task_definition}

Historically, audio diarization~\cite{reynolds2005approaches} has referred to the broad task of segmenting a recording and categorizing those segments (for example, into silence, noise, music, commercials, speech, etc.). The popularity it gained in the early 2000s has waned and few works are framed under that task nowadays. Speaker diarization corresponds to the more specific task of segmenting a recording in terms of speech segments and assigning speaker labels. Speaker diarization (SD) is usually referred to as ``who spoke when'' and gained popularity in the current definition also during the early 2000s, mainly due to the National Institute of Standards and Technology (NIST) in their series of Rich Transcriptions (RT) evaluations and several European projects~\cite{anguera2012speaker}. 

Examples of both tasks are presented in Figure~\ref{fig:diarization_example} over the same recording. 
Note that for SD, the goal is to distinguish speakers and using relative speaker labels such as `Spk1' and `Spk2' is enough. Still, for some applications, the role (i.e. presenter, announcer, guest, etc.) or specific speaker labels (i.e. Alex, Bob, Carol, etc.) are necessary and in such cases, the SD labels are post-processed by a speaker identification system. Since only speaker segments are of interest, anything else is simply not labeled. Speech segments can be bounded by non-speech segments or speaker change points, as observed in Figure~\ref{fig:diarization_example}.

\begin{figure}[ht]
    \centering
    \begin{tikzpicture}
        \filldraw[fill=white!50] (0,2.5) rectangle (1,3.25) node[pos=.5] {sil};
        \filldraw[fill=green!50] (1,2.5) rectangle (2.5,3.25) node[pos=.5] {music};
        \filldraw[fill=purple!50] (2.5,2.5) rectangle (4.5,3.25) node[pos=.5] {speech};
        \filldraw[fill=orange!50] (4.5,2.5) rectangle (8,3.25) node[pos=.5] {commercial};
        \filldraw[fill=purple!50] (8,2.5) rectangle (12.5,3.25) node[pos=.5] {speech};
        \filldraw[fill=white!50] (12.5,2.5) rectangle (13,3.25) node[pos=.5] {sil};
        \filldraw[fill=purple!50] (13,2.5) rectangle (15,3.25) node[pos=.5] {speech};
        
        \filldraw[fill=white!50] (0,1) rectangle (2.5,1.75) node[pos=.5] {};
        \filldraw[fill=blue!50] (2.5,1) rectangle (4.5,1.75) node[pos=.5] {Spk1};
        \filldraw[fill=orange!50] (4.5,1) rectangle (8,1.75) node[pos=.5] {Spk2};
        \filldraw[fill=blue!50] (8,1) rectangle (10.5,1.75) node[pos=.5] {Spk1};
        \filldraw[fill=red!50] (10.5,1) rectangle (12.5,1.75) node[pos=.5] {Spk3};
        \filldraw[fill=white!50] (12.5,1) rectangle (13,1.75) node[pos=.5] {};
        \filldraw[fill=blue!50] (13,1) rectangle (15,1.75) node[pos=.5] {Spk1};
    \end{tikzpicture}
    \caption{Audio diarization and speaker diarization example. ``sil'' means silence. Both sequences correspond to the same recording. }
    \label{fig:diarization_example}
\end{figure}
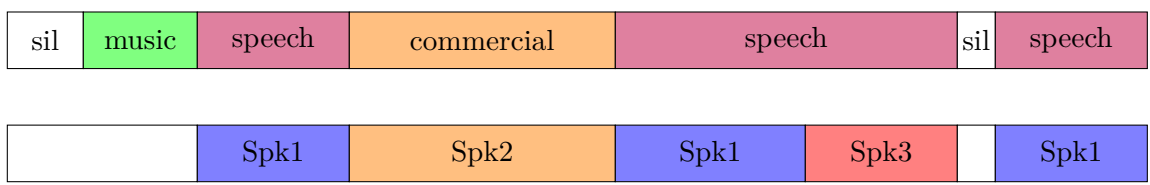

In this work, we focus on the most general scenario for speaker diarization, where no information is given about the input recording. Some examples of a priori information can be 
\begin{itemize}
    \item number of speakers
    \item language
    \item scenario/source (i.e. telephone conversation, meeting, broadcast news) which can help in choosing specific pre-processing techniques, infer the number of speakers or what type of speech to expect (formal, spontaneous, etc.)
    \item audio samples of speakers which can help to look for those speakers using anchor models
\end{itemize}

A priori information can make the diarization task more specific and potentially simpler. However, in this work, we will not work under the assumption of given information as we look for robust models that can deal with a myriad of situations.

Two extensively studied tasks are related to diarization: voice activity detection (VAD), sometimes also referred to as speech activity detection (SAD), and overlapped speech detection (OSD). The first one has historically been considered an external task for the sake of reducing the load on the diarization system. VAD is also of interest for other tasks such as automatic speech recognition (ASR) or speaker recognition (SR), and thus, efforts have been put into improving VAD systems in general and not only for diarization. OSD has usually been considered an external task; although much more related to diarization. Nevertheless, both are part of diarization for the scope of this work.

\section{Evaluation}\label{sec:evaluation}

\subsection{Ground truth labels}\label{sec:ground_truth_labels}
In order to evaluate a diarization system, just as with any other task, the annotations produced by a system need to be compared against ground truth ones. However, ground truth labels are generated following different criteria depending on the dataset. To illustrate the procedure of generating diarization labels, we utilize a few corpora as examples: 

For the First DIHARD Challenge\footnote{\url{https://dihardchallenge.github.io/dihard1/}}, manual annotations were first produced, then refined using forced alignment (where pauses longer than 200\,ms were used to split segments) and then checked and corrected by annotators~\cite{ryant2018first}. This is a very expensive annotation pipeline since human annotators need to listen to a recording several times to produce quality annotations. 

A more automated approach was utilized when collecting and annotating VoxConverse~\cite{chung20_interspeech}, where videos were available and voice activity detection was correlated with mouth movement of in-screen speakers to assign speaker labels to most speech segments. Speaker embeddings of off-screen speech segments were obtained and compared with those of already determined speakers in order to assign identities to those segments. This was followed by manual inspection to identify common errors and instructions were given to manual annotators to correct them following certain guidelines (i.e. pauses longer than 250\,ms were used to split segments, mark segments with 0.1\,s precision, etc.).

As observed in these two examples, different criteria can be used to produce the reference annotations. 
Related to the precision of the reference annotations is also the discussion of calculating metrics using a forgiveness collar or not. Using a collar means that small differences (below a certain threshold) between the reference and system annotations will not be considered for calculating errors. Different corpora have chosen different conventions; for example, DIHARD does not use a collar (meaning that manual annotations are fully trusted), while VoxConverse considers a 250\,ms collar. It can be argued that such a collar can account for inconsistencies in the annotation pipeline. However, as we have stated in~\cite{landini2022bayesian}, even if human annotations have errors, it does not mean that speech that could be evaluated should always be discarded around each speaker change point. The only thing that the annotation errors imply is that diarization systems have a non-zero lower bound on the diarization error. Moreover, not all metrics can, a priori, be calculated with a collar and introducing them creates different instances of a given metric.

\subsection{Metrics}\label{sec:metrics}

Since speaker labels generated by diarization systems do not need to match the naming convention of the reference ones (e.g. `Spk1', `Spk2', etc.), a 1:1 mapping between the system-generated speaker clusters and the reference ones is obtained using the Hungarian method~\cite{kuhn1955hungarian} for finding the solution of a bipartite graph. Given this optimal mapping, errors are calculated to determine how good the system output is with regard to the reference.

The most widely used performance metric for diarization is the diarization error rate (DER) proposed by NIST in their series of RT evaluations during the 2000s~\cite{NISTRT}. DER is defined by

\begin{equation}
    DER=\frac{SER+FA+Miss}{Total \ speech},
\end{equation}
\noindent where:
\begin{itemize}
    \item $SER$ stands for speaker error, the amount of time that speech is attributed to incorrect speakers (i.e. confusion of speakers)
    \item $FA$ is false alarm, the amount of time that non-speech regions are incorrectly attributed to a speaker (or time when overlapped speech of $K$ speakers is found in speech regions with $N<K$ reference speakers)
    \item $Miss$ stands for missed speech, the amount of time that speech is not attributed to any speaker (or time when overlapped speech of $K$ speakers is found for regions with $N>K$ reference overlapping speakers)
    \item $Total \ speech$ is the total amount of speech, accounting also for speaker overlaps.
\end{itemize}

An example of how DER is calculated can be seen in Figure~\ref{fig:DER_example} where the best matching \{Interviewer - Spk1, Interviewee1 - Spk2, Interviewee2 - <>\} is denoted by the colors. For more details about DER, please refer to the RT-09 evaluation plan\footnote{\url{https://web.archive.org/web/20100606092041if\_/http://www.itl.nist.gov/iad/mig/tests/rt/2009/docs/rt09-meeting-eval-plan-v2.pdf}}.

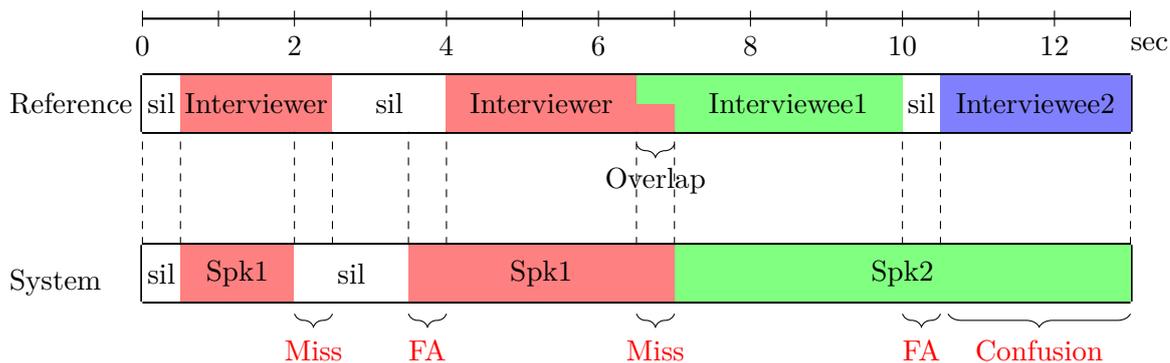
\begin{figure}[ht]
    \centering
    \begin{tikzpicture}
        \draw[thick, -] (0,4) -- (13,4);
        \foreach \x in {0,1,2,3,4,5,6,7,8,9,10,11,12,13}
        \draw (\x cm,4cm+3pt) -- (\x cm,4cm-3pt);
        
        \draw[ultra thick] (0,4) node[below=3pt,thick] {0} node[above=3pt] {};
        \draw[ultra thick] (2,4) node[below=3pt,thick] {2} node[above=3pt] {};
        \draw[ultra thick] (4,4) node[below=3pt,thick] {4} node[above=3pt] {};
        \draw[ultra thick] (6,4) node[below=3pt,thick] {6} node[above=3pt] {};
        \draw[ultra thick] (8,4) node[below=3pt,thick] {8} node[above=3pt] {};
        \draw[ultra thick] (10,4) node[below=3pt,thick] {10} node[above=3pt] {};
        \draw[ultra thick] (12,4) node[below=3pt,thick] {12} node[above=3pt] {};
        \draw[ultra thick] (13.25,4) node[below=3pt,thick] {sec} node[above=3pt] {};

        \node[text width=3.5cm] at (0,2.875) {Reference};

        \draw[ultra thick, -] (0,2.5) -- (13,2.5);
        \draw[ultra thick, -] (0,2.5) -- (0,3.25);
        \draw[ultra thick, -] (0,3.25) -- (13,3.25);
        \draw[ultra thick, -] (13,2.5) -- (13,3.25);
        
        \fill[fill=white!50] (0,2.5) rectangle (0.5,3.25) node[pos=.5] {sil};
        \fill[fill=red!50] (0.5,2.5) rectangle (2.5,3.25) node[pos=.5] {Interviewer};
        \fill[fill=white!50] (2.5,2.5) rectangle (4,3.25) node[pos=.5] {sil};
        \fill[fill=red!50] (4,2.5) rectangle (6.5,3.25) node[pos=.5] {Interviewer};
        \fill[fill=red!50] (6.48,2.5) rectangle (7,2.875) node[pos=.5] {};
        \fill[fill=green!50] (6.5,2.875) rectangle (7.02,3.25) node[pos=.5] {};
        \fill[fill=green!50] (7,2.5) rectangle (10,3.25) node[pos=.5] {Interviewee1};
        \fill[fill=white!50] (10,2.5) rectangle (10.5,3.25) node[pos=.5] {sil};
        \fill[fill=blue!50] (10.5,2.5) rectangle (13,3.25) node[pos=.5] {Interviewee2};

        \node[text width=3.5cm] at (0,0.5) {System};

        \draw[ultra thick, -] (0,0.25) -- (13,0.25);
        \draw[ultra thick, -] (0,0.25) -- (0,1);
        \draw[ultra thick, -] (0,1) -- (13,1);
        \draw[ultra thick, -] (13,0.25) -- (13,1);
        
        \fill[fill=white!50] (0,0.25) rectangle (0.5,1) node[pos=.5] {sil};
        \fill[fill=red!50] (0.5,0.25) rectangle (2,1) node[pos=.5] {Spk1};
        \fill[fill=white!50] (2,0.25) rectangle (3.5,1) node[pos=.5] {sil};
        \fill[fill=red!50] (3.5,0.25) rectangle (7,1) node[pos=.5] {Spk1};
        \fill[fill=green!50] (7,0.25) rectangle (13,1) node[pos=.5] {Spk2};

        \draw[dashed] (0,1) -- (0,2.5);
        \draw[dashed] (0.5,) -- (0.5,2.5);
        \draw[dashed] (2,1) -- (2,2.5);
        \draw[dashed] (2.5,1) -- (2.5,2.5);
        \draw[dashed] (3.5,1) -- (3.5,2.5);
        \draw[dashed] (4,1) -- (4,2.5);
        \draw[dashed] (6.5,1) -- (6.5,2.5);
        \draw[dashed] (7,1) -- (7,2.5);
        \draw[dashed] (10,1) -- (10,2.5);
        \draw[dashed] (10.5,1) -- (10.5,2.5);
        \draw[dashed] (13,1) -- (13,2.5);

        \draw [decorate,decoration={brace,amplitude=5pt,mirror,raise=4ex}]
  (6.5,3) -- (7,3) node[midway,yshift=-3em]{Overlap}{};
        \draw [decorate,decoration={brace,amplitude=5pt,mirror,raise=4ex}]
  (2,0.75) -- (2.5,0.75) node[midway,yshift=-3em]{\textcolor{red}{Miss}}{};
        \draw [decorate,decoration={brace,amplitude=5pt,mirror,raise=4ex}]
  (3.5,0.75) -- (4,0.75) node[midway,yshift=-3em]{\textcolor{red}{FA}}{};
        \draw [decorate,decoration={brace,amplitude=5pt,mirror,raise=4ex}]
  (6.5,0.75) -- (7,0.75) node[midway,yshift=-3em]{\textcolor{red}{Miss}}{};
        \draw [decorate,decoration={brace,amplitude=5pt,mirror,raise=4ex}]
  (10,0.75) -- (10.5,0.75) node[midway,yshift=-3em]{\textcolor{red}{FA}}{};
        \draw [decorate,decoration={brace,amplitude=5pt,mirror,raise=4ex}]
  (10.6,0.75) -- (13,0.75) node[midway,yshift=-3em]{\textcolor{red}{Confusion}}{};
        
    \end{tikzpicture}
    \caption{Speaker diarization evaluation example in terms of DER. Total missed speech: 1\,s, total FA speech: 1\,s, speaker confusion: 2.5\,s, total reference speech: 11\,s; therefore, $DER (\%)=100 \cdot \frac{1 + 1 + 2.5}{11} = 40.9\%$}
    \label{fig:DER_example}
\end{figure}

Due to inherent inconsistencies in the annotations introduced by the annotators when finding the precise boundaries of segments, NIST proposed using an optional 0.25\,s forgiveness collar around each boundary of a speech reference segment where speech is not scored. Using a collar means that small differences (below a certain threshold) between the reference and system annotations will not be considered for calculating errors.

For the Second DIHARD Speaker Diarization Challenge~\cite{ryant19_interspeech}, the organizers introduced the Jaccard error rate (JER) based on the Jaccard index~\cite{jaccard1901etude}. JER has become a popular secondary metric after DER. 

As explained in~\cite{ryant19_interspeech}, to obtain JER, the Hungarian algorithm is used as for the DER. Then, the JER for the reference speaker \emph{ref} is obtained as
\begin{equation}
    JER_{ref}=\frac{FA_{ref}+Miss_{ref}}{Total \ speech_{ref}},
\end{equation}

where
\begin{itemize}
    \item $Total \ speech_{ref}$ is the duration of the union of reference and system speaker segments; if the reference speaker was not paired with a system speaker, it is the speech time of the reference speaker
    \item $FA_{ref}$ is the total system speaker time not attributed to the reference speaker; if the reference speaker was not paired with a system speaker, it is 0
    \item $Miss_{ref}$ is the total reference speaker time not attributed to the system speaker; if the reference speaker was not paired with a system speaker, it is equal to $Total \ speech$.
\end{itemize}

Then, JER is the average of the $N$ speaker-specific Jaccard error rates: 
\begin{equation}
    JER = \frac{1}{N} \sum\limits_{ref} JER_{ref}
\end{equation}

\begin{figure}[ht]
    \centering
    \begin{tikzpicture}
        \draw[thick, -] (0,4) -- (13,4);
        \foreach \x in {0,1,2,3,4,5,6,7,8,9,10,11,12,13}
        \draw (\x cm,4cm+3pt) -- (\x cm,4cm-3pt);
        
        \draw[ultra thick] (0,4) node[below=3pt,thick] {0} node[above=3pt] {};
        \draw[ultra thick] (2,4) node[below=3pt,thick] {2} node[above=3pt] {};
        \draw[ultra thick] (4,4) node[below=3pt,thick] {4} node[above=3pt] {};
        \draw[ultra thick] (6,4) node[below=3pt,thick] {6} node[above=3pt] {};
        \draw[ultra thick] (8,4) node[below=3pt,thick] {8} node[above=3pt] {};
        \draw[ultra thick] (10,4) node[below=3pt,thick] {10} node[above=3pt] {};
        \draw[ultra thick] (12,4) node[below=3pt,thick] {12} node[above=3pt] {};
        \draw[ultra thick] (13.25,4) node[below=3pt,thick] {sec} node[above=3pt] {};

        \node[text width=3.5cm] at (0,2.875) {Reference};

        \draw[ultra thick, -] (0,2.5) -- (13,2.5);
        \draw[ultra thick, -] (0,2.5) -- (0,3.25);
        \draw[ultra thick, -] (0,3.25) -- (13,3.25);
        \draw[ultra thick, -] (13,2.5) -- (13,3.25);
        
        \fill[fill=white!50] (0,2.5) rectangle (0.5,3.25) node[pos=.5] {sil};
        \fill[fill=red!50] (0.5,2.5) rectangle (2.5,3.25) node[pos=.5] {Interviewer};
        \fill[fill=white!50] (2.5,2.5) rectangle (4,3.25) node[pos=.5] {sil};
        \fill[fill=red!50] (4,2.5) rectangle (6.5,3.25) node[pos=.5] {Interviewer};
        \fill[fill=red!50] (6.48,2.5) rectangle (7,2.875) node[pos=.5] {};
        \fill[fill=green!50] (6.5,2.875) rectangle (7.02,3.25) node[pos=.5] {};
        \fill[fill=green!50] (7,2.5) rectangle (10,3.25) node[pos=.5] {Interviewee1};
        \fill[fill=white!50] (10,2.5) rectangle (10.5,3.25) node[pos=.5] {sil};
        \fill[fill=blue!50] (10.5,2.5) rectangle (13,3.25) node[pos=.5] {Interviewee2};

        \node[text width=3.5cm] at (0,0.5) {System};

        \draw[ultra thick, -] (0,0.25) -- (13,0.25);
        \draw[ultra thick, -] (0,0.25) -- (0,1);
        \draw[ultra thick, -] (0,1) -- (13,1);
        \draw[ultra thick, -] (13,0.25) -- (13,1);
        
        \fill[fill=white!50] (0,0.25) rectangle (0.5,1) node[pos=.5] {sil};
        \fill[fill=red!50] (0.5,0.25) rectangle (2,1) node[pos=.5] {Spk1};
        \fill[fill=white!50] (2,0.25) rectangle (3.5,1) node[pos=.5] {sil};
        \fill[fill=red!50] (3.5,0.25) rectangle (7,1) node[pos=.5] {Spk1};
        \fill[fill=green!50] (7,0.25) rectangle (13,1) node[pos=.5] {Spk2};

        \draw[dashed] (0,1) -- (0,2.5);
        \draw[dashed] (0.5,) -- (0.5,2.5);
        \draw[dashed] (2,1) -- (2,2.5);
        \draw[dashed] (2.5,1) -- (2.5,2.5);
        \draw[dashed] (3.5,1) -- (3.5,2.5);
        \draw[dashed] (4,1) -- (4,2.5);
        \draw[dashed] (6.5,1) -- (6.5,2.5);
        \draw[dashed] (7,1) -- (7,2.5);
        \draw[dashed] (10,1) -- (10,2.5);
        \draw[dashed] (10.5,1) -- (10.5,2.5);
        \draw[dashed] (13,1) -- (13,2.5);

        \draw [decorate,decoration={brace,amplitude=5pt,mirror,raise=4ex}]
  (6.5,3) -- (7,3) node[midway,yshift=-3em]{Overlap}{};
        \draw [decorate,decoration={brace,amplitude=5pt,mirror,raise=4ex}]
  (2,0.75) -- (2.5,0.75) node[align=center,midway,xshift=-0.5em,yshift=-3.5em,text width=3cm]{\textcolor{red}{Miss \\ Interviewer}}{};
        \draw [decorate,decoration={brace,amplitude=5pt,mirror,raise=4ex}]
  (3.5,0.75) -- (4,0.75) node[align=center,midway,xshift=0.5em,yshift=-3.5em, text width=3cm]{\textcolor{red}{FA \\ Interviewer}}{};
        \draw [decorate,decoration={brace,amplitude=5pt,mirror,raise=4ex}]
  (6.5,0.75) -- (7,0.75) node[align=center,midway,yshift=-3.5em,text width=3cm]{\textcolor{red}{Miss \\ Interviewee1}}{};
        \draw [decorate,decoration={brace,amplitude=5pt,mirror,raise=4ex}]
  (10,0.75) -- (13,0.75) node[align=center,midway,yshift=-3.5em, text width=3cm]{\textcolor{red}{FA \\ Interviewee1}}{};
        \draw [decorate,decoration={brace,amplitude=5pt,mirror,raise=4ex}]
  (10.6,-0.5) -- (13,-0.5) node[align=center,midway,yshift=-3.5em,text width=3cm]{\textcolor{red}{Miss \\ Interviewee2}}{};
        
    \end{tikzpicture}
    \caption{Speaker diarization evaluation example in terms of JER. $JER_{Interviewer} = \frac{0.5+0.5}{5.5} = 0.18$, $JER_{Interviewee1} = \frac{3+0.5}{6.5} = 0.54$, $JER_{Interviewee2} = \frac{0+2.5}{2.5} = 1$; therefore, $JER (\%) = 57.3\%$.}
    \label{fig:JER_example}
\end{figure}
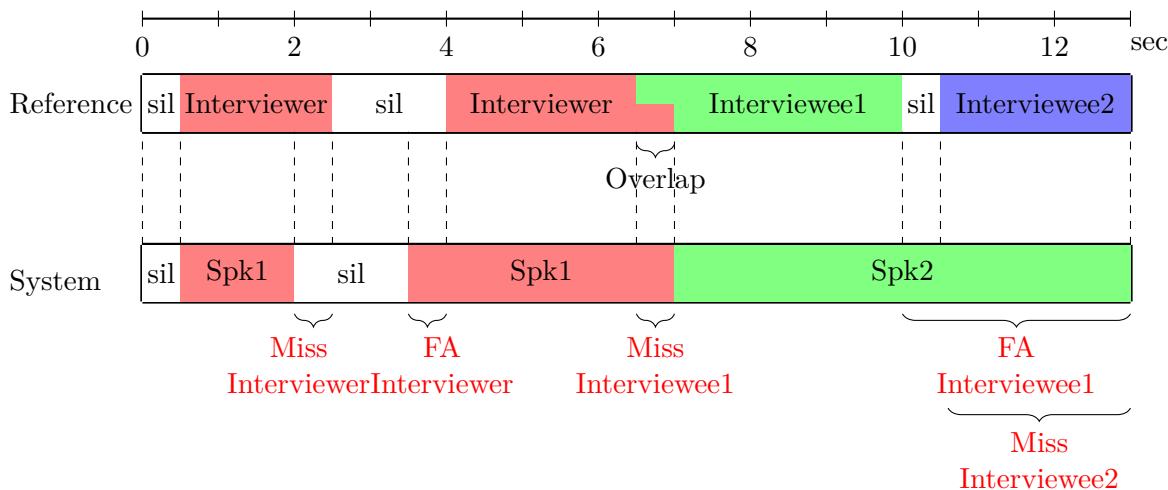

An example of how JER is calculated can be seen in Figure~\ref{fig:JER_example}.
Although JER highly correlates with DER, it gives equal weight to all speakers in the recording regardless of whether they contribute largely to the total amount of speech or not. This can be interesting in cases where some speaker(s) dominate the recording but all speakers are relevant. For more details about JER, please refer to the Second DIHARD Challenge evaluation plan\footnote{\url{https://dihardchallenge.github.io/dihard2/docs/second\_dihard\_eval\_plan\_v1.1.pdf}}.

Currently, the most popular tools for evaluation are \textit{dscore}\footnote{\url{https://github.com/nryant/dscore}} and \textit{pyannote.metrics}~\cite{pyannote.metrics} which include, besides DER and JER, several other clustering performance metrics and the possibility of defining one's own metrics.

\section{Relevance of the task}\label{sec:relevance}

Although speaker diarization has received comparatively less attention from the speech community than other tasks such as speaker recognition or speech recognition, it has several applications. In this section, we briefly describe the possible applications of diarization and show its relevance even in tandem with other tasks.

In general, the application of diarization is related to providing rich meta-data in terms of speaker segments which allow indexing of audiovisual resources with speakers and structured search and access to resources. There are several domains to which it is of interest to apply diarization. We briefly list here some of the most common ones as reviewed in~\cite{tranter2006overview,anguera2012speaker,moattar2012review}:

\begin{itemize}
    \item broadcast news, debates, television shows, movies
    \item recorded meetings (which becomes particularly interesting nowadays as many meetings are being held remotely)
    \item telephone conversations
    \item clinical interviews
    \item recorded lectures or conferences
\end{itemize}

However, SD can be combined with other tasks, including the most studied ones: ASR and SR. SD in combination with ASR, determine ``who said what'' and solve the ``cocktail party problem''~\cite{watanabe20b_chime,cornell23_chime}. Sometimes SR has to be performed on recordings with several speakers and, in such cases, diarization is key to allow finding the speaker of interest as in a recent speaker recognition evaluation (SRE)~\cite{NISTSRE2018evalplan}. However, there are some other applications for diarization such as speaker indexing and retrieval, speaker counting, or improving ASR systems by allowing speaker-specific adaptation.

Other tasks such as target speaker extraction (TSE) or enhancement can benefit from diarization. Speaker turns given by a diarization system can be used to inform a TSE model about segments of a speaker of interest which can be used as enrollment. This has been explored in~\cite{zmolikovaphd,zmolikova2023neural}.

In source separation, given a recording that is the result of multiple overlapping sources (i.e. speakers), the goal is to produce several recordings each containing only the sound of a particular source. This task is related to speaker diarization in the sense that the turns of different speakers can be inferred from its output. However, it is a more complex task because it not only recognizes the different speakers but also has to extract their speech to produce different outputs. Diarization can aid such systems by providing the number of speakers in the recording and their segments so that the task can resemble TSE, a priori, a simpler task~\cite{boeddeker2018front,raj2021integration}.

\chapter{Modular diarization systems}
\label{sec:cascaded_diarization}

Some of the first works related to diarization started to appear around the year 2000~\cite{garofolo2004rich,fiscus2006rich}. Works published in the late 90s~\cite{saunders1996real,gauvain1998partitioning,liu1999fast,siegler1997automatic} focusing on telephone conversations or broadcast news formulated the task as ``partitioning'' or ``segmentation and clustering'', among others. But it was the series of Rich Transcription Evaluations~\cite{NISTRT} organized by NIST between 2002 and 2009 that first defined the diarization task and set a framework for comparing different approaches for diarization. The focus was initially on broadcast news data and later on meeting data. Several European projects focused on meeting data during the first decade of 2000 and also shifted the attention to the challenges related to the meeting scenario~\cite{anguera2012speaker}. The works related to diarization mostly focused on those types of recordings in the following years until approximately 2017-2018 with the emerging DIHARD challenge. Since this work focuses on off-line diarization, i.e. diarization when the whole recording is available at evaluation time, we only cover previous work in this line. 

Until the recent advent of end-to-end neural diarization models, most diarization systems were composed of several subsystems that solved different subtasks, sometimes related to specific setups (e.g. multi-channel recordings). In this chapter, we briefly cover the history of the cascaded approach, comment on its strengths and weaknesses and finally describe the clustering-based method studied in this thesis.

\section{Structure of a standard cascaded system}\label{sec:cascaded_structure}

\subsection*{Overview}

Throughout the years, most competitive diarization systems have used several of the blocks depicted in Figure~\ref{fig:blocks_history}, applied one after the other in a ``cascaded'' fashion. A common categorization for diarization systems is between bottom-up and top-down. Bottom-up refers to starting from short segments of speech and clustering them until finding just a few speakers. Top-down consists of starting with the full recording as a whole and partitioning it into smaller parts that in the end will correspond to a few speakers. The former approach has been more thoroughly studied (Figure~\ref{fig:blocks_history} represents this one) but a comparison between the two can be found in~\cite{anguera2012speaker,evans2012comparative}. 

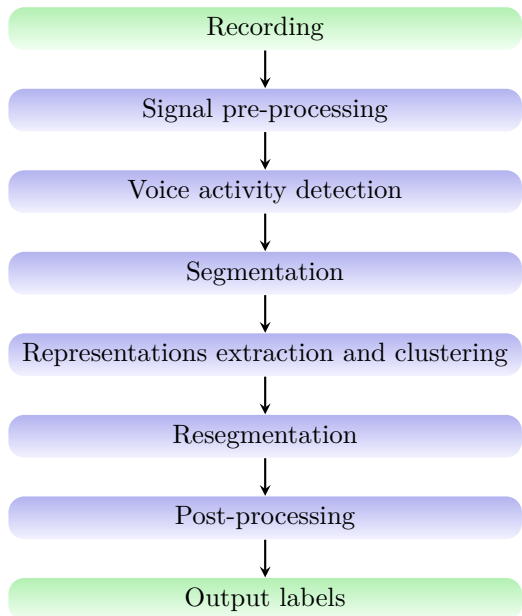
\begin{figure}[ht]
    \centering
    \tikzstyle{block} = [rectangle, 
        minimum height=0.5cm, 
        minimum width=0.8cm, align=center,
        draw=none, shade,
        top color=black!20!blue!30,
        bottom color=blue!5,
        rounded corners=6pt] 
    \tikzstyle{arrow} = [single arrow, draw]
    
    \tikzstyle{block2} = [rectangle, 
        minimum height=0.5cm, 
        minimum width=0.8cm, align=center,
        draw=none, shade,
        top color=black!20!green!30,
        bottom color=green!5,
        rounded corners=6pt] 
    \tikzstyle{arrow} = [single arrow, draw]
    
    \hspace*{-0.5cm}
    \begin{tikzpicture}[auto, >=stealth, node distance=0.4cm and 0.4cm, arr/.style={->,thick}, line/.style={thick}, font=\small]
    
        \node (rec) [block2, text width=6.5cm] {Recording};
        \node (Prep) [block, text width=6.5cm, below=0.5cm of rec] {Signal pre-processing};
        \node (VAD) [block, text width=6.5cm, below=0.5cm of Prep] {Voice activity detection};
        \node (segmentation) [block, text width=6.5cm, below=0.5cm of VAD] {Segmentation};
        \node (clustering) [block, text width=6.5cm, below=0.5cm of segmentation] {Representations extraction and clustering};
        \node (resegmentation) [block, text width=6.5cm, below=0.5cm of clustering] {Resegmentation};
        \node (postprocessing) [block, text width=6.5cm, below=0.5cm of resegmentation] {Post-processing};
        \node (out) [block2, text width=6.5cm, below=0.5cm of postprocessing] {Output labels};
        
        \draw[arr] (rec.south) -- (Prep.north);
        \draw[arr] (Prep.south) -- (VAD.north);
        \draw[arr] (VAD.south) -- (segmentation.north);
        \draw[arr] (segmentation.south) -- (clustering.north);
        \draw[arr] (clustering.south) -- (resegmentation.north);
        \draw[arr] (resegmentation.south) -- (postprocessing.north);
        \draw[arr] (postprocessing.south) -- (out.north);
    \end{tikzpicture}

    \caption{Typical components in a cascaded speaker diarization system (marked in blue).}
    \label{fig:blocks_history}
\end{figure}

We describe below each of the modules mentioned in Figure~\ref{fig:blocks_history}. Note that many other possible sub-modules as well as more details about each of the steps are omitted as this is a high-level overview. More details can be found in~\cite{tranter2006overview} for systems until the year 2006 and in~\cite{anguera2012speaker} for systems created until 2012. A more thorough description and review of literature can be found in~\cite{moattar2012review} where even non-standard approaches are described. More recent approaches that make use of neural networks (NN) are presented in~\cite{park2022review}. 

\subsection*{Signal pre-processing}
In the context of diarization, signal pre-processing refers to applying methods that enhance speech and facilitate the ultimate task. While speech pre-processing originally focused on spectral-based methods, nowadays approaches rely on neural-network-based methods. For diarization, the most common methods include denoising and dereverberation.

Even though telephone conversations and broadcast news can benefit from such pre-processing, this step became more relevant for recorded meetings. While the former are generally recorded with close-talk microphones, the latter usually present setups with far-field microphones where the effect of reverberation or noises made by the participants make for a more challenging scenario~\cite{anguera2012speaker}. Furthermore, meetings offer in some cases recordings with multiple microphones~\cite{janin2004icsi, mccowan2005ami, mostefa2007chil} and making use of the signal from all of them can provide better performance~\cite{pardo2006speaker}. Different approaches have been used to deal with multiple microphones but the most popular is acoustic beamforming~\cite{anguera2007acoustic}.

With the more challenging scenarios presented in DIHARD challenge~\cite{ryant2018first}, preprocessing has become quite relevant and the best-performing systems in the challenge made use of at least one type of enhancement method~\cite{sell2018diarization,diez2018but,sun2018speaker,landini2020but,wang2021ustc}.

\subsection*{Voice activity detection}
The task of VAD is to find the segments of voice or speech to be labeled in the diarization output. This means discarding segments of silence, noise and possibly music while keeping all segments of speech. VAD is of interest not only for diarization but for many other tasks such as SR or ASR. Thus, large efforts have been made to improve the capabilities of VAD systems~\cite{ramirez2007voice}. Some of the most popular approaches are based on energy or spectrum~\cite{anguera2005robust,istrate2005nist}, or Gaussian mixture models (GMM), eventually aided by hidden Markov models (HMM), trained with acoustic features such as Mel frequency cepstral coefficients (MFCC)~\cite{wooters2004towards,reynolds2004lincoln}. 

More recently, systems based on NN have outperformed previous ones~\cite{ryant2013speech,gelly2017optimization,sun2018speaker}. However, one of the main problems with these systems is that the domains of the training data (telephone, meetings, etc.) bias the model to work well in some scenarios while failing in others.
In general, if the model allows to choose different detection thresholds, a lower miss rate of VAD is preferred over a lower false alarm rate since speech discarded in this step is usually not recovered later. Something to consider is that different domains (i.e. meetings, telephone, etc.) present speech with different characteristics and different proportions. For example, in a telephone conversation, speakers usually take turns with very little silence in between whereas in a household day-long recording, speech is generally less common than other sounds or simply silence. Systems are commonly tuned or trained with some application in mind and using a configuration optimal for a certain domain can produce sub-optimal results on recordings of a different domain. 

Furthermore, the annotation criteria and guidelines can vary among different corpora. This means for instance that (sung) background music can be labeled as silence in certain contexts. Moreover, speech from speakers that are not of main interest can be marked as non-speech (think of a recording in a restaurant where people at the next table have a conversation). At the same time, voiced sounds such as cough or sneeze from a speaker of interest can be considered ``voice'' even if they are actually not speech. In spite of this, the terms VAD and SAD are usually used interchangeably.

This task is so crucial that many challenges have proposed two tracks: one where diarization is evaluated as a whole and another where oracle (extracted from the manual annotations used for evaluations) VAD labels are allowed in order to evaluate the rest of the steps assuming perfect VAD.

Nowadays, most NN-based systems can perform reasonably good VAD except in very challenging conditions, such as dinner-party scenarios or extremely noisy backgrounds.

\subsection*{Segmentation}
Segmentation, sometimes also referred to as speaker change point detection (SCD), finds the points in the recording that separate audio sources (including different speakers). 
While longer segments allow to have more information to perform clustering in the next step, it is important that the segments do not contain speech from different speakers as this can create problems when trying to find which segments correspond to the same speaker. Therefore, SCD deals with the trade-off between long speech segments versus having pure segments.

This step can be combined with VAD in the case that silence or noise are considered audio sources. However, it is common to make use of the VAD output given by the module previously described. The most popular approaches used for SCD are based on comparing adjacent windows of MFCC (or similar) features. 
Historically, the most common technique consisted of taking a segment of speech, dividing it in half to obtain two sub-windows and then deciding if it was more likely that each sub-window would be represented by different distributions or that the whole window would be represented by a single distribution. The most popular approaches assumed Gaussian distributions and shifted the window to calculate a metric for every frame. Usually, windows were of fixed length but some works have explored windows of increasing length~\cite{chen1998speaker}. In either case, the change detection was based on thresholding the metric at local maxima. The most popular distances were based on the Bayesian information criterion (BIC) in~\cite{chen1998speaker}, based on the Kullback-Leibler divergence in~\cite{siegler1997automatic}, and on using the generalized log-likelihood ratio (GLR) in~\cite{jin2004speaker}.

More recently, some works using NN have focused on SCD: using
deep neural networks (DNN)~\cite{gupta2015speaker},
convolutional neural networks (CNN)~\cite{hruz2017convolutional},
bidirectional long short-term memory (BLSTM) networks~\cite{bredin2017tristounet,yin2017speaker,yin2018neural}, 
long short-term memory networks (LSTM) with acoustic features and word embeddings as input~\cite{india2017lstm}, 
LSTMs with attention mechanism using text as input~\cite{meng2017hierarchical}, 
LSTMs with speaker embeddings as input~\cite{sari2019pre}, and
CNNs with speaker embeddings as input~\cite{aronowitz2020context}. 

However, many current systems do not make use of any SCD module since iterative segmentation-clustering or the use of speaker embeddings computed on fixed segments allows for a good performance without the SCD step. In particular, the use of uniform segmentation to extract speaker embeddings (to be discussed in the following section) has become the standard approach in state-of-the-art systems.

\subsection*{Representation extraction and clustering}
This step refers to two of the most crucial steps in the pipeline which usually act in tandem. Originally, speech segments were represented by acoustic features such as MFCC, but more carefully designed representations (i.e. embeddings) are preferred nowadays. The speaker embeddings used for diarization, described below, are usually adopted from the technologies used for speaker verification. The clustering builds on top of these representations and assigns them to different speaker identities in a conversation.

The clustering step has the objective of providing one (global) cluster per speaker for the whole recording. The predominant approach for doing so is agglomerative hierarchical clustering (AHC) which consists of the following steps: first, pair-wise distances are computed between each pair of initial clusters (simply segments); then, the closest clusters according to the estimated distances are merged and the distances are updated given the new setup. These operations are carried out iteratively until a stopping criterion is met. 

For calculating the distances, the segments have been represented with Gaussian distributions
~\cite{moh2003towards} or GMMs~\cite{moraru2003elisa} computed on acoustic features such as MFCCs. Different distance metrics have been used as well as different stopping criteria and some of them followed the same approaches described for segmentation. Among distance metrics, the most popular have been GLR and BIC for which the stopping criterion is based on comparing the statistics of two clusters versus a single cluster after merging, thus allowing for fast clustering.

Other approaches model speakers by maximum-a-posteriori adapting a GMM-based universal background model (UBM) to speaker clusters~\cite{betser2004speaker,reynolds2004lincoln}. Also making use of BIC, both approaches use a fixed threshold as the stopping criterion, which needs to be tuned on development data of similar characteristics as the test set. 

An alternative to AHC is spectral clustering~\cite{ning2006spectral} in combination with a clustering method such as k-means. Like AHC, spectral clustering also relies on pair-wise distances between pairs of segments but uses the eigen-vectors of the Laplacian matrix associated with the affinity matrix to transform the space where clustering will be performed. With this approach, a common method to determine the number of clusters is to compute the largest eigen-gap, that is the maximum difference between adjacent (after sorting) eigen-values of the Laplacian~\cite{park2019auto}.

Several approaches have used HMMs to cluster representations and model diarization but they will be covered in more detail in Section~\ref{sec:diarization_bhmm} as they are related to one of the main research lines in this work.

For nearly a decade, joint factor analysis (JFA)~\cite{kenny2005joint,kenny2007joint} and its simplified variant, i-vectors~\cite{dehak2010front}, were the predominant models for speaker verification. Both the JFA and i-vector extractor models utilize a GMM to model the speaker-specific distribution of spectral features, such as MFCCs. However, the GMM parameters are constrained to live in a low-dimensional, high-variability subspace. Only a low-dimensional latent vector (i-vector or, as originally named in JFA, speaker factors) needs to be estimated from speaker-specific data to represent the speaker-specific distribution. This i-vector (or vector of speaker factors) can then be extracted from (or estimated on) a segment of speech, serving as an embedding that represents the speaker of that segment. 

Soon after their introduction, these concepts were applied to diarization as well: i-vectors were used for diarization in~\cite{shum2011exploiting} and, just like for SR, they pushed the field forward. Several works~\cite{dupuy2012vectors,shum2012use,stafylakis2012mean,rouvier2012global} extended this idea either in combination with resegmentation or with improved clustering. 
Along with the introduction of such representations, the use of uniform segmentation using overlapping segments (e.g. 2\,s-long segments every 0.5\,s) started to be more common, replacing more complex segmentation approaches. This represented a ``back to the basics'' trend since uniform segmentation had been commonly used 20 years before~\cite{siu1992unsupervised,wilcox1994segmentation}.

Again, following the advances in speaker recognition, probabilistic linear discriminative analysis (PLDA) was also used as similarity metric for i-vectors in the context of diarization~\cite{prazak2011speaker}. In particular, \cite{sell2014speaker} paved the way for the state-of-the-art diarization systems for the following years. 
Several efforts were made to extend and improve the embeddings/representations in this approach, such as replacing unsupervised GMMs with supervised DNN senone posteriors to calculate i-vectors~\cite{sell2015speaker} and introducing a prior for the number of speakers~\cite{sell2016priors} but the main break-through occurred when the first successful deep neural network embeddings appeared~\cite{hershey2016deep,snyder2016deep}. These embeddings were named \emph{x-vectors} and the first diarization work using them was~\cite{garcia2017speaker} resulting in a new state-of-the-art performance at the time. Another type of speaker embeddings, \emph{d-vectors}~\cite{variani2014deep}, were used for diarization in~\cite{wang2018speaker} together with spectral clustering to push the state-of-the-art further. \cite{garcia2017speaker} followed~\cite{sell2014speaker} with the exception that instead of a PLDA model, a distance metric was learned with a neural network; although, with training criteria similar to discriminatively trained PLDAs. After being a central part of the best-performing systems~\cite{sell2018diarization,diez2018but} in the First DIHARD Challenge~\cite{ryant2018first}, a recipe in Kaldi~\cite{povey2011kaldi} was made available and this approach has become a staple in competitive diarization systems being even used as the baseline of subsequent diarization challenges~\cite{ryant19_interspeech,watanabe20b_chime,ryant21_interspeech,nagrani2020voxsrc}. 

Advances regarding NN-based speaker embeddings tailored for diarization and NN-based clustering methods are discussed in Section~\ref{sec:nn_overview}.

In spite of the efforts, the most common approach nowadays consists of extracting x-vectors using fixed segmentation (1.5\,s-long segments every 0.5\,s or 0.25\,s) and clustering them with AHC or spectral clustering. Yet, this is an area of wide interest as it represents the core of modular systems. 

\subsection*{Resegmentation}
Resegmentation refers to an optional step in which the boundaries of the clustering output are refined, sometimes followed by new segmentation, embedding extraction, and re-clustering stage. The most common approach uses Viterbi decoding (with or without iterations); this step is closely related to the HMM-based approach covered in Section~\ref{sec:diarization_bhmm}. Other approaches based on sliding windows were proposed~\cite{adam2002new} in which windows for the speakers are estimated and their boundaries re-estimated based on some distance measure. Given initial segments that represent speakers, the distances between them and new candidate windows give possible points for resegmentation. However, this approach has not been largely explored. 

The clustering and resegmentation steps were used iteratively in~\cite{gauvain1998partitioning,barras2004improving} and later in~\cite{shum2012use} with more sophisticated clustering methods.

\subsection*{Post-processing}
As in many speech processing tasks, one common procedure to leverage the performance of individual systems is system combination or fusion. Diarization is not the exception in this regard and there have been attempts to combine the outputs of different systems such as using cluster voting~\cite{tranter2005two} or merging outputs~\cite{meignier2006step} considering the agreement of the models. Another possibility is hybridization (or ``piped'' systems)~\cite{meignier2006step} which uses the output of one system to initialize another one. 

More recently, an approach following a technique commonly used in ASR: recognizer output voting error reduction (ROVER) has been proposed~\cite{stolcke2019dover} under the name of DOVER where voting between different systems is done to fuse the outputs of several systems to generate a final decision. DOVER was extended to handle systems with overlapping segments in DOVER-Lap~\cite{raj2021dover} and has become relevant in recent challenges.

\subsection*{Overlapped speech detection}
A different type of post-processing that deserves attention on its own is dealing with segments of overlapped speech, where two or more speakers speak simultaneously. 
Overlapped speech is present in general in all types of recordings but with different proportions. Given that most modular systems only output one speaker per frame, the penalization on recordings with high levels of overlap is considerable.

Some efforts were made to analyze the nature of overlapped speech~\cite{shriberg2001observations,cetin2006speaker}. One of the first works to include overlap processing for diarization~\cite{otterson2007efficient} proposed the heuristic of picking the closest speaker (in time) as the second speaker. Although it provided a reasonable approach for assigning a secondary speaker, their analysis assumed oracle overlap detection and using real detection systems largely degraded the results. In general, the impact of OSD is two-fold~\cite{boakye2008two,huijbregts2009speech,zelenak2010overlap}: (1) detected overlap segments can be removed before doing clustering in order to consider only pure segments to obtain cleaner speaker models. (2), it can be used at the end for post-processing; that is, labeling segments for more than one speaker given the clustering output.
The earliest successful works in OSD~\cite{boakye2008overlapped,zelenak2010overlap,boakye2011improved} used an HMM with three states (non-speech, speech, and overlapped speech) with GMMs as models and MFCCs as features (and spatial features in the case of~\cite{zelenak2010overlap}). The task of overlapped speech detection has, however, been explored more in recent years:
The most prolific works have used NNs for the purpose of overlap detection. \cite{geiger2013detecting} made use of an LSTM to detect overlap and an HMM (states for non-speech, speech, and overlapped speech) to post-process the per-frame labels. \cite{andrei2017detecting,kunevsova2019detection} made use of a CNN to detect overlap while~\cite{hagerer2017enhancing,bullock2020overlap} used a BLSTM. \cite{raj2021multi} used a time-delay NN (TDNN) and BLSTM together with an HMM. \cite{jung21_interspeech} used a convolutional recurrent neural network architecture to leverage both local and sequential information.

However, one of the main problems for training NNs for the task is the lack of training data. Moreover, the quality of the annotations in overlapping segments is not always satisfactory, especially for short segments. For this reason, works sometimes resort to artificially created overlaps~\cite{hagerer2017enhancing,kunevsova2019detection}.

Recent papers have also paid attention to the problem of handling overlap once it is detected. For instance, \cite{bullock2020overlap} used the output of the variational Bayes HMM framework for diarization (to be discussed in Section~\ref{sec:diarization_bhmm}) by considering the first and second most likely speakers given by the model as present speakers when the OSD finds an overlapped speech segment. \cite{raj2021multi} extended the spectral clustering technique to return two labels for segments recognized by the OSD system. 

Although some authors have used linguistic cues~\cite{wlodarczak2012temporal,geiger2013using,yella2013improved,chowdhury2015annotating} together with standard features for OSD, acoustic features still provide most of the relevant information with current techniques. Other approaches such as using the amount of detected silence~\cite{yella2012speaker} or analyzing the formants and harmonic structure of speech~\cite{shokouhi2015robust} to find patterns corresponding to overlapping speakers have been proposed but their performance was worse than that of NN-based systems. Some have focused on learning speaker embeddings used for clustering jointly with VAD and OSD~\cite{akira2018joint} but without much success.

Related to overlapped speech detection is the field of source separation which intends to create a separate output stream (signal or spectrogram) for each of the sources (speakers). Although some works~\cite{chen2017progressive,chen2018multi,yoshioka18_interspeech} focused on this task in similar conditions as we do, the task itself is different (to separate rather than to detect) and focuses on ASR as the down-stream task. Thus, we do not cover it here.

Given the quality of VAD systems, speaker embeddings and clustering methods, most sources of errors in diarization performance nowadays are related to overlapped speech. Even more, the performance of OSD systems is in most cases far from satisfactory. In general, excellent precision is preferred over good recall since false alarms directly increase the diarization errors. Yet, correctly found overlap segments do not always translate into decreased error due to the complicated task of assigning the correct speaker. 

\section{Diarization based on Bayesian hidden Markov model (VBx)}\label{sec:diarization_bhmm}

This section covers the work done regarding the Bayesian hidden Markov model for diarization and the configuration of a full system with this approach. This system falls into the category of clustering of embeddings and, being a more sophisticated approach than other simpler clustering methods, it has allowed for state-of-the-art performance on different sets for considerable time. Nowadays, it is still used as a contrastive baseline system in many publications.

\subsection{Background}\label{sec:bhmm_background}

Hidden Markov models~\cite{rabiner1986introduction} have been used for speech processing for a very long time. In particular, an ergodic HMM presents a suitable framework for diarization where states represent speakers, the transitions between states represent the speaker turns in a conversation and each state has a speaker model which is assumed to generate the observed data.

Some of the first works in diarization used HMMs~\cite{sugiyama1993speech,wilcox1994segmentation,sonmez1999speaker} although at that time they referred to the task as ``speaker tracking'' or simply ``speaker segmentation''. The main problem with those approaches was that the number of speakers in the conversation had to be known in advance. An extension to tackle that problem consisted in an evolutive HMM~\cite{meignier2000evolutive,meignier2001hmm,meignier2006step,fredouille2008new} where new states were added iteratively and the Viterbi decoding ran after each change of HMM configuration to see if the new model fitted the observed data better. Another approach~\cite{ajmera2002unknown} started from a large number of segments (many more than the number of expected speakers) and ``merged'' states as long as the new configuration fitted the data better than the previous one.

However, more relevant work for this chapter was proposed in~\cite{valente2004variational,valente2005variational} where, for the first time, variational Bayesian (VB) inference was used together with an HMM for diarization. The idea was to use an ergodic HMM modeling the speakers (states) with GMMs. We do not describe the VB inference here as it is a central part of the model we use in this work and it is covered in section~\ref{sec:bhmm_description}.

Alternatives using a hierarchical Dirichlet process HMM (HDP-HMM)~\cite{teh2004sharing} and a sticky HDP-HMM were presented in~\cite{fox2007sticky} using also a Bayesian framework in the context of diarization. 
With the HDP-HMM, the model allowed potentially an infinite number of speakers, since the number of states is inferred naturally. As presented by the authors, this approach has the problem that it does not model the persistence of states adequately. This means that the approach gives a higher posterior probability to models with a higher number of states and with rapid switching between them. In order to fix this, they included a parameter of self-transition bias with its corresponding prior so that it is part of the whole Bayesian framework. Their extension, called sticky HDP-HMM, presented a fully Bayesian framework (i.e. priors were assumed over all the parameters) to model the diarization problem. Unlike in~\cite{valente2005variational}, where a variational expectation-maximization-like algorithm was used to obtain the parameters, in~\cite{fox2007sticky}, Gibbs sampling was used to obtain the model parameters. The sticky HDP-HMM approach was shown to attain state-of-the-art performance when proposed. Although a very principled way to tackle the problem of diarization, the solution requires quite a complex model and a large number of Gibbs sampling iterations to reach good performance.

Another Bayesian approach was proposed in~\cite{kenny2008bayesian,kenny2010diarization} where a Bayesian GMM based on JFA was used for diarization. At that time, JFA was the common framework for speaker verification and using eigen-voices imposed a more powerful prior on the parameters of the GMMs than the priors used in~\cite{valente2005variational}. 
One of the problems of the approach presented in~\cite{kenny2008bayesian} is that the turn durations are not modeled so the natural solution was to extend the model with an HMM.

This Bayesian hidden Markov model (BHMM) was used in ~\cite{sell2015diarization} (and released later by the author\footnote{\url{http://http://speech.fit.vutbr.cz/software/vb-diarizationeigenvoice-and-hmm-priors}}) reaching new state-of-the-art results. The BHMM framework was presented as resegmentation: after obtaining a carefully tuned diarization output by means of AHC of i-vectors, the BHMM applied at MFCC level (i.e. 10\,ms) was used to improve the boundaries of the segments and to potentially correct the mistakes made by AHC. A more thorough analysis of the BHMM approach was then presented in~\cite{diez2018speaker,diez2019analysis} where it was shown that it can reach excellent performance even if randomly initialized (instead of using the AHC output).

This model was further adapted to work with x-vectors as input instead of MFCCs in~\cite{diez2019bayesian} resulting in a more light-weight approach (given by the lower frequency of x-vectors in comparison with MFCCs, and states modeled by single Gaussians rather than GMMs). However, this newer model is more powerful due to the discriminative capabilities of the x-vectors, and the pre-trained PLDA model that is incorporated into the BHMM, to properly model between- and within-speaker variabilities. This approach forms the core of VBx, the model we describe next.

\subsection{Model description}\label{sec:bhmm_description}

The model that we describe here corresponds to the latest version of the BHMM which works at x-vector level as described in~\cite{landini2022bayesian}. Often we refer to this model as VBx\footnote{Variational Bayes diarization working with x-vectors.}. Although many aspects are shared with its predecessor version working at MFCC level, we omit those differences and focus only on the latest version. More details about the previous version can be found in~\cite{diez2018speaker,diez2019analysis}.

\subsection*{Overview}
The model works on the basis of a VAD segmentation which can be given by a system or by oracle labels, i.e. extracted from the reference annotations. On the speech portions, uniform segmentation is used to define the segments where speaker embeddings (x-vectors) are to be extracted.

The diarization model assumes that the input sequence of x-vectors is generated by an HMM with speaker-specific state distributions. Building on a common back-end approach for comparing x-vectors, the state distributions are derived from a PLDA model pre-trained on a large number of speaker-labeled x-vectors. 

The HMM is ergodic with one-to-one correspondence between the speakers and states. This allows us to model transitions between any pair of speakers, as can be expected in a conversation. As will be discussed in section~\ref{sec:bhmm_weaknesses}, this approach does not consider overlapped speech and in the case of having it in the recording, the model will output only one speaker for the corresponding embeddings.

Let $\XX=\{\xx_1, \xx_2,...,\xx_T\}$ be the sequence of observed x-vectors and $\ZZ=\{z_1, z_2,...,z_T\}$ the corresponding sequence of discrete latent variables defining the hard alignment of x-vectors to HMM states. In our notation, $z_t=s$ indicates that the speaker (HMM state) $s$  is responsible for generating observation $\xx_t$. To address the speaker diarization task, the speaker distributions represented by latent vectors $\mathbf{y}_s$ and latent variables $z_t$ are jointly estimated given an input sequence $\XX$. The diarization output is then given by the most likely sequence $\ZZ$, which encodes the alignment of x-vectors to speakers.

\subsection*{HMM topology}\label{sec:topology}
The HMM topology and transition probabilities model the speaker turn durations. Figure~\ref{HMMmodel1state} shows an example of the HMM topology for $S=3$ speakers.  The transition probabilities are set as follows: we transition back to the same speaker/state with probability $P_{loop}$. Note that this probability is not learned (unlike in~\cite{fox2007sticky}) but it is a tunable parameter in the model.
The remaining probability $(1-P_{loop})$ is the probability of changing the speaker, which corresponds to the transition to the non-emitting node in Figure \ref{HMMmodel1state}. 
From the non-emitting node, we immediately transition to one of the speaker states with probability $\pi_s$.\footnote{For convenience, we allow to re-enter the same speaker as it leads to simpler update formulae.} Therefore, the probability of leaving one speaker and entering another speaker $s$ is $(1-P_{loop})\pi_s$.
To summarize, the probability of transitioning from state $s'$  to state $s$ is
\begin{equation}
p(s|s')=(1-P_{loop})\pi_s+\delta(s=s') P_{loop}, 
\end{equation}
where $\delta(s=s')$ equals 1 if we go to the same state ($s=s'$) and is $0$ otherwise.

The non-emitting node in Figure \ref{HMMmodel1state} is also the initial state of the model. Therefore, the probabilities $\pi_s$ also control the selection of the initial HMM state (i.e. the state generating the first observation).
These probabilities $\pi_s$ are inferred (jointly with the variables $\mathbf{y}_s$ and $z_t$) from the input conversation. 

Thanks to the automatic relevance determination principle~\cite{Bishop2006} stemming from the Bayesian model, zero probabilities will be learned for the $\pi_s$ corresponding to redundant speakers, which effectively drops such speakers from the HMM model. 
Typically, the HMM is initialized with a larger number of speakers (see section \ref{sec:bhmm_experiments}) and we make use of this behavior to drop the redundant speakers (i.e. to estimate the number of speakers in the conversation).

\begin{figure}[htb]
\centering
\begin{tikzpicture}
\tikzstyle{main}=[circle, minimum size = 6mm, thick, draw =black!80, node distance = 10mm]
\tikzstyle{small}=[circle, minimum size = 1mm, thick, draw =black!80, node distance = 8mm]
\tikzstyle{nonr}=[circle, thick, draw =black!80, node distance = 10mm]
\tikzstyle{connect}=[-latex, thick]
\tikzstyle{box}=[rectangle, draw=black!100]
  \node[main, fill = white!100,draw=black!80] (s1) [] {$\mathbf{s}_{1}$};
  \node[small] (s) [fill = black!100,below= of s1] {};
  \node[main] (s2) [below right=of s] {$\mathbf{s}_{2}$ };
  \node[main] (s3) [below left=of s] { $\mathbf{s}_3$};
  \node[nonr, fill = white!100,draw=none] (z1) [above right =of s1] {};
  \node[nonr, fill = white!100,draw=none] (z2) [above right =of s2] {};
  \node[nonr, fill = white!100,draw=none] (z3) [above left =of s3] {};
  \path (s1) edge [connect,  out=-120,in=135] node[left] {$1-P_\text{loop}$} (s)
		(s2) edge [connect,  out=100,in=-10] node[right=3pt] {$1-P_\text{loop}$} (s)
		(s3) edge [connect,  out=5,in=-100]  node[below  =13pt] {$1-P_\text{loop}$}(s)
        (s1) edge [loop, connect, out=70,in=110, looseness=5] node[ right=3pt]{$P_{loop}$} (s1)
        (s2) edge [loop, connect,  out=-70,in=-30, looseness=5] 
        node[right=3pt]{$P_{loop}$} (s2)
        (s3) edge [loop, connect, out=-160,in=-120, looseness=5] node[left=3pt]{$P_{loop}$} (s3)
        (s) edge [connect,  out=45,in=-60] node[right] {$\pi_1$} (s1)
		(s) edge [connect,  out=-80,in=170] node[right=1pt] {$\pi_2$} (s2)
		(s) edge [connect,  out=190,in=80]  node[left=3pt] {$\pi_3$}(s3);
\end{tikzpicture}
\caption{HMM model for 3 speakers with 1 state per speaker, with a dummy non-emitting (initial) state. Figure adopted from \cite{diez2019analysis}.}
 \label{HMMmodel1state}
\end{figure}
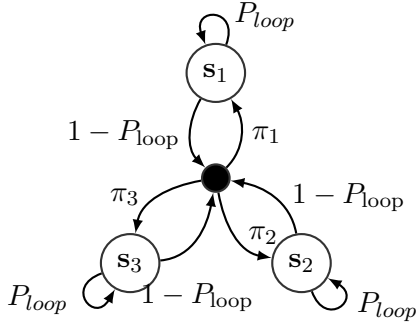

\subsection*{Speaker-specific distributions}\label{sec:speakerdist}
The speaker-specific distributions are derived from a PLDA which is a standard model used for comparing x-vectors in speaker verification~\cite{kenny10PLDA_HTP}. 
Here, only a simplified variant of PLDA is considered, which is often referred to as \textit{two-covariance model}~\cite{brummer2010spkpartitioning}.

This model assumes that the distribution of x-vectors specific to speaker $s$ is Gaussian $\mathcal{N}(\hat{\mathbf{x}}_t;\hat{\mathbf{m}}_s,\boldsymbol{\Sigma}_{w})$, where $\boldsymbol{\Sigma}_{w}$ is the within-speaker covariance matrix shared by all speaker models, and $\hat{\mathbf{m}}_s$ is the speaker-specific mean.
Speaker means are further assumed to be Gaussian distributed $\mathcal{N}(\hat{\mathbf{m}}_s; \mathbf{m}, \boldsymbol{\Sigma}_{b})$, where $\mathbf{m}$ is the global mean and $\boldsymbol{\Sigma}_{b}$ is the between-speaker covariance matrix.
In general, $\boldsymbol{\Sigma}_{w}$ and $\boldsymbol{\Sigma}_{b}$ can be full covariance matrices. However, to further simplify and speed up the inference in our model, we assume that the x-vectors are linearly transformed into a space where $\boldsymbol{\Sigma}_{b}$ is diagonal and $\boldsymbol{\Sigma}_{w}$ is identity. This can be achieved as follows:

Let $\hat{\XX}$ be the matrix of original (untransformed) x-vectors that the parameters of the original PLDA model $\mathbf{m},$ $\boldsymbol{\Sigma}_{w}$ and $\boldsymbol{\Sigma}_{b}$  were estimated on.
The x-vectors that are used as input for the diarization algorithm are obtained as 
\begin{equation}
\label{eq:transformxvec}
    \XX=(\hat{\XX}-\mathbf{m})\mathbf{E},
\end{equation} 
\noindent where $\mathbf{E}$ is the matrix that transforms the x-vectors into the desired space.
This matrix can be obtained by solving the standard generalized eigen-value problem
\begin{equation}
    \boldsymbol{\Sigma}_{b} \mathbf{E} = \boldsymbol{\Sigma}_{w} \mathbf{E} \boldsymbol{\Phi}, 
\end{equation}
\noindent where $\mathbf{E}$ is the matrix of eigen-vectors and $\boldsymbol{\Phi}$ is the diagonal matrix of eigen-values, which is also the between-speaker covariance matrix in the transformed space.
Note that the eigen-vectors $\mathbf{E}$ are, in fact, bases of linear discriminant analysis (LDA) estimated directly from the PLDA model parameters. Therefore, if we construct $\boldsymbol{\Phi}$ only using $R$ largest eigen-values and assemble $\mathbf{E}$ only using the corresponding eigen-vectors, \eqref{eq:transformxvec} further performs LDA dimensionality reduction of x-vectors to $R$-dimensional space. We use $R$ as one of the hyper-parameters of the VBx method. In \eqref{eq:transformxvec}, we have also subtracted the global mean from the original x-vectors to have the new set of x-vectors zero-centered.

In summary, the PLDA model compatible with the new set of x-vectors $\XX$ suggests that speaker-specific means are distributed as 
\begin{equation}
p(\mathbf{m}_s) = \mathcal{N}(\mathbf{m}_s; \mathbf{0}, \boldsymbol{\Phi}).
\end{equation}

For convenience and for compatibility with the notation introduced in \cite{diez2019analysis}, we further re-parametrize the speaker means with speaker-specific latent variables $\mathbf{y}_s$ as: 
\begin{equation}
\mathbf{m}_s=\mathbf{V} \mathbf{y}_s,
\label{eq:PLDA}
\end{equation}
\noindent where $\VV=\boldsymbol{\Phi}^{\frac{1}{2}}$ is a diagonal matrix and $\mathbf{y}_s$ is a standard normal distributed random variable
\begin{equation}
\label{eq:normalys}
    p(\yy_s)=\mathcal{N}(\yy_s;\mathbf{0},\mathbf{I}).
\end{equation}
The speaker-specific distribution of x-vectors is then
\begin{equation}
p(\mathbf{x}_t|\mathbf{y}_s) = \mathcal{N}(\mathbf{x}_t ; \mathbf{V} \mathbf{y}_s, \mathbf{I}),
\label{eq:datcondspkSIMP}
\end{equation}
\noindent where $\mathbf{I}$ is identity matrix.

In our diarization model, we use \eqref{eq:datcondspkSIMP} to model the speaker (HMM state) distributions. 
This distribution is fully defined only in terms of the speaker vector $\yy_s$ (and the pre-trained matrix $\VV$ shared by all the speakers).
The speaker vector $\yy_s$ is treated as a latent variable with standard normal prior~\eqref{eq:normalys}, which is why the BHMM model is called \textit{Bayesian}\footnote{However, unlike other ``fully Bayesian'' HMM implementations~\cite{fox2007sticky,BealVBLowerBThesis}, we do not impose any prior on the transition probabilities.}.
This way, the full PLDA model is incorporated into the BHMM in order to properly model between- and across-speaker variability. Therefore, the model is capable of discriminating between speakers just like the PLDA model when used for speaker verification.

\subsection*{Bayesian HMM}\label{sec:BayesianHMM}
To summarize, our complete  model for SD is a Bayesian HMM, which is defined in terms of the state-specific distributions (or so-called output probabilities)
\begin{equation}
p(\mathbf{x}_t|z_t=s) =  p(\mathbf{x}_t|s) =  p(\mathbf{x}_t|\mathbf{y}_s)
 \end{equation}
and the transition probabilities
\begin{equation}
p(z_t=s|z_{t-1}=s')  = p(s|s')
 \end{equation}
described in the two previous sections.
By abuse of notation, $p(z_1|z_0)$ will correspond to the initial state probability $p(z_1{=}s)=\pi_s$ in the following formulae.

The complete model can be also defined in terms of the joint probability of the observed and latent random variables (and their factorization) as
\begin{align}
\label{jointprobeq}
 p  (\XX,\ZZ,\YY)&= p(\XX|\ZZ,\YY) p(\ZZ) p(\YY) =
\prod_t p\left(\xx_t|z_t\right)
\prod_t p\left(z_t|z_{t-1}\right)
\prod_s  p\left(\yy_s\right),
\end{align}
where $\YY=\{\yy_1, \yy_2,...,\yy_S\}$ is the set of all the speaker-specific latent variables. The corresponding graphical model can be seen in Figure~\ref{fig:bayesian_net}.

The model assumes that each x-vector sequence corresponding to an input conversation is obtained using the following generative process:
\begin{mdframed}[
    align=center,
    linecolor=black,
    linewidth=0.6pt,
    userdefinedwidth=0.3\columnwidth,
]
\begin{algorithmic}
  \For{$s=1..S$}
      \State $\mathbf{y}_s \sim \mathcal{N}(0,\mathbf{I})$
  \EndFor
  \For{$t=1..T$}
  \State $z_t \sim p(z_t|z_{t-1})$
  \State $\mathbf{x}_t \sim p(\mathbf{x}_t|z_t)$
  \EndFor
\end{algorithmic}
\end{mdframed}

\begin{figure}[ht]
\centering
    \begin{tikzpicture}[x=1.7cm,y=1.8cm]
      \node[obs]                   (X_1) {$\mathbf{x}_1$} ; %
      \node[latent, above=of X_1, yshift=-0.75cm]  (Z_1) {$z_1$} ; %
      
      \node[obs, right=of X_1, xshift=-0.75cm]    (X_2) {$\mathbf{x}_2$} ; %
      \node[latent, above=of X_2, yshift=-0.75cm]  (Z_2) {$z_2$} ; %
    
      \node[const, right=of Z_2, xshift=-0.75cm]   (etc1)  {$\ \ \ldots \ \ $} ; %
      \node[const, right=of X_2, xshift=-0.75cm]   (etc2)  {$\ \ \ldots \ \ $} ; %
      
      \node[obs, right=of etc2, xshift=-0.75cm]    (X_T) {$\mathbf{x}_T$} ; %
      \node[latent, above=of X_T, yshift=-0.75cm]  (Z_T) {$z_T$} ; %
    
      \node[const, above=of Z_2, yshift=-0.75cm, xshift=1cm]   (pi)  {$\boldsymbol{\pi}_{}$} ; %
      \node[latent, below=of X_2, yshift=0.75cm, xshift=1cm]  (Y_s) {$\mathbf{y}_s$} ; %
    
      \edge {pi}  {Z_1} ; %
      \edge {Z_1} {X_1} ; %
      \edge {Y_s} {X_1} ; %
    
      \edge {pi}  {Z_2} ; %
      \edge {Z_1} {Z_2} ; %
      \edge {Z_2} {X_2} ; %
      \edge {Y_s} {X_2} ; %
    
      \edge {Z_2} {etc1} ; %
      
      \edge {pi}  {Z_T} ; %
      \edge {etc1} {Z_T} ; %
      \edge {Z_T} {X_T} ; %
      \edge {Y_s} {X_T} ; %
    \end{tikzpicture}

    \caption{Bayesian network describing the model.}
    \label{fig:bayesian_net}
\end{figure}
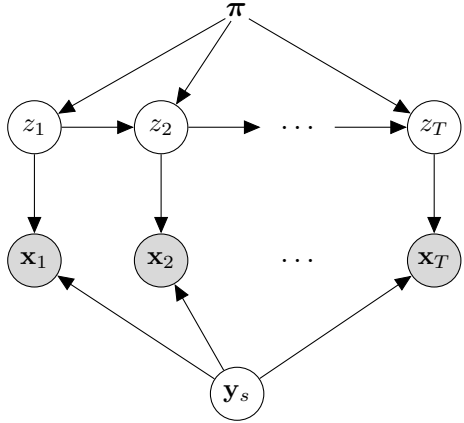

\subsection*{Diarization inference}\label{sec:inference}
The diarization problem consists in finding the assignment of embeddings to speakers, which is represented by the latent sequence $\ZZ$. In order to find the most likely sequence $\ZZ$, we need to infer the posterior distribution  $p(\ZZ|\mathbf{X})=\int p(\ZZ,\mathbf{Y}|\mathbf{X}) d\mathbf{Y}$. 
Unfortunately, the evaluation of this integral is intractable, and therefore, we approximate it using variational Bayes inference~\cite{Bishop2006}, where the distribution $p(\ZZ,\YY|\mathbf{X})$ is approximated by $q(\ZZ,\YY)$. We use the mean-field approximation~\cite{Bishop2006,kenny2008bayesian} assuming that the approximate posterior distribution factorizes as
\begin{equation}
q(\ZZ,\YY) = q(\ZZ)q(\YY).
\label{eq:meanfieldfactorization}
\end{equation}
The particular form of the approximate distributions $q(\ZZ)$ and $q(\YY)$ directly follows from the optimization described below.

We search for such $q(\ZZ,\YY)$ that minimizes the Kullback-Leibler divergence \\ $D_{KL}(q(\ZZ,\YY)\|p(\mathbf{Z},\mathbf{Y}|\mathbf{X}))$, which is equivalent to maximizing the standard VB objective -- the evidence lower bound objective (ELBO)~\cite{Bishop2006}\footnote{In some works, the variational free energy is optimized but that is equivalent to optimizing ELBO.}
\begin{equation}\label{ELBO}
\begin{split}
\mathcal{L}\left(q(\mathbf{\ZZ,\YY})\right)= E_{q(\ZZ,\YY)}\left\lbrace \ln\left(\frac{p(\XX,\YY,\ZZ)}{q(\YY,\ZZ)}\right)\right\rbrace.
\end{split}
\end{equation}
Using the factorization~\eqref{eq:meanfieldfactorization}, the ELBO can be split into three terms
\begin{equation}\label{fetchELBO}
\hat{\mathcal{L}}\left(q(\ZZ,\YY)\right)= F_A E_{q(\ZZ,\YY)}\left[ \ln p(\XX|\YY,\ZZ)\right] + F_B E_{q(\YY)}\left[ \ln\frac{p(\YY)}{q(\YY)}\right] + E_{q(\ZZ)}\left[ \ln\frac{p(\ZZ)}{q(\ZZ)}\right], 
\end{equation}
where the first term is the expected log-likelihood of the observed x-vector sequence $\XX$ and the second and third terms are Kullback-Leibler divergences $D_{KL}(q(\YY)\|p(\YY))$ and $D_{KL}(q(\ZZ)\|p(\ZZ))$ regularizing the approximate posterior distributions $q(\YY)$ and $q(\ZZ)$ towards the priors $p(\YY)$ and $p(\ZZ)$. In~\eqref{fetchELBO}, we modified the ELBO by scaling the first two terms by constant factors $F_A$ and $F_B$.\footnote{Note that similar scaling factor for the third term would be redundant as only the relative scale of the three factors is relevant for the optimization.} The theoretically correct values for these factors leading to the original ELBO~\eqref{fetchELBO} are $F_A=F_B=1$. However, choosing different values gives us finer control over the inference, which can be used to improve diarization performance. 

$F_A$ is the acoustic scaling factor. Choosing values between 0 and 1 scales the first term in~\ref{fetchELBO}, counteracting the HMM's (incorrect) assumption of statistical independence between observations. $F_B$ is the speaker regularization coefficient which penalizes the complexity of the speaker models, thus keeping the posteriors of the speaker latents close to the priors. Values higher than 1 result in the model dropping more speakers in the context of the VB inference.
There is a certain level of interplay between these hyperparameters and, given a specific application scenario, a particular combination of $F_A$ and $F_B$ can allow the model to find a better solution. For further details on the specific effects these scaling factors have in the inference and their interplay, we refer the reader to~\cite{diez2019analysis}.

As described above, we search for the approximate posterior $q(\ZZ,\YY)$ that maximizes the ELBO~\eqref{fetchELBO}. In the case of the mean-field factorization~\eqref{eq:meanfieldfactorization}, we proceed iteratively by finding the $q(\YY)$ that maximizes the ELBO given fixed $q(\ZZ)$ and vice versa. 
The updates and ELBO evaluation are described below.
We omit the derivations but the interested reader can find them in~\cite{DiezVBxreport21}. Note, that an initial sequence $\ZZ$ is necessary to start the iterative procedure to obtain the approximate posterior $q(\ZZ,\YY)$. While the initial assignment of observations to states $\ZZ$ can be obtained randomly, we observed that in practice, a more sensible assignment can lead to better results as will be discussed in Section~\ref{sec:configuration}.

\subsubsection*{Updating $q(\YY)$}
\label{update_q_y}
Given a fixed $q(\ZZ)$, the distribution over $\YY$ that maximizes the ELBO is 
\begin{equation}
q^*(\YY) = \prod_s q^*(\yy_s),
\end{equation}
where the speaker-specific approximate posteriors
\begin{equation}
q^*(\yy_s)= \mathcal{N}\left(\mathbf{y}_s|\boldsymbol{\alpha}_s,\mathbf{L}_s^{-1}\right)
\label{YVBupdate}
\end{equation}
are Gaussians with the mean vector and precision matrix
\begin{align}
\boldsymbol{\alpha}_s=\frac{F_A}{F_B}\mathbf{L}_s^{-1}\sum_t \gamma_{ts}\boldsymbol{\rho}_t\label{YVBupdate2_alpha}\\ \mathbf{L}_s=\mathbf{I}+\frac{F_A}{F_B}\left(\sum_t \gamma_{ts}\right)\boldsymbol{\Phi},
\label{YVBupdate2_L}
\end{align}
\noindent where 
\begin{equation}
\label{eq:rho}
    \boldsymbol{\rho}_t= \VV^T \xx_t.
\end{equation}
In this update formula, $\gamma_{ts}=q(z_{t}=s)$ is the marginal approximate posterior derived from the current estimate of the distribution $q(\ZZ)$ (see below), which can be interpreted as the responsibility of speaker $s$ for generating observation $\xx_t$ (i.e. defines a soft alignment of x-vectors to speakers). 

It is worth noting that $\boldsymbol{\Phi}$ is a diagonal matrix. Therefore, also matrix $\mathbf{L}_s$ is diagonal, and its inversion and application in \eqref{YVBupdate2_alpha} become trivial.

\subsubsection*{Updating $q(\ZZ)$}
We never need to infer the complete distribution over all the possible alignments of observations to speakers $q(\ZZ)$. When updating $q(\YY)$ using~\eqref{YVBupdate2_alpha} and~\eqref{YVBupdate2_L}, we only need the marginals $\gamma_{ts}=q(z_{t}=s)$. Therefore, when updating $q(\ZZ)$, we can directly search for the responsibilities $\gamma_{ts}$ that correspond to the distribution $q^*(\ZZ)$ maximizing the ELBO given a fixed $q(\YY)$. Similar to the standard HMM training, such responsibilities can be calculated efficiently using a forward-backward algorithm as
\begin{equation}
\gamma_{ts}=\frac{A(t,s)B(t,s)}{\overline{p}(\XX)},
\label{spkrespons}
\end{equation}
where the forward probability
\begin{equation}
\begin{split}
\label{forward}
A(t,s)=\bar{p}(\xx_t|s)\sum_{s'}A(t-1,s')p(s|s')
\end{split}
\end{equation}
is recursively evaluated by progressing forward in time for ${t{=}1..T}$ starting with $A(0,s)=\pi_s$. Similarly,
\begin{equation}
\begin{split}
\label{backward}
B(t,s)=\sum_{s'}B(t+1,s')\bar{p}(\xx_{t+1}|s')p(s'|s)
\end{split}
\end{equation}
is the backward probability evaluated using backward recursion for times $t=T..1$ starting with $B(T,s)=1$. 
\begin{equation}
\label{totalA}
\overline{p}(\XX)=\sum_s A(T,s)
\end{equation}
is the total forward probability and
\begin{align}
\label{almostHMMp}
\nonumber
\ln \overline{p}(\xx_t|s) = & \ F_A \left[  \boldsymbol{\alpha}_s^T\boldsymbol{\rho}_t-\frac{1}{2}\tr\left(\boldsymbol{\Phi}\left[\mathbf{L}_s^{-1}+\boldsymbol{\alpha}_s\boldsymbol{\alpha}_s^T\right]\right) -\frac{D}{2}\ln 2\pi -\frac{1}{2}\xx_t^T\xx_t \right] \\
= & \ F_A \left[  \boldsymbol{\alpha}_s^T\boldsymbol{\rho}_t-\frac{1}{2}\boldsymbol{\phi}^T\left[\boldsymbol{\lambda}_s+\boldsymbol{\alpha}_s^2\right]  -\frac{D}{2}\ln 2\pi -\frac{1}{2}\xx_t^T\xx_t \right]
\end{align}

\noindent is the expected log-likelihood of observation $\xx_t$ given a speaker $s$ taking into account its uncertainty $q(\yy_s)$. The second line of \eqref{almostHMMp} corresponds to an efficient evaluation of this term, where vector $\boldsymbol{\phi}$ is the diagonal of the diagonal matrix $\boldsymbol{\Phi}$, vector $\boldsymbol{\lambda}$ is the diagonal of the diagonal matrix $\mathbf{L}_s^{-1}$ and the square in $\boldsymbol{\alpha}_s^2$ is element-wise. Note also that the terms $-\frac{D}{2}\ln 2\pi -\frac{1}{2}\xx_t^T\xx_t$ in \eqref{almostHMMp} are not only constant over VB iterations but also constant for different speakers $s$. As a consequence, the contribution of these terms cancels in \eqref{spkrespons}, and therefore, they do not have to be calculated at all.

\subsubsection*{Updating $\pi_s$}
\label{sec:updatepi}Finally, the speaker priors $\pi_s$ are updated as maximum likelihood type II estimates \cite{Bishop2006}: Given fixed $q(\YY)$ and $q(\ZZ)$, we search for the values of $\pi_s$ that maximize the ELBO \eqref{fetchELBO}, which gives the following fixed point iteration update formula
\begin{equation}
\pi_s \propto\ \gamma_{1s}+ \frac{(1{-}P_{loop})\pi_s}{\overline{p}(\XX)}\sum_{t=2}^T \sum_{s'} A(t{-}1,s') p(\xx_t|s) B(t,s)
\label{thepiupdate}
\end{equation}
with the constraint $\sum_s \pi_s=1$.
As described in section \ref{sec:topology}, this update tends to drive the $\pi_s$ corresponding to ``redundant speakers'' to zero values, which effectively drops them from the model and selects the right number of speakers in the input conversation.\footnote{This is known as automatic relevance determination principle \cite{Bishop2006}}

\subsubsection*{Evaluating the ELBO}
The convergence of the iterative VB inference can be monitored by evaluating the ELBO objective. For the Bayesian HMM, the ELBO can be efficiently evaluated (see page 95 of \cite{BealVBLowerBThesis}) as
\begin{equation}
\hat{\mathcal{L}} = \ln\overline{p}(\XX) +\sum_s \frac{F_B}{2} \left(R+\ln|\mathbf{L}_s^{-1}|-\tr(\mathbf{L}_s^{-1})-\boldsymbol{\alpha}_s^T\boldsymbol{\alpha}_s\right),
\label{lowerbound}
\end{equation}
where $R$ is the dimensionality of the x-vectors. Note, that since $\mathbf{L}_s$ is a diagonal matrix, $\ln|\mathbf{L}_s^{-1}|$ can be calculated just as the sum of the log of the elements in the diagonal.
This way of evaluating the ELBO is very practical as the term $\overline{p}(\XX)$ from~\eqref{totalA} is obtained as a byproduct of ``updating $q(\ZZ)$'' using the forward-backward algorithm. On the other hand, the ELBO can be evaluated using~\eqref{lowerbound} only right after the $q(\ZZ)$ update. Thus, the improvements in the ELBO obtained from $q(\YY)$ or $\pi_s$ updates cannot be monitored, which might be useful for debugging purposes. 

The complete VB inference consisting of iterative updates of $q(\YY)$, $q(\ZZ)$ and parameters $\pi_s$ is summarized in the following algorithm:

\begin{mdframed}[
    align=center,
    linecolor=black,
    linewidth=0.6pt,
    userdefinedwidth=0.9\columnwidth,
]\begin{algorithmic}
  \State Initialize all $\gamma_{ts}$ (e.g. randomly or using another diarization method)
  \Repeat  
    \State   Update $q(\mathbf{y}_s)$ for $s{=}1..S$ using \eqref{YVBupdate}
    \For{$t=1..T$}
    \State Calculate $A(t,s)$ for $s{=}1..S$ using \eqref{forward}
    \EndFor
    \For{$t=T..1$}  
    \State Calculate $B(t,s)$ for $s{=}1..S$ using  \eqref{backward}
    \EndFor
    \State Update $\gamma_{ts}$ for $t{=}1..T, s{=}1..S$ using \eqref{spkrespons}
    \State Update $\pi_s$ for $s{=}1..S$ using \eqref{thepiupdate}
    \State Evaluate ELBO $\hat{\mathcal{L}}$ using \eqref{lowerbound}
    \Until convergence of $\hat{\mathcal{L}}$
\end{algorithmic}
\end{mdframed}

\section{Experiments}\label{sec:bhmm_experiments}

\subsection{Model configuration}
\label{sec:configuration}
This section presents results with a cascaded system based on clustering of x-vectors by means of VBx. However, the system consists of a few blocks as depicted in Figure~\ref{fig:blocks_vbx_system}. Voice activity detection is initially applied and, on the speech segments, a fixed segmentation is used where each x-vector is extracted from 1.5\,s windows every 0.25\,s. Towards the end of each segment, some x-vectors might be estimated on shorter-than-1.5\,s windows. Then, the BHMM framework is applied using AHC as initialization where the AHC threshold is set the same for all datasets and low enough to undercluster (i.e. produce more clusters than speakers). Finally, the output of an overlapped speech detection system is used to add overlapping speakers since the clustering step inherently assigns a single speaker to each x-vector. Each of the modules is described in detail next. Since we present results on telephone conversation datasets as well as wide-band corpora, two similar models are described, one for each data type.

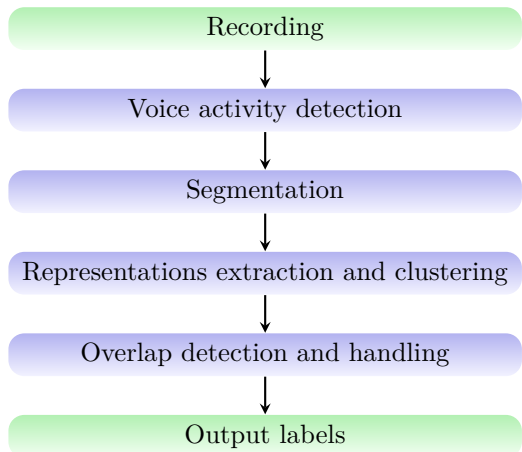
\begin{figure}[ht]
    \centering
    \tikzstyle{block} = [rectangle, 
        minimum height=0.5cm, 
        minimum width=0.8cm, align=center,
        draw=none, shade,
        top color=black!20!blue!30,
        bottom color=blue!5,
        rounded corners=6pt] 
    \tikzstyle{arrow} = [single arrow, draw]
    
    \tikzstyle{block2} = [rectangle, 
        minimum height=0.5cm, 
        minimum width=0.8cm, align=center,
        draw=none, shade,
        top color=black!20!green!30,
        bottom color=green!5,
        rounded corners=6pt] 
    \tikzstyle{arrow} = [single arrow, draw]
    
    \hspace*{-0.5cm}
    \begin{tikzpicture}[auto, >=stealth, node distance=0.4cm and 0.4cm, arr/.style={->,thick}, line/.style={thick}, font=\small]
    
        \node (rec) [block2, text width=6.5cm] {Recording};
        \node (VAD) [block, text width=6.5cm, below=0.5cm of rec] {Voice activity detection};
        \node (segmentation) [block, text width=6.5cm, below=0.5cm of VAD] {Segmentation};
        \node (clustering) [block, text width=6.5cm, below=0.5cm of segmentation] {Representations extraction and clustering};
        \node (postprocessing) [block, text width=6.5cm, below=0.5cm of clustering] {Overlap detection and handling};
        \node (out) [block2, text width=6.5cm, below=0.5cm of postprocessing] {Output labels};
        
        \draw[arr] (rec.south) -- (VAD.north);
        \draw[arr] (VAD.south) -- (segmentation.north);
        \draw[arr] (segmentation.south) -- (clustering.north);
        \draw[arr] (clustering.south) -- (postprocessing.north);
        \draw[arr] (postprocessing.south) -- (out.north);
    \end{tikzpicture}

    \caption{Components in the speaker diarization system utilized.}
    \label{fig:blocks_vbx_system}
\end{figure}

\subsubsection{Voice activity detection}
\label{sec:vad}

Two VAD systems are used and the best-performing on each corpus development set is chosen for such corpus. On one side, a model based on TDNN and statistics pooling is utilized. The model is widely used and available in Kaldi\footnote{~\url{http://kaldi-asr.org/models/m4}}. It is trained on the alignments provided by an HMM-GMM model with three classes: speech, non-speech and noise. The training data consist of Fisher parts 1 and 2~\cite{cieri2004fisher}. To produce the silence/speech decisions, an HMM is built to impose minimum silence and speech durations, and the TDNN's outputs are converted into likelihoods to perform Viterbi decoding with the HMM and produce the final VAD outputs. This system should, a priori, have better performance on narrow-band data.

The second model is based on a recurrent neural network (RNN)~\cite{gelly2017optimization}. It is widely used and available in pyannote 2.1~\cite{bredinpyannote}. The architecture comprises SincNet~\cite{ravanelli2018speaker} (formed by convolutional blocks), two stacked bidirectional LSTM layers, and two linear layers before the classification head into two outputs (silence/speech). 

\subsubsection{x-vector extractor and PLDA}
\label{sec:xvector}

As described in the previous section, VBx diarization relies on a pre-trained x-vector extractor and a PLDA model. Since we report results on both 16\,kHz recordings and 8\,kHz telephone recordings, two x-vector extractors and the corresponding PLDA models were needed, one for each condition. These were trained in cooperation with Phonexia and the complete PyTorch~\cite{paszke2019pytorch} recipe for x-vector extractor and PLDA training is available at \url{https://github.com/phonexiaresearch/VBx-training-recipe}.

\paragraph{x-vector extractor architecture}
Both 8\,kHz and 16\,kHz x-vector extractors use the same deep neural network architecture based on ResNet101~\cite{he2016deep, zeinali2019but}.
In both cases, the neural network inputs are 64 log Mel filter bank features extracted every 10\,ms using a 25\,ms window. The two x-vector extractors differ only in the frequency ranges spanned by the Mel filters, which are 20-3700\,Hz and 20-7700\,Hz for the 8\,kHz and 16\,kHz systems, respectively. The x-vector extractor architecture is summarized in Table~\ref{tab:resnet101}. The first 2D convolutional layer operates on the $64\times M$ matrix of log Mel filter bank features, where $M$ is the number of frames in the input segments. For training, we use 4\,s segments (i.e. $M=400$).  The following layers are standard ResNet blocks \cite{he2016deep}. As in the original x-vector architecture~\cite{snyder2018x}, the statistical pooling layer is used to aggregate information over the whole speech segment (i.e. mean and standard deviation of activations is calculated over the time dimension).
After the pooling layer, a linear transformation is used to reduce the dimensionality to obtain the $256$-dimensional x-vectors.
\begin{table}[!th]
\caption{\label{tab:resnet101} The structure of the proposed ResNet101 architecture. The first dimension of the input shows the size of the filterbank and the second dimension indicates the number of frames.}
  \centering
    \setlength\tabcolsep{4pt}
    \begin{tabular}{l l l l}
    \toprule
    Layer & Structure & Stride & Output \\
    \midrule
    Input & - & - & $64 \times $M$ \times 1$ \\
    Conv2D-1 & $3 \times 3, 32$ & 1 & $64 \times $M$ \times 32$ \\
    \midrule
    ResNetBlock-1 & $\myatopthree{1 \times 1, 32}{3 \times 3, 32}{1 \times 1, 128} \times 3$ & 1 & $64 \times $M$ \times 128$ \\
    ResNetBlock-2 & $\myatopthree{1 \times 1, 64}{3 \times 3, 64}{1 \times 1, 256} \times 4$ & 2 & $32 \times $M/2$ \times 256$ \\
    ResNetBlock-3 & $\myatopthree{1 \times 1, 128}{3 \times 3, 128}{1 \times 1, 512} \times 23$ & 2 & $16 \times $M/4$ \times 512$ \\
    ResNetBlock-4 & $\myatopthree{1 \times 1, 256}{3 \times 3, 256}{1 \times 1, 1024} \times 3$ & 2 & $8 \times $M/8$ \times 1024$ \\
    \midrule
    Statistics Pooling & - & - & $16 \times 1024$ \\
    Flatten & - & - & $16384$ \\
    Linear & - & - & $256$ \\
   \bottomrule 
   \end{tabular}
\end{table}

The x-vector extractors are trained using stochastic gradient descent and additive angular margin loss~\cite{deng2019arcface} with speaker identities as class labels. We ramp up the margin during the first two epochs (pass through the training data) and then train the neural network for another epoch with a fixed margin $m=0.2$.
 
\paragraph{x-vector extractor training data}
\label{training_data}
The 16\,kHz x-vector extractor is trained using data from VoxCeleb1~\cite{nagrani17_interspeech}, VoxCeleb2~\cite{chung18b_interspeech} and CN-CELEB~\cite{fan2020cn}. Table~\ref{tab:train_data_16k} details number of speakers and hours in each set. The energy-based VAD from Kaldi~\cite{povey2011kaldi} toolkit is used to remove silence frames. Speakers with less than 2 recordings are discarded. Further, we drop utterances with less than 4 seconds of speech. This way, about 4\% of speech data is discarded. Data augmentation is performed the same way as in the SRE16 Kaldi recipe\footnote{\url{https://github.com/kaldi-asr/kaldi/tree/master/egs/sre16/v2}}, resulting in four additional copies of the data with artificially added noise, music, or reverberation. 
Training examples are randomly sampled from the training data. This way we extract about 89 million examples (original and augmented 4s segments), which cover more than 60\% of the speech from the training corpora.

\begin{table}[!th]
\caption{\label{tab:train_data_16k} Number of speakers and the total amount of training data used for the 16\,kHz x-vector extractor.}
  \centerline{
  \setlength\tabcolsep{4pt}
    \begin{tabular}{l l l}
    \toprule
    Set & Speakers & Hours\\
    \midrule
    VoxCeleb1 & 1211 & 323\\
    VoxCeleb2 & 5994 & 2290\\
    CN-CELEB & 973 & 264\\
    \midrule
    Total & 8178 & 2877\\
   \bottomrule 
   \end{tabular}
}       
\end{table}

To train the 8\,kHz x-vector extractor, the same data sets are used as in the 16\,kHz case. Additionally, the following data sets were used: 
Mixer collection (NIST SRE 2004-2010), Switchboard, and DeepMine~\cite{zeinali2018deepmine}. See Table~\ref{tab:train_data_8k} for details. Any wide-band data used were downsampled to 8\,kHz and passed through a telephone codec (selected randomly from AMR, G.711, G.726, GSM-FR, and GSM-EFR). The same data selection and augmentation was used as for the 16\,kHz case.
Note that about 30\% of DeepMine data were discarded as this dataset contains many utterances with less than 4 seconds of speech (mostly phrases for text-dependent speaker verification). 

\begin{table}[!th]
\caption{\label{tab:train_data_8k} Number of speakers and the total amount of training data for the 8\,kHz x-vector extractor.}
  \centerline{
  \setlength\tabcolsep{4pt}
    \begin{tabular}{l l l}
    \toprule
    Set & Speakers & Hours\\
    \midrule
    VoxCeleb1 & 1211 & 323\\
    VoxCeleb2 & 5994 & 2290\\
    CN-CELEB & 973 & 264\\
    \midrule
    Mixer collection & 4254 & 3805\\
    Switchboard & 2591 & 1170\\
    DeepMine & 1858 & 688\\
    \midrule
    Total & 16881 & 8540\\
   \bottomrule 
   \end{tabular}
}       
\end{table}

\paragraph{PLDA training}
The 8\,kHz and 16\,kHz PLDA models are trained on the same data as the corresponding x-vector extractors. For this purpose, one x-vector is extracted from each individual recording (e.g. one cut from a YouTube video in the case of the VoxCeleb data). The length of such recordings can range from 4\,s to several minutes. Note that the PLDA trained on such x-vectors is later used in VBx to operate on x-vectors extracted from much shorter 1.5\,s segments.

\subsubsection{Overlap detection and handling}
\label{sec:osd_handling}

To detect overlapped speech segments, the model shared in pyannote~\cite{Bredin2020} was used. Alike their VAD system, the architecture comprises SincNet~\cite{ravanelli2018speaker} (formed by convolutional blocks), two stacked bidirectional LSTM layers, and two linear layers before the classification head into two outputs. 

Once the overlap segments are detected, given the diarization labels provided by VBx, second speaker labels are assigned on overlap segments. This is done using a heuristic~\cite{otterson2007efficient} which uses the closest-in-time speaker.

\subsection{Evaluation Data}
\label{sec:vbx_evaluation_data}
To evaluate the quality of the model, we considered several standard datasets in the field comprising different characteristics in terms of acoustics and style. Information about the datasets can be seen in Table~\ref{tab:datasets_information}.

\begin{table}[H]
    \caption{DER evaluation collar, types of microphone, and characteristics of each evaluation dataset.}
    \label{tab:datasets_information}
    \setlength{\tabcolsep}{4pt} 
    \centering
    \begin{tabular}{l|cll}
    \toprule
    Dataset & \multicolumn{1}{c}{collar (s)} & Microphone & Characteristics \\ 
    \midrule
   AISHELL-4 & 0 & array & Discussions in Mandarin in different rooms \\
   AliMeeting & 0 & array \& headset & Meetings in Mandarin in different rooms \\
   AMI & 0 & array \& headset & Meetings in English in different rooms \\
   Callhome & 0.25 & telephone & Conversations with family or friends \\
   CHiME6 & 0.25 & array & Dinner parties in home environments\\
   DIHARD2 & 0 & varied & Wide variety of domains\\
   DIHARD3 full & 0 & varied & Wide variety of domains \\
   DipCo & 0.25 & array & Dinner party sessions in the same room \\
   Mixer6 & 0.25 & varied & Interviews and calls in English \\
   MSDWild & 0.25 & varied & Videos of daily casual conversations \\
   RAMC & 0 & mobile phone & Phone calls in Mandarin \\
   VoxConverse & 0.25 & varied & Wide variety of videos (different languages) \\
    \bottomrule
  \end{tabular}
\end{table}

For each one, we utilized official or community-established partitions as described next. The training set is used to tune the hyperparameters of VBx using discriminatively trained VBx (DVBx)~\cite{klement2023discriminative} which presents a discriminative approach to learn such hyperparameters. The development set is used as validation set during that procedure.

\begin{itemize}
    \item AISHELL-4~\cite{fu21b_interspeech} is an open-source dataset collected with microphone arrays in a meeting scenario with Mandarin speakers. We use the train/evaluation split provided.
    \item AliMeeting~\cite{yu2022m2met}, similarly to AISHELL-4, is an open-source dataset collected with microphone arrays in a meeting scenario with Mandarin speakers. However, headset recordings are also available so we report results on both far-field and near-field scenarios using the train/evaluation/test split provided. Unlike in the M2MET Challenge, for which the dataset was introduced, we do not use oracle VAD labels for our systems.
    \item AMI~\cite{carletta2005ami,kraaij2005ami} is a multi-modal open-source dataset collected with microphone arrays and headsets in a meeting scenario. The language spoken is English and even though it was collected almost 20 years ago, it is still relevant these days. We use the full-corpus-ASR partition into train/dev/test and the diarization annotations of the ``only words'' setup described in~\cite{landini2022bayesian}\footnote{\url{https://github.com/BUTSpeechFIT/AMI-diarization-setup}}.
    \item Callhome~\cite{przybocki2001nist}, the speaker segmentation data from the 2000 NIST Speaker Recognition Evaluation dataset, usually referred to as ``Callhome'', has become the de facto telephone conversations evaluation set for diarization. We report results using the standard Callhome partition\footnote{Sets listed in \url{https://github.com/BUTSpeechFIT/CALLHOME\_sublists}}.
    \item CHiME6~\cite{watanabe20b_chime} consists of colloquial conversations in a dinner party scenario among 4 participants speaking English collected with far-field microphones in different parts of the rooms. The corpus is characterized by very high levels of overlapped speech and long recordings, making it one of the most challenging scenarios at the time of writing. We use the official partition (train/dev/eval) and annotations from CHiME7 challenge~\cite{cornell23_chime}. 
    \item DIHARD2~\cite{ryant2019second} which was used for the Second DIHARD Challenge and is composed of recordings from a variety of sources with scenarios ranging from read books, radio interviews, courtrooms to YouTube videos. We use the official partition.
    \item DIHARD3~\cite{ryant21_interspeech} was used for the Third DIHARD Challenge and follows a similar distribution of scenarios. We use the official partition and the ``full'' set, which includes a large portion of telephone conversations, in order to have a more distinct corpus with regard to DIHARD2. 
    \item DipCo~\cite{segbroeck20_interspeech} consists of English-spoken dinner party sessions collected with far-field microphones. We use the official partition (dev/eval) and annotations from CHiME7 challenge~\cite{cornell23_chime}. 
    \item Mixer6~\cite{brandschain2010mixer} comprises recordings of interviews and telephone calls as well as prompt reading (with a single speaker) collected with different microphones placed around the room. We use the official partition (train/dev/eval) and annotations from CHiME7 challenge~\cite{cornell23_chime} but, given that the train part has only one speaker per recording, we only use the dev and eval parts. 
    \item MSDWild~\cite{liu22t_interspeech} is an open-source dataset containing recordings from public videos in different real-world situations in different languages. We use the official partition into few.train/many.val/few.val as train/dev/test following other works.
    \item RAMC~\cite{yang22h_interspeech} is an open-source corpus of conversational speech in Mandarin collected with mobile phones. We use the official partition.
    \item VoxConverse~\cite{chung20_interspeech} was proposed for the VoxSRC challenge and consists of videos of different natures collected from YouTube. We use the official partition into dev/test and the latest annotations\footnote{Version 0.3 in \url{https://github.com/joonson/voxconverse/tree/master}}.
\end{itemize}

In general, we prioritize the evaluation of diarization without forgiveness collar and we use 0.25\,s collar in the cases where challenge organizers propose it or it is customary (and therefore needed for comparison with other works). More information about each set can be found in Tables~\ref{tab:datasets_stats1} and \ref{tab:datasets_stats2}. It should be noted that if a dataset did not have the three splits, then the development set was used as training set and the test set as validation set so those results should be taken with a grain of salt.

As the models we evaluate are trained on single-channel data, we mix all channels in the microphone array (far-field) or headsets (near-field) when the datasets contain microphone array data. For DipCo, only far-field channels are used.

\begin{table}[H]
    \caption{Number of files, minimum and maximum number of speakers per recording, and number of hours per partition of each evaluation dataset.}
    \label{tab:datasets_stats1}
    \setlength{\tabcolsep}{3pt} 
    \centering
    \begin{tabular}{l|ccc|ccc|ccc}
    \toprule
    \multirow{2}{*}{Dataset} & \multicolumn{3}{c|}{train} & \multicolumn{3}{c|}{development} & \multicolumn{3}{c}{test} \\ 
    & \#files & \#spk & \#\,h & \#files & \#spk & \#\,h & \#files & \#spk & \#\,h \\
    \midrule
   AISHELL-4 & 191 & 3-7 & 107.53 & -- & -- & -- & 20 & 5-7 & 12.72 \\
   AliMeeting & 209 & 2-4 & 111.36 & 8 & 2-4 & 4.2 & 20 & 2-4 & 10.78 \\
   AMI & 136 & 3-5 & 80.67 & 18 & 4 & 9.67 & 16 & 3-4 & 9.06 \\
   Callhome & -- & -- & -- & 249 & 2-7 & 8.70 & 250 & 2-6 & 8.55 \\
   CHiME6 & 14 & 4 & 35.68 & 2 & 4 & 4.46 & 4 & 4 & 10.05 \\
   DIHARD2 & -- & -- & -- & 192 & 1-10 & 23.81 & 194 & 1-9 & 22.49 \\
   DIHARD3 full & -- & -- & -- & 254 & 1-10 & 34.15 & 259 & 1-9 & 33.01 \\
   DipCo & -- & -- & -- & 5 & 4 & 2.73 & 5 & 4 & 2.6 \\
   Mixer6 & 243 & 1 & 183.09 & 59 & 2 & 44.02 & 23 & 2 & 6.02  \\
   MSDWild & 2476 & 2-7 & 66.1 & 177 & 3-10 & 4.1 & 490 & 2-4 & 9.85 \\
   RAMC & 289 & 2 & 149.65 & 19 & 2 & 9.89 & 43 & 2 & 20.64 \\
   VoxConverse & -- & -- & -- & 216 & 1-20 & 20.3 & 232 & 1-21 & 43.53 \\
    \bottomrule
  \end{tabular}
\end{table}

\begin{table}[H]
    \caption{Percentages of silence (sil), segments with a single speaker (1-spk), and overlapped speech (ov) per partition for each evaluation dataset.}
    \label{tab:datasets_stats2}
    \setlength{\tabcolsep}{6pt} 
    \centering
    \begin{tabular}{l|ccc|ccc|ccc}
    \toprule
    \multirow{2}{*}{Dataset} & \multicolumn{3}{c|}{train} & \multicolumn{3}{c|}{development} & \multicolumn{3}{c}{test} \\ 
    & sil & 1-spk & ov & sil & 1-spk & ov & sil & 1-spk & ov \\
    \midrule
   AISHELL-4 & 9.97 & 80.92 & 9.11 & -- & -- & -- & 9.54 & 85.99 & 4.47 \\
   AliMeeting & 7.12 & 66.22 & 26.66 & 7.7 & 72.36 & 19.94 & 8.04 & 73.24 & 18.72 \\
   AMI & 16.52 & 72.31 & 11.17 & 22.01 & 61.14 & 16.85 & 14.74 & 67.9 & 17.36 \\
   Callhome & -- & -- & -- & 10.23 & 74.49 & 15.28 & 10.55 & 74.46 & 14.99 \\
   CHiME6 & 23.52 & 52.85 & 23.63 & 12.58 & 43.66 & 43.76 & 19.96 & 50.95 & 29.09 \\
   DIHARD2 & -- & -- & -- & 24.26 & 68.33 & 7.41 & 25.92 & 67.48 & 6.6 \\
   DIHARD3 full & -- & -- & -- & 20.19 & 71.23 & 8.58 & 20.9 & 71.68 & 7.42 \\
   DipCo & -- & -- & -- & 7.91 & 66.38 & 25.71 & 9.41 & 65.7 & 24.89 \\
   Mixer6 & 75.07 & 24.96 & 0 & 3.05 & 76.19 & 20.76 & 2.4 & 83.66 & 13.94 \\
   MSDWild & 9.88 & 80.1 & 10.02 & 14.63 & 70.72 & 14.65 & 10.51 & 78.37 & 11.12 \\
   RAMC & 16.15 & 83.84 & 0.01 & 17.21 & 82.78 & 0.01 & 17.39 & 82.6 & 0.01 \\
   VoxConverse & -- & -- & -- & 6.83 & 89.68 & 3.49 & 10.45 & 86.82 & 2.73 \\
    \bottomrule
  \end{tabular}
\end{table}

\subsection{Results}

The results in this section are presented mainly to serve as baseline for Chapter~\ref{sec:conclusion}. For this reason, there is no particular analysis of the configurations in VBx nor the other steps of the pipeline. However, the interested reader can refer to:
\begin{itemize}
    \item \cite{landini2020but} for comparisons regarding the impact of the PLDA model and the OSD step in the context of DIHARD2
    \item \cite{diez2020optimizing} for analyses with respect to the segmentation and the initialization in the context of DIHARD2
    \item \cite{landini2021but} for comparisons between using PLDA and heavy-tailed PLDA across domains in DIHARD3
    \item \cite{landini2021analysis} for an in-depth analysis of every step of a modular system in the context of VoxConverse
    \item \cite{landini2022bayesian} for an extensive comparison of VBx with the state-of-the-art (at the time of publication) for AMI, Callhome and DIHARD2
\end{itemize}

It should be noted that the above-mentioned publications presented results with full modular pipelines focused on single datasets. In this section, we utilize updated sub-modules with the same pipeline across 12 different datasets. We expect these results will also serve as detailed baselines for future diarization publications.

Given the acoustic characteristics of different corpora, for the thesis, we evaluated two systems in terms of sampling rate: 8\,kHz and 16\,kHz comprising the most common scenarios at the time of writing. 

In terms of VAD, as mentioned in Section~\ref{sec:vad}, Kaldi ASpIRE and pyannote 2.1 models were evaluated with default parameters on signals at 8\,kHz and 16\,kHz. Recordings that were originally sampled at 16\,kHz were downsampled to obtain their 8\,kHz versions and vice-versa. For each condition (dataset and sampling rate), the best of the two VADs was chosen depending on the performance on the development set. The performance on the test sets is presented in Table~\ref{tab:VAD_results}. While precision and recall are detection performance metrics of the models in general, the missed and FA errors give a better hint about the effect the VAD can have on the final diarization performance.

For most of the datasets, the performances for different sampling rates are very similar. The only cases with relevant differences are CHiME6, DipCo and to a lesser extent Mixer6, for which the higher sampling rate provides advantages. These corpora, together with AMI array are those with the higher VAD errors. This is no surprise since they all contain data recorded with far-field microphones which are intrinsically more difficult and where the information carried in higher frequencies can be more helpful. In all cases, the error is very high due to the high percentage of missed speech, corresponding to low recall. It should be noted that one possibility to improve the performance on such sets would be to fine-tune the models or tune their detection thresholds using a training or development set of matching characteristics.

\begin{table}[H]
    \caption{VAD performance on the corresponding test set for each dataset in terms of missed speech, false alarm, precision and recall, all in \%.}
    \label{tab:VAD_results}
    \setlength{\tabcolsep}{2.5pt} 
    \centering
    \begin{tabular}{l|lcccc|lcccc}
    \toprule
    \multirow{2}{*}{Dataset} & \multicolumn{5}{c|}{8\,kHz} & \multicolumn{5}{c}{16\,kHz}  \\ 
    & VAD & Miss & FA & Prec. & Reca. & VAD & Miss & FA & Prec. & Reca. \\
    \midrule
   AISHELL-4 & Kaldi & 5.55 & 1.36 & 98.4 & 93.9 & Kaldi & 5.41 & 1.39 & 98.4 & 94.0 \\
   AliMeeting far & Kaldi & 5.39 & 3.03 & 96.6 & 94.1 & Kaldi & 5.36 & 3.04 & 95.4 & 94.2 \\
   AliMeeting near & Kaldi & 1.81 & 2.67 & 96.9 & 97.9 & Kaldi & 1.74 & 2.69 & 96.9 & 98.0 \\
   AMI array & pyannote & 13.85 & 1.34 & 98.0 & 82.8 & pyannote & 13.85 & 1.34 & 98.0 & 82.8 \\
   AMI headset & pyannote & 2.5 & 3.03 & 96.3 & 96.9 & pyannote & 2.5 & 3.03 & 96.3 & 96.9 \\
   Callhome & Kaldi & 1.5 & 6.61 & 93.0 & 98.3 & Kaldi & 1.47 & 6.62 & 93.0 & 98.4 \\
   CHiME6 & Kaldi & 54.07 & 1.92 & 93.0 & 32.1 & pyannote & 25.73 & 0.94 & 98.3 & 67.7 \\
   DIHARD2 & pyannote & 5.13 & 2.95 & 95.9 & 93.1 & pyannote & 5.13 & 2.95 & 95.9 & 93.1 \\
   DIHARD3 full & pyannote & 3.3 & 2.55 & 96.8 & 95.8 & pyannote & 3.3 & 2.55 & 96.8 & 95.8 \\
   DipCo & pyannote & 40.08 & 0.57 & 98.9 & 55.8 & pyannote & 28.88 & 0.86 & 98.6 & 68.1 \\
   Mixer6 & pyannote & 24.19 & 0.37 & 99.5 & 75.2 & pyannote & 17.52 & 0.43 & 99.5 & 82.0 \\
   MSDWild & pyannote & 2.65 & 5.38 & 94.2 & 97.0 & pyannote & 2.65 & 5.38 & 94.2 & 97.0 \\
   RAMC & Kaldi & 0.33 & 9.24 & 89.9 & 99.6 & Kaldi & 0.32 & 9.27 & 89.9 & 99.6 \\
   VoxConverse & pyannote & 1.35 & 3.22 & 96.5 & 98.5 & pyannote & 1.35 & 3.22 & 96.5 & 98.5 \\
    \bottomrule
  \end{tabular}
\end{table}

For OSD, only the model in pyannote with default parameters was used and its performance on test sets is presented in Table~\ref{tab:OSD_results}. 
In this case, the differences in performance between the two sampling rates are very small. As compared to VAD, the classification performance for OSD is substantially worse. The task itself is more difficult (even for humans) so more errors are expected to a certain extent. Like with the VAD, far-field sets have higher errors which is expected since they present more challenging conditions. In most cases, most of the mistakes are related to a low recall. However, a low recall is preferred to a low precision in a modular pipeline since false alarms count immediately as diarization errors and being more conservative is preferred.

In VBx diarization itself, as denoted in \eqref{fetchELBO}, the hyperparameters $F_A$ and $F_B$ have an effect on the inference and using values different from the theoretical $F_A=1$ and $F_B=1$ can lead to better performance. Different values can be explored exhaustively as done for AMI, Callhome and DIHARD2 in \cite{landini2022bayesian} but this is a time-consuming task. Recent work~\cite{klement2023discriminative} has tackled this issue with a discriminative training framework where the hyperparameters (and potentially the PLDA parameters) of VBx can be learned automatically. We used this framework, which extends the work in this thesis, to find suitable $F_A$ and $F_B$ values for the rest of the corpora as presented in \cite{landini2023diaper}.

\begin{table}[H]
    \caption{OSD performance on the corresponding test set for each dataset in terms of missed speech, false alarm, precision and recall, all in \%.}
    \label{tab:OSD_results}
    \setlength{\tabcolsep}{6pt} 
    \centering
    \begin{tabular}{l|cccc|cccc}
    \toprule
    \multirow{2}{*}{Dataset} & \multicolumn{4}{c|}{8\,kHz} & \multicolumn{4}{c}{16\,kHz}  \\ 
    & Miss & FA & Prec. & Reca. & Miss & FA & Prec. & Reca. \\
    \midrule
   AISHELL-4 & 4.12 & 0.24 & 59.0 & 7.9 & 3.78 & 0.35 & 66.4 & 15.5 \\
   AliMeeting far & 15.06 & 0.58 & 86.3 & 19.6 & 13.56 & 0.91 & 85.1 & 27.6 \\
   AliMeeting near & 6.9 & 0.86 & 92.4 & 60.2 & 7.83 & 0.81 & 92.2 & 54.8 \\
   AMI array & 8.77 & 0.48 & 86.1 & 25.2 & 8.77 & 0.48 & 86.1 & 25.2 \\
   AMI headset & 5.62 & 0.89 & 87.3 & 52.1 & 5.62 & 0.89 & 87.3 & 52.1 \\
   Callhome & 9.39 & 1.65 & 77.1 & 37.2 & 9.23 & 1.87 & 75.3 & 38.3 \\
   CHiME6 & 27.76 & 0.19 & 86.6 & 4.1 & 26.82 & 0.26 & 89.1 & 7.4 \\
   DIHARD2 & 5.05 & 0.79 & 66.1 & 23.4 & 5.05 & 0.79 & 66.1 & 23.4 \\
   DIHARD3 full & 4.6 & 0.99 & 74.1 & 38.0 & 4.6 & 0.99 & 74.1 & 38.0 \\
   DipCo & 24.0 & 0.42 & 67.6 & 3.6 & 23.4 & 0.61 & 71.0 & 6.0 \\
   Mixer6 & 12.53 & 9.11 & 13.4 & 10.1 & 11.99 & 12.41 & 13.6 & 14.0 \\
   MSDWild & 7.58 & 2.33 & 61.9 & 33.4 & 7.58 & 2.33 & 61.9 & 33.4 \\
   RAMC & 0.01 & 1.49 & 0.0 & 14.1 & 0.01 & 1.6 & 0.0 & 14.4 \\
   VoxConverse & 1.29 & 0.87 & 62.5 & 52.9 & 1.29 & 0.87 & 62.5 & 52.9 \\
    \bottomrule
  \end{tabular}
\end{table}

\begin{table}[H]
    \caption{Diarization results on the corresponding test set for each dataset in terms of DER and its three components and the mean speaker counting error. For convenience, the DER forgiveness collar is also presented.}
    \label{tab:VBx_results}
    \setlength{\tabcolsep}{2.5pt} 
    \centering
    \begin{tabular}{lc|c|ccc|c|c|ccc|c}
    \toprule
    \multirow{2}{*}{Dataset} & DER & \multicolumn{5}{c|}{8\,kHz model} & \multicolumn{5}{c}{16\,kHz model}  \\ 
    & Collar & DER & Miss & FA & Conf. & MSCE & DER & Miss & FA & Conf. & MSCE \\
    \midrule
   AISHELL-4 & 0 & 14.46 & 10.34 & 1.69 & 2.43 & 0.15 & 15.84 & 9.98 & 1.8 & 4.06 & 0.45 \\
   AliMeeting far & 0 & 29.6 & 21.93 & 3.13 & 4.55 & 0.1 & 28.84 & 20.64 & 3.40 & 4.80 & 0.15 \\
   AliMeeting near & 0 & 23.49 & 12.38 & 3.29 & 7.82 & 0.05 & 22.59 & 13.19 & 3.26 & 6.14 & 0.1 \\
   AMI array & 0 & 34.14 & 26.12 & 1.93 & 6.08 & 0.44 & 34.61 & 26.12 & 1.93 & 6.56 & 0.5 \\
   AMI headset & 0 & 22.23 & 14.06 & 3.73 & 4.45 & 0.31 & 22.42 & 14.06 & 3.73 & 4.63 & 0.31 \\
   Callhome & 0.25 & 13.62 & 7.69 & 1.92 & 4.02 & 0.31 & 26.65 & 8.08 & 1.72 & 16.85 & 0.8 \\
   CHiME6 & 0.25 & 84.01 & 77.84 & 0.91 & 5.26 & 2.25 & 70.42 & 47.72 & 0.91 & 21.79 & 3 \\
   DIHARD2 & 0 & 27.47 & 14.13 & 4.44 & 8.9 & 1.02 & 26.59 & 14.14 & 4.5 & 7.95 & 1.01 \\
   DIHARD3 full & 0 & 20.49 & 10.13 & 3.96 & 6.4 & 0.6 & 20.28 & 10.15 & 3.99 & 6.15 & 0.6 \\
   DipCo & 0.25 & 56.19 & 50.77 & 0.72 & 4.69 & 0.8 & 49.22 & 40.66 & 1.05 & 7.51 & 2 \\
   Mixer6 & 0.25 & 38.09 & 28.14 & 8.66 & 1.29 & 0.13 & 35.6 & 21.74 & 12.43 & 1.44 & 0.04 \\
   MSDWild & 0.25 & 18.81 & 5.99 & 2.55 & 10.28 & 1.14 & 16.86 & 5.81 & 2.75 & 8.29 & 1.05 \\
   RAMC & 0 & 18.33 & 0.41 & 12.98 & 4.94 & 0.05 & 18.19 & 0.39 & 13.15 & 4.65 & 0 \\
   VoxConverse & 0.25 & 6.69 & 1.6 & 2.02 & 3.08 & 1.11 & 6.12 & 1.6 & 2.02 & 2.51 & 1.12 \\
    \bottomrule
  \end{tabular}
\end{table}

Table~\ref{tab:VBx_results} presents the final results when applying the whole pipeline to each of the datasets. The mean speaker counting error (MSCE) is defined as $MSCE = \frac{1}{R} \sum_{r=1}^R | C_r - \widehat{C}_r |$, where $C_r$ and $\widehat{C}_r$ are the actual and estimated number of speakers in recording $r$, and $R$ is the total number of recordings in the test set. 

As expected, the 16\,kHz model has an edge over the 8\,kHz one in most cases in terms of DER, given that most corpora are originally recorded with that sampling rate. The clear exception is Callhome comprising telephone conversations on 8\,kHz for which the wide-band model has almost twice the error. The other exception is AISHELL-4 for which there is some advantage when using the 8\,kHz model. At the same time, the MSCE presents an opposite trend, with the 8\,kHz model usually performing a bit better. In any case, the MSCE is in general quite low for both models in most datasets.

The final errors can also be partly traced to the performance of the VAD and OSD modules. For corpora with very high percentages of overlapped speech such as CHiME6 and DipCo, we can see that the DERs are the highest among all sets. For these corpora, the VAD errors are high showing that the dinner-party scenarios are challenging even for a relatively simple task. These are two of the sets that present far-field conditions. Besides them, AliMeeting far and AMI array share the same condition and their errors are among the highest. Finally, Mixer6 presents very high DER but this is mainly due to the bad performance in terms of OSD for this set.

\section{Strengths of VBx}\label{sec:bhmm_strengths}
One aspect is that the method relies on x-vectors as features for performing clustering. 
Current state-of-the-art systems for speaker verification rely on this type of features and large efforts are being made by the community to optimize and improve their quality. VBx ingests current state-of-the-art x-vectors but it can also use any new version of such embeddings, benefiting from any improvements that they can bring.

Another aspect is that the speaker models are derived from a PLDA model trained with large amounts of x-vectors from thousands of speakers which makes the speaker models in VBx very robust.

As shown in the experiments, the model can learn the number of speakers in the recording by itself and with great accuracy. Other approaches fail on this or need to use a different sub-module to estimate the number of speakers while the Bayesian nature of VBx can handle this elegantly. Furthermore, the model is principled as it relies on a probabilistic modeling of the diarization task. This is in contrast to other approaches that rely heavily on ad-hoc tuning of thresholds.

Moreover, the model itself does not require large amounts of data with diarization annotations to perform well. This contrasts NN-based approaches, which are usually very data-hungry and more computationally expensive.

Finally, the main advantage of a cascaded model is that it is relatively easy to replace sub-modules. For example, one can focus on improving the VAD module with little effect on other parts of the system but impact in the final performance.

\section{Weaknesses of VBx}\label{sec:bhmm_weaknesses}
One of the problems is that a system that makes use of VBx for clustering has also other modules to deal with VAD, OSD and potentially yet other modules such as enhancement or resegmentation. This makes running such a system somewhat cumbersome as several sub-systems have to be run sequentially. A more elegant approach would be a single model that can take the signal as the input and produce directly the diarization output.

Although some works have proposed the idea of having a silence state in the HMM, in our case, we would need to ensure that the x-vector extractor and PLDA can model silence. This is normally not done for speaker verification and it would potentially require special treatment for silence in the extractor. Furthermore, the VAD performance can be definitely better with an NN-based model trained specifically for the task.

Since the x-vector extractor model is trained for speaker verification, the training samples are usually longer than the 1.5\,s used for diarization. Moreover, this length for the segments is already quite long and it is likely that speech from more than one speaker is included in that window, especially for colloquial conversations with short turns. In order to deal with this, an extractor would need to be trained with shorter segments (which are by nature less reliable) to be able to deal better with short segments. 

Related to this, whenever there is overlapped speech, the embedding is extracted necessarily with speech from more than one speaker. However, the x-vector extractor is trained with single-speaker speech and it is possible that x-vectors do not model overlapped speech well. In this direction, there have been recent efforts~\cite{jung2023search} to train extractors with augmentations suited for diarization which have shown promising results by incorporating more than one speaker in the segments.

Using an x-vector extractor that can handle overlaps is only half of the solution as modifications in the HMM would be required to handle such overlaps. This aspect is related to one clear weakness of the approach: the presented model cannot deal with overlapped speech; yet, most of the errors correspond currently to overlap segments. Nevertheless, a recent extension~\cite{delcroix23_interspeech} has made use of VBx to reconcile decisions between short segments made by an end-to-end model that handles overlap, showing that it might be possible to adapt the approach adequately.

\section{Final remarks}

This chapter described the standard cascaded framework for diarization models. A brief historical overview of each sub-module was presented and special emphasis was put on the Bayesian hidden Markov model scheme for clustering, commenting on the initial models that inspired the current version and describing its formulation in detail. Making use of x-vectors as input, VBx was evaluated on standard tasks with data from different domains.

Finally, the strengths and weaknesses of the model were discussed. In spite of its problems, the ease of use of VBx recipe\footnote{\url{https://github.com/BUTSpeechFIT/VBx}} and strong results allowed it to be used as the baseline in numerous publications and even in some of the most recent challenges~\cite{yu2022m2met,grauman2022ego4d}.

\chapter{End-to-end diarization systems}
\label{sec:e2e_diarization}

The main weakness of cascaded methods for diarization is their inherent multi-modular nature. Not only the inference can be cumbersome and potentially slow but also each subsystem is trained independently with different objectives. Therefore, having a single model to handle the whole problem is desirable.

In this chapter, we discuss the best-performing discriminatively trained diarization systems with special emphasis on end-to-end neural diarization (EEND), for being the most promising framework lately. Then, given the need of these models for large amounts of annotated training data, we discuss an alternative approach to create ``simulated conversations'' (in contrast to the original ``simulated mixtures'') and show its advantages. Finally, we present DiaPer, a model that builds on the EEND framework.

\section{Neural networks for diarization}\label{sec:nn_overview}
Neural networks have been used in the context of diarization for some time. Given the most common diarization pipeline described in Figure~\ref{fig:blocks_history}, the first applications of NN focused on individual steps. We have covered examples for pre-processing, VAD, segmentation and overlap detection in section~\ref{sec:cascaded_structure}. In this section, we cover specific works for speaker embeddings extraction and decoding of speaker activities which span the most recent efforts in these directions. The lists in Sections \ref{sec:nn_embeddings_and_affinity} and \ref{sec:nn_speaker_activities}  do not pretend to be exhaustive\footnote{An extensive and frequently updated list of works can be found in \url{https://github.com/DongKeon/Awesome-Speaker-Diarization}} but rather cover the works that have shown remarkable results at their time or that presented a new, and promising, line of research. In particular, Section~\ref{sec:e2e_tsvad} covers target speaker voice activity detection (TS-VAD) which processes the recordings in an end-to-end fashion but relies on a previously trained speaker embedding extractor and an initial diarization output given by another model.

Although end-to-end diarization models have only been developed in the last few years, they have attracted a broad interest. Section~\ref{sec:e2e_EEND} covers the main works in the end-to-end neural diarization framework. EEND has been the first fully end-to-end model capable of dealing with the permutation problem intrinsic to speaker diarization and has paved the road for most of the recent developments in this field.

While both EEND and TS-VAD have their advantages, they also struggle to handle certain scenarios. Different extensions have been proposed and some of them can be categorized as ``two-stage'' models where the first stage usually performs EEND on short segments of the recording and the second stage can be a standard clustering method (such as AHC or spectral clustering) or a neural-based clustering approach. We discuss them in Section~\ref{sec:e2e_twostage}.

\subsection{Neural networks for speaker embeddings and affinity matrices}
\label{sec:nn_embeddings_and_affinity}
Since the advent of neural speaker embeddings (d-vectors~\cite{variani2014deep} or x-vectors~\cite{snyder2018x}), we could say that all competitive diarization systems nowadays make use of an NN. NN-based embeddings capture speaker information that allows for better segment representations which lead to better performance. However, many approaches have focused on modifying the speaker representations or the affinity matrix (obtained by comparing every embedding against each other). These are, in turn, used by a clustering method for the purpose of diarization using different techniques: 
\begin{itemize}
    \item \cite{yella2014artificial} is one of the first works in this direction where an NN is trained to discriminate speakers. Then, the output of an intermediate layer (`bottleneck') is used as features in an HMM-GMM system to perform clustering. The main novelty in this work was the combination of acoustic features with discriminatively trained representations in the context of a diarization system.
    \item \cite{garcia2017speaker} proposed using DNN-based embeddings in combination with AHC to produce diarization outputs. Even though the term ``x-vector'' was not coined yet, the embeddings were obtained with the same strategy so this is the first work to have performed clustering of x-vectors for diarization. Analogously, \cite{wang2018speaker} present the first application of d-vectors for diarization by clustering them with spectral clustering.
    \item \cite{lin19_interspeech} proposed to produce the affinity matrix using a BLSTM. This is accomplished by feeding the BLSTM with sequences where the $i^{th}$ frame is concatenated to each frame in the original sequence. The network is trained to learn the similarity between frame $i$ and each other frame using binary cross-entropy (BCE) with ``same'' versus ``different'' classes. The approach showed excellent results at the time and, in spite of the limitations of LSTMs with long sequences, the BLSTM-based affinity matrix performed significantly better than a PLDA-based one on long utterances. A similar idea is presented in~\cite{Lin2020selfattentive}, where self-attention is used to produce the similarity matrix in the same fashion as with the BLSTM and leads to better results. 
    \item \cite{wang2020speaker} used graph neural networks (GNN) to refine the input embeddings. The GNNs (which work as an encoder) are trained so that the affinity matrix of the modified embeddings resembles the ground truth adjacency matrix, using a binary classification loss for linkage prediction for all pairs of segments inside each session. When using spectral clustering with the processed embeddings, they manage to obtain significant improvements over the original representations. This shows the potential for transforming embeddings normally trained for speaker recognition towards diarization-specific applications. \cite{wang2023community} also utilize GNNs but instead of refining the embeddings in a global fashion, they refine clusters of embeddings (denoted by their corresponding subgraphs) one at a time. Then, the refined subgraphs are merged and the new whole graph is partitioned to obtain speaker clusters. Unlike~\cite{wang2020speaker}, the clustering is performed in the graph space and it is shown to perform considerably better than other clustering methods such as AHC or spectral clustering.
    \item Another line of works~\cite{singh20_interspeech,singh2021self,singh2021selfsup} also explores refinement of the affinity matrix and graph-based clustering. In~\cite{singh20_interspeech}, a triplet loss scheme is used to train a DNN that refines the cosine similarity-based affinity matrix for AHC. In~\cite{singh2021self}, path integral clustering (a graph-structural agglomerative clustering algorithm) is used to define the clusters resulting in better performance than AHC. In~\cite{singh2021selfsup} the similarities are given by a PLDA model whose parameters are updated as part of the training process to improve the performance with respect to cosine similarity.
    \item \cite{kwon21b_interspeech} explore two techniques: dimensionality reduction of embeddings (by means of training an auto-encoder) and iterative affinity matrix refinement (by means of an attention-based scheme), in the context of AHC and spectral clustering. The dimensionality reduction does not affect spectral clustering but it improves the performance with AHC while the opposite happens for the affinity matrix refinement. This shows that the local hierarchies built with AHC can often be given by irrelevant information and transforming the embeddings, even if unsupervisedly, can alleviate this issue. Alternately, spectral clustering is highly affected by noises in the affinity matrix and several refinement operations~\cite{wang2018speaker} are usually applied to improve the quality of the matrix.
    \item One of the limitations when clustering embeddings lies in the need for long enough segments to extract good representations but short enough segments to model short speaker turns. To address this, one line of research has combined embeddings extracted with different window sizes and shifts. In~\cite{park2021multi}, the affinity matrices of three different configurations are processed with an NN and the resulting matrix is used for spectral clustering. In~\cite{kwon2022multi}, embeddings from different resolutions are paired with vectors that denote which scale they were extracted with (analogously to positional encoding in Transformer models) and they are processed with an attention mechanism to obtain similarities between the embeddings (through the inherent attention weights). The resulting affinity matrix is finally used for clustering with spectral clustering.

    \item \cite{jung2023search} argue that speaker embedding extractors are normally trained with a speaker recognition application in mind where there is speech from a single speaker. In diarization, it is likely that embeddings will be extracted from speech corresponding to more than one speaker so they analyze augmentation schemes for training the extractor, where different levels of overlapped speech are introduced or where speaker changes occur. It is shown that training the embedding extractors using these augmentation strategies improves the performance of a clustering-based diarization system. Furthermore, they also present a strategy to create target and non-target trials (including speaker changes and overlap) that are better proxies to evaluate the performance of a speaker embedding extractor for the purpose of diarization (instead of speaker recognition, that is usually the target application when training embedding extractors). Given the importance that embedding extractors still have for diarization, this is a promising line of research that could potentially lead to extractors more tailored for diarization.

\end{itemize}

\subsection{Neural networks for decoding speaker activities}
\label{sec:nn_speaker_activities}
A different line of work to the ones described above has focused on replacing the standard clustering approach by an NN-based solution. In some cases, the assignment of speaker labels to the embeddings was treated as a sequence-to-sequence problem modeled with BLSTMs \cite{zhang2022towards} and with an encoder-decoder Transformer-based architecture \cite{li2021discriminative}. In this last case, while remarkable results were obtained in comparison with spectral clustering, using augmentation strategies is crucial to make this approach successful. 

One particular model, named unbounded interleaved-state recurrent neural network~\cite{zhang2019fully} handles segmentation, embedding extraction and clustering steps together. It works in an online manner and can deal with an unbounded number of speakers. Although they present remarkable results, their full recipe\footnote{\url{https://github.com/google/uis-rnn}} is not shared as it relies on internal components of Google's infrastructure. 

Some approaches have focused on clustering or diarization using NNs but from the perspective of source separation~\cite{hershey2016deep,von2019all}. Given that they focus on a different task which usually requires clean speech from each speaker, we do not discuss them further here.

Approaches focusing on clustering with an NN, just like any other clustering method for diarization, cannot deal naturally with overlapping segments.
Another application of NN consists in decoding speaker activity probabilities for each speaker, eventually handling overlapped speech. TS-VAD follows this strategy and is described next.

\subsubsection{Target speaker voice activity detection}
\label{sec:e2e_tsvad}

Target speaker voice activity detection ~\cite{medennikov2020target} was developed in the context of the CHiME-6 Challenge in 2020 and its performance was so remarkably better than any other system that they won the challenge by a large margin. This approach makes use of speaker embeddings (i-vectors) from each speaker as references to find where those speakers appear in the recording, thus the name. TS-VAD relies on an initial diarization output used to estimate one i-vector per speaker. Then, acoustic features processed with CNN layers and BLSTM projection layers are used to detect if there is voice activity for each of the speakers using the i-vectors as anchor speaker references, as shown in Figure~\ref{fig:tsvad}. 
The outputs are per-speaker speech activity probabilities. However, they can benefit from post-processing too: \cite{medennikov2020target} evaluate a few options such as tuning the detection threshold, applying a median filter, removing short pauses and Viterbi decoding using an HMM with a few states. Eventual re-estimation of the i-vectors is needed to improve the performance but the approach manages to perform very well, especially in scenarios with large amounts of overlapped speech such as CHiME-6. 
Nevertheless, one of the main drawbacks of this approach as originally proposed is that it requires a priori knowledge of the number of speakers. 

\begin{figure}[ht]
    \centering
    \includegraphics[width=0.8\textwidth]{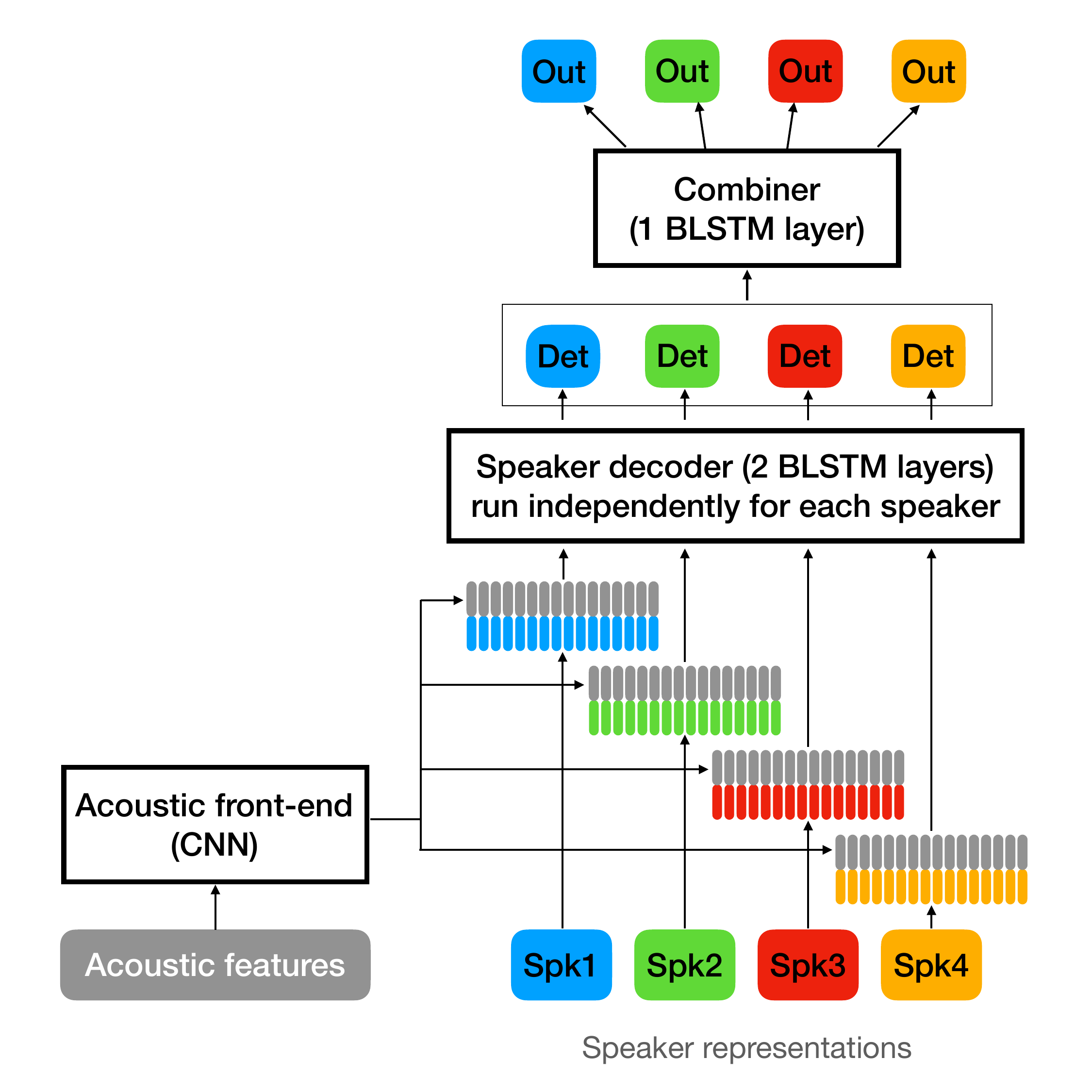}
    \caption{Diagram of the original target speaker voice activity detection approach.}
    \label{fig:tsvad}
\end{figure}

TS-VAD was extended in \cite{wang2021ustc} to deal with an unknown number of speakers by the winning team of the Third DIHARD Challenge~\cite{ryant21_interspeech} by adding (other, non-appearing in the conversation) or removing (least talkative) anchor speaker references. In \cite{he21c_interspeech}, an upper bound on the number of speakers was used so that the model can drop some of the representations. 
Adaptive neural speaker diarization using memory-aware multi-speaker embedding~\cite{he2023ansd} presents an approach inspired by TS-VAD which can handle an undefined number of speakers. It is based on speaker templates encoded in a memory module and an attention mechanism using the frames of individual speakers according to the initial diarization. The resulting speaker embeddings for each of the observed speakers are further reinforced to be similar to i-vectors, x-vectors, or the concatenation of both. This model is iteratively applied by fine-tuning it on simulated mixtures obtained from the single-speaker segments according to the initial diarization output, thus providing extra gains.

The modifications proposed by \cite{cheng2023multi,cheng2023target} can handle even more speakers (although always capped by some pre-defined maximum). The main changes consist in using NN-based representations instead of i-vectors. \cite{cheng2023target} use Conformer \cite{gulati20_interspeech} instead of only CNN to produce per-frame representations. Furthermore, the BLSTM projection is replaced by an attention-based decoder.
In~\cite{wang2022cross}, the channels are processed with cross-channel attention instead of CNN with attention. A recent extension~\cite{wu2023semi} uses cross-channel attention in the context of CHiME-7~\cite{cornell23_chime} and obtains remarkable results. They also perform semi-supervised training of the model using pseudo-labels generated for unlabeled in-domain data to improve further.

In~\cite{cheng2023target}, the speaker embeddings are obtained with a ResNet-based architecture incorporated into the whole model and the speaker activity module (originally a BLSTM) is replaced by an attention-based module which not only considers across-time but also across-speaker context.

A model similar to TS-VAD is presented in \cite{du2021speaker,du2022speaker} where the speaker activities returned by the model are not treated independently but a power set encoding is used to produce overlap-aware outputs.

\subsection{End-to-end neural diarization}\label{sec:e2e_EEND}

\subsubsection{EEND and SA-EEND to handle a fixed number of speakers}

End-to-end neural diarization was originally proposed in \cite{fujita19_interspeech} as a single neural-based system that can handle diarization without the need for other modules.  The diarization problem is defined as a per-speaker-per-time-frame binary classification problem. Given a sequence of observations (features) $\textbf{X} \in \mathbb{R}^{T \times F}$ where $T$ denotes the sequence length and $F$ the feature dimensionality, and labels $\textbf{Y} \in \{0, 1\}^{T \times S}$ where $y_{t,s} = 1$ if speaker $s$ is active at time $t$ and 0 otherwise; the most probable labels $\bar{\textbf{Y}}$ are estimated among all possible labels as
\begin{equation}
    \bar{\textbf{Y}} = \underset{\textbf{Y} \ \in \ \mathcal{Y}}{\arg \max} \ P(\textbf{Y} | \textbf{X})
\end{equation}
using the following factorization: 
\begin{equation}
\label{eq:cond_indep_assumption}
    P(\textbf{Y} | \textbf{X}) = \prod_{t}^T \prod_{s}^S P(y_{t,s} | \textbf{X}).
\end{equation}

This formulation handles silence and overlaps between speakers naturally since silence at time $t$ corresponds to the case where $y_{t,s} = 0 \ \forall s \in S$ while overlap happens at time $t$ if $y_{t,s_1} = 1$ and $y_{t,s_2} = 1$ with $s_1 \neq s_2$. The model predicts the posteriors $P(y_{t,s} | \textbf{X})$ and is trained to maximize the probability of the reference labels.

One of the main features of the EEND framework is the permutation-invariant training (PIT) loss borrowed from source separation~\cite{yu2017permutation} which allows to train the model to distinguish speakers without assigning specific identities. If $\mathbf{\hat{y}}_{t} \in \{0, 1\}^S$ are the posteriors estimated by the model for time $t$, the loss is defined as
\begin{equation}
    \label{eq:diar_loss}
    \mathcal{L}_d(\textbf{Y}, \hat{\textbf{Y}}) = \frac{1}{T S} \ \underset{\phi \in perm(S)}{\min} \sum_{t}^T BCE(\mathbf{y}_t^{\phi}, \hat{\mathbf{y}}_t),
\end{equation}

where $\mathbf{y}_t^{\phi}$ is the $\phi$-th permutation of the ground truth speaker labels at time $t$ and BCE denotes the binary cross entropy
\begin{equation}
    \label{eq:BCE}
    BCE(\mathbf{y}_t, \hat{\mathbf{y}}_t) = \sum_s^S -y_{t,s} \log \hat{y}_{t,s} - (1 - y_{t,s}) \log (1 - \hat{y}_{t,s}).
\end{equation}

Note that the permutation of speaker labels is the same for all frames.

Together with this loss, an auxiliary deep clustering loss~\cite{hershey2016deep} was proposed but not used in the following works for not showing improvements.

The first version of EEND consisted of two BLSTM layers followed by a linear layer with $S$ outputs and a sigmoid function to produce the posteriors as shown in Figure~\ref{fig:eend}. This model was followed by the so-called SA-EEND~\cite{fujita2019end} where the BLSTM layers were replaced by one linear layer and several multi-head self-attention blocks as shown in Figure~\ref{fig:sa-eend}. SA-EEND showed improvements over the BLSTM-based one. However, one strong limitation of these models was that they could only handle a fixed number of speakers (limited to 2 in the original publications) per recording.

\begin{figure*}[t!]
    \centering
    \begin{subfigure}[t]{0.35\textwidth}
        \centering
        \caption{EEND in its original version. The branch for the clustering loss is omitted.}
        \label{fig:eend}
        \centering
        \includegraphics[width=\textwidth]{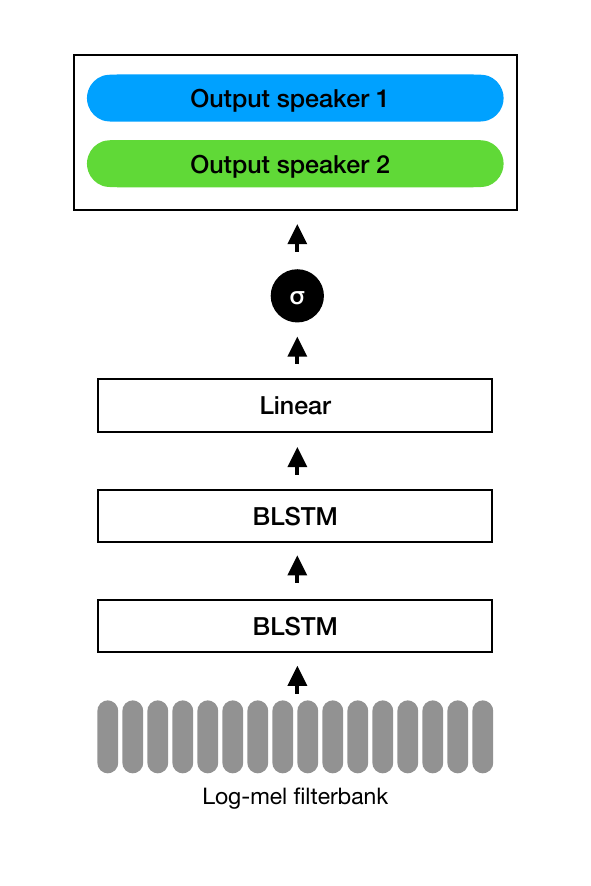}
    \end{subfigure}%
    ~ 
    \begin{subfigure}[t]{0.66\textwidth}
        \centering
        \caption{EEND with self-attention encoder blocks.}
        \label{fig:sa-eend}
        \centering
        \includegraphics[width=\textwidth]{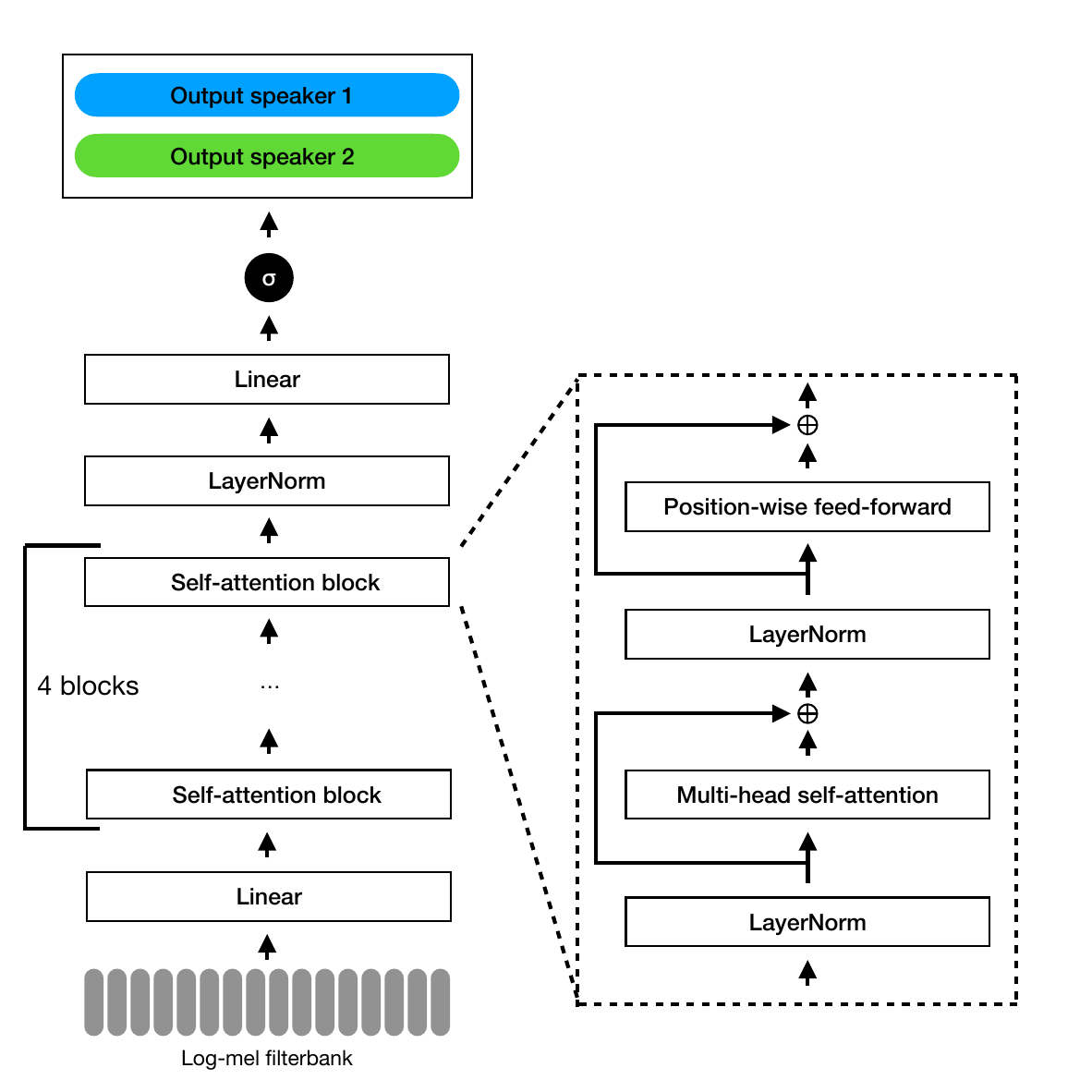}
    \end{subfigure}
    \caption{End-to-end neural diarization for a fixed number of speakers.}
\end{figure*}

\subsubsection{SC-EEND and EEND-EDA to handle variable numbers of speakers}
\label{sec:e2e_EEND_EDA}

The following works focused on extending the EEND approaches to handle variable numbers of speakers per recording. Speaker-wise chain EEND (SC-EEND)~\cite{fujita2020neural} presents an extension where, instead of assuming conditional independence (as in \eqref{eq:cond_indep_assumption}), using the probabilistic chain rule, the posterior probability is expressed as
\begin{equation}
    P(\textbf{Y} | \textbf{X}) = \prod_{s} P(\mathbf{y}_{s} | \mathbf{y}_1, \ldots, \mathbf{y}_{s-1}, \textbf{X}).
\end{equation}

The speech activities of the speakers are sequentially decoded using also the activities of previously estimated speakers. A neural network based on LSTMs outputs the activities of one speaker at a time until the current speaker has no speech. When training this model, one issue arises since training with PIT implies that there is no particular order for decoding the speakers. However, SC-EEND decodes the activities of the current speaker given the activity of the previously-decoded one. To reduce the impact of errors during training, teacher-forcing is employed (i.e. the ground truth activities are given instead of the ones decoded by the model). After decoding the activities of a speaker, one could pair them with the reference labels that lead to the lowest error to know which reference speaker's activities to feed to the next step but this leads to bad performance. A better strategy consists in decoding all speakers activities one by one, then finding the optimal match with the reference labels (as usual when using PIT) and then utilizing teacher-forcing with the correct order of speakers. This method has been extended to perform multitasking with VAD and/or OSD~\cite{takashima2021end}.

Another alternative to handle variable numbers of speakers was almost simultaneously proposed and, given its superior performance, has made the speaker-wise chain decoding to fall into disuse. The alternative model, making use of encoder-decoder-based attractors, was named EEND-EDA~\cite{horiguchi20_interspeech}. Following the SA-EEND approach, frame embeddings are obtained based on self-attention, but, instead of having fixed outputs for a limited number of speakers, the frame embeddings are passed through an encoder to produce a representation of the whole sequence which, in turn, is passed through a decoder to obtain one attractor per speaker, as shown in Figure~\ref{fig:eend-eda}. A unidirectional LSTM layer is used as encoder so that if $(\mathbf{e}_t)_{t=1}^T$ is the sequence of frame embeddings, 
\begin{equation}
    \mathbf{h}_0, \mathbf{c}_0 = Encoder(\mathbf{e}_1, \ldots, \mathbf{e}_T)
\end{equation}
is the representation of the whole sequence given by the last hidden state $\mathbf{h}_0 \in \mathbb{R}^D$ and last cell state $\mathbf{c}_0 \in \mathbb{R}^D$ of the LSTM where $D$ is the dimension of frame embeddings and attractors. Note that the output of the LSTM is not used in any way in the encoder. Furthermore, in practice, it has been observed that randomly shuffling the input of the encoder provides better results. This clearly goes against the intrinsic ability of LSTMs to model time sequences and shows their inability to handle long sequences. 

The LSTM decoder is used to produce one speaker representation per speaker seen in the sequence. These representations, called attractors, $\mathbf{a}_s \in \mathbb{R}^D$ are iteratively decoded one at a time as follows
\begin{equation}
    \mathbf{h}_s, \mathbf{c}_s, \mathbf{a}_s = Decoder(\mathbf{h}_{s-1}, \mathbf{c}_{s-1}, \mathbf{0}),
\end{equation}
where $\mathbf{0} \in \mathbb{R}^D$ is a zero vector passed as input to the decoder LSTM. The initial hidden and cell states ($\mathbf{h}_0$ and $\mathbf{c}_0$) correspond to those given by the encoder. 

\begin{figure*}[t!]
    \centering
    \centering
    \includegraphics[width=\textwidth]{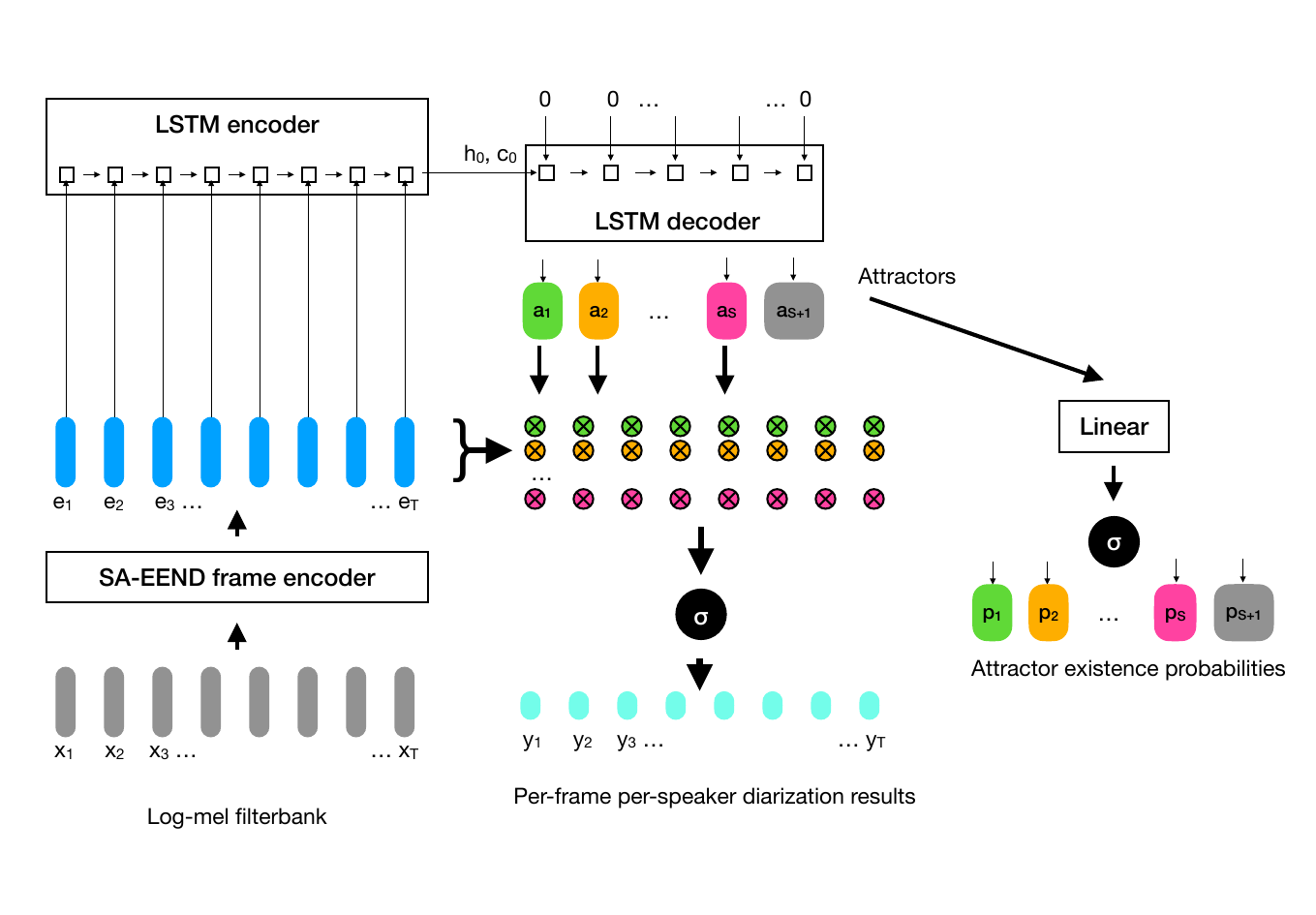}
    \caption{End-to-end neural diarization with encoder-decoder attractors (to handle a flexible number of speakers).}
    \label{fig:eend-eda}
\end{figure*}

This approach allows for decoding any number of attractors. If one knows how many speakers appear in the sequence, one can decode that number of attractors; however, in a less restricted scenario, the model needs to know which attractors to consider as valid for the diarization output. To determine so, an attractor existence probability is obtained for  attractor $s$ as
\begin{equation}
    p_s = \sigma(\mathbf{w}^\top \mathbf{a}_s + \mathbf{b}),
\end{equation}
where $\mathbf{w}$ and $\mathbf{b}$ are trainable weights and bias of a fully-connected layer and $\sigma$ is the logistic sigmoid function.

Given that the attractors are decoded in order, only the first $S$ will be deemed valid. During training, reference labels for the attractor existence $l_s$ are defined as 
\begin{equation}
    l_s = 
    \begin{cases}
        1 & \text{if } s \leq S \\
        0 & \text{S+1}
    \end{cases},
\end{equation}
where $S$ is the reference number of speakers in the sequence.

The parameters used to classify the attractors as valid or invalid are updated by means of the attractor existence loss 
\begin{equation}
    \label{eq:att_loss}
    \mathcal{L}_a(\mathbf{p}, \mathbf{l}) = BCE(\mathbf{p}, \mathbf{l}),
\end{equation}
where $\mathbf{p} = [ p_1, \ldots, p_{S+1} ]$ and $\mathbf{l} = [ l_1, \ldots, l_{S+1} ]$ are the predicted posterior attractor existence probabilities and references, and BCE is the binary cross entropy. The model is, therefore, trained to mark as valid the first $S$ attractors and the $(S+1)^{th}$ as invalid.

During inference, the estimated number of speakers $\hat{S}$ is obtained as
\begin{equation}
    \hat{S} = \max \{s | s \in \mathbb{Z}_0^+ \ , \ p_s \geq \tau\},
\end{equation}
where $\tau$ is a given threshold, e.g. $0.5$.

Having obtained attractors for all speakers $\textbf{A} = [\mathbf{a}_1, \ldots, \mathbf{a_S}] \in \mathbb{R}^{D \times S}$, and given the frame embeddings for the whole sequence $\textbf{E} = [\mathbf{e}_1, \ldots, \mathbf{e}_T] \in \mathbb{R}^{D \times T}$, the per-frame per-speaker activity posterior probabilities are obtained as
\begin{equation}
    \hat{y}_{t,s} = \sigma(\mathbf{e}_t^\top \mathbf{a}_s).
\end{equation}

The model is trained using the following combined loss
\begin{equation}
    \label{eq:total_loss}
    \mathcal{L} = \mathcal{L}_d + \alpha \mathcal{L}_a,
\end{equation}
with $\alpha$ usually set to $1$. While the original implementation for a fixed number of speakers would compute the loss for all permutations of reference and estimated speakers, this approach does not scale for larger numbers of speakers. Faster implementations have been explored based on the optimal solution of a bipartite graph~\cite{lin2020optimal,landini2023multi}, usually solved with the so-called Hungarian algorithm.

\subsubsection{EEND extensions}
\label{sec:eend_extensions}

Several extensions have been proposed based on the EEND framework. Some works have focused on extensions to the online scenario: limiting the context of the encoder in EEND-EDA~\cite{han2021bw}, using a speaker-tracing buffer for a fixed number of speakers~\cite{xue2021online} and a flexible number of speakers~\cite{xue21d_interspeech}. Others have used EEND in combination with the standard cascaded framework: \cite{horiguchi2021end} used EEND as a post-processing of a clustering-based method to find the overlaps between each pair of speakers. An extension to make use of multiple microphones is presented in \cite{horiguchi2022multi} exploring a replacement of the self-attention layers by spatio-temporal encoders (where self-attention is applied across channels and across frames sequentially) or the alternative co-attention encoders (where across frame self-attention is applied on each channel independently and then multi-head attention using information of all channels is used to condense the information). In \cite{horiguchi2023mutual}, the co-attention alternative is further used in a knowledge distillation scheme~\cite{hinton2014distilling} so a single-channel model and a multi-channel model are used iteratively as student and teacher models to improve their performances.

Conformer~\cite{gulati20_interspeech} was used to replace the self-attention layers of SA-EEND in \cite{liu21j_interspeech} and of EEND-EDA in \cite{leung21_interspeech}. Both works have also replaced the feature subsampling used on the input by convolutional subsampling and \cite{leung21_interspeech} introduced a convolutional upsampling to return more fine-grained outputs. Given the quadratic complexity of attention on the sequence length, a linear approximation was evaluated in the context of EEND-EDA in \cite{izquierdodelalamo22_iberspeech}. 

Given the residual connections in between the self-attention layers in EEND, some works have explored auxiliary losses calculated after each of the layers in the context of SA-EEND~\cite{yu2022auxiliary} and in the context of EEND-EDA~\cite{fujita2023neural} but both in the 2-speaker scenario. 

Using the same framework for training, based on BCE and permutation invariant loss, there have been other works that train end-to-end models such as \cite{maiti2021end} where time-delay convolutional blocks and self-attention layers with a linear attention approximation are combined to provide global context. Two extra losses are explored: one that reinforces diarization in a local context and another that reinforces speaker recognition capabilities in the embeddings. Multi-tasking reinforcing speaker recognition has also been explored in other works~\cite{rybicka2022end,pan2022towards}. EEND has also been aided by features extracted from an ASR model in \cite{khare2022asr} and in a unified framework to perform diarization and separation in \cite{maiti2023eend}.

One of the drawbacks of end-to-end models is the need for a vast amount of data annotated for diarization purposes. Some works~\cite{takashima21_interspeech,coria2023continual} have focused on this problem from the perspective of creating pseudo-labels on non-annotated data and iteratively re-train a seed model on those data. In \cite{cord2023frame}, a pre-trained speaker recognition model is used in the EEND framework to take advantage of the speaker ID labels. First, a student model (same architecture as the pre-trained teacher model) is trained so that per-frame representations resemble those of the teacher obtained over a longer speech segment. The goal is that the student produces speaker embeddings more frequently (and with finer granularity) than the pre-trained speaker embedding extractor. Then, these speaker embeddings with higher resolution are used as input for EEND, replacing the filterbank features.

One of the limitations of EEND-EDA is the LSTM-based encoder-decoder mechanism. In order to obtain better results, the inputs of the LSTM encoder are randomly shuffled, clearly removing the temporal information. \cite{pan2022towards} proposed an alternative where the input of the LSTM encoder is not shuffled and the LSTM decoder incorporates an attention mechanism over the frame embeddings returned by the frame encoder. Instead of using zero vectors as input for the decoder, the input is obtained as a weighted sum of the frame encoder outputs, thus providing the decoder with better cues. A similar idea is explored in \cite{broughton23_interspeech} where the decoder is fed with summary representations calculated together with the embeddings produced by the conformer-based frame encoder.

\begin{figure}[H]
    \centering
    \begin{subfigure}[b]{0.5\textwidth}
        \centering
        \includegraphics[width=\linewidth]{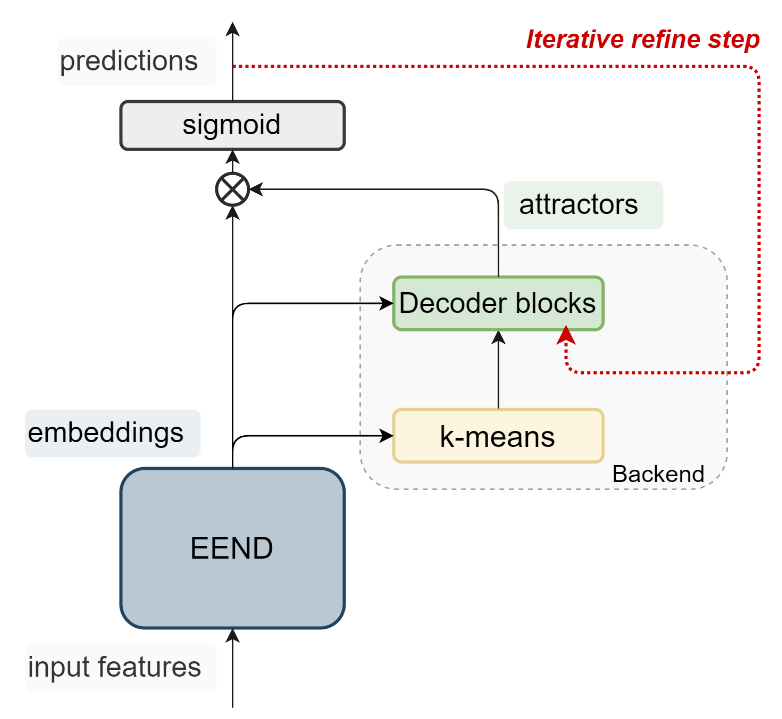}
        \caption{General diagram of the model. The k-means module is used only as initialization.}
        \label{fig:rybicka_diagram} 
    \end{subfigure}%
    ~
    \begin{subfigure}[b]{0.45\textwidth}
        \centering
        \includegraphics[width=\linewidth]{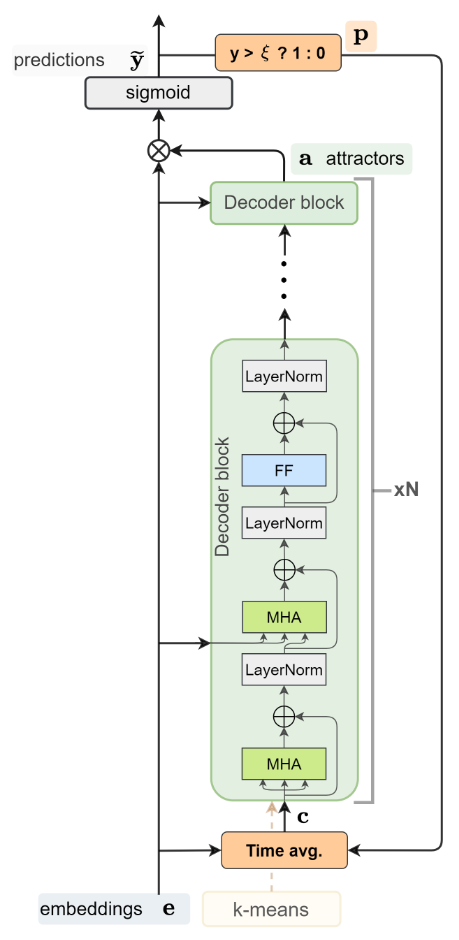}
        \caption{Detailed diagram of the backend block.}
        \label{fig:rybicka_decoder} 
    \end{subfigure}
    \caption{Model diagram for obtaining non-autoregressive attention-based attractors as presented in \cite{rybicka2022end}. Figure adopted from \cite{rybicka2022end}.}
    \label{fig:rybicka}
\end{figure}

\paragraph{Non-autoregressive attractor estimation}
\label{sec:nonautoregressive}
Given the limitations of the LSTM-based encoder and decoder to obtain attractors, some works have explored non-autoregressive approaches for obtaining attractors with attention-based schemes.
In this subsection, we cover in more detail previous works in this line since they are more relevant to our model, presented in Section~\ref{sec:e2e_diaper}.

The first of these works \cite{rybicka2022end}, replaced the LSTM-based encoder-decoder by two layers of cross-attention decoder as shown in Figure~\ref{fig:rybicka}. In this work, initial attractors are estimated using k-means on the frame embeddings to cluster them to the number of speakers in the recording (e.g. 2). These initial attractors are then iteratively transformed using multi-head cross-attention blocks where the frame embeddings serve as keys and values and the attractors serve as queries. The queries of the first block are obtained as the average of frame embeddings weighted by predicted posterior activities and the following blocks simply refine them attending to the frame embeddings. It is shown that this method can improve by performing a few refinement iterations (i.e. by predicting activities in one iteration and using them to initialize the queries in the next iteration).

\begin{figure}[H]
    \centering
    \includegraphics[width=0.8\textwidth]{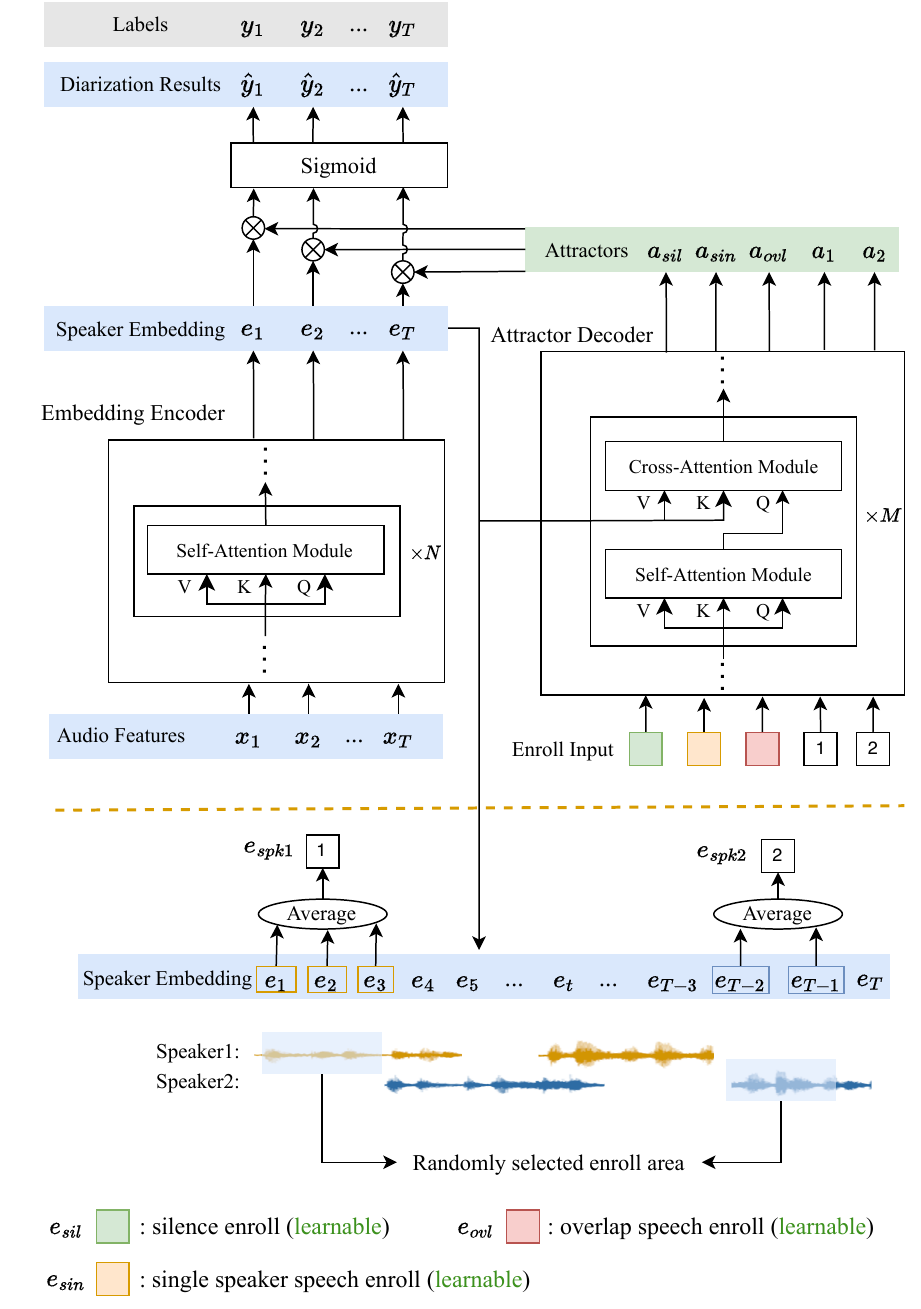}
    \caption{Model diagram for attention-based target-speaker attractors as presented in \cite{chen23n_interspeech}. Note that the figure states that the segment is selected randomly but other options are possible. Figure adopted from \cite{chen23n_interspeech}.}
    \label{fig:att_eda}
\end{figure}

In \cite{fujita2023neural}, the LSTM-based encoder-decoder is also replaced by a cross-attention decoder; however, in this case, the set of initial queries (which are transformed into attractors) is not defined by the output of the model but they are learnable parameters.

\cite{chen23n_interspeech} combine the former approaches two as shown in Figure~\ref{fig:att_eda}. The LSTM-based encoder-decoder is also replaced by layers of cross-attention decoder and three of the initial queries are fixed (but learned during training): they represent ``silence'', ``single speaker'' and ``overlap'' while the other $S$ queries represent each of the speakers in the recording. In the first pass, only the fixed queries are used and then the speaker queries are estimated iteratively one at a time from the frame embeddings. Different strategies are considered for defining the speaker queries from frames labeled as ``single speaker'': selecting the first segment of speech, selecting the segment randomly, and using spectral clustering on the frame embeddings and picking a segment of the majoritarian cluster. Of these, spectral clustering provides the best results. The weighted average of the selected frames is computed with weights given by the predicted posterior activities to produce the new initial speaker query. The new set of queries is refined through a few cross-attention layers in order to produce the final attractors. These attractors are used to obtain the speech activity posteriors similar to EEND-EDA. This procedure is repeated for each speaker, one at a time until all ``single speaker'' segments are consumed by some attractor. It should be noted that the inference procedure with this method is more complicated than in the original EEND-EDA due to the iterative procedure to estimate first silence, single speaker and overlap attractors and then each of the speakers iteratively.

\subsection{Two-stage models}
\label{sec:e2e_twostage}

EEND models usually have problems handling long recordings or recordings with many speakers. Two-stage models allow for taking advantage of the good performance of EEND to handle silence and overlaps in short segments by using it as first stage. Then, the second stage model can handle more speakers and joins the information from short segments of a potentially very long recording. Three lines of work can be distinguished regarding the second stage method and they are described next.

\subsubsection{Neural- and clustering-based models}
This line combines end-to-end models (used on chunks of a few seconds each) and clustering (to find the inter-chunk correspondence between speakers). These approaches have the advantage of dealing with overlapped speech and have good granularity in the decisions (given by the end-to-end part). Furthermore, they can deal with long recordings as well as be able to better estimate the number of speakers (given by the clustering part).

It was in \cite{kinoshita2021integrating,kinoshita21_interspeech} where EEND was run for the first time in chunks to produce diarization outputs and then the decisions reconciled by clustering in a method coined EEND+vector clustering (EEND-VC). Together with the speech activities, speaker representations for each of the speakers in the chunk were produced by means of a weighted average of frame-level embeddings using the speech activities as weights. Then, these speaker representations were used to find the inter-chunk agreement between speaker labels, imposing the restriction that different labels in a given chunk are not merged. Furthermore, speaker representations were encouraged to have small intra-speaker and large inter-speaker distances. Even though the EEND part is limited in terms of the number of speakers it can handle, if the chunks are short enough, the number of speakers is in most cases bounded and the performance is not largely affected. The clustering step was performed with constrained k-means or AHC in \cite{kinoshita2021integrating,kinoshita21_interspeech} respectively.

An extension was proposed in \cite{kinoshita2022tight} where the clustering was performed with an infinite Gaussian mixture model where the parameters were estimated with variational Bayes inference, allowing for better performance on recordings with higher numbers of speakers. 
In \cite{kinoshita22_interspeech}, the focus was on how to better split recordings. They detected the beginnings of utterances rather than arbitrarily separating the recording into chunks and used Graph-PIT~\cite{neumann21_interspeech} to handle overlapping segments.
More recently in~\cite{delcroix23_interspeech}, the VBx framework was used for clustering where the HMM states did not only consider one speaker but all possible pairs of speakers, in order to handle overlaps given by the EEND part of the system.

EEND+clustering was explored for the online scenario in ~\cite{coria2021overlap}. The most recent recipe of the pyannote toolkit~\cite{bredinpyannote} is also based on this framework. EEND is applied on 5\,s chunks (where, in practice, more than three speakers almost never appear) but at a resolution of 0.016\,s, instead of the usual 0.1\,s used in EEND systems, thus allowing for more accurate time boundaries. Using single-speaker audio (given by the diarization output in the chunk) one speaker embedding per speaker is extracted using ECAPA-TDNN~\cite{desplanques20_interspeech}. The speaker embeddings of different chunks are clustered by means of AHC and the local diarization decisions are combined to provide whole-utterance diarization annotations.

The EEND framework was utilized in both stages in \cite{horiguchi2021towards} where a variant of EEND-EDA was used in a global and a local context. In this model, EEND-EDA was first run on 5\,s chunks to produce diarization outputs as well as the corresponding attractors. Then, to find the inter-chunk agreement between the local attractors, they were passed through a transformer decoder where they functioned as queries and the frame embeddings were keys and values. This aimed to convert all local attractors into global attractors which were then clustered by means of spectral clustering with a k-means variant that accounted for constraints. An online version of this approach was presented in \cite{horiguchi2022online}.

\subsubsection{Neural- and TS-VAD-based models}

\cite{wang2022incorporating} presented an architecture where a ResNet-based model was pre-trained for speaker recognition and used to produce per-frame features consumed by both EEND and TS-VAD. The EEND part of the model was used to produce the diarization initialization normally required by TS-VAD. Using the activities returned by the EEND module as weights, the per-frame embeddings were weight-averaged to define the speaker embeddings needed by TS-VAD.

One of the most successful approaches up to date in this line was presented in~\cite{wang2023told}. The first stage consists of EEND-EDA with the addition of an LSTM after the dot-product between frame embeddings and attractors to make use of contextual information of the speakers activities at the frame level. Instead of treating each speaker independently, as usually done in EEND, they use power set encoding (PSE) labels to model the output. PSE reinforces the dependencies between speakers by modeling up to three simultaneous (overlapping) speakers. Given the diarization output provided by the first stage, x-vectors are extracted with a pre-trained ResNet-based model. These x-vectors (one per speaker) are obtained using the frames where the speaker was active according to the first stage. The x-vectors, functioning as speaker embeddings are in turn compared with per-frame embeddings of high resolution to produce the final outputs. Unlike in TS-VAD, the comparison between frame embeddings and speaker profiles is not done with BLSTMs. Instead, they utilize two options to obtain scores for each speaker and each frame. The first option is to use the concatenated speaker and frame embeddings transformed with feed-forward layers independently of each other. In the second option, the speaker embedding is concatenated to all frame embeddings and the sequence of extended frames is transformed with self-attention layers to provide context-dependent activities for the given speaker. The similarity scores given by these two subsystems for all frames and speakers are then processed with an LSTM to deal with context and produce the final PSE classes.

\subsection{Other}
Some models do not fall into the categories above but rather present a combination of them or an alternative approach.

In \cite{wang2023target}, a model that combines ideas from EEND-EDA and TS-VAD was presented. The dot-product that usually compares frame embeddings and attractors in EEND-EDA was replaced by an approach similar to TS-VAD. Frame embeddings and attractors were stacked and run through the speaker activity detection module. Originally BLSTM-based in~\cite{medennikov2020target}, the speaker activity module was replaced (similarly to~\cite{cheng2023target}) by a transformer-based architecture and enhanced by considering joint-speaker-detection rather than treating each speaker independently.

DIVE~\cite{zeghidour2021dive} presented an end-to-end approach that iteratively finds the speakers in a recording, similarly to~\cite{fujita2020neural}. In \cite{fujita2020neural}, each speaker activity is obtained with an LSTM-based decoder and the previous speakers' activities are given to the next LSTM pass. Given the limitations of LSTMs to model long sequences, DIVE (independently) classifies each frame into one of four classes: ``single novel speaker is active'', ``single already selected speaker is active'', ``overlapped speech'' or ``silence''. This approach decodes activities for each speaker one at a time iteratively. The classifier consists of linear layers followed by nonlinearities and it receives as input the frame embedding, the speaker vector corresponding to the current speaker and a representation for each of the previously selected speakers. Unlike LSTMs, this approach allows for better handling of long sequences. Results for two speaker conversations are remarkable but no analysis has been presented for more speakers.

Another type of models are those focused on speaker-attributed ASR. While these works do not directly tackle the task of speaker diarization, they are closely related.
\begin{itemize}
    \item Serialized output training (SOT)~\cite{kanda20b_interspeech} models multi-speaker ASR. Speech tokens (i.e. words or sub-words) of different speakers are sorted by time-stamp and ``speaker-change'' tokens are introduced in between to denote that the tokens correspond to different speakers. Given the serialized ASR labels including the new ``speaker-change'' token, an end-to-end ASR model is trained as usual. This way, each word can be assigned a speaker label, providing word-level diarization. Several works~\cite{kanda20_interspeech,kanda21b_interspeech,kanda22b_interspeech} have extended this approach using a speaker encoder and a decoder to produce speaker embeddings associated to the tokens which are clustered to produce token-level diarization.
    \item Turn-to-diarize~\cite{xia2022turn} presents a similar idea where an ASR model is trained to also predict ``speaker-change'' tokens. Given the timestamps of these change tokens, a single d-vector is extracted for each of the segments. Spectral clustering is then used to assign speaker labels to each of the segments. Unlike the SOT works, providing timestamps implies that this model effectively produces diarization outputs as previously described methods.
\end{itemize}

A specific category of its own is formed by the models in the SUPERB benchmark~\cite{yang21c_interspeech}. Given the advent of self-supervisedly trained models for generating speech representations, such as wav2vec 2.0~\cite{baevski2020wav2vec}, HuBERT~\cite{hsu2021hubert} or WavLM~\cite{chen2022wavlm}, SUPERB presents a standard framework for comparison. Among the many speech-related tasks, speaker diarization is included. SUPERB focuses on the 2-speaker scenario and the model is a simple single LSTM layer. It should be noted that in this framework, the self-supervised model is normally frozen and only a rather simple task-specific model is trained.

Although many works on speech-related tasks have used models like wav2vec 2.0 or HuBERT as feature extractors (without following the SUPERB framework), this path has not been widely explored for diarization. The only work in this direction is based on WavLM~\cite{chen2022wavlm}, where the representations are processed with EEND-VC to produce remarkable results. On diarization-related tasks, in \cite{kunevsova2023multitask}, simple classifiers are trained using the features from wav2vec 2.0 for speaker change detection, voice activity detection and overlapped speech detection.

\subsection{Summary}

In the last few years, a large bulk of work in the diarization field has focused on end-to-end models, mainly under the EEND (fully end-to-end systems) TS-VAD (non-fully end-to-end systems) frameworks. The latter leverages speaker embedding extractors, deeply studied in the context of speaker recognition and even partially for diarization. However, their nature partly resembles modular systems and brings back their disadvantages such as having to run multiple modules. 

EEND systems, on the other hand, propose a single system for the task but they struggle to find the correct number of speakers when there are many of them, and face problems with long recordings. Furthermore, they require large amounts of training data, which has to be generated artificially, and do not perform as well on wide-band data as they do on telephone conversations. Thus, good quality training data plays a major role in the performance of these systems. Nevertheless, the research in this area has been limited. Section~\ref{sec:e2e_data} focuses on this aspect of EEND models.

With the superior performance of attention-based models over recurrent ones in several speech tasks, it is surprising that EEND-EDA is still one of the most sought-after methods for handling multiple speakers in a fully end-to-end system. Recent works have focused on non-autoregressive variants for attractor calculation but are mostly limited to only two speakers or using iterative procedures. Furthermore, all of them follow a common pattern. Section~\ref{sec:e2e_diaper} describes a model that expands on these ideas.

\section{Synthetic training data generation}\label{sec:e2e_data}
End-to-end diarization models have shown remarkable improvements over cascaded ones in certain scenarios. However, one of the disadvantages of end-to-end models is that they require large amounts of training data while manually annotated data for diarization are notoriously lacking. 

When presenting the first version of EEND, \cite{fujita19_interspeech} proposed a strategy for constructing simulated mixtures using telephone conversations from different collections and this strategy has been used (with both telephony data or read books) to create simulated mixtures by mixing speakers from different recordings. However, little analysis was presented about how the simulations were devised nor what impact the used data augmentations have. Furthermore, the mixtures do not resemble real conversations, especially when more than two speakers are included. 

This chapter describes the original approach~\cite{fujita19_interspeech} for creating

training data to which, respecting the original terminology, we will refer to as ``simulated mixtures'' (SM). Next, we describe our approach, denoted ``simulated conversations'' (SC), which was specifically developed to better resemble real conversations. The proposal of SC represents one of the contributions of the thesis. Not only we showed improvements over SM but other works have adopted the approach and shown that it is an important step to reach state-of-the-art results.

\begin{figure}[ht]
    \centering
    \includegraphics[width=\textwidth]{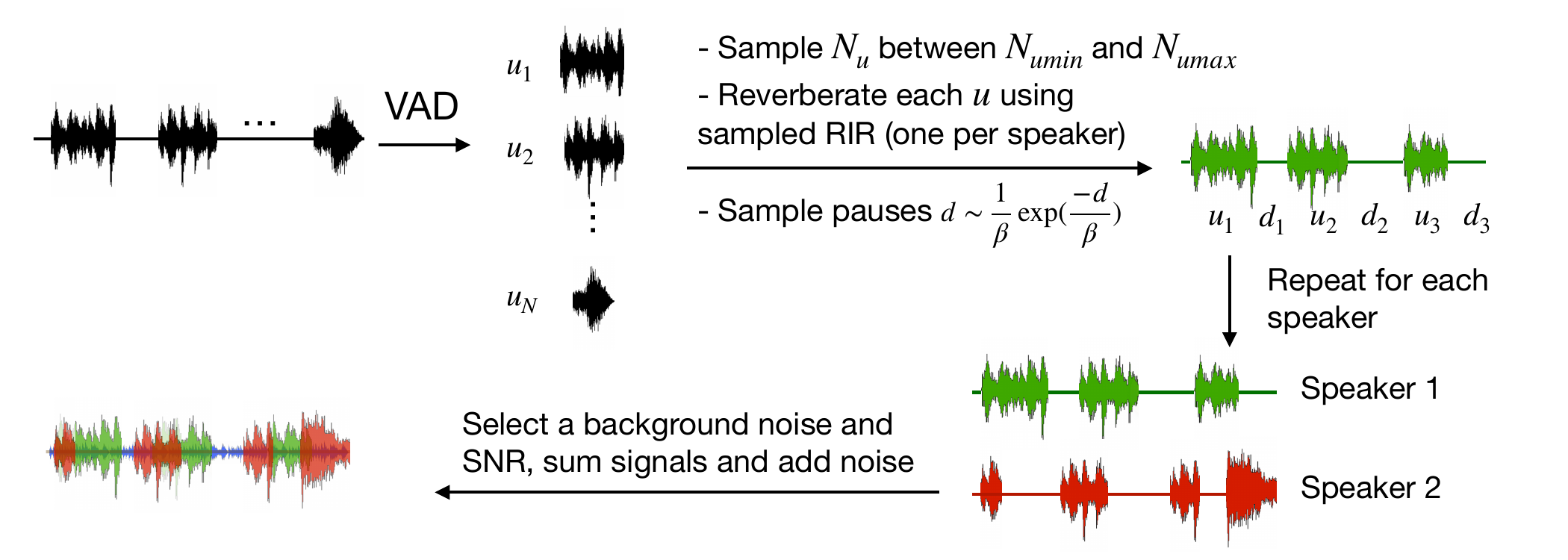}
    \caption{Procedure for generating simulated mixtures.}
    \label{fig:SM}
\end{figure}

\subsection{From simulated mixtures \ldots}
Diarization training data consist of audio recordings and their corresponding speaker segment annotations. In order to create the mixtures with their corresponding annotations, \cite{fujita19_interspeech} have used a VAD system run on each side of each conversation in a pool of thousands of telephone conversations. Assuming a single speaker per telephone channel, this means that speech segments and their speaker labels can be gathered from the pool to construct mixtures.

To create a mixture, as many speakers as wanted in the mixture ($N_{spk}$) are sampled from the total pool. The pool is represented by $\mathcal{U} = \{ U_s \}_{s \in \mathcal{S}}$ where $U_s$ is an utterance of speaker $s$ formed by all the segments denoted by the VAD that belong to that speaker in an original utterance\footnote{Note that \cite{fujita19_interspeech} use ``utterances'' to refer to each of the segments of speech in a real conversation.}. For each selected speaker, one of their utterances is randomly sampled and $N_u$ consecutive segments are selected randomly from it. 
$N_u$ is a random number in between $N_{umin}$ and $N_{umax}$ (parameters of the mixture creation procedure). 
Pauses are introduced in between the selected segments of a speaker to simulate turns in a conversation, where the length of the pause is sampled from an exponential distribution with parameter $\beta$. The resulting audio defines a channel for that speaker in the mixture. This process is repeated for all speakers in the mixture and finally, the channels are summed to obtain a single utterance. This procedure is enriched by adding background noises and reverberating each speaker channel with a room impulse response (RIR) randomly sampled from a set of predefined rooms. The whole procedure is depicted in Figure~\ref{fig:SM} and described in Algorithm~\ref{alg:mixture}. Note that since the channel of each speaker is generated independently, depending on the choice of $\beta$, the output can contain a high percentage of overlapped speech (usually higher than in real conversations). 
This is even more accentuated when the number of speakers is more than two, i.e. long overlaps between three or more speakers are not uncommon in the simulated mixtures while they rarely occur in real conversations.

\begin{algorithm}
\caption{Mixture simulation (from \cite{fujita19_interspeech})}\label{alg:mixture}
\hspace*{\algorithmicindent} \textbf{Input:} $\mathcal{S}$, $\mathcal{N}$, $\mathcal{I}$, $\mathcal{R}$ \Comment{\footnotesize{Set of speakers, noises, RIRs, and SNRs}} \\
\hspace*{\algorithmicindent} \hspace*{\algorithmicindent} $\mathcal{U} = \{ U_s \}_{s \in \mathcal{S}}$ \Comment{\footnotesize{Set of utterances}} \\
\hspace*{\algorithmicindent} \hspace*{\algorithmicindent} $N_{\text{spk}}$ \Comment{\footnotesize{\#speakers per mixture}} \\
\hspace*{\algorithmicindent} \hspace*{\algorithmicindent} $N_{\text{umax}}$, $N_{\text{umin}}$ \Comment{\footnotesize{Maximum and minimum \#segments per speaker}} \\
\hspace*{\algorithmicindent} \hspace*{\algorithmicindent} $\beta$ \Comment{\footnotesize{Average interval}} \\
\hspace*{\algorithmicindent} \textbf{Output:} $\mathbf{y}$ \Comment{\footnotesize{Mixture}} 
\begin{algorithmic}
    \State Sample a set of $N_{\text{spk}}$ speakers $\mathcal{S}'$ from $\mathcal{S}$ 
    \State $\mathcal{X} \gets \emptyset$ \Comment{\footnotesize{Set of $N_{\text{spk}}$ speakers' signals}}
    \ForAll{$s \in \mathcal{S}'$}
        \State $\mathbf{x}_s \gets 0s$ \Comment{\footnotesize{Channel for speaker $s$}}
        \State Sample $\mathbf{i}$ from $\mathcal{I}$ \Comment{Select RIR}
        \State Sample $N_u$ from $\{N_{\text{umin}}, \ldots, N_{\text{umax}} \}$
        \State $G \gets getNConsecutiveSegments(U_s, N_u)$ \Comment{Randomly}
        \State $pos \gets 0$
        \For{$u = 1$ to $N_u$}
            \State Sample $d \sim \frac{1}{\beta} \exp{\Big(-\frac{d}{\beta}\Big)}$ \Comment{\footnotesize{Sample pause length}}
            \State $pos \gets pos + d$
            \State $\mathbf{x}_s.addFromPosition(pos, G[u] * \mathbf{i})$ \Comment{Insert segment}
        \EndFor
        \State $\mathcal{X}$.add($\mathbf{x}_s$)
    \EndFor
    \State $L_{max} = \max_{\mathbf{x} \in \mathcal{X}} |\mathbf{x}|$
    \State $\mathbf{y} \gets \sum_{\mathbf{x} \in \mathcal{X}} \Big( \mathbf{x} \oplus \mathbf{0}^{L_{max} - |\mathbf{x}|} \Big)$ \Comment{Fill with 0's to match lengths}
    \State Sample $\mathbf{n}$ from $\mathcal{N}$ \Comment{Background noise}
    \State Sample $r$ from $\mathcal{R}$ \Comment{SNR}
    \State Determine a mixing scale $p$ from $r$, $\mathbf{y}$, and $\mathbf{n}$
    \State $\mathbf{n}' \gets$ repeat $\mathbf{n}$ until reaching the length of $\mathbf{y}$
    \State $\mathbf{y} \gets \mathbf{y} + p \cdot \mathbf{n}'$
\end{algorithmic}
\end{algorithm}

\subsection{\ldots to simulated conversations}

One of the main concerns with SM is that each speaker in the mixture is treated independently. Although the lengths of the pauses are randomly drawn from an exponential distribution, a sensible choice as it represents the inter-arrival times between independent events (the speech segments of the speaker), this does not resemble the dynamics of a real conversation where speakers do not take turns independently but collaboratively. For this reason, the proposed approach to create SC~\cite{landini22_interspeech} uses the following statistics on frequencies, lengths of pauses and overlaps from real conversations:

\begin{itemize}
    \item Number of times that two consecutive segments are of the same speaker and separated by a pause and a histogram defining $D_{=\text{speaker}}$, the distribution of pause lengths between segments of the same speaker.
    \item Number of times $ds$ that two consecutive segments are of different speakers and separated by a pause and a histogram defining $D_{\neq\text{speaker}}$, the distribution of pause lengths between segments of different speakers.
    \item Number of times $ov$ that two consecutive segments overlap\footnote{If annotations are correct, they are always from different speakers.} and a histogram defining $D_{\text{overlap}}$, the distribution of overlap lengths.
    
    \item $p = \frac{ds}{ds + ov}$ is the probability of having a pause between two segments of different speakers. Note that the probability of having overlap between two segments of different speakers is $1-p$.
\end{itemize}

\begin{algorithm}
\caption{Conversation simulation}\label{alg:conversation}
\hspace*{\algorithmicindent} \textbf{Input:} $\mathcal{S}$, $\mathcal{N}$, $\mathcal{I}$, $\mathcal{R}$ \Comment{\footnotesize{Set of speakers, noises, RIRs, and SNRs}} \\
\hspace*{\algorithmicindent} \hspace*{\algorithmicindent} $\mathcal{U} = \{ U_s \}_{s \in \mathcal{S}}$ \Comment{\footnotesize{Set of utterances}} \\
\hspace*{\algorithmicindent} \hspace*{\algorithmicindent} $N_{\text{spk}}$ \Comment{\footnotesize{\#speakers per conversation}} \\
\hspace*{\algorithmicindent} \hspace*{\algorithmicindent} $p, D_{=\text{speaker}}, D_{\neq\text{speaker}}, D_{\text{overlap}}$ \Comment{Estimated distributions}
\begin{algorithmic}
\State $G \gets \{\}$ \Comment{\footnotesize{Dictionary with list of segments per speaker}}
\State Sample a set of $N_{\text{spk}}$ speakers $\mathcal{S}'$ from $\mathcal{S}$ 
\ForAll{$s \in \mathcal{S}'$}
    \State Sample $u$ from $U_s$ without replacement
    \State Sample $\mathbf{i}$ from $\mathcal{I}$ \Comment{RIR}
    \State $u' \gets u * \mathbf{i}$ \Comment{Reverberate all segments in the utterance}
    \State $G[s] \gets u'$
\EndFor
\State $L \gets$ Randomly interleave $G$ into a single list
\State $\mathbf{y}.addFromPosition(0, L[1]$) \Comment{Start signal with first segment}
\State $pos \gets \text{len}(L[1])$
\For{$t = 2$ to $\text{len}(L)$}
    \If{Speaker$(L[t-1]) = $ Speaker$(L[t])$}
        \State Sample $gap$ from $D_{=\text{speaker}}$
    \ElsIf{Sample of $Bernoulli(p)$ is 1}
        \State Sample $gap$ from $D_{\neq\text{speaker}}$
    \Else
        \State Sample $-gap$ from $D_{\text{overlap}}$
    \EndIf
    \State $\mathbf{y}.addFromPosition(pos + gap, L[t])$
    \State $pos \gets pos + gap + \text{len}(L[t])$
\EndFor
\State Sample $\mathbf{n}$ from $\mathcal{N}$ \Comment{Background noise}
\State Sample $r$ from $\mathcal{R}$ \Comment{SNR}
\State Determine a mixing scale $p$ from $r$, $\mathbf{y}$, and $\mathbf{n}$
\State $\mathbf{n}' \gets$ repeat $\mathbf{n}$ until reaching the length of $\mathbf{y}$
\State $\mathbf{y} \gets \mathbf{y} + p \cdot \mathbf{n}'$
\end{algorithmic}
\end{algorithm}

The proposed approach is described in Algorithm~\ref{alg:conversation} and depicted in Figure~\ref{fig:SC}. In this case, utterances are sampled without replacement. When creating a large set of simulated conversations, this ensures that an utterance is not used more than once\footnote{In practice, the code is prepared to run until exhausting all utterances and re-start again for a specific number of times or until generating a specific amount of audio.}. Furthermore, for a given utterance all segments are used as part of a single SC. In contrast, in the original approach, only a subset of the segments from each original conversation is used at a time. 

The segments (defined by the VAD labels) of the selected utterances (one per speaker) are randomly interleaved as shown in Figure~\ref{fig:SC_shuffle}. The speaker labels of all segments of all speakers are listed and shuffled. Then, for each label, the segments of the corresponding speaker are picked in order. This guarantees that the per-speaker order is kept while assigning random order to the speaker turns of the different speakers. Then, for each pair of consecutive segments after interleaving them, a gap is defined depending on the nature of the two segments as shown in Algorithm~\ref{alg:conversation}. The lengths of pauses or overlaps are sampled from the estimated distributions. In the pseudo-code, there is a single channel initialized with $0$'s and segments are added to it in the obtained order by means of $out.addFromPosition(pos, in)$ which adds the signal $in$ onto $out$ starting from position $pos$.

\begin{figure}[ht]
    \centering
    \includegraphics[width=\textwidth]{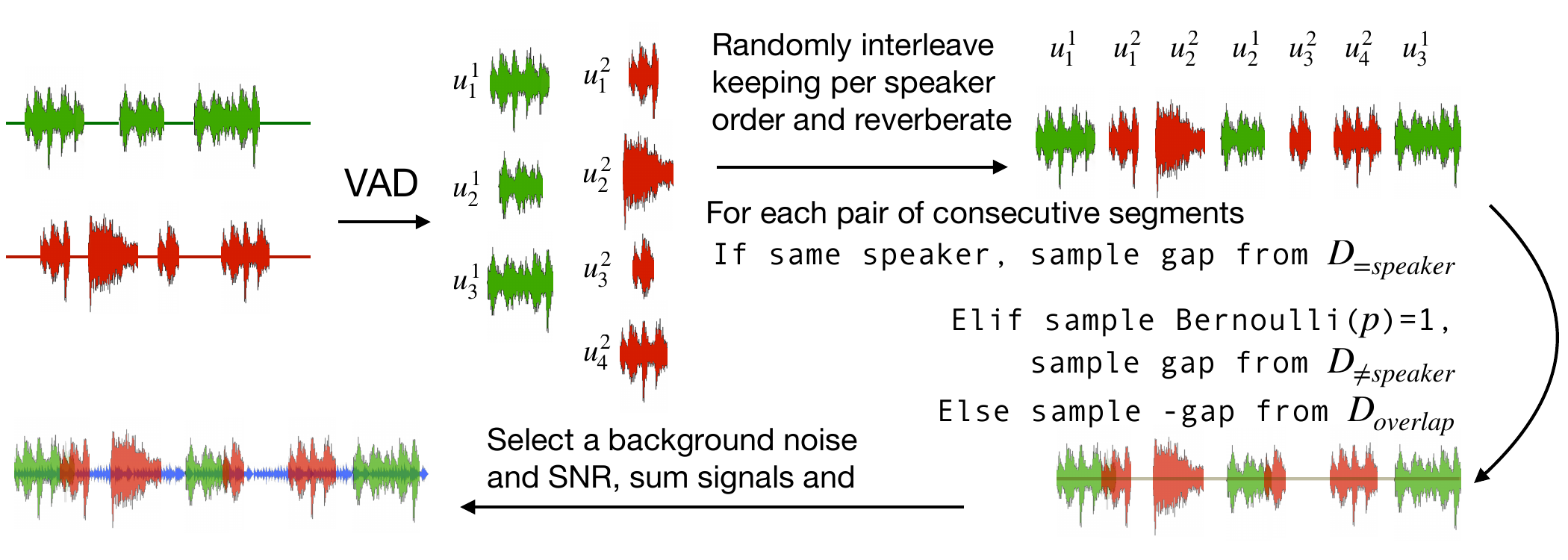}
    \caption{Procedure for generating simulated conversations.}
    \label{fig:SC}
\end{figure}

\begin{figure}[ht]
    \centering
    \includegraphics[width=\textwidth]{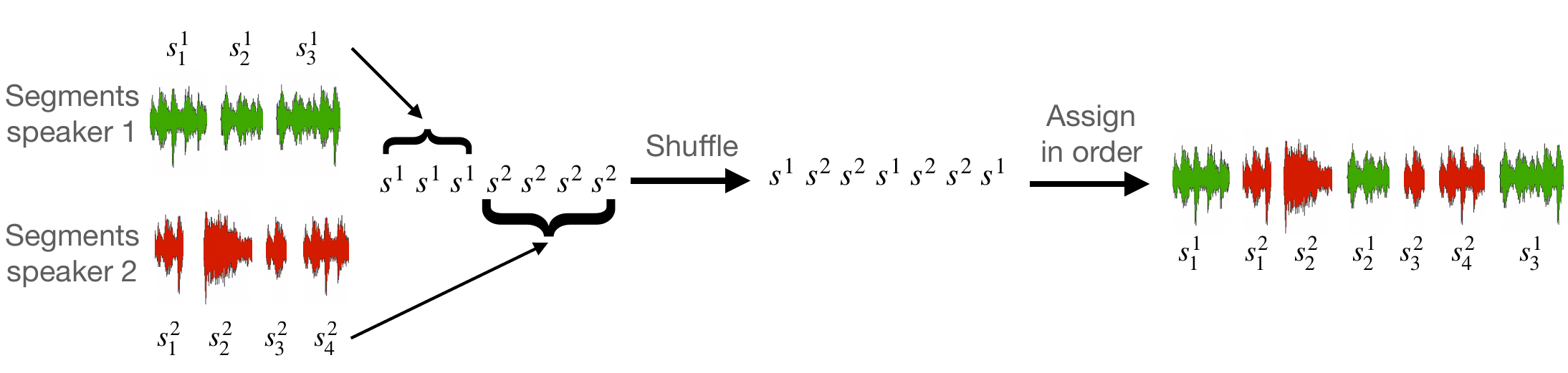}
    \caption{Procedure for interleaving segments when generating simulated conversations.}
    \label{fig:SC_shuffle}
\end{figure}

It should be noted that an alternative method was devised concurrently in \cite{yamashita22_odyssey} where the turn-taking categories (turn-hold, turn-switch, interruption, backchannel) were taken into account to define simulated data. This approach also considers the interactions between speakers for defining the turns so that a Markov process expresses the transitions and the probabilities of each type of transition are estimated on real conversational data.
Although the method has also a theoretical background and a more complex scheme, the results are not better than those obtained with our proposed approach. Furthermore, the analysis is only focused on the two-speaker telephone scenario.

\subsection{SM versus SC}
\label{sec:sm_vs_sc}

For the sake of the analysis, we consider the addition of background noises and reverberation (already present in~\cite{fujita19_interspeech}) as augmentations. For analyzing the effect of using or not using each of these augmentations, different training datasets are generated for each setup. The same EEND model is trained on each one independently and its performance is evaluated on real data. 
Besides those two (already present in the original approach), we explore a third one which aims to make the proportions in energy levels between speakers of a created conversation to resemble those seen in real conversations. This augmentation estimates the energy levels of all utterances to be used in a single conversation, takes as reference the maximum of those and scales the remaining following a ratio sampled from energy relations estimated on real conversations. However, this approach did not impact the performance.

For the sake of making the comparison between the two approaches (SM and SC) as fair as possible, several aspects of the data preparation follow those of the original SM approach~\cite{fujita19_interspeech}:

\begin{itemize}
    \item The set of utterances used: comprises Switchboard-2 (phases I, II, III)~\cite{graff1998switchboard,graff1999switchboard,graff2002switchboard}, Switchboard Cellular (parts 1 and 2)~\cite{graff2001switchboard,graff2004switchboard}, and NIST Speaker Recognition Evaluation datasets (from years 2004, 2005, 2006, 2008)~\cite{nist20112006test,nist20122006test,nist20112008train,nist20112008test,nist20112005test,nist20112006train,nist20062004,nist20112005}. All the recordings are sampled at 8\,kHz and, out of 6381 speakers, 90\% are used for creating training data.
    \item The VAD used to obtain time annotations is based on time-delay neural networks and statistical pooling\footnote{\url{http://kaldi-asr.org/models/m4}}.
    \item The set of background noises and mechanism for augmenting data: 37 noises labeled as ``background'' in the MUSAN collection~\cite{snyder2015musan} are added to the signal scaled with a signal-to-noise-ratio (SNR) selected randomly from \{5, 10, 15, 20\} dB.
    \item The set of RIRs used to reverberate utterances: an RIR is sampled from the collection introduced in~\cite{ko2017study} and used to reverberate the utterances of each speaker with 0.5 probability.
\end{itemize}

In order to shed some light on how the proposed SCs resemble real conversations more than SMs, some statistics about the recordings are presented in Table~\ref{tab:datasets_stats} where we see that SC is more similar to real sets in terms of percentages of silence, speech from a single speaker and overlap. We selected a subset of the SC to match the number of hours used with SM in previous works. The code for generating SC can be found in \url{https://github.com/BUTSpeechFIT/EEND\_dataprep}.

\begin{table}[ht]
\centering
\caption{Statistics for synthetic and real datasets. All listed sets have only 2 speakers per recording.}
\label{tab:datasets_stats}
\setlength{\tabcolsep}{3pt} 
  \begin{tabular}{@{}
                  l
                  S[table-format=6.0]
                  S[table-format=3.2]
                  S[table-format=4.2]|
                  S[table-format=2.2]
                  S[table-format=2.2]
                  S[table-format=2.2]
                  @{}}
  \toprule
          & 
         & 
        \multicolumn{1}{c}{\begin{tabular}{@{}c@{}}Average\end{tabular}} &
        \multicolumn{1}{c}{\begin{tabular}{@{}c@{}}Total \end{tabular}} &
        \multicolumn{3}{|c}{Average (\%)} \\
        Dataset       & 
        \multicolumn{1}{c}{\#Files} & 
        \multicolumn{1}{c}{\begin{tabular}{@{}c@{}}duration (s)\end{tabular}} &
        \multicolumn{1}{c}{\begin{tabular}{@{}c@{}}audio (h)\end{tabular}} &
        \multicolumn{1}{|c}{\begin{tabular}{@{}c@{}}Silence\end{tabular}} & 
        \multicolumn{1}{c}{\begin{tabular}{@{}c@{}}1-speaker\end{tabular}} &
        \multicolumn{1}{c}{\begin{tabular}{@{}c@{}}Overlap\end{tabular}} \\
        
  \midrule
  SM (train $\beta=2$) & 100000 & 89.30 & 2480.62 & 18.61 & 49.62 & 31.77 \\ 
  SC (train) & 25074 & 356.15 & 2480.56 & 12.80 & 78.83 & 8.37 \\
  \midrule
  Callhome Part 1 (2 speakers) & 155 & 74.02 & 3.19 & 9.05 & 77.90 & 13.05 \\
  Callhome Part 2 (2 speakers) & 148 & 72.14 & 2.97 & 9.84 & 78.34 & 11.82 \\
  DIHARD3 full CTS dev & 61 & 599.95 & 10.17 & 10.56 & 77.27 & 12.17 \\
  DIHARD3 full CTS eval & 61 & 599.95 & 10.17 & 10.89 & 78.62 & 10.49 \\
   \bottomrule
  \end{tabular}
\end{table}

\subsubsection{Comparison for two-speaker conversations}
\label{sec:2spkexp}

The experiments were performed using the self-attention EEND with encoder-decoder attractors for showing superior performance in previous works. It should be noted that for experiments with 2 speakers per recording, one could use SA-EEND but we use EEND-EDA (Section~\ref{sec:e2e_EEND_EDA}) to be consistent with the analyses in the next sections dealing with variable numbers of speakers. In all cases, the architecture used was exactly the same as that described in~\cite{horiguchi20_interspeech}. For the sake of making the code more efficient, we used our PyTorch implementation\footnote{\url{https://github.com/BUTSpeechFIT/EEND}}. 15 consecutive 23-dimensional log-scaled Mel-filterbanks (computed over 25\,ms every 10\,ms) are stacked to produce 345-dimensional features every 100\,ms. These are transformed by 4 self-attention encoder blocks (with 4 attention heads each) into a sequence of 256-dimensional embeddings. These are then shuffled in time and fed into the LSTM-based encoder-decoder module which decodes as many attractors as speakers are predicted (2 in these experiments). The speech activity probabilities for each speaker (represented by an attractor) at each time (represented by an embedding) are obtained by thresholding the output of the sigmoid function applied to the product of the attractors and frames. 

During inference time, per-speaker per-time predicted probabilities $\hat{Y}$ are thresholded at 0.5 to determine speech activities. To ease the comparison, this parameter was not tuned in any experiment but tuning it could lead to better results. 
Each training was run for 100 epochs on a single GPU. The batch size was set to 64 or 32 with 100000 or 200000 minibatch updates of warm up respectively. Following~\cite{horiguchi20_interspeech}, the Adam optimizer~\cite{kingma2014adam} was used and scheduled with noam~\cite{vaswani2017attention}. 

Since there can be a mismatch between the simulated training data and the real application data, fine-tuning (FT) is performed on a small set of in-domain data. For this step, 100 epochs were run and the Adam optimizer was used with learning rate $10^{-5}$. For the inference as well as for obtaining the model from which to fine-tune, the models from the last 10 epochs were averaged. During training and fine-tuning phases, batches are formed by sequences of stacked 500 Mel-filterbank outputs, corresponding to 50\,s. During inference, the full recordings are fed to the network one at a time. 

Diarization performance is evaluated in terms of diarization error rate as defined in Section~\ref{sec:metrics}.

\paragraph{SC augmentations analysis}

We evaluated all combinations of the augmentations, i.e. reverberating and adding noises. For this analysis, statistics to generate SC were estimated on DIHARD3 CTS development. Figure~\ref{fig:v1_callhome_part2_2spk} presents results on recordings with 2 speakers from Callhome Part 2 comparing the best-published result with the original Chainer implementation~\cite{horiguchi20_interspeech}, our runs with the Chainer implementation, our PyTorch implementation with SM and the PyTorch implementation with the different SC options. Our runs with Chainer correspond with the result in~\cite{horiguchi20_interspeech}. 
The results with PyTorch are slightly worse, note that there might be small implementation differences between Chainer and PyTorch. Also, the same hyper-parameters tuned for training with Chainer were used with the PyTorch implementation to ease the comparison and adjusting them could lead to further improvement.

Using any of the SC options for generating training data outperforms SM. Moreover, the performance of the models trained using the best SC option (SC 2) is close to the results obtained with the models trained on SM after fine-tuning on the real sets. 
The models trained on the best SC option can, in turn, be further improved by means of fine-tuning, significantly outperforming those trained on SM, showing that a model trained on better quality SC data not only performs better but can also still be improved by fine-tuning to real data.

When comparing the four augmentation options, the main gains are observed when adding background noises; as expected, this simulates noisier conditions and adds variability to the set. Reverberating the signals with the default parameters actually harms the performance. This could be because each speaker is reverberated with different acoustic conditions, possibly making the task easier for the model. We hypothesize that the effect might be different in other scenarios such as meetings or interviews where speakers do not have their own close-talk microphones, as is the case in telephone conversations. Furthermore, the choice of RIRs could be narrowed down to take into account only rooms that resemble those in real applications. These aspects need to be further studied, especially for the more diverse scenarios seen in wide-band data. Results in the following sections are obtained using the best result with SC 2 and SM (P).

\begin{figure}[ht]
  \centering
  \includegraphics[width=0.8\linewidth]{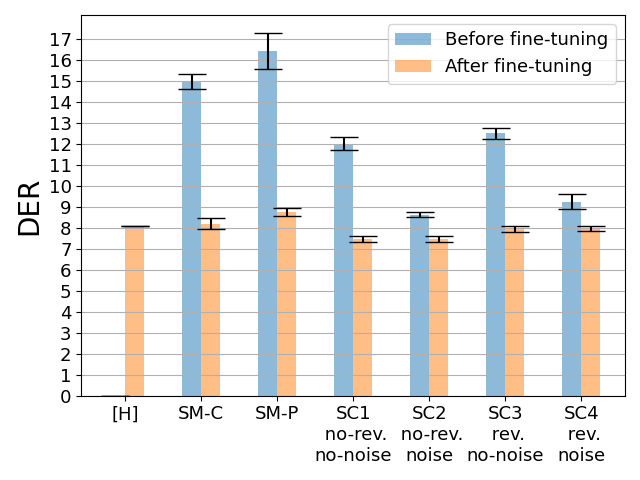}
  \caption{DER (\%) comparison for SM (C stands for Chainer and P for PyTorch) and SC options on CH2-2spk and fine-tuning to CH1-2spk. [H] corresponding to \cite{horiguchi20_interspeech} is a single run. All other experiments were repeated 5 times (with different initializations) and we show means and error bars.}
  \label{fig:v1_callhome_part2_2spk}
\end{figure}

\paragraph{DER breakdown analysis}

Table~\ref{tab:DER_breakdown} presents a detailed comparison of the errors from SM (P) and SC 2. In terms of VAD, both systems perform very similarly before and after fine-tuning. However, larger differences can be seen in terms of OSD. When using SC for training, the model makes considerably fewer false alarms for OSD, especially before fine-tuning. This can be explained by the larger percentage of overlap seen in SM (Table~\ref{tab:datasets_stats}). 

\begin{table}[ht]
  \caption{Error analysis before and after fine-tuning on recordings with 2 speakers of CH2-2spk in terms of DER and its three components: false alarm (FA), missed speech (Miss) and confusion error (Conf.), VAD and OSD.}
  \label{tab:DER_breakdown}
  \setlength{\tabcolsep}{6pt} 
  \centering
  \begin{tabular}{@{}
                  l
                  S[table-format=2.2] |
                  S[table-format=1.2]
                  S[table-format=1.2]
                  S[table-format=1.2] |
                  S[table-format=1.2]
                  S[table-format=1.2] |
                  S[table-format=1.2]
                  S[table-format=2.2]
                  @{}}
    \toprule
    &  & \multicolumn{3}{c|}{DER breakdown} & \multicolumn{2}{c|}{VAD} & \multicolumn{2}{c}{OSD} \\
    System & \multicolumn{1}{c|}{DER} & \multicolumn{1}{c}{Miss} & \multicolumn{1}{c}{FA} & \multicolumn{1}{c|}{Conf.} & \multicolumn{1}{c}{Miss} & \multicolumn{1}{c|}{FA} & \multicolumn{1}{c}{Miss} & \multicolumn{1}{c}{FA} \\
    \midrule
    SM (P) & 15.09 & 2.83 & 8.24 & 4.01 & 0.48 & 7.70 & 4.84 & 10.71 \\
    \hspace{0.2cm}+ FT & 8.44 & 5.23 & 2.32 & 0.90 & 3.39 & 4.06 & 6.20 & 4.22 \\
    \midrule
        SC 2 & 8.64 & 3.11 & 4.84 & 0.69 & 0.49 & 7.53 & 4.68 & 9.03 \\
    \hspace{0.2cm}+ FT & 7.28 & 4.72 & 1.98 & 0.58 & 3.23 & 4.03 & 6.02 & 3.82 \\

    \bottomrule
  \end{tabular}
\end{table}

\paragraph{SC generation statistics and fine-tuning analysis}

To analyze the effect of the set used to estimate the statistics for creating SC data, we made a comparison using either DH-dev or CH1. As seen in Table~\ref{tab:statistics_source}, the effect of the set used for the estimation of the statistics is not large. Both datasets present several recordings that amount to a few hours of speech which is enough to have a reasonable amount of data to estimate the statistics. At the same time, the fine-tuning step improves the performance on both datasets suggesting that real conversations still differ from SC, leaving room for improving the quality of SC.

\begin{table}[ht]
  \caption{DER (\%) for models trained with SC 2 using statistics estimated on DH-dev or CH1. Fine-tuning for each test set is done in the corresponding dev set. Numbers in gray denote results where the test data were used for training. In underlined results, test data were used to compute the statistics.}
  \label{tab:statistics_source}
  \setlength{\tabcolsep}{6pt} 
  \centering
  \begin{tabular}{@{}
                  l |
                  S[table-format=2.2] 
                  S[table-format=2.2] |
                  S[table-format=2.2]
                  S[table-format=2.2] 
                  @{}}
    \toprule
    & \multicolumn{2}{c|}{Callhome 2 speakers} & \multicolumn{2}{c}{DIHARD3 CTS full} \\
    System & \multicolumn{1}{c}{Part 1} & \multicolumn{1}{c|}{Part 2} & \multicolumn{1}{c}{dev} & \multicolumn{1}{c}{eval} \\
    \midrule
    DH stats & \textcolor{black}{8.16} & 8.64 & \textcolor{gray}{\underline{23.47}} & 22.06 \\
    \hspace{0.2cm}+ FT & \textcolor{gray}{6.20} & 7.28 & \textcolor{gray}{16.99} & 17.00 \\
    \midrule
    CH stats & \textcolor{gray}{\underline{8.26}} & 8.73 & \textcolor{black}{22.29} & 21.53 \\
    \hspace{0.2cm}+ FT & \textcolor{gray}{6.13} & 7.28 & \textcolor{gray}{17.38} & 17.14 \\
    \bottomrule
  \end{tabular}
\end{table}

\paragraph{Amount of data analysis}
\label{sec:dataamount}

One aspect to analyze is the effect that the amount of training data has on EEND performance when using SC. Figure~\ref{fig:confbands} shows that the performance degrades considerably before fine-tuning when training with \textit{as little as} 310 hours. However, using 1240 hours (half of the amount used in all other experiments) already allows for a similar performance as using more data, e.g. 4961\,h or 7442\,h. Moreover, using these larger training sets allows the model after fine-tuning to reach as little as 7.03\% DER on average and 6.8\% DER in the best run.

\begin{figure}[ht]
  \centering
  \includegraphics[width=0.8\linewidth]{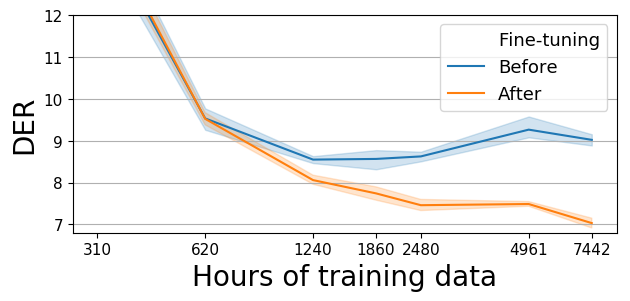}
  \caption{DER (\%) on CH2-2spk when training with different numbers of hours of SC 2. Each experiment is repeated 5 times (with different initializations) to obtain the mean and confidence intervals.}
  \label{fig:confbands}
\end{figure}

\paragraph{Comparison with previous results}

In this paragraph, we compare our results to the best results published with the same architecture. Table~\ref{tab:final_results} shows that when training with SC and without fine-tuning, it is possible to attain similar performance as when training with SM after fine-tuning on Callhome. This gain is smaller in the case of DIHARD CTS, where fine-tuning has a larger effect. One of the aspects to consider with the architecture used is that the input is downsampled so that one output every 100\,ms is produced. When evaluating with a 0\,s forgiveness collar, such as in DIHARD, this severely impacts the results. Table~\ref{tab:final_results} presents results when evaluating at the standard downsampling of one output every 100\,ms and 50\,ms\footnote{This is done only at inference time (not during training), the system is exactly the same and cannot address the issue completely. Having finer granularity considerably increases the inference time and memory requirements.}, showing that a considerable error reduction is possible after fine-tuning when producing outputs with finer time resolution.

\begin{table}[ht]
\centering
\caption{DER (\%) for different systems. Results on DIHARD3 CTS are presented when the input is downsampled to one every 100\,ms (standard) and 50\,ms.}
\label{tab:final_results}
\setlength{\tabcolsep}{6pt} 
  \begin{tabular}{@{}
                  l |
                  S[table-format=2.2]
                  S[table-format=2.2] |
                  S[table-format=2.2]
                  S[table-format=2.2] |
                  S[table-format=2.2]
                  S[table-format=2.2]
                  @{}}
  \toprule
            & \multicolumn{2}{c|}{CH 2 speakers} & \multicolumn{4}{c}{DIHARD3 CTS full} \\
    System & \multicolumn{1}{c}{Part 1} & \multicolumn{1}{c|}{Part 2} & \multicolumn{2}{c}{dev} & \multicolumn{2}{c}{eval} \\
    
  \midrule
  \cite{horiguchi20_interspeech} & \text{--} & 8.07 & \multicolumn{2}{c}{\text{--}} & \multicolumn{2}{c}{\text{--}} \\
  \midrule
  & & & \multicolumn{4}{c}{Downsample rate (ms)} \\
  & & & \text{50} & \text{100} & \text{50} & \text{100} \\
  \midrule
  SM (P) & 13.62 & 15.09 & 25.36 & 25.46 & 22.16 & 23.58 \\
  \hspace{0.2cm}+ FT & \textcolor{gray}{7.61} & 8.44 & \textcolor{gray}{12.97} & \textcolor{gray}{17.98} & 11.99 & 17.44 \\
  \midrule
  SC 2 & 8.16 & 8.64 & \textcolor{gray}{\underline{21.16}} & \textcolor{gray}{\underline{23.47}} & 19.81 & 22.06 \\
  \hspace{0.2cm}+ FT & \textcolor{gray}{6.2} & 7.28 & \textcolor{gray}{11.68} & \textcolor{gray}{16.99} & 11.20 & 17.00 \\
  \bottomrule
  \end{tabular}
\end{table}

\subsubsection{Comparison for wide-band data}
\label{sec:sc_wideband}

Most works with end-to-end models so far have focused on the telephony scenario where large amounts of data are available and where each speaker is recorded in a separate channel. This allows to create diarization segment annotations by simply running a VAD system on each channel. This scenario presents certain advantages such as (usually) one speaker per channel, conversational speech, and relatively similar channel characteristics across recordings. This last aspect allows for a relatively easy amalgamation of different recordings when creating synthetic data where the channel differences are less relevant than the speaker ones and, thus, the model can focus on learning to separate speakers rather than channels. 

Contrarily, wide-band data can have more variability in terms of recording devices, settings and types of interactions. However, it is rare to have a single speaker per recording, even less in a conversational scenario. Thus, applying the same techniques for creating synthetic data becomes more challenging since the mismatch between (artificial) training and (authentic) test data can become much larger.
There have been efforts~\cite{leung2021end,liu21j_interspeech,kinoshita2021integrating,maiti2021end} in generating training data for EEND using wide-band recordings from LibriSpeech~\cite{panayotov2015librispeech} mostly based on SM. 

Although there are numerous wide-band collections with thousands of hours of speech, they contain normally more than one speaker in the same channel. This makes it impossible to use the same strategy as with telephony data for deriving the speaker turns. Instead, it is necessary to have some kind of segmentation already available that contains information about the speakers. Next, we describe the freely available datasets that we used to create wide-band SC in this work.

\begin{itemize}[label={$\bullet$}, topsep=0pt, itemsep=0pt, leftmargin=10pt]
    \item LibriSpeech~\cite{panayotov2015librispeech} consists of 1000 hours of read English speech from almost 2500 speakers. Each recording is expected to contain speech of a single speaker; thus, the original strategy of running VAD to obtain segmentation is possible. Recordings are of good quality and without background noises meaning that channel characteristics are relatively similar across recordings. However, all speech is read and not conversational.
    \item VoxCeleb2~\cite{chung18b_interspeech} consists of more than 2400 hours of recordings from more than 6000 speakers speaking mostly English. Originally prepared as a training set for speaker recognition systems, the recordings are partially annotated. This means that for the speakers of interest, some of their segments are labeled. Thus, it is possible to derive speech segments for a given speaker without the need for any VAD system. At the same time, these annotations are automatically generated, possibly introducing errors such as including small excerpts from other speakers. Furthermore, the speech is collected ``in the wild'' so the recordings can have different noises. We observed better results if recordings with a poor SNR were filtered out.
    \item VoxPopuli~\cite{wang-etal-2021-voxpopuli} consists of recordings from the European Parliament in different languages. For this work, we used the subset for which transcriptions exist. Each recording is expected to contain speech from a single speaker and the transcription timestamps are used to derive the segmentation used for SC. Since the annotations were automatically derived, in some rare cases there were short excerpts from other speakers. This subset contains approximately 2700 hours from 2600 speakers. The recordings correspond to speaker turns during plenary sessions. Therefore, this corpus presents speech that is not read nor conversational but has exclusive turns and spontaneous speech. Recordings are normally of good quality and without background noises.
\end{itemize}

As in the telephony scenario, SCs were augmented with background noises from the MUSAN collection~\cite{snyder2015musan} scaled with an SNR selected randomly from \{5, 10, 15, 20\} dB. Room impulse responses and leveling of relative energy between speakers were evaluated as mentioned in~\cite{landini22_interspeech} but the performance was about the same so we present results only when adding background noises.

\paragraph{Evaluation sets}

Different collections were used for evaluation. For the wide-band scenario, only two-speaker recordings were used\footnote{The multi-speaker scenario is studied in Section~\ref{sec:e2e_diaper}.} in order to simplify the conditions and have clear comparisons with the, already studied, telephone scenario. We evaluate results on the four domains from the Third DIHARD Challenge~\cite{ryant21_interspeech} that satisfy such condition (Table~\ref{tab:evaluation_datasets} summarizes the information about each subset):

\begin{itemize}
    \item CTS, consisting of previously unpublished telephone conversations from the Fisher collection. Both development and evaluation sets consist of 61 10-minute recordings each (full set). Originally 8\,kHz signals, they were upsampled to 16\,kHz for the challenge.
    \item Clinical, consisting of Autism Diagnostic Observation Schedule (ADOS) interviews conducted to identify whether a child fits the clinical diagnosis for autism recorded using a ceiling-mounted microphone. In the full set of DIHARD3, the development set contains 48 recordings lasting 320\,s on average and the evaluation set contains 51 recordings lasting 307\,s on average.
    \item Maptask, consisting of interactions of speakers during a map task in which one participant communicates instructions to the other participant to draw a path in a map. Each speaker is recorded independently with a close-talk microphone and the channels are summed to obtain a mono recording. The development set contains 23 recordings lasting 395\,s on average and the evaluation set contains 19 recordings lasting 393\,s on average.
    \item Sociolinguistic (lab), consisting of interviews recorded under quiet conditions in a controlled environment with a single omnidirectional microphone. The development and evaluation sets contain 16 and 12 recordings respectively lasting around 10 minutes each.
\end{itemize}

\begin{table}[H]
    \caption{Evaluation sets for wide-band scenario. For AMI, numbers do not correspond to dev but to the train set.}
    \label{tab:evaluation_datasets}
    \setlength{\tabcolsep}{4pt} 
    \centering
    \begin{tabular}{llcccc}
    \toprule
    \multirow{2}{*}{Name} & \multirow{2}{*}{Type} & \multicolumn{2}{c}{\#files} & \multicolumn{2}{c}{Avg. length (s.)} \\
    & & dev & eval & dev & eval \\
    \midrule
    CTS & telephone conversations & 61 & 61 & 600 & 600 \\
    Clinical & interviews with children & 48 & 51 & 320 & 307 \\
    Maptask & fast-paced interactions & 23 & 19 & 395 & 393 \\
    Socio lab & interviews with adults & 16 & 12 & 600 & 600 \\
    AMI & meetings (2-spk) & 804 & 93 & 1190 & 1137 \\
    VoxConverse & broadcast (2-spk) & 44 & 31 & 280 & 520 \\
    
    \bottomrule
  \end{tabular}
\end{table}

Another domain of interest for diarization is meetings. However, there is a lack of datasets in a meeting-like scenario with only two speakers. Given that many end-to-end diarization works still focus on the two-speaker scenario, we considered of relevance to derive from AMI meetings~\cite{carletta2005ami} all possible subsets of two speakers for each conversation and make them publicly available\footnote{\url{https://github.com/BUTSpeechFIT/AMI_2speaker_subset}}. For each recording, all pairs of speakers were drawn and, for each pair, all speech where another speaker spoke was removed from the waveforms using reference diarization annotations of the ``only words'' setup described in~\cite{landini2022bayesian}\footnote{Available in \url{https://github.com/BUTSpeechFIT/AMI-diarization-setup}.}. Then, for each original conversation with four speakers, six \textit{conversations} of two speakers were created. We evaluate results on Mix-Headset audios (AMI H) and the beamformed microphone array N1 (AMI A), where BeamformIt~\cite{anguera2007acoustic} is applied using the specific AMI setup.

Finally, to add more diversity, recordings with two speakers from VoxConverse~\cite{chung20_interspeech} were used as these come from varied broadcast sources and present different characteristics from those covered in previously mentioned sets. Following VoxConverse evaluation setup, a forgiveness collar of 0.25\,s was used to compute DER while all the sets mentioned above were scored with forgiveness collar 0\,s. In all cases, all speech (including overlap) was evaluated.

\paragraph{Results}

One of the main aspects when generating synthetic training data is the choice of recordings to use. Figure~\ref{fig:audio_sources} presents a comparison of SM and SC using different sets, namely telephone conversations, LibriSpeech downsampled to 8\,kHz (mimicking a telephone channel scenario), LibriSpeech, VoxCeleb2 and VoxPopuli. 16\,kHz evaluation data were downsampled to run the inference with 8\,kHz models and 8\,kHz data were upsampled (with empty upper spectrum) to run the inference with 16\,kHz models.

Even though some clear differences exist before fine-tuning, such as SC LibriSpeech performing the best for Maptask and both AMI sets, differences are reduced after fine-tuning. As observed in~\cite{landini22_interspeech}, differences of up to 0.5\%\,DER can easily be attributed to chance; thus, here there are no statistically significant ``winners'' in most cases.
Some of the patterns worth mentioning are that 8\,kHz models perform the worst for Maptask and the AMI sets, suggesting that information in higher frequencies is more relevant in these sets; and that using VoxCeleb2 recordings allows for better performance in VoxConverse, probably due to the similarities between both sets, formed by diverse recordings collected from YouTube.

Unlike the effect seen in the telephony scenario where SCs are clearly superior and the need for fine-tuning is reduced substantially\footnote{Results on CTS when training with SM telephone are 4.9\% and 5.2\% relatively worse than when training with SC telephone before and after fine-tuning respectively.}~\cite{landini22_interspeech}, when working with different wide-band datasets, fine-tuning still plays a major role and even SMs allow for similar performance. Unfortunately, this shows that the main challenge in wide-band scenarios is, until now, not the realism or naturalness of the synthetic data in terms of turns but rather differences in the channel between source and test data, quality of speaker annotation, and conversational nature of data or a combination of all these. 

\begin{figure*}
    \begin{subfigure}[b]{\textwidth}
        \centering
        \includegraphics[width=\linewidth]{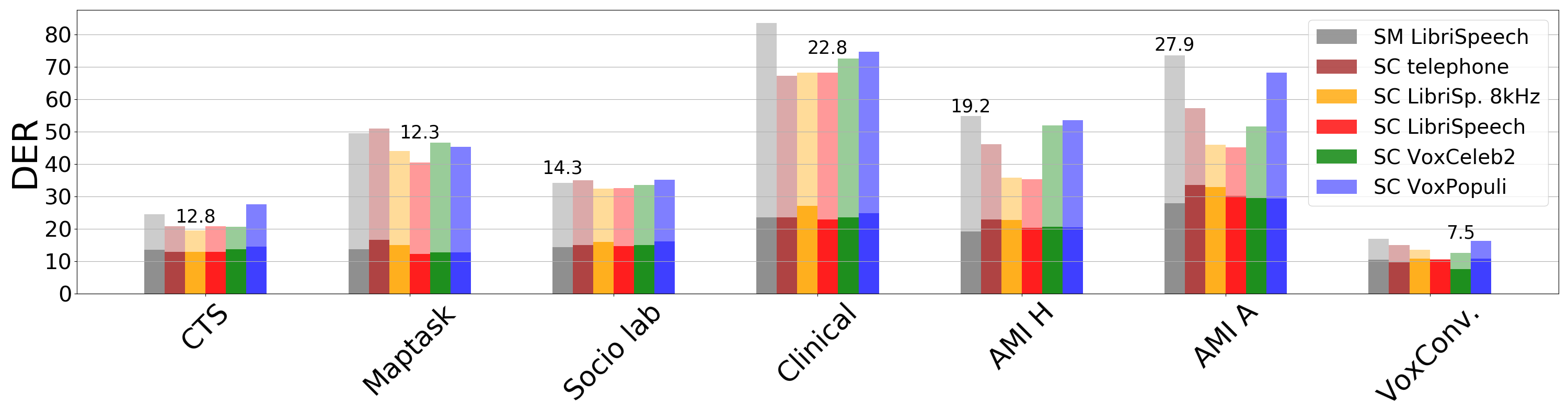}
        \caption{Comparison of type of recordings used to generate SC (or SM): LibriSpeech, telephone (Switchboard and SRE), VoxCeleb2 and VoxPopuli. Numbers mark the best result among bars.}
        \label{fig:audio_sources} 
    \end{subfigure}
    
    \begin{subfigure}[b]{\textwidth}
        \centering
        \includegraphics[width=\linewidth]{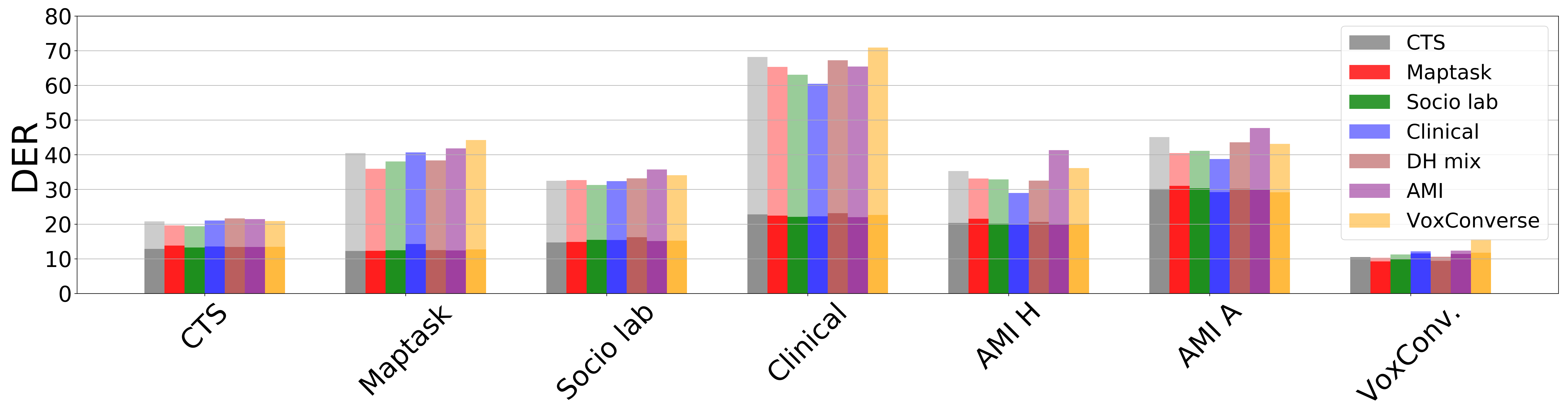}
        \caption{Comparison of sources for estimating the statistics used to generate SC with LibriSpeech recordings.}
        \label{fig:stats_sources} 
    \end{subfigure}
    \caption{Results on the evaluation/test sets of different subsets of DIHARD3, AMI mix-headset (H) and beamformed array (A) and 2-speaker recordings from VoxConverse. Light shade colors show results before fine-tuning and darker colors refer to results after fine-tuning to a matching development set.}
\end{figure*}

The mechanism proposed for generating SC makes use of statistics about pauses and overlaps observed in real conversations. We explored different sets for extracting such statistics: CTS, Maptask, Socio lab, Clinical, all the aforementioned together (in the core setup, to have them equally represented), AMI and VoxConverse for generating SC using audios from LibriSpeech. As shown in Figure~\ref{fig:stats_sources}, there are no significant differences.
While these findings do not imply that it is not possible to produce synthetic data of superior quality that will permit better performance without the need for fine-tuning, we intend to share our results with the community simply to shed light on the directions that have already been explored. There are still other aspects to consider such as improving data augmentation mechanisms like using more realistic background noises, reverberation or loudness levels that match the application scenarios. For instance, CTS, Maptask and AMI H have speech recorded with close-talk microphones, and Socio lab, Clinical and AMI A were recorded with far-field microphones. These aspects were not particularly considered in our analysis but they might have a strong influence. We generated SC from specific domains, i.e., from Maptask recordings only, and evaluated the performance on the same domain but the performance was considerably inferior; probably due to the limited number of speakers. Even more, choosing recordings with similar channel characteristics or choosing speakers with similar voices to create SC could be mechanisms for creating more challenging training data that would allow for better performance on real scenarios.

Finally, we hope that by sharing recipes\footnote{\url{https://github.com/BUTSpeechFIT/EEND_dataprep}} for generating training data using public datasets and the models trained on publicly available data, the community will have easier access to baselines that otherwise require expensive computations.

\subsubsection{Comparison for a flexible number of speakers in the telephony scenario}
\label{sec:sc_morespeakers}

Having seen the advantages of SC over SM in the constrained scenario of 2 speakers per recording, the next analysis extends the comparison for more speakers. Using the model described in~\ref{sec:2spkexp}, the same setup was followed for SM and SC: the initial training using 2 speakers per recording is run for 100 epochs with 50\,s sequences, the adaptation to simulated training data with more speakers is run for 100 epochs on smaller sets or 25 epochs on (approximately four times) larger sets both with 50\,s sequences. Finally, the fine-tuning step is run for 100 epochs using sequences of 200\,s.

\paragraph{Evaluation sets}

To evaluate the performance when training EEND-EDA with more than two speakers, the 2000 NIST Speaker Recognition Evaluation~\cite{przybocki2001nist} dataset, usually referred to as ``Callhome''~\cite{NISTSRE2000evalplan} was used.
We report results using the standard Callhome partition\footnote{\url{https://github.com/BUTSpeechFIT/CALLHOME_sublists}}. We will refer to the parts as CH1 and CH2. 
The numbers of recordings with different amounts of speakers are listed in Table~\ref{tab:callhome_speakers}. 

\begin{table}[H]
    \caption{Numbers of recordings per list for Callhome parts 1 and 2. Percentages over the total are expressed between brackets.}
    \label{tab:callhome_speakers}
    \setlength{\tabcolsep}{6pt} 
    \centering
    \begin{tabular}{lcccccc}
    \toprule
    No. speakers & 2 & 3 & 4 & 5 & 6 & 7 \\ 
    \midrule
    CH1 & 155 (62.3\%) & 61 (24.5\%) & 23 (9.2\%) & 5 (2\%) & 3 (1.2\%) & 2 (0.8\%) \\
    CH2 & 148 (59.2\%) & 74 (29.6\%) & 20 (8\%) & 5 (2\%) & 3 (1.2\%) & 0 (0\%) \\
    
    \bottomrule
  \end{tabular}
\end{table}

\paragraph{Training sets}

Different training sets were compared besides the standard set of SM used in \cite{horiguchi20_interspeech} with 1 to 4 speakers per recording and 100k mixtures for each number of speakers. SC sets are listed in Table~\ref{tab:multidatasets_stats}. SC \textbf{2}-4\,spk contains recordings with 2, 3 and 4 speakers following the proportions of 50\%, 25\%, and 25\% respectively in terms of hours while SC 2-\textbf{4}\,spk follows the proportions 25\%, 25\%, and 50\%. The SC 2-7\,spk follows the proportions of speakers seen in Callhome Part 1 (shown in Table~\ref{tab:callhome_speakers}).

\begin{table}[ht]
\centering
\caption{Statistics for synthetic and real datasets.}
\label{tab:multidatasets_stats}
\setlength{\tabcolsep}{6pt} 
  \begin{tabular}{@{}
                  l
                  S[table-format=6.0]
                  S[table-format=3.2]
                  S[table-format=5.2]
                  S[table-format=2.2]
                  S[table-format=2.2]
                  S[table-format=2.2]
                  @{}}
  \toprule
          & 
         & 
        \multicolumn{1}{c}{\begin{tabular}{@{}c@{}}Average\end{tabular}} &
        \multicolumn{1}{c}{\begin{tabular}{@{}c@{}}Total \end{tabular}} &
        \multicolumn{3}{c}{Average (\%)} \\
        Dataset       & 
        \multicolumn{1}{c}{\#Files} & 
        \multicolumn{1}{c}{\begin{tabular}{@{}c@{}}duration (s)\end{tabular}} &
        \multicolumn{1}{c}{\begin{tabular}{@{}c@{}}audio (h)\end{tabular}} &
        \multicolumn{1}{c}{\begin{tabular}{@{}c@{}}Silence\end{tabular}} & 
        \multicolumn{1}{c}{\begin{tabular}{@{}c@{}}1-speaker\end{tabular}} &
        \multicolumn{1}{c}{\begin{tabular}{@{}c@{}}Overlap\end{tabular}} \\
        
  \midrule
  SM 2\,spk & 100000 & 89.30 & 2480.62 & 18.61 & 49.62 & 31.77 \\ 
  SC 2\,spk & 25074 & 356.15 & 2480.56 & 12.80 & 78.83 & 8.37 \\
  \midrule
  SC \textbf{2}-4\,spk & 20449 & 436.62 & 2480.14 & 11.31 & 77.86 & 10.83 \\
  SC 2-\textbf{4}\,spk & 17717 & 503.96 & 2480.17 & 10.49 & 77.15 & 12.36 \\
  SC 2-7\,spk & 21652 & 412.41 & 2480.39 & 11.71 & 78.09 & 10.2 \\
  SM 1-4\,spk & 400000 & 139.46 & 15495.00 & 20.06 & 50.79 & 29.15 \\
  SC 1-4\,spk & 99373 & 359.47 & 9922.67 & 12.76 & 78.49 & 8.76 \\
  \midrule
  Callhome Part 1 & 249 & 125.80 & 8.70 & 10.23 & 74.49 & 15.28 \\
  Callhome Part 2 & 250 & 123.18 & 8.55 & 10.55 & 74.46 & 14.99 \\
  \bottomrule
  \end{tabular}
\end{table}

\paragraph{Results}

\begin{table}[ht]
    \caption{DER (\%) comparison on Callhome Part 2. FT stands for fine-tuning.}
    \label{tab:multispeaker}
    \setlength{\tabcolsep}{6pt} 
    \centering
    \begin{tabular}{@{}
                  l
                  l |
                  S[table-format=2.2] |
                  S[table-format=2.2]
                  S[table-format=2.2]
                  S[table-format=2.2] 
                  S[table-format=2.2]
                  S[table-format=2.2] 
                  @{}}
    \toprule
    & System & All & \multicolumn{1}{c}{2-spk} & \multicolumn{1}{c}{3-spk} & \multicolumn{1}{c}{4-spk} & \multicolumn{1}{c}{5-spk} & \multicolumn{1}{c}{6-spk} \\
    \midrule
    1 & SM 2 spk & 28.67 & 16.85 & 27.46 & 40.4 & 52.94 & 50.51 \\
    2 & \hspace{0.2cm}+ adapt SM 1-4 spk & 26.14 & 16.28 & 24.67 & 34 & 50.15 & 53.24 \\
    3 & \hspace{1cm}+ FT CH1 & 17.45 & 8.38 & 16.14 & 24.26 & 36.75 & 46.79 \\ 
    
    \cline{2-8}
    
    4 & SM 1-4 spk & 27 & 16.1 & 25.44 & 37.56 & 46.52 & 54.92 \\
    5 & \hspace{1cm}+ FT CH1 & 25.78 & 13.92 & 24.77 & 34.62 & 43.77 & 66.37 \\
    \midrule
    
    6 & SC 2 spk & 20.86 & 8.48 & 21.07 & 29.56 & 45.61 & 49.2 \\
    7 & \hspace{0.2cm}+ adapt SC 1-4 spk & 16.18 & 8.95 & 13.78 & 21.22 & 37.35 & 46.32 \\
    8 & \hspace{1cm}+ FT CH1 & 16.07 & 10.03 & 14.35 & 19.3 & 30.67 & 46.94 \\ 
    
    9 & \hspace{0.2cm}+ adapt SC \textbf{2}-4 spk & 17.52 & 9.07 & 15.11 & 23.3 & 38.55 & 54.03 \\
    
    10 & \hspace{0.2cm}+ adapt SC 2-\textbf{4} spk & 17.09 & 9.55 & 14.69 & 22.5 & 34.96 & 51.46 \\
    
    11 & \hspace{0.2cm}+ adapt SC 2-7 spk & 17.49 & 9.16 & 15.43 & 24.18 & 39.17 & 45.41 \\ 
    \cline{2-8}
    
    12 & SC 1-4 spk & 19.9 & 10.2 & 17.79 & 26.49 & 42.67 & 58.09 \\ 
    13 & \hspace{1cm}+ FT CH1 & 21.24 & 15.45 & 18.89 & 25.17 & 38.16 & 49.54 \\ 
    
    \cline{2-8}
    
    14 & SC \textbf{2}-4 spk & 21.23 & 11.44 & 20.03 & 29.46 & 41.77 & 47.44 \\
    15 & \hspace{1cm}+ FT CH1 & 20.59 & 13.32 & 18.89 & 25.94 & 39.93 & 45.82 \\ 
    \cline{2-8}
    
    16 & SC 2-\textbf{4} spk & 20.39 & 11.45 & 18.67 & 27.75 & 42.16 & 47.11 \\
    17 & \hspace{1cm}+ FT CH1 & 20.2 & 14.39 & 17.67 & 24.51 & 42.75 & 42.38 \\ 
    \cline{2-8}
    
    18 & SC 2-7 spk & 24.98 & 16.55 & 23.14 & 31.57 & 48.76 & 50.04 \\
    19 & \hspace{1cm}+ FT CH1 & 24.9 & 16.27 & 23.29 & 31.42 & 47.36 & 51.32 \\
    
    \bottomrule
  \end{tabular}
\end{table}

Table ~\ref{tab:multispeaker} presents a comparison when different sets are used for training, adapting (to more speakers) and fine-tuning (to Callhome Part 1). The first 5 rows show the performance when using SM. When following the same scheme with SC (rows 6-8 and 12, 13), the model trained on 2 speakers is already considerably better but adapting to more speakers pushes the performance further. The fine-tuning step provides only small gains, reducing the dependence on this step. Analogously, if the model is directly trained on the 1-4 sets, SC (rows 12, 13) allow for much better performance than SM (rows 4, 5). However, this set is so biased towards fewer speakers that the fine-tuning only helps for 4 or more speakers while harming for 2 and 3 which actually represent the majority of the recordings.

We explored using other SC multi-speaker sets with different proportions of recordings. Training on a set with a higher proportion of recordings with 4 than 2 speakers (row 10) is beneficial since the resulting model can deal better with recordings with more speakers. Training with a set that follows the same proportion of speakers seen in CH1 (row 11) does not bring considerable gains. However, it should be noted that the training set is rather small and it is possible that learning to deal with more speakers requires a larger training set. Similar behaviors are seen when training the model directly on these sets (rows 12-19).

Overall, the best results are obtained when first training only on 2 speakers and then using the 1-4 speakers set for adaptation which is considerably larger than the other ones. This suggests that such model can take advantage of larger sets and that it would be beneficial to produce training data on the fly in order to encourage larger variability in the training set.

\subsection{Summary}

In this section, we discussed the generation of synthetic training data and its effects on an end-to-end model's performance. The original approach for generating simulated mixtures and the proposed approach to create simulated conversations were described showing how the results of the latter resemble real conversations better than the former. Both were compared in the telephony scenario for conversations with two or more speakers and it was shown that the simulated conversations provide advantages; mainly the improved performance and less need for a fine-tuning stage in the training procedure.

The analysis was extended to wide-band data in different domains. However, it was shown that the main challenges are not in terms of how realistic are the interactions but most likely in the acoustic aspects of the recordings. This is also related to the increased intrinsic variability observed in wide-band data as well as in the available datasets that can be used to create synthetic data. These two aspects need to be addressed in the future to be able to take full advantage of synthetic training data, crucial for the good performance of end-to-end models. 

Possible ideas to improve simulated conversations were also discussed throughout the section, mostly related to the ``hard'' aspects of the simulated data, like the channel. However, ``soft'' facets which have not been addressed can be very relevant to improve the quality of simulated conversations. For example, back-channels are used by speakers to indicate agreement or understanding during a conversation. When randomly interleaving speech segments to create simulated conversations, back-channels can be placed in appropriate positions in the conversation but, more often than not, they simply appear in wrong places. In order to adequately position back-channel segments in a simulated conversation, deeper understanding of the context will be needed and transcriptions (manual or automatic) could help. Then, segments could be treated differently if they correspond to back-channels, in order to place them only during or in between another speaker's segments.

Another common aspect in natural speech are changes in levels of energy or tone towards the end of a turn, indicating another speaker that it is their turn. Improving the quality of simulated conversations regarding turn-taking cues could also lead to more natural interactions. It has been shown~\cite{brusco2023automatic} that turn-taking cues can be automatically labeled with relatively high accuracy and these annotations could be used as part of the process that creates synthetic data. For example, given a segment with certain characteristics like a `smooth switch', a segment from a different speaker should follow, thus restricting the pool of segments to choose from for the next segment.

Entrainment refers to the similarities in the communication behavior between participants such as matching prosody or stress patterns. When generating simulated conversations and matching speakers on different conversations the level of expected entrainment will be very low since the speakers did not really interact with each other. Improving in this direction is considerably more difficult than in the points aforementioned mainly because matching certain characteristics requires modifying (or generating new) speech. 

In order to address this, and given the recent developments on voice conversion and text-to-speech~\cite{shen2023naturalspeech,wang2023neural}, generating natural-sounding scripted conversations is more feasible than it was only a couple of years ago. Using transcriptions from real conversations and generating speech with different speaker voices could lead to simulated conversations that address several of the points raised above. Even more, current generative text tools are capable of creating very diverse dialogues and this could increase the variability of the content. The overall scheme to generate such data would consist of:
\begin{itemize}
    \item Collect and/or generate transcriptions of conversations.
    \item Extract speaker profile embeddings for speakers in a large pool.
    \item Using voice conversion over an existing recording or text-to-speech with a transcription, and a speaker profile, generate a speech segments of that speaker in a given conversation.
    \item Combine segments corresponding to the same conversation for different subsets of speakers.
\end{itemize}

Naturally, aspects like back-channels, turn-taking cues and entrainment would be handled in those steps depending on the quality of the voice conversion and text-to-speech systems. 

This discussion does not pretend to be exhaustive regarding improving synthetic data but rather paths that were not explored in our previous work and which we believe have potential to bridge the gap between real and simulated conversations.

\section{DiaPer: End-to-end neural diarization with Perceiver-based attractors}
\label{sec:e2e_diaper}

\subsection{Motivation and background}

As discussed in Section~\ref{sec:e2e_EEND}, the EEND framework is the most prolific single-stage end-to-end model, showing remarkable performance. EEND-EDA, although in theory capable of dealing with flexible numbers of speakers, has certain limitations. One of them is that the model is strongly biased towards the number of speakers per recording seen during training and has difficulties handling conversations with many speakers. Even more, it is necessary to shuffle the frames at the input of the attractor encoder to obtain the best performance showing that the choice of LSTM as encoder is suboptimal.

Several works have proposed modifications to the EEND-EDA model, as discussed in Section~\ref{sec:eend_extensions}. Three of them~\cite{rybicka2022end,fujita2023neural,chen23n_interspeech}, in particular, present a non-autoregressive attractor-decoder framework based on attention as shown in Figures~\ref{fig:rybicka_generalized},~\ref{fig:fujita_generalized},~\ref{fig:chen_generalized}. The three models are presented under the same general framework where some blocks are common to all of them. All share a similar frame encoder based on self-attention layers to produce the frame embeddings and in all cases, the frame-speaker activities are obtained by applying sigmoid on the dot product between frame embeddings and attractors. Thus, the main differences stem from how the attractors are obtained.

\begin{figure}
    \centering
    \begin{minipage}[b]{.35\textwidth}
        \centering
        \includegraphics[width=\linewidth]{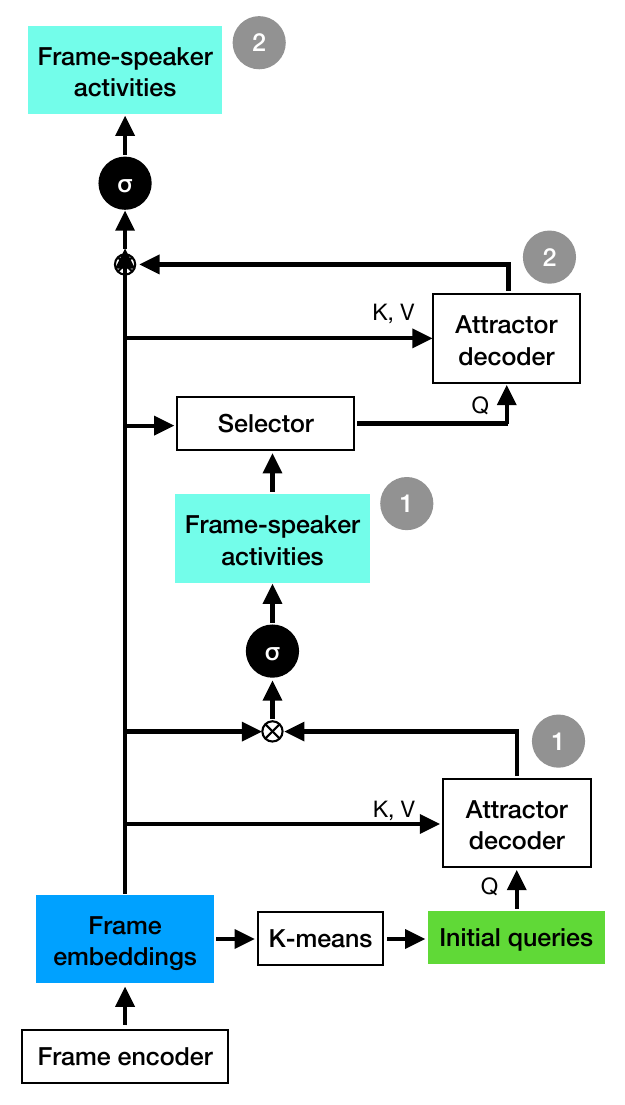}
        \captionsetup{justification=centering}
        \caption{Diagram for \\ \cite{rybicka2022end}.}
        \label{fig:rybicka_generalized} 
    \end{minipage}%
    \hfill
    \begin{minipage}[b]{.25\textwidth}
        \centering
        \includegraphics[width=\linewidth]{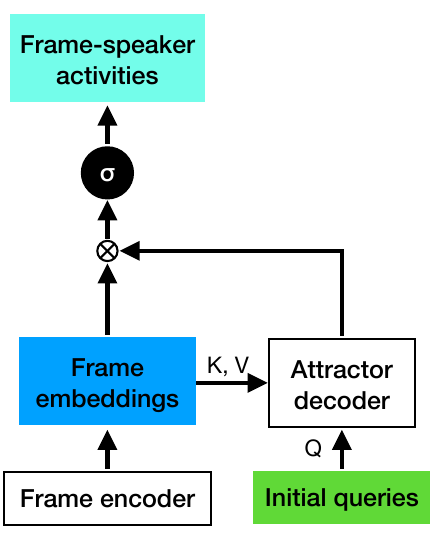}
        \caption{Diagram for \cite{fujita2023neural}.}
        \label{fig:fujita_generalized} 
    \end{minipage}%
    \hfill
    \begin{minipage}[b]{.35\textwidth}
        \centering
        \includegraphics[width=\linewidth]{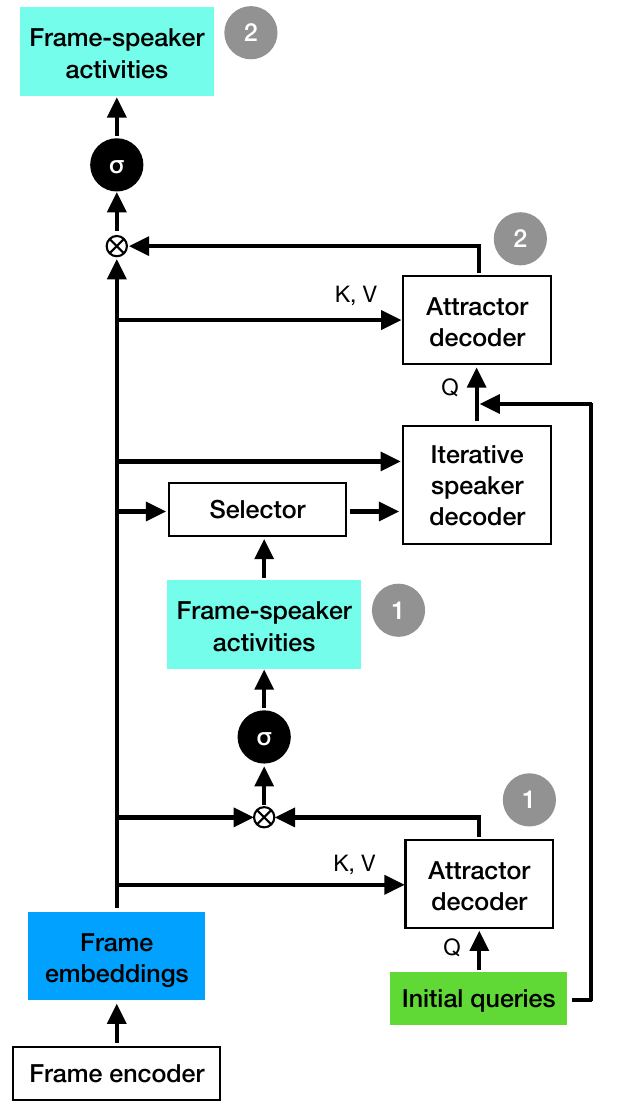}
        \captionsetup{justification=centering}
        \caption{Diagram for \\ \cite{chen23n_interspeech}.}
        \label{fig:chen_generalized}
    \end{minipage}
\end{figure}

The approach in Figure~\ref{fig:rybicka_generalized}~\cite{rybicka2022end} utilizes transformer decoder blocks where frame embeddings function as keys and values and the attractors are used as queries and refined by each decoder block. The attractors in the first block are obtained as weighted averages of the frame embeddings by the probability of the speaker being active in that frame (given by the diarization output of the previous iteration). This procedure is repeated iteratively to refine the attractors. However, an initial set of queries is necessary for the first iteration and they utilize the centroids obtained after using k-means clustering over frame embeddings. K-means is used to obtain a pre-defined number of clusters, thus this approach handles a fixed number of speakers. The attractor decoder comprises two sequential decoder blocks (as denoted in Figure~\ref{fig:rybicka_decoder}). In the figure, the gray ``1'' denotes the first iteration when k-means is used, and ``2'' refers to the second pass when the attractors are obtained with the weighted average (performed by the Selector). Further iterations are not represented but follow the path denoted by ``2''.

In Figure~\ref{fig:fujita_generalized}~\cite{fujita2023neural}, a simpler scheme is presented where the initial queries are learnable parameters learned during training. It also utilizes a single transformer decoder block formed by only a cross-attention module where frame embeddings function as keys and values. Note that this approach does not iterate.

Figure~\ref{fig:chen_generalized} combines ideas of the two above. \cite{chen2023attention} utilize six sequential transformer decoder blocks where frame embeddings function as keys and values and the queries are separated into two groups: Three learnable queries represent non-speech, single-speaker speech and overlap respectively and are denoted by the initial queries in the figure. As explained in Section~\ref{sec:nonautoregressive} and shown in Figure~\ref{fig:att_eda}, queries for the speakers are defined iteratively one speaker at a time. The selector chooses single-speaker speech frames from a speaker (randomly or clustering frames first with SC and selecting a segment of the majoritarian cluster). The frames selected from the new speaker are averaged to produce one attractor. The initial queries and this preliminary attractor are then refined throughout the decoder blocks to produce the final attractors. This procedure is repeated until all ``single speaker'' segments are consumed by some attractor. The gray ``1'' denotes the first step where only categories are decoded and ``2'' denotes the second step where the attractor decoder is used to obtain the final frame-speaker activities. 

The first two approaches have demonstrated their capabilities only in the two-speaker scenario where there is no need to find the number of speakers in the recording and where the architecture can be crafted to handle that specific number. The extension to more speakers is definitely possible but follow-up works have not yet been published. The third one presents results with a flexible number of speakers but relies on an autoregressive scheme since the speakers are iteratively decoded in the second step.
At the same time, these approaches present particular interpretations of a more generic architecture: the Perceiver~\cite{jaegle2021perceiver} which iteratively refines a set of latents (queries in cross-attention) informed by the input sequence (keys and values in cross-attention) but in a non-autoregressive framework.

The model we propose in this work generalizes some of the ideas described above and directly tackles the problem of handling several speakers using Perceivers to obtain attractors in an EEND-based framework. We name this approach DiaPer: end-to-end neural Diarization with Perceiver-based attractors.

\subsection{Model description}

The Perceiver~\cite{jaegle2021perceiver} is a Transformer~\cite{vaswani2017attention} variant that employs cross-attention to project the variable-size input onto a fixed-size set of latent representations, as shown in Figure~\ref{fig:Perceiver}. These latents are transformed by a chain of self-attention and cross-attention blocks. By encoding the variable-size input (``Byte array'' in the figure) into the fixed-size latent space (``Latent array'' in the figure), the Perceiver reduces the quadratic complexity of the Transformer to linear. In this work, we utilize the Perceiver framework to encode speaker information into the latent space and then derive attractors from the latents. Perceiver allows us to handle a variable number of speakers per conversation while addressing some of the limitations of EDA with a fully non-autoregressive (and iteration-free) scheme. 

\begin{figure}[ht]
  \centering
  \includegraphics[width=\linewidth]{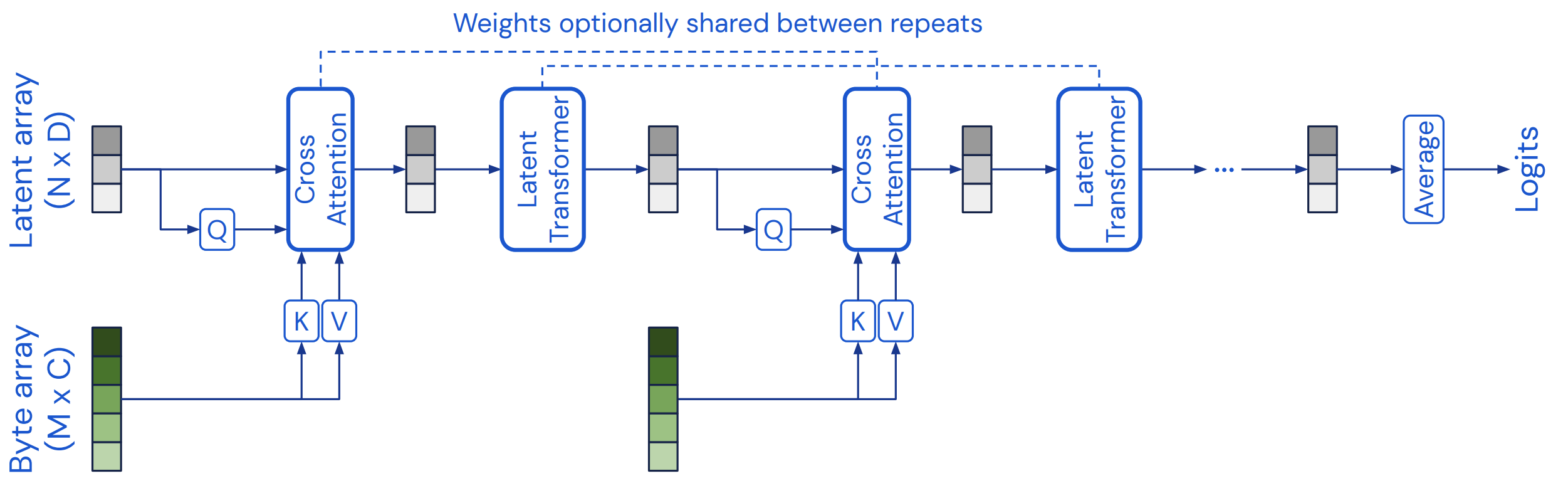}
  \caption{Perceiver architecture as presented in \cite{jaegle2021perceiver}. Figure adopted from \cite{jaegle2021perceiver}.}
  \label{fig:Perceiver}
\end{figure}

\begin{figure}[ht]
  \centering
  \includegraphics[width=0.7\linewidth]{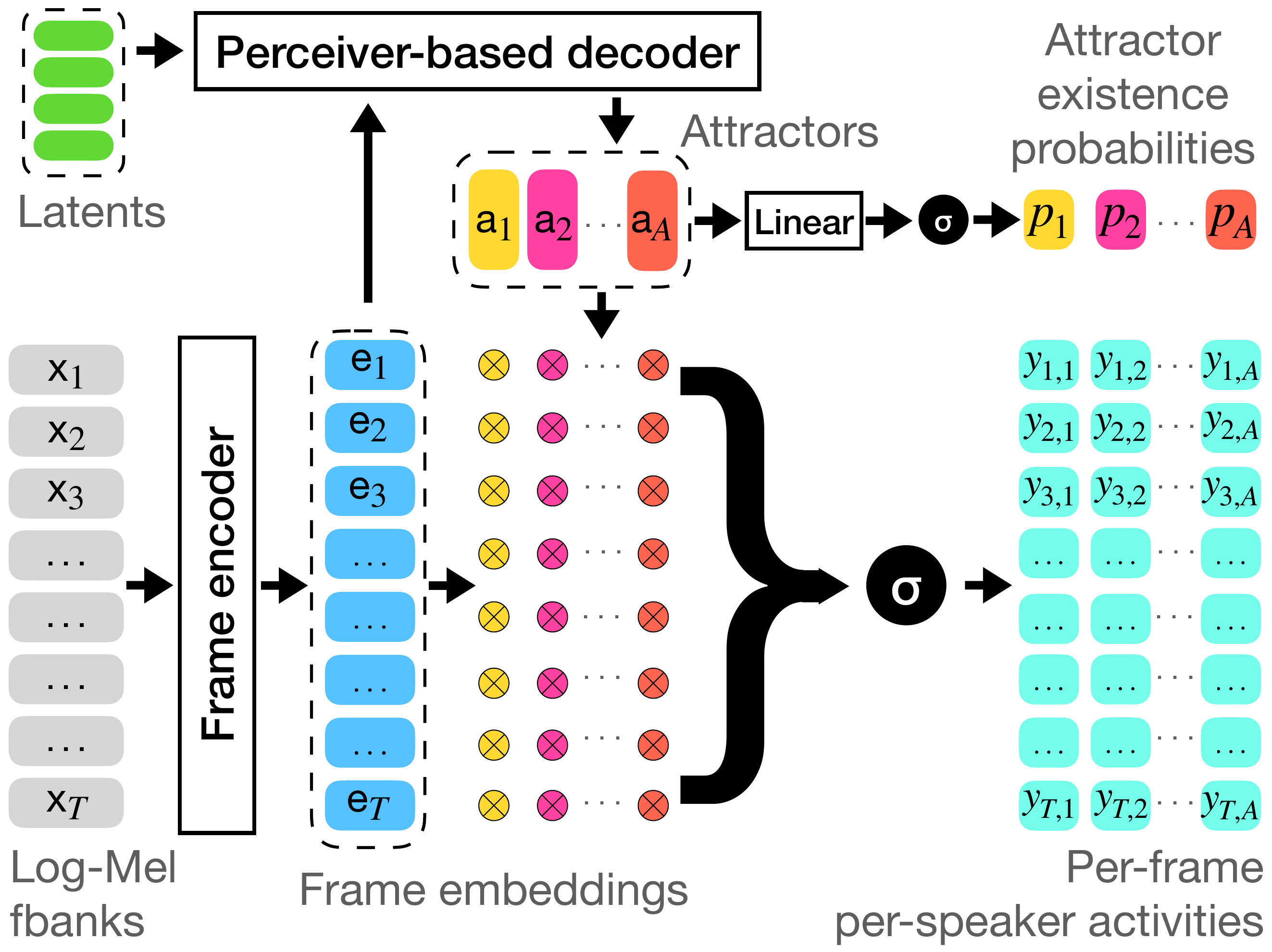}
  \caption{DiaPer diagram.}
  \label{fig:DiaPer}
\end{figure}

DiaPer shares many facets with other EEND models, such as defining diarization as a per-speaker-per-time-frame binary classification problem. 
Given a sequence of observations (features) $\textbf{X} \in \mathbb{R}^{T \times F}$ where $T$ denotes the sequence length and $F$ the feature dimensionality, the model produces $\hat{\textbf{Y}} \in (0, 1)^{T \times S}$ which represent the speech activity probabilities of the $S$ speakers for each time-frame. Just like with EEND-EDA, the model is trained so that $\hat{\textbf{Y}}$ matches the reference labels $\textbf{Y} \in \{0, 1\}^{T \times S}$ where $y_{t,s} = 1$ if speaker $s$ is active at time $t$ and 0 otherwise. The main difference between EEND-EDA and DiaPer is in how the attractors are obtained given the frame embeddings. As shown in Figure~\ref{fig:DiaPer}, DiaPer makes use of Perceiver to obtain the attractors instead of the LSTM-based encoder-decoder.

The main two modules in DiaPer are the frame encoder and the attractor decoder. As shown in Figures~\ref{fig:Frame_Encoder_new_simple} and \ref{fig:encoder_layer} and proposed in~\cite{fujita2019end}, the frame encoder receives the sequence of frame features $\textbf{X}$ and transforms them with a few chained self-attention layers $\textbf{E} = FrameEncoder(\textbf{X})$ to obtain the frame embeddings $\textbf{E} \in \mathbb{R}^{T \times D}$, where $D$ is the dimension of the embeddings. 

The attractor decoder receives the frame embeddings and produces attractors $\textbf{A} = PerceiverDecoder(\textbf{E})$ with $\textbf{A} \in  \mathbb{R}^{A \times D}$\footnote{In practice, $A$ equals to the number of speakers $S$.} which are in turn compared with the frame embeddings to determine which speaker is active at each time-frame: $\hat{\textbf{Y}} = \sigma(\textbf{E} \textbf{A}^\top)$.

\begin{figure}
    \centering
    \begin{minipage}[b]{.35\textwidth}
        \centering
        \includegraphics[width=\linewidth]{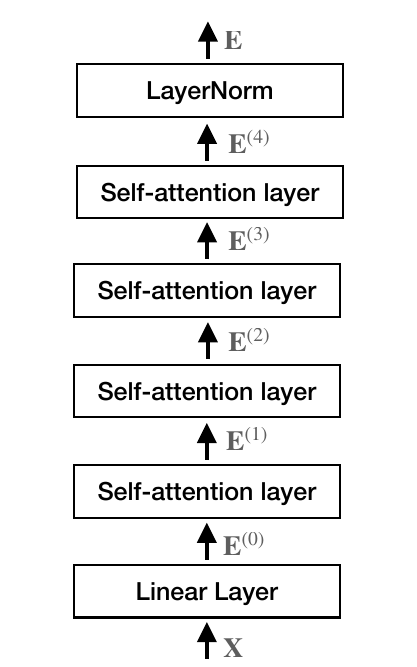}
        \captionsetup{justification=centering}
        \caption{Frame encoder scheme.}
        \label{fig:Frame_Encoder_new_simple} 
    \end{minipage}%
    \hfill
    \begin{minipage}[b]{.35\textwidth}
        \centering
        \includegraphics[width=\linewidth]{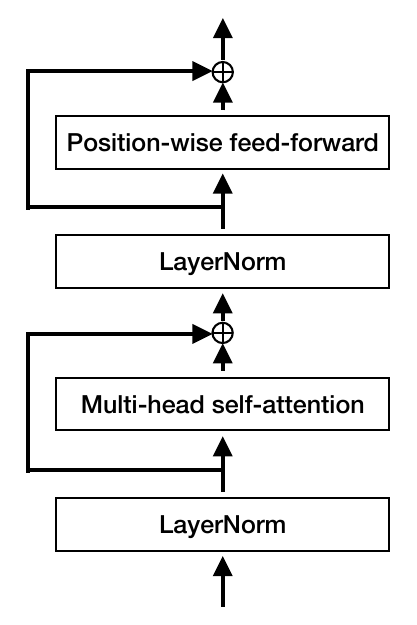}
        \caption{Self-attention layer.}
        \label{fig:encoder_layer} 
    \end{minipage}
\end{figure}

To explain how the frame encoder and the attractor decoder operate, we will make use of two functions: multi-head attention (MHA) described in Algorithm~\ref{alg:mha} and feed-forward layer (FF) described in Algorithm~\ref{alg:ff}. Note that multi-head cross-attention corresponds to $MHA(Q, K, V)$ when $Q \neq K = V$ and multi-head self-attention (MHSA) corresponds to $MHA(Q, K, V)$ when $Q = K = V$. Both operations (namely MHA and FF) operate using internal parameters so one can think of using them later as instances. For example, $MHA^{(i)}$ will refer to the $i$\textsuperscript{th} instance with its own set of parameters.

\begin{algorithm}
\caption{Multi-head attention}\label{alg:mha}
\textbf{Parameters:} $H \in \mathbb{N}$, $\textbf{W}_h^{Q} \in \mathbb{R}^{D \times d}, \textbf{W}_h^{K} \in \mathbb{R}^{D \times d}, \textbf{W}_h^{V} \in \mathbb{R}^{D \times d}$, $\textbf{W}^{O} \in \mathbb{R}^{(H \cdot d) \times D}$ \\
where $H$ is the number of heads (with $1 \leq h \leq H$), $\textbf{W}_h^{Q}$, $\textbf{W}_h^{K}$, $\textbf{W}_h^{V}$ are the query, key and value projection matrices for the $h$\textsuperscript{th} head, $\textbf{W}^{O}$ is the output projection matrix, and $d = \frac{D}{H}$ is the dimension of each head. \\
\textbf{Input:} $\mathbf{Q} \in \mathbb{R}^{q \times D}, \mathbf{K} \in \mathbb{R}^{T \times D}, \mathbf{V} \in \mathbb{R}^{T \times D}$ \\
\textbf{Output:} $MHA(\mathbf{Q}, \mathbf{K}, \mathbf{V}) \in \mathbb{R}^{q \times D}$ \\
\begin{algorithmic}
\ForAll{$h \in 1 \ldots H$}
    \State $C_h \gets Softmax\Big( \frac{\mathbf{Q} \textbf{W}_h^{Q} (\mathbf{K} \textbf{W}_h^{K})^\top}{\sqrt{d}} \Big) (\mathbf{V} \textbf{W}_h^{V})$
\EndFor
\State $MHA(\mathbf{Q}, \mathbf{K}, \mathbf{V}) \gets Concatenate(C_1, \ldots, C_H) \textbf{W}^{O}$
\end{algorithmic}
\end{algorithm}

\begin{algorithm}
\caption{Feed-forward}\label{alg:ff}
\textbf{Parameters:} $\mathbf{W}_1 \in \mathbb{R}^{D \times D_{ff}}, \mathbf{b}_1 \in \mathbb{R}^{D_{ff}}, \mathbf{W}_2 \in \mathbb{R}^{D_{ff} \times D}, \mathbf{b}_2 \in \mathbb{R}^{D_{ff}}$  \\
where $\textbf{W}_1$, $\textbf{W}_2$, $\textbf{b}_1$, $\textbf{b}_2$ are the weights and biases, $\mathbf{1}$ is an all-one vector, $ReLU(\cdot)$ is the rectified linear unit activation function. $D$ is the dimension of the input and $D_{ff}$ is the dimension of the internal linear layer.\\
\textbf{Input:} $\mathbf{E} \in \mathbb{R}^{T \times D}$ \\
\textbf{Output:} $FF(\mathbf{E}) \in \mathbb{R}^{T \times D}$ \\
\begin{algorithmic}
\State $FF(\mathbf{E}) \gets ReLU(\mathbf{E} \mathbf{W}_1 + \mathbf{1} \mathbf{b}_1^{\top}) \mathbf{W}_2 + \mathbf{1} \mathbf{b}_2^{\top}$
\end{algorithmic}
\end{algorithm}

To describe the model, we need to define $FrameEncoder$ and $PerceiverDecoder$ and the loss(es) used to train it. We will first focus on a minimal version of a working model and then cover some extensions that lead to better performance.

The frame encoder is in charge of transforming the initial input features into deeper and more contextualized representations from which (a) the attractors will be estimated, and (b) the frame-wise activation of each speaker will be determined. Several encoder layers are used to extract such representations as shown in Figure~\ref{fig:Frame_Encoder_new_simple}. More formally, the $FrameEncoder$ can be described as:
\begin{flalign}
    &\textbf{e}_t^{(0)} = \textbf{W}_{in} \textbf{x}_t + \textbf{b}_{in}& \\
    &\textbf{E}^{(0)} = [\textbf{e}_1^{(0)}, \ldots, \textbf{e}_T^{(0)}]\\
    &\bar{\textbf{E}}^{(l)} = LN(\textbf{E}^{(l-1)})& \\
    &\hat{\textbf{E}}^{(l)} = LN(\bar{\textbf{E}}^{(l)} + MHA^{(l)}(\bar{\textbf{E}}^{(l)}, \bar{\textbf{E}}^{(l)}, \bar{\textbf{E}}^{(l)}))& \\
    &\check{\textbf{E}}^{(l)} = \hat{\textbf{E}}^{(l)} + FF^{(l)}(\hat{\textbf{E}}^{(l)})& \\
    &\textbf{E}^{(l)} = ConditionBlock(\check{\textbf{E}}^{(l)})& \label{eq:frame_encoder_layer}\\
    &FrameEncoder(\textbf{X}) = \textbf{E}^{(L)}&
\end{flalign}
where $LN$ stands for layer normalization, and $L$ is the number of layers in the frame encoder (with $1 \leq l \leq L$). $ConditionBlock$ is optional and can be ignored at this point (it will be defined in \eqref{eq:conditionblock}).

\begin{figure}[ht]
  \centering
  \includegraphics[width=0.6\linewidth]{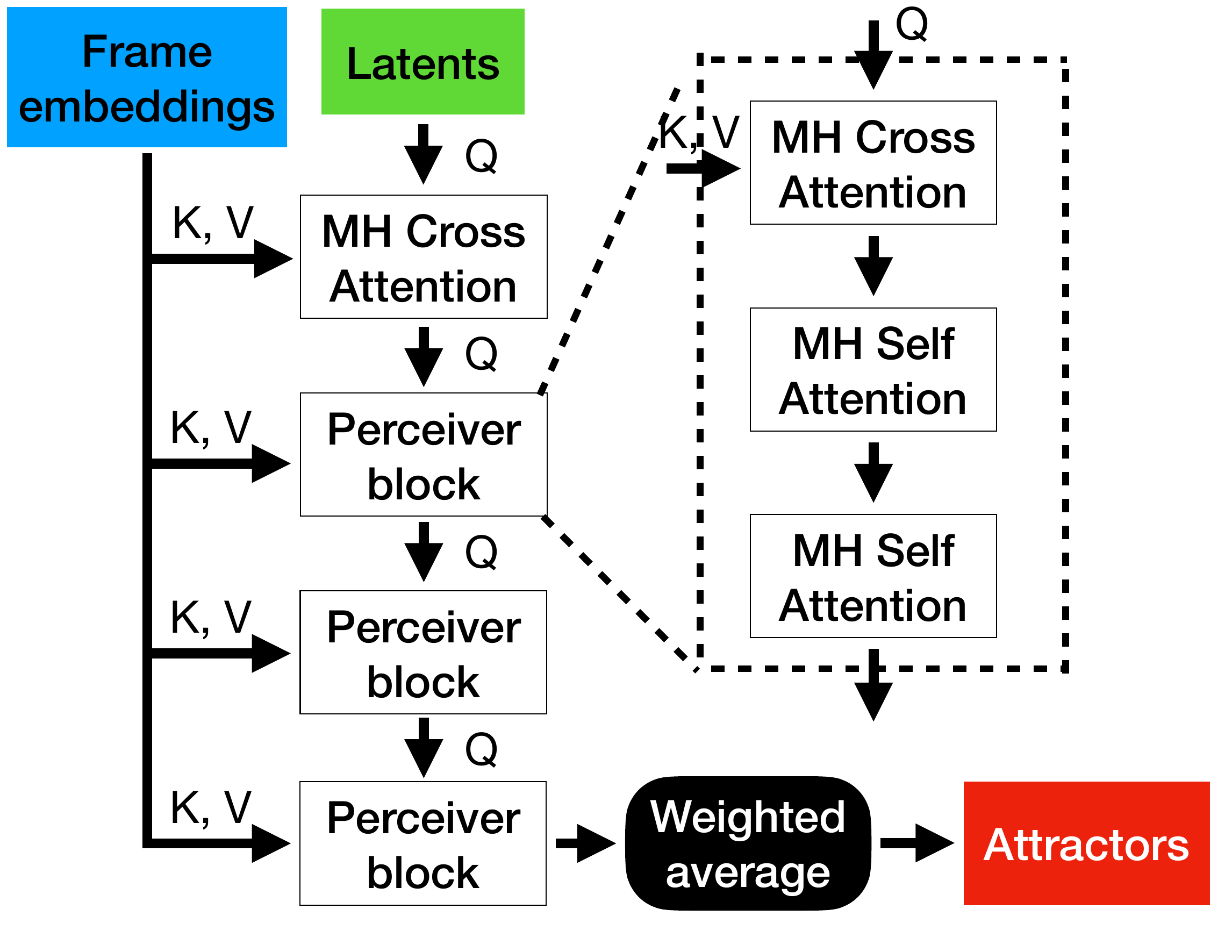}
  \caption{Scheme of Perceiver decoder.}
  \label{fig:perceiver_decoder}
\end{figure}

The decoder makes use of a chain of a few Perceiver blocks as depicted in Figure~\ref{fig:perceiver_decoder}. The set of learnable latents is transformed by each block utilizing the frame embeddings as keys and values. One could have an equal number of latents and attractors, in which case the latents are an initial representation transformed by the Perceiver blocks to obtain the attractors. In practice, we observed that this leads to instability in the training and that obtaining the attractors as linear combinations of a larger set of (transformed) latents performed better. More formally, the $PerceiverDecoder$ is defined as:
\begin{flalign}
    &\textbf{L}^{(0)} = MHA^{(0)}(\textbf{L}, \textbf{E}^{(L)}, \textbf{E}^{(L)})& \\
    &\bar{\textbf{L}}^{(b)} = MHA^{(b)}(\textbf{L}^{(b-1)}, \textbf{E}^{(L)}, \textbf{E}^{(L)})& \label{eq:cross_attention} \\
    &\hat{\textbf{L}}^{(b)} = MHA^{(b)_2}(\bar{\textbf{L}}^{(b)}, \bar{\textbf{L}}^{(b)}, \bar{\textbf{L}}^{(b)})& \\
    &\textbf{L}^{(b)} = MHA^{(b)_1}(\hat{\textbf{L}}^{(b)}, \hat{\textbf{L}}^{(b)}, \hat{\textbf{L}}^{(b)})& \\
    &\textbf{A} = PerceiverDecoder(\textbf{E}^{(L)}) = \textbf{W} \textbf{L}^{(B)},&
\end{flalign}
where $\textbf{L} \in \mathbb{R}^{L \times D}$ is the set of learnable latents, $\textbf{L}^{(b)}$ are the transformed latents after the $b$\textsuperscript{th} Perceiver block, $B$ is the number of Perceiver blocks in the decoder (with $1 \leq b \leq B$), $\textbf{W} \in \mathbb{R}^{A \times L}$ is the matrix that linearly combines latents to obtain attractors, and $\mathbf{A}$ represents the decoded attractors.

DiaPer decodes always the same fixed number of attractors, denoted by $A$. As mentioned above, the attractors are obtained as linear combinations of the latents. Therefore, the original latents are encouraged to represent information about the speakers in a general manner so that these representations can be transformed (through cross- and self-attention) given a particular input sequence in order to capture the characteristics of the speakers in the utterance. 
In standard scaled dot-product attention~\cite{vaswani2017attention}, the softmax is applied on the time-axis to normalize the attention weights along the sequence length before multiplying with the values. In Perceiver, cross- and self-attention on the latents are intertwined. When doing cross-attention (i.e. in \eqref{eq:cross_attention}), we observed slightly better performance if the softmax was applied to normalize across latents rather than along the sequence length, i.e. each frame embedding is ``probabilistically'' assigned to each latent using weights that sum up to one. This and other decisions are compared in the experimental section.

As usual for EEND-based models, the diarization loss $\mathcal{L}_d$ is calculated as 
\begin{equation}
    \label{eq:diarization_loss}
    \hat{\mathcal{L}}_d(\textbf{Y}, \hat{\textbf{Y}}) = \frac{1}{T S} \ \underset{\phi \in perm(S)}{\min} \sum_{t}^T BCE(\mathbf{y}_t^{\phi}, \hat{\mathbf{y}}_t),
\end{equation}
where considering all reference labels permutations denotes PIT loss.

Similarly to EEND-EDA, to determine which attractors are valid, an attractor existence loss $\hat{\mathcal{L}}_a$ is calculated as 
\begin{equation}
    \label{eq:attractors_loss}
    \hat{\mathcal{L}}_a(\mathbf{r}, \mathbf{p}) = BCE(\mathbf{r}^{\phi}, \mathbf{p}),
\end{equation}
using the same permutation given by $\hat{\mathcal{L}}_d$, where $\mathbf{p} = [ p_1, \ldots, p_{A} ]$ are the attractor posterior existence probabilities and $\mathbf{r}^{\phi} = [ r_1, \ldots, r_{A} ]$ are the reference presence labels $r_i \in \{0, 1\}$ for $1 \leq i \leq A$. However, unlike EEND-EDA where the loss only considers the number of speakers in the utterance plus one (for the first invalid attractor), in DiaPer, the loss is calculated for all outputs in the model (independently of the number of speakers in the utterance).

In order to encourage the model to utilize all latents, an extra ``entropy term'' $\mathcal{L}_e$ is added to the loss so that the weights that define the linear combination of latents do not become extreme values (i.e. no latent has a very high weight, therefore making all others very small):
\begin{equation}
    \mathcal{L}_e = \sum_{a=1}^{A} mean(Softmax(\textbf{w}_a) \cdot \log Softmax(\textbf{w}_a)),
\end{equation}
where $\textbf{w}_a \in \mathbb{R}^{L}$ is the row of $\textbf{W}$ corresponding to attractor $a$.

The final loss to be optimized is $\hat{\mathcal{L}} = \hat{\mathcal{L}}_d + \hat{\mathcal{L}}_a + \mathcal{L}_e$. However, we introduced two modifications.

First, in a similar way as presented in~\cite{fujita2023neural}, conditioning on frame-speaker activities is included to each layer in the frame encoder: As shown in Figure~\ref{fig:conditioning}, intermediate attractors are calculated given the frame embeddings of each frame encoder layer. The intermediate attractors are then weighted by intermediate frame activities, transformed and added into the frame embedding space to produce the conditioning. The conditioning block can be expressed as
\begin{flalign}
    & \textbf{A}^{l} = PerceiverDecoder(\hat{\mathbf{E}^{(l)}}) & \label{eq:partial_attractors}\\
    & \hat{\mathbf{Y}}^{l} = \sigma(\hat{\textbf{E}^{(l)}} \textbf{A}^{l\top}) & \label{eq:partial_activities_l}\\
    & ConditionBlock(\hat{\textbf{E}^{(l)}}) = \check{\textbf{E}}^{(l)} + \hat{\mathbf{Y}}^{l} \textbf{A}^{l} \textbf{W}_c, & \label{eq:conditionblock}
\end{flalign}
where $\textbf{A}^{l}$ are the attractors corresponding to the intermediate frame embeddings for layer $l$ in the frame encoder, $\hat{\mathbf{Y}}^{l}$ are the posterior frame-speaker activities when using the frame embeddings for layer $l$ in the frame encoder, and $\textbf{W}_c \in \mathbb{R}^{D \times D}$ is a learnable transformation matrix that also weighs the importance of the conditioning. Figure~\ref{fig:Frame_Encoder_extended_new} represents the new frame encoder including the conditioning blocks.

\begin{figure}
    \centering
    \begin{minipage}[b]{.45\textwidth}
        \centering
        \includegraphics[width=\linewidth]{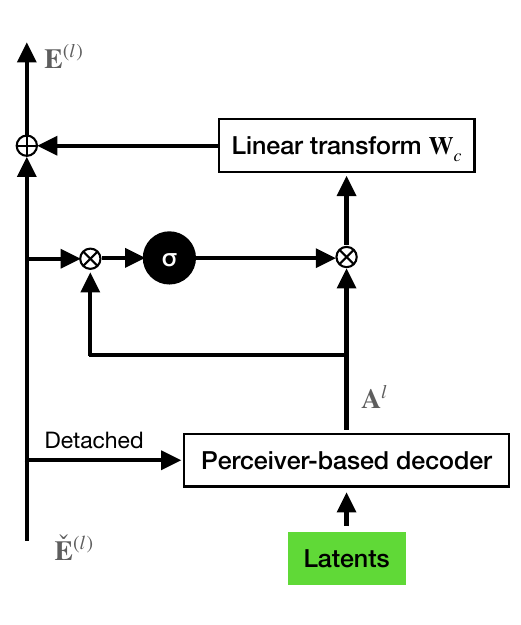}
        \caption{Conditioning scheme.}
        \label{fig:conditioning} 
    \end{minipage}
    \hfill
    \begin{minipage}[b]{.35\textwidth}
        \centering
        \includegraphics[width=\linewidth]{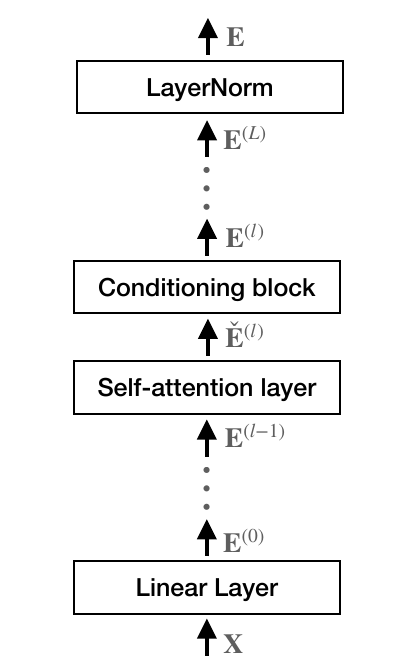}
        \captionsetup{justification=centering}
        \caption{Frame encoder scheme.}
        \label{fig:Frame_Encoder_extended_new} 
    \end{minipage}%
\end{figure}

The second modification consists in introducing auxiliary losses, inspired by other works ~\cite{yu2022auxiliary,fujita2023neural,jeoung2023improving}.
The main idea is to reinforce the diarization and attractor losses after each layer in the frame encoder and after each Perceiver block in the attractor decoder. Using $\textbf{A}^{l}$ from \eqref{eq:partial_attractors}, $\hat{\mathcal{L}}_a(\mathbf{r}, \mathbf{p}^{l})$ is calculated where $\mathbf{p}^{l}$ are the attractor existence probabilities for $\textbf{A}^{l}$; and $\hat{\mathcal{L}}_d(\textbf{Y}, \hat{\textbf{Y}}^{l})$ is calculated using the intermediate speech activities $\hat{\textbf{Y}}^{l}$ for layer $l$ as defined in \eqref{eq:partial_activities_l}. 

Analogously but for the Perceiver blocks in the attractor decoder, $\textbf{A}^{b} = \textbf{W} \textbf{L}^{(b)}$ are the attractors for the intermediate latents (after the $b$\textsuperscript{th} block) from which it is possible to evaluate $\hat{\mathcal{L}}_a(\mathbf{r}, \mathbf{p}^{b})$. Then, $\hat{\textbf{Y}}^{b} = \sigma(\textbf{E} \textbf{A}^{b\top})$ represents the intermediate speech activities and it is possible to evaluate $\hat{\mathcal{L}}_d(\textbf{Y}, \hat{\textbf{Y}}^{b})$ for block $b$. Note that in this case, the final frame embeddings $\textbf{E}$ are used while for $\hat{\mathbf{Y}}^{l}$, the attractors are decoded for the intermediate frame embeddings $\textbf{E}^{(l)}$. Other combinations, such as using $\textbf{A}^{L}$ instead of $\textbf{A}^{l}$ for $\hat{\textbf{Y}}^{l}$, would be possible but they were not studied.

The averages of the intermediate (all but the last one) losses over frame encoder layers and over Perceiver blocks are summed to the losses $\hat{\mathcal{L}}_d(\textbf{Y}, \hat{\textbf{Y}})$ and $\hat{\mathcal{L}}_a(\mathbf{r}, \mathbf{p})$ which use ``final'' attractors and ``final'' frame embeddings.
Then, $\mathcal{L}_d$ and $\mathcal{L}_a$ are obtained as
\begin{flalign}
    \mathcal{L}_d = \hat{\mathcal{L}}_d(\textbf{Y}, \hat{\textbf{Y}}) + \frac{1}{L-1} \sum_{l=1}^{L-1} \hat{\mathcal{L}}_d(\textbf{Y}, \hat{\textbf{Y}}^l) + \frac{1}{B-1} \sum_{b=1}^{B-1} \hat{\mathcal{L}}_d(\textbf{Y}, \hat{\textbf{Y}}^b) \\
    \mathcal{L}_a = \hat{\mathcal{L}}_a(\mathbf{r}, \mathbf{p}) + \frac{1}{L-1} \sum_{l=1}^{L-1} \hat{\mathcal{L}}_a(\mathbf{r}, \mathbf{p}^l) + \frac{1}{B-1} \sum_{b=1}^{B-1} \hat{\mathcal{L}}_a(\mathbf{r}, \mathbf{p}^b).
\end{flalign}

The final loss to be optimized is $\mathcal{L} = \mathcal{L}_d + \mathcal{L}_a + \mathcal{L}_e$.

One of the major disadvantages when using a non-autoregressive decoder is that the number of elements to decode (attractors in this case) has to be set in advance and this imposes a limit on the architecture. However, unlike the original versions of EEND, we do not focus on a scenario with a specific number of speakers but rather set the model to have a maximum number of attractors $A$ large enough to handle several scenarios. This is done in one way or another in all methods that handle ``flexible'' numbers of speakers. For example, in practice with EEND-EDA, one decodes a-maximum-expected-number-of-speakers attractors and later verifies which ones are valid rather than iteratively decode and check for the validity of each individual attractor. DiaPer decodes always the same number of attractors and, like in EEND-EDA~\cite{horiguchi20_interspeech}, a linear layer plus sigmoid determine which attractors are valid, i.e. correspond to a speaker in the conversation.\footnote{The effect of this is discussed in Section~\ref{sec:predictors_necessary}.}

\subsection{Experiments}

\subsubsection{Training data}
Given the conclusions of Section~\ref{sec:e2e_data}, SC are utilized to train the models. Different sets of SC were generated. To train 8\,kHz models, 10 sets were created, each with a different number of speakers per SC (ranging from 1 to 10) and each containing 2500\,h of audio. Utterances from the following sets were used: Switchboard-2 (phases I, II, III)~\cite{graff1998switchboard,graff1999switchboard,graff2002switchboard}, Switchboard Cellular (parts 1 and 2)~\cite{graff2001switchboard,graff2004switchboard}, and NIST Speaker Recognition Evaluation datasets (from years 2004, 2005, 2006, 2008)~\cite{nist20062004,nist20112005,nist20112005test,nist20112006test,nist20112006train,nist20122006test,nist20112008train,nist20112008test}. All the recordings are sampled at 8\,kHz and, out of 6381 speakers, 90\% are used for creating training data.
The Kaldi ASpIRE VAD\footnote{\url{http://kaldi-asr.org/models/m4}} is used to obtain time annotations (in turn used to produce reference diarization labels).
To augment the training data, we use 37 noises from MUSAN~\cite{snyder2015musan} labeled as ``background''. They are added to the signal respecting a signal-to-noise ratio selected randomly from \{5, 10, 15, 20\} dB.

In order to train 16\,kHz models, a similar strategy was followed to also generate SC with different numbers of speakers ranging from 1 to 10 per conversation, all containing 2500\,h of audio. Instead of telephone conversations, utterances were taken from LibriSpeech~\cite{panayotov2015librispeech} which consists of 1000 hours of read English speech from almost 2500 speakers. More details can be found in Section~\ref{sec:sc_wideband}. The same VAD as described above was used to produce annotations and equivalent background noises were used, but in 16\,kHz.

Statistics about the whole sets and the 2-speaker ones can be seen in Table~\ref{tab:datasets_stats_SC}. Naturally, SC with more speakers tend to have more overlap.
Since the algorithm for creating SC uses the whole recordings, and LibriSpeech recordings are usually longer than telephone conversations, this results in longer utterances for SC LibriSpeech. The longer turns in LibriSpeech also explain why SC generated with such data have proportionally less silence. However, the amount of audio is about the same for similar conditions (i.e. same number of speakers).

\begin{table}[H]
\setlength{\tabcolsep}{3pt} 
\centering
\caption{Statistics for synthetic datasets. }
\label{tab:datasets_stats_SC}
\setlength{\tabcolsep}{5pt} 
  \begin{tabular}{@{}
                  lccc|ccc
                  @{}}
  \toprule
        \multirow{2}{*}{Dataset}       & 
        \multirow{2}{*}{\#files} & Total & Average &
        \multicolumn{3}{c}{Average \%} \\
        & & audio (h) & dur. (s) & silence & 1-speaker & overlap \\
  \midrule
  SC telephone 2 speakers & 24967 & 2481 & 357.68 & 12.78 & 78.87 & 8.35 \\
  SC telephone 1-10 speakers & 148732 & 24807 & 600.45 & 9.69 & 76.15 & 14.16 \\
  SC LibriSpeech 2 speakers & 7158 & 2480 & 1247.51 & 10.09 & 82.48 & 7.43 \\
  SC LibriSpeech 1-10 speakers & 35558 & 23385 & 2367.57 & 7.23 & 79.51 & 13.24 \\
   \bottomrule
  \end{tabular}
\end{table}

\subsubsection{Models}
As the main baseline for this work, we utilize end-to-end neural diarization with encoder-decoder attractors (EEND-EDA)~\cite{horiguchi20_interspeech} which is the most popular EEND approach that can handle multiple speakers and was used in previous sections. Its architecture was exactly the same as that described in~\cite{horiguchi20_interspeech} and we used our PyTorch implementation\footnote{\url{https://github.com/BUTSpeechFIT/EEND}}. 15 consecutive frames of 23-dimensional log Mel-filterbanks (computed over 25\,ms every 10\,ms) are stacked to produce 345-dimensional features every 100\,ms. These are transformed by the frame encoder, comprised of 4 self-attention encoder blocks (with 4 attention heads each) into a sequence of 256-dimensional embeddings. These are then shuffled in time and fed into the LSTM-based encoder-decoder module that decodes attractors, which are deemed as valid if their existence probability is above a certain threshold. The sigmoid function applied on the dot product between attractors and frame embeddings is used to obtain speech activity probabilities for each speaker (represented by a valid attractor) at each time step (represented by an embedding). 

Part of the setup for DiaPer is shared with the baseline, namely the input features, the frame encoder configuration (except in experiments where the number of layers was changed). In the mechanism for determining attractor existence, DiaPer also utilizes a linear layer followed by sigmoid applied on the attractors but they are processed independently. This is in contrast with EEND-EDA where the order of the attractors matters as they are decoded until the existence probablity is below a certain threshold.

Following standard practice with EEND models, the training scheme starts with training the model first on synthetic training data and proceeds with fine-tuning (FT) using a small development set of real data of the same domain as the test set. In the experiments with more than two speakers, a model initially trained on synthetic data with two speakers per recording is first adapted to a synthetic set with a variable number of speakers, and finally fine-tuned to a development set. 

\subsubsection{Evaluation data}

For the first parts of the analysis, where we show the architectural choices of DiaPer and where we compare DiaPer with EEND-EDA, we utilize two 2-speaker conversational recordings. For that, we use the setup from Section~\ref{sec:2spkexp} based on Callhome 2-speaker subsets and DIHARD3 CTS full. We also utilize the Callhome corpus with multiple speakers.

In order to compare DiaPer and EEND-EDA in terms of finding the right numbers of speakers, a set of SC is used consisting of 10 recordings for each number of speakers from 1 to 10, totalling 100 recordings.

Finally, for the analysis on wide-band sets, we use the same corpora as used for the modular system (Section~\ref{sec:vbx_evaluation_data}).

\subsubsection{Training}

Most trainings were run on a single GPU. The batch size was set to 32 with 200000 minibatch updates of warm-up. Following~\cite{horiguchi20_interspeech}, the Adam optimizer~\cite{kingma2014adam} was used and scheduled with noam~\cite{vaswani2017attention}. For a few trainings with variable number of speakers where 4 GPUs were used, the batchsize and warm-up steps were adapted accordingly (i.e. divided by 4). Other hyperparameters (e.g. dropout, learning rate) can be seen in the training configuration files shared in the repository.

For FT on a development set, the Adam optimizer was used. Both EEND-EDA and DiaPer were fine-tuned with learning rate $10^{-5}$ for Callhome 2 speakers and with $10^{-4}$ for whole Callhome and DIHARD3 CTS. For the wide-band datasets, DiaPer was fine-tuned on the training set using learning rate $10^{-6}$ until the performance on the development set stopped improving; or, in case there was no official training set available, the fine-tuning was performed on the development set till not further improvement on the test set.

During training (with 2-speaker SC), adaptation (with a variable number of speakers SC), and FT (with in-domain data), batches were formed by sequences of 600 Mel-filterbank outputs, corresponding to 1 minute, unless specified otherwise (i.e. the analysis in Section~\ref{sec:wideband}). These sequences are randomly selected from the generated SC\footnote{An attentive reader will notice that it might not be possible to see as many as 10 speakers in 1 minute, this is addressed in the experimental section.}. During inference, the full recordings are fed to the network one at a time. In all cases, when evaluating a given epoch, the parameters from the checkpoints of the previous 10 epochs are averaged to run the inference.

To compare EEND-EDA and DiaPer on equal ground, we train both models for the same number of epochs, evaluate them after regular intervals and choose the best-performing on the development set. For comparisons on 2-speaker scenarios of Callhome, each model is trained for 100 epochs on telephony SC. Every 10 epochs, the parameters of the 10 previous checkpoints are averaged and performance is evaluated on the CH1-2spk set to determine the best one. The performance of such model is reported on the CH2-2spk set and DIHARD3 CTS full eval before and after FT.

When doing adaptation to more speakers for comparison on Callhome, the best-performing 2-speaker model as described above is selected as initialization. The adaptation to an SC set with different numbers of speakers per recording is run for 75 epochs. Using a sliding window, the parameters of the last 10 models are averaged every 5 epochs and performance is evaluated on CH1 to determine the best one.  The performance of such model is reported on CH2. This model is also used as initialization when doing FT to a development set. To avoid selecting results on the test sets, all fine-tunings are run for 20 epochs and the parameters of the last 10 epochs are averaged to produce the final model.

For comparisons on the variety of wide-band sets, three variants of DiaPer are trained. An 8\,kHz model following a similar approach as described above: trained for 100 epochs on SC of 2 speakers created with telephony speech and then adapted to the SC with 1-10 speakers set for 100 epochs. The 16\,kHz model is trained in the same manner but using SC generated from LibriSpeech. Two flavors of this ``wide-band DiaPer'' are used, one with 10 attractors and another with 20 attractors to analyze the impact on datasets with several speakers. For the comparisons on wide-band sets, results are also shown without and with FT.

\subsubsection{Metrics}
Diarization performance is evaluated in terms of DER as defined by NIST~\cite{NISTRT} and computed using dscore\footnote{\url{https://github.com/nryant/dscore}}.
During inference time, the model outputs are thresholded at 0.5 to determine speech activities. For evaluation sets where a forgiveness collar is used for DER, a median filter with window 11 (corresponding to 1.1\,s) is applied as post-processing over the speech activities. If the forgiveness collar is 0\,s, no filtering is applied and, instead of running the inference with 10 frames subsampling in the frame encoder, 5 frames only are subsampled as this provides a better resolution (50\,ms vs. 100\,ms) in the output. However, due to the high memory consumption, when processing very long files for CHiME6, a subsampling of 15 frames had to be used.
To analyze the model quality in terms of finding the correct number of speakers, confusion matrices for correct/predicted numbers of speakers are presented for SC with 10 recordings for each number of speakers from 1 to 10.

\subsubsection{Selection of parameters}
In order to shed some light on the influence of different aspects of the architecture in DiaPer, we present first a comparison of the performance when varying some key elements. We start from the best configuration we found, namely: 3 Perceiver blocks in the attractor decoder, 128 latents, 4 self-attention layers in the frame-encoder and 128-dimensional latents, frame embeddings and attractors. This configuration is marked with a gray background in the comparisons. The models are trained on 2-speaker SC and no FT is applied.

Table~\ref{tab:perceiver_blocks_comparison} shows the impact of the number of Perceiver blocks in the attractor decoder. Out of the configurations explored, having 3 blocks presents the best performance.

\begin{table}[H]
  \centering
  \caption{Comparison on CH1-2spk when varying the number of Perceiver blocks in the attractor decoder.}
  \label{tab:perceiver_blocks_comparison}
  \setlength{\tabcolsep}{4pt} 
  \begin{tabular}{@{}
                  l
                  S[table-format=1.2] 
                  S[table-format=1.2] 
                  S[table-format=1.2] 
                  S[table-format=1.2] 
                  S[table-format=1.2] 
                  @{}}
  \toprule
   \# Blocks & \multicolumn{1}{c}{1} & \multicolumn{1}{c}{2} & \multicolumn{1}{c}{\cellcolor{lightgray}3} & \multicolumn{1}{c}{4} & \multicolumn{1}{c}{5} \\ 
    \midrule
   DER (\%) & 8.27 & 8.41 & \cellcolor{lightgray}7.96 & 8.44 & 8.09 \\
   \# Parameters (M) & 3.1 & 3.7 & \cellcolor{lightgray}4.3 & 4.9 & 5.5\\
  \bottomrule
  \end{tabular}
\end{table}

Table~\ref{tab:number_latents_comparison} shows how the number of latents can affect the performance. Differences are small for all amounts equal to or below 256, even with as few as 8. Nevertheless, given that the number of parameters is very similar for any configuration, we keep 128 latents as having more could ease the task when more speakers appear in a recording.

\begin{table}[H]
  \centering
  \caption{Comparison on CH1-2spk when varying the number of latents.}
  \label{tab:number_latents_comparison}
  \setlength{\tabcolsep}{4pt} 
  \begin{tabular}{@{}
                  l
                  S[table-format=1.2]  
                  S[table-format=1.2]
                  S[table-format=1.2]
                  S[table-format=1.2]
                  S[table-format=1.2]
                  S[table-format=1.2]
                  S[table-format=1.2]
                  @{}}
  \toprule
  \# Latents & \multicolumn{1}{c}{8} & \multicolumn{1}{c}{16} & \multicolumn{1}{c}{32} & \multicolumn{1}{c}{64} & \multicolumn{1}{c}{\cellcolor{lightgray} 128} & \multicolumn{1}{c}{256} & \multicolumn{1}{c}{512} \\
  \midrule
  DER (\%) & 8.15 & 8.14 & 8.29 & 8.10 & \cellcolor{lightgray}7.96 & 8.10 & 8.54 \\
  \# Parameters (M) & 4.29 & 4.29 & 4.29 & 4.30 & \cellcolor{lightgray}4.31 & 4.32 & 4.36 \\
  \bottomrule
  \end{tabular}
\end{table}

Table~\ref{tab:number_layers_frameencoder_comparison} presents a comparison when varying the number of layers in the frame encoder. Standard SA-EEND and EEND-EDA use 4 and some works have used 6 layers. In the case of DiaPer, we do not observe large differences in the performance and obtain the best performance with 4.

\begin{table}[H]
  \centering
  \caption{Comparison on CH1-2spk when varying the number of layers in frame encoder.}
  \label{tab:number_layers_frameencoder_comparison}
  \setlength{\tabcolsep}{4pt} 
  \begin{tabular}{@{}
                  l
                  S[table-format=1.2]  
                  S[table-format=1.2]  
                  S[table-format=1.2]  
                  S[table-format=1.2] 
                  @{}}
  \toprule
        \# Layers & \multicolumn{1}{c}{3} & \multicolumn{1}{c}{\cellcolor{lightgray}4} & \multicolumn{1}{c}{5} & \multicolumn{1}{c}{6} \\
        \midrule
        DER (\%) & 8.18 & \cellcolor{lightgray}7.96 & 8.33 & 8.31 \\
        \# Parameters (M) & 3.7 & \cellcolor{lightgray}4.3 & 4.9 & 5.5\\
  \bottomrule
  \end{tabular}
\end{table}

Finally, Table~\ref{tab:latent_dimension_comparison} shows the impact of the model dimensions on the performance. Increasing the dimensionality of latents, frame embeddings and attractors further than 128 does not show improvements in terms of DER but increases the number of model parameters significantly. Figure~\ref{fig:latents_dimensions_part1} shows performance throughout the epochs for the development set. It is clear how more dimensions allow for a faster convergence; however, more than 128 do not provide gains in terms of final performance. In addition, more dimensions make the training less stable: using 512 would always lead to instability. Configurations with less than 128 dimensions (64 and 32) can improve further and after 200 epochs reduce the DER by about 1 point but still with worse final results than other configurations. These findings show that reasonable performances can be achieved even with more lightweight versions of DiaPer.

\begin{figure}[H]
    \centering
    \includegraphics[width=0.7\linewidth]{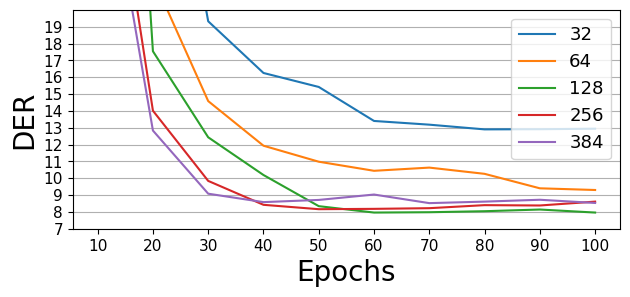}
    \caption{Performance on CH1-2spk for different model dimensions (latents, frame embeddings and attractors).}
    \label{fig:latents_dimensions_part1}
\end{figure}

\begin{table}[H]
  \centering
  \caption{Comparison on CH1-2spk when varying the model dimension (latents, frame embeddings and attractors).}
  \label{tab:latent_dimension_comparison}
  \setlength{\tabcolsep}{4pt} 
  \begin{tabular}{@{}
                  l
                  S[table-format=2.2]  
                  S[table-format=2.2]  
                  S[table-format=2.2]  
                  S[table-format=2.2]  
                  S[table-format=2.2]
                  @{}}
  \toprule
        Dimensions & \multicolumn{1}{c}{32} & \multicolumn{1}{c}{64} & \multicolumn{1}{c}{\cellcolor{lightgray}128} & \multicolumn{1}{c}{256} & \multicolumn{1}{c}{384}\\ 
    \midrule
        DER (\%) & 12.90 & 9.30 & \cellcolor{lightgray}7.96 & 8.16 & 8.52 \\
        \# Parameters (M) & 0.7 & 1.6 & \cellcolor{lightgray}4.3 & 12.9 & 26.6 \\
  \bottomrule
  \end{tabular}
\end{table}

\subsubsection{Ablation analysis}
Different decisions were made when developing DiaPer and some have a big impact on the performance. Table~\ref{tab:ablation_comparison} presents a comparison of DiaPer in the best configuration shown above and when removing some of the operations performed during training. The first one refers to the normalization of the loss (see \eqref{eq:diarization_loss}) by the reference number of speakers.
DiaPer always outputs $A$ attractors and the loss is calculated for all of them, even if only training with 2-speakers SC. If the loss is not normalized by the number of speakers, the model tends to find less speech, increasing the missed speech rate considerably. 

Another ablation targets the frame encoder conditioning shown in Figure~\ref{fig:Frame_Encoder_extended_new}. Similarly to \cite{fujita2023neural}, where the scheme was introduced, removing it worsens the performance by around 0.5\% DER. Comparable degradation is observed by removing the loss reinforcements in both frame encoder and Perceiver blocks.

Finally, the attention normalization in the cross-attention calculations inside the Perceiver blocks is performed across latents in DiaPer. If done across time, as it is usually done, the performance is slightly worse. We have also explored using across-time normalization in half of the heads and across-latents in the other half but the performance was not better than using across-latents in all heads.

\begin{table}[H]
  \centering
  \caption{DER (\%) on CH1-2spk with different ablations.}
  \label{tab:ablation_comparison}
  \setlength{\tabcolsep}{4pt} 
  \begin{tabular}{@{}
                  l
                  S[table-format=2.2] 
                  @{}}
  \toprule
  DiaPer & 7.96 \\
  \midrule
  Without normalization of loss per \#speakers & 11.10 \\
  Without frame encoder conditioning & 8.55 \\
  Without intermediate loss in frame encoder & 8.53 \\
  Without intermediate loss in Perceiver blocks & 8.43 \\
  Perceiver cross-attention across time (instead of latents) & 8.07 \\
  \bottomrule
  \end{tabular}
\end{table}

While publications always focus on the positive aspects of the models, we believe there is substantial value in sharing those options that were explored and did not provide gains. Among them were:

\begin{itemize}[leftmargin=0.4cm]
    \item using absolute positional encoding when feeding the frame embeddings into the attractor decoder (no improvement). Perhaps due to the training vs. test mismatch regarding input length.
    \item using specaugment for data augmentation (no improvement).
    \item following \cite{maiti2021end,rybicka2022end}, adding a speaker recognition loss to reinforce speaker discriminative attractors (slightly worse results). However, this was only explored in the scenario with 2 speakers and the effect might be different for recordings with more speakers.
    \item following \cite{wang2023told}, including an LSTM-based mechanism to model output speaker activities through time (worse performance).
    \item modeling silence with a specific attractor (worse performance).
    \item length normalization of frame embeddings and attractors before performing dot-product to effectively compute cosine similarity (worse performance).
    \item using cross-attention to compare frame embeddings and attractors instead of dot-product (worse performance).
    \item as analyzed in \cite{du2021speaker,wang2023told,plaquet23_interspeech}, using power set encoding to model the diarization problem instead of per-frame per-speakers activities (worse performance). We believe that the reason for this approach not to work with DiaPer is that, when handling many speakers, the number of classes in the power set is too high and most of them are not well represented. This approach has much more potential in scenarios with limited numbers of of speakers as shown in~\cite{plaquet23_interspeech}.
\end{itemize}
Implementations of most of these variants can be found in our public repositories in \url{https://github.com/BUTSpeechFIT/DiaPer} to enable others to easily experiment with them.

\subsubsection{Two-speaker telephone conversations}
Even though DiaPer is specifically designed for the scenario with multiple speakers, as it is common practice, in this section we first present results for the 2-speaker telephone scenario. It should be noted that both EEND-EDA and DiaPer, when trained only with 2-speaker SC, learn to only output activities for 2 speakers, even if they are prepared to handle a variable number of them. Figure~\ref{fig:telephone_two_speakers} compares the performance on two sets before and after FT to the in-domain development set. Both EEND-EDA and DiaPer were trained on the same data with 5 different parameter initializations to produce the error bars. Results show that DiaPer can reach better performance on both datasets, both with and without FT.

\begin{figure}[H]
    \centering
    \begin{subfigure}[b]{0.5\textwidth}
        \centering
        \includegraphics[width=\linewidth]{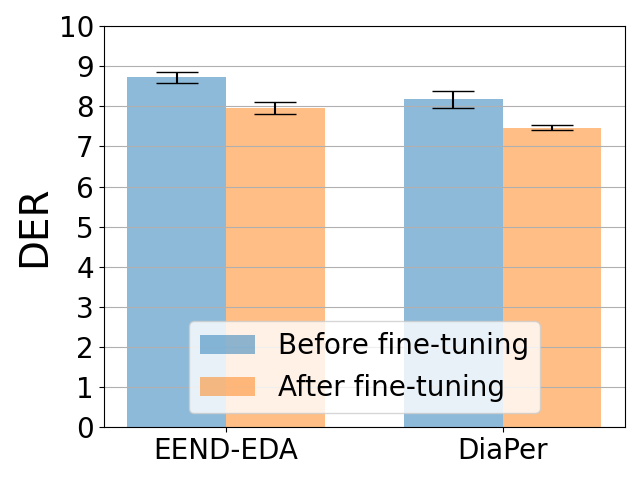}
        \caption{CH Part 2 (2 speakers).}
    \end{subfigure}%
    \hfill
    \begin{subfigure}[b]{0.5\textwidth}
        \centering
        \includegraphics[width=\linewidth]{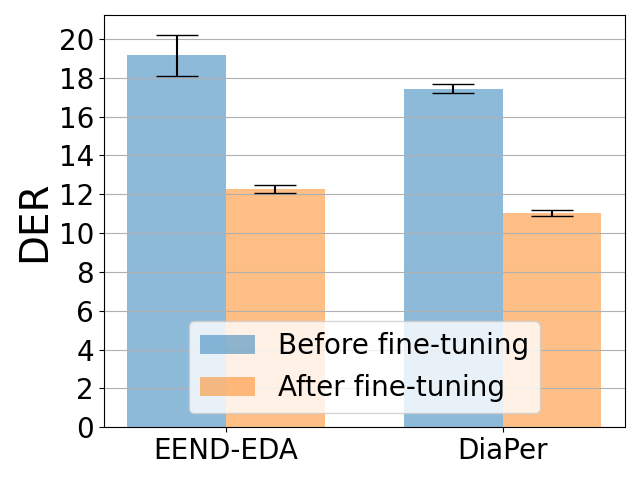}
        \caption{DH3 CTS full eval.}
    \end{subfigure}
    \caption{DER (\%) for telephone recordings of Callhome and DIHARD3 conversational telephone speech (CTS) with 2 speakers.}
    \label{fig:telephone_two_speakers}
\end{figure}

Table~\ref{tab:ch2two_comparison} presents a comparison between EEND-EDA and DiaPer without and with FT. We can see that even though DiaPer is devised to perform better when handling multiple speakers, it can reach a better performance than its EEND-EDA counterpart on the 2-speaker telephone scenario, with fewer parameters.

\begin{table}[H]
    \caption[capt]{DER (\%) comparison on CH2-2spk. EEND-EDA and DiaPer are trained with SC with 2 speakers per conversation. For each method (EEND-EDA and DiaPer), the best model on CH1-2spk out of 5 runs is selected.}
    \label{tab:ch2two_comparison}
    \setlength{\tabcolsep}{6pt} 
    \centering
    \begin{tabular}{@{}
                  l 
                  c
                  c
                  S[table-format=2.2] 
                  S[table-format=2.2]
                  @{}}
    \toprule
    \multirow{2}{*}{System} & \#Param. & Data & \multicolumn{1}{c}{No} & \multicolumn{1}{c}{With} \\
    &  (Million) & (kHour) & \multicolumn{1}{c}{FT} & \multicolumn{1}{c}{FT} \\
    \midrule
    EEND-EDA & 6.4 & 2.5 & 8.77 & 7.96 \\
    DiaPer & 4.6 & 2.5 & 8.05 & 7.51 \\
    \bottomrule
  \end{tabular}
\end{table}

Figure~\ref{fig:inferencetime} presents a comparison between the standard EEND-EDA baseline and DiaPer inference times. Although DiaPer is slower for very short recordings, it can run faster than the standard EEND-EDA when processing several-minute recordings. 
This speed-up is mainly given by the more light-weight nature of DiaPer which results in a faster frame encoder processing, which dominates the computation time versus the attractor decoding in both models. To have a fair comparison, ``EEND-EDA small'' denotes a version of EEND-EDA where the model dimension matches that of DiaPer in its best configuration (128-dimensional frame embeddings and attractors). This corresponds to using the same frame encoder in both models and we can see that DiaPer is slightly slower due to more computations in the attractor decoder. It should be noted that EEND-EDA small performs slightly worse than EEND-EDA in terms of DER. This was not the case for DiaPer and with the smaller configuration, we are able to obtain better DER performance than EEND-EDA and run faster.

\begin{figure}
    \centering
    \includegraphics[width=0.7\linewidth]{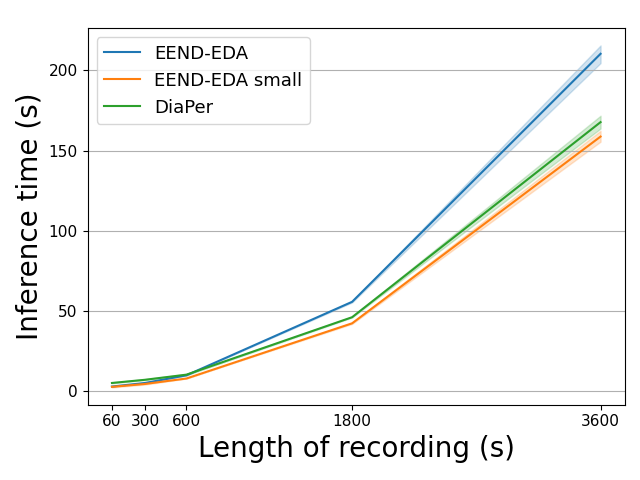}
    \caption{Inference time for EEND-EDA and DiaPer for recordings from 1 minute to 1 hour running 5 times each inference with a downsampling factor of 10. In black is the percentage of time taken by DiaPer wrt EEND-EDA. Ran on Intel(R) Xeon(R) CPU E5-2680 v4 @ 2.40GHz.}
    \label{fig:inferencetime}
\end{figure}

\subsubsection{Multiple-speakers telephone conversations}
Figure~\ref{fig:telephone_more_speakers} presents the comparison for recordings with multiple numbers of speakers where EEND-EDA and DiaPer are trained on the same data. Once again, DiaPer presents significant advantages over EEND-EDA both before and after fine-tuning to the development set. Table~\ref{tab:ch2_per_speaker} shows the DER for different numbers of speakers per conversation where gains are observed in almost all cases. The largest differences are for recordings with more speakers, suggesting the superiority of DiaPer in handling such situations.

\begin{figure}[H]
    \centering
    \includegraphics[width=0.55\linewidth]{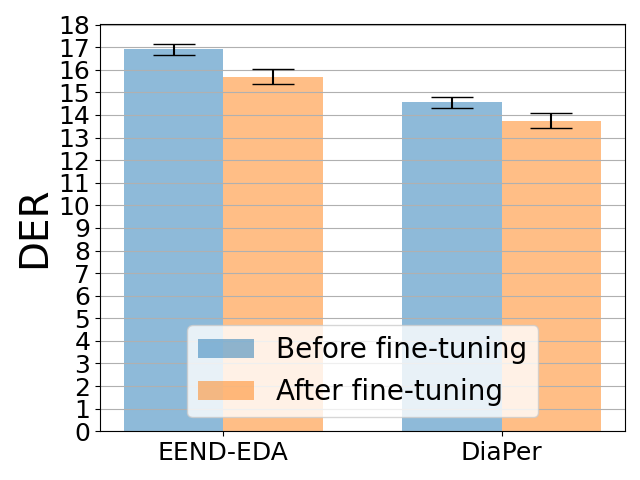}
    \caption{DER (\%) for CH Part 2 with varying number of speakers.}
    \label{fig:telephone_more_speakers}
\end{figure}

\begin{table}[H]
    \caption{DER (\%) comparison on CH2. For each method (EEND-EDA and DiaPer), the best model on CH1 out of 5 runs is selected.}
    \label{tab:ch2_per_speaker}
    \setlength{\tabcolsep}{5pt} 
    \centering
    \begin{tabular}{@{}
                  l |
                  S[table-format=2.2] |
                  S[table-format=2.2]
                  S[table-format=2.2]
                  S[table-format=2.2] 
                  S[table-format=2.2]
                  S[table-format=2.2] 
                  @{}}
    \toprule
    System & All & \multicolumn{1}{c}{2-spk} & \multicolumn{1}{c}{3-spk} & \multicolumn{1}{c}{4-spk} & \multicolumn{1}{c}{5-spk} & \multicolumn{1}{c}{6-spk} \\
    \midrule
    EEND-EDA & 16.70 & 8.99 & 13.84 & 24.57 & 33.10 & 46.25 \\
    \hspace{0.25cm}+ FT CH1 & 15.29 & 7.54 & 14.01 & 20.84 & 33.34 & 41.36 \\ 
    \midrule
    
    DiaPer & 14.86 & 9.10 & 12.70 & 19.18 & 29.52 & 41.81 \\
    \hspace{0.25cm}+ FT CH1 & 13.60 & 7.39 & 12.08 & 19.62 & 30.25 & 28.84 \\ 
    \bottomrule
  \end{tabular}
\end{table}

Finally, Table~\ref{tab:ch2_der_detail} shows the comparison of DER components. It can be observed that 
without fine-tuning, DiaPer does not improve the confusion error of EEND-EDA but rather fixes missed and false alarm (FA) speech. A closer look at the inherent VAD and OSD performances of the two models allows us to see that DiaPer improves considerably the OSD recall with similar OSD precision. Therefore, most of the improvement is related to more accurate overlapped speech detection. Nevertheless, it should be pointed out that precision and recall slightly above 50\% are still very low. There is clearly large room for improving the performance in this aspect.

\begin{table}[H]
    \caption{Comparison on CH2. For each method (EEND-EDA and DiaPer), the best model on CH1 out of 5 runs is selected. DER and its three components, and precision and recall for VAD and OSD performance are reported.}
    \label{tab:ch2_der_detail}
    \setlength{\tabcolsep}{3.5pt} 
    \centering
    \begin{tabular}{@{}
                  l |
                  S[table-format=2.2] |
                  S[table-format=1.2]
                  S[table-format=1.2]
                  S[table-format=1.2] |
                  S[table-format=2.1]
                  S[table-format=2.1] |
                  S[table-format=2.1]
                  S[table-format=2.1] 
                  @{}}
    \toprule
    \multirow{2}{*}{System} & \multicolumn{1}{c|}{DER} & \multicolumn{1}{c}{Miss} & \multicolumn{1}{c}{FA} & \multicolumn{1}{c|}{Conf.} & \multicolumn{2}{c|}{VAD} & \multicolumn{2}{c}{OSD} \\
     & \multicolumn{1}{c|}{(\%)} & \multicolumn{1}{c}{(\%)} & \multicolumn{1}{c}{(\%)} & \multicolumn{1}{c|}{(\%)} & \multicolumn{1}{c}{P (\%)} & \multicolumn{1}{c|}{R (\%)} & \multicolumn{1}{c}{P (\%)} & \multicolumn{1}{c}{R (\%)} \\
    \midrule
    EEND-EDA & 16.70 & 7.08 & 4.88 & 4.73 & 93.3 & 97.6 & 50.0 & 41.9 \\
    \hspace{0.25cm}+ FT CH1 & 15.29 & 8.24 & 2.61 & 4.44 & 95.8 & 94.5 & 63.8 & 38.3 \\ 
    \midrule
    DiaPer & 14.86 & 6.16 & 3.90 & 4.80 & 93.1 & 98.1 & 51.5 & 52.1 \\
    \hspace{0.25cm}+ FT CH1 & 13.60 & 7.80 & 2.06 & 3.74 & 95.4 & 95.3 & 64.1 & 44.8 \\ 
    \bottomrule
  \end{tabular}
\end{table}

\begin{figure}[H]
    \centering
    \begin{subfigure}[b]{0.4\textwidth}
        \centering
        \includegraphics[width=\linewidth]{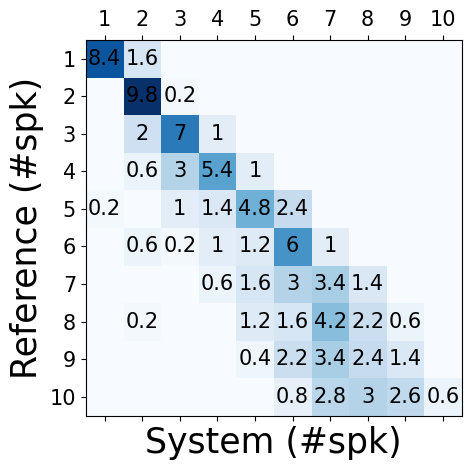}
        \caption{EEND-EDA.}
    \end{subfigure}
    \hfill
    \begin{subfigure}[b]{0.4\textwidth}
        \centering
        \includegraphics[width=\linewidth]{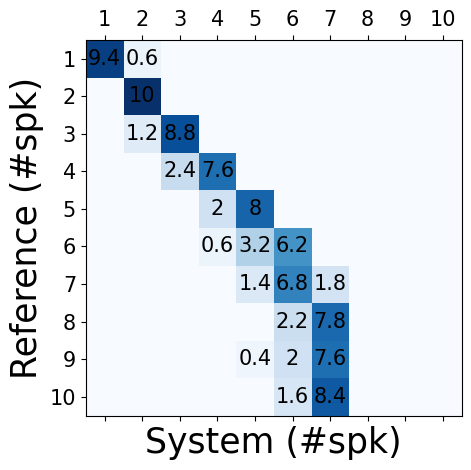}
        \caption{DiaPer.}
    \end{subfigure}
        
    \begin{subfigure}[b]{0.4\textwidth}
        \centering
        \includegraphics[width=\linewidth]{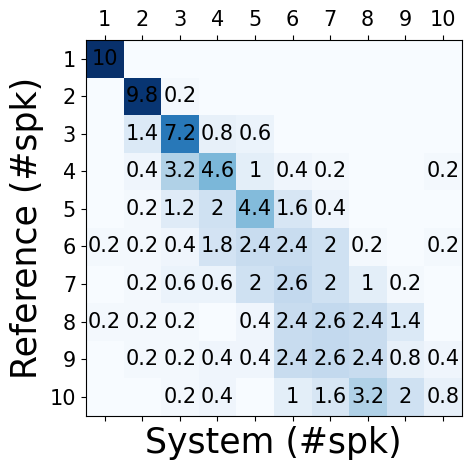}
        \caption{EEND-EDA.}
    \end{subfigure}
    \hfill
    \begin{subfigure}[b]{0.4\textwidth}
        \centering
        \includegraphics[width=\linewidth]{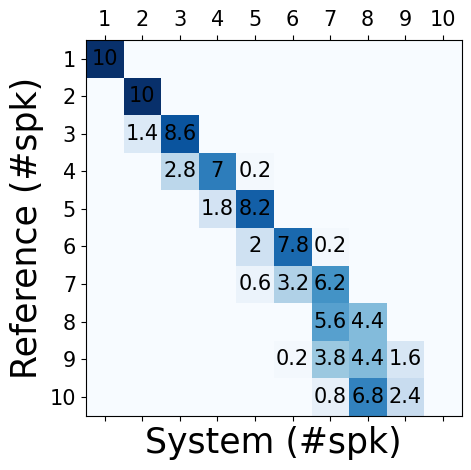}
        \caption{DiaPer.}
    \end{subfigure}
    
    \caption{Confusion matrices averaging five models evaluated on SC when adapted for 50 epochs with 2-7 speakers (top) and 1-10 speakers (bottom).}
    \label{fig:confmatrixavg}
\end{figure}

EEND-EDA has been shown to have problems handling several speakers (i.e. not being able to find more than the number seen in training and significantly miscalculating the number of speakers when more than 3 are present in a conversation)~\cite{horiguchi2021towards,horiguchi2022encoder}. To compare DiaPer's performance in this sense, we trained 5 of both such models with the same procedure but different initialization seeds and evaluated them on a set of 100 SC with 10 recordings for each number of speakers from 1 to 10. Confusion matrices between the numbers of real (reference) speakers and the numbers found by the system were calculated for each model. The averages of such confusion matrices for the 5 DiaPer and 5 EEND-EDA models are presented in Figure~\ref{fig:confmatrixavg}. Although both EEND-EDA and DiaPer are trained on the same data with only up to 7 speakers per SC (matrices above), EEND-EDA is able to find more speakers. Yet, DiaPer is considerably more accurate for SC with up to 6 speakers. When both EEND-EDA and DiaPer are trained with up to 10 speakers per SC (matrices below), we can see that DiaPer is still considerably more accurate. However, its performance is limited when the number of speakers is 8 or more.

\begin{figure}[H]
    \centering
    \begin{subfigure}[b]{0.4\textwidth}
        \centering
        \includegraphics[width=\linewidth]{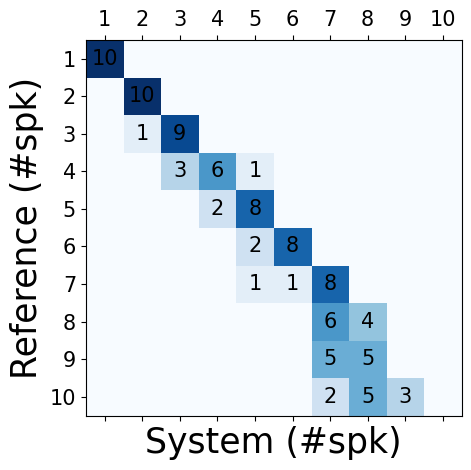}
        \caption{1 minute, 50 epochs.}
    \end{subfigure}
    \hfill
    \begin{subfigure}[b]{0.4\textwidth}
        \centering
        \includegraphics[width=\linewidth]{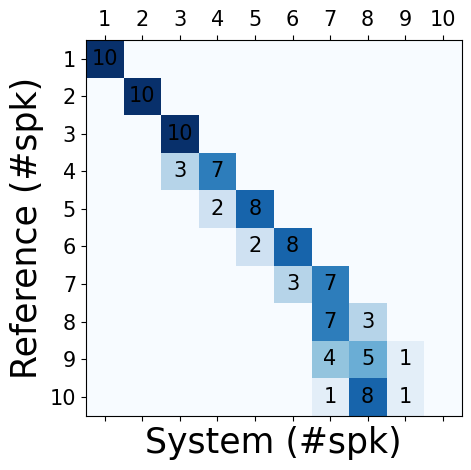}
        \caption{1 minute, 100 epochs.}
    \end{subfigure}
    
    \begin{subfigure}[b]{0.4\textwidth}
        \centering
        \includegraphics[width=\linewidth]{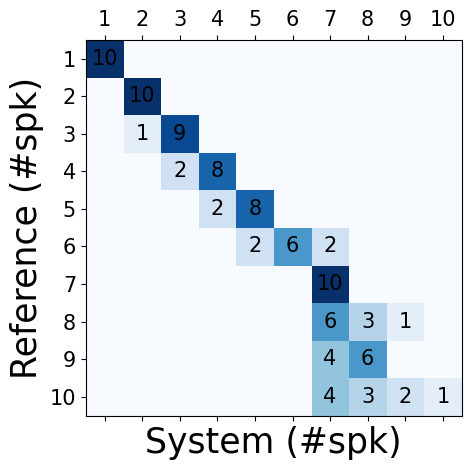}
        \caption{4 minutes, 50 epochs.}
    \end{subfigure}
    \hfill
    \begin{subfigure}[b]{0.4\textwidth}
        \centering
        \includegraphics[width=\linewidth]{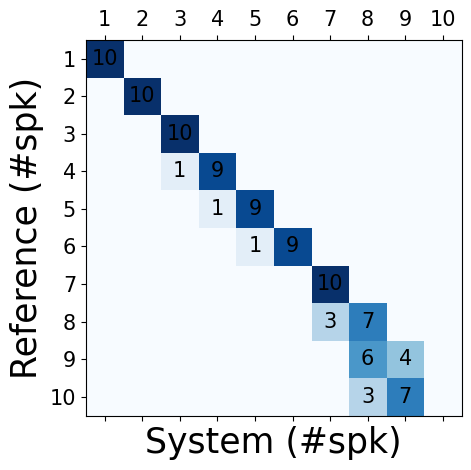}
        \caption{4 minutes, 100 epochs.}
    \end{subfigure}
    
    \caption{Confusion matrices for DiaPer adapted to telephony SC with 1 to 10 speakers per recording using different sequence lengths to create the batches: 1 minute (top) and 4 minutes (bottom).}
    \label{fig:confmatrixminutes}
\end{figure}

One element to consider is that all the models above were trained and adapted using batches of 1-minute-long sequences. The 1-minute training examples were randomly selected from the simulated conversations so it is possible that not all speakers in the conversations appeared in each sample. Furthermore, it is less likely for 10 speakers in a simulated conversation to be heard in only one minute. For this reason, we also performed adaptation of one model using 4-minute-long sequences. While sequences of 1 minute have on average 3.6 speakers, sequences of 4 minutes have 5.2, allowing the model to see a higher number of speakers per training sample on average during training. A comparison is presented in Figure~\ref{fig:confmatrixminutes} after 50 and 100 epochs training with 1 and 4 minutes sequences. A slight improvement is observed when using 4 minutes after 50 epochs but such advantage increases after 100 epochs. Nevertheless, it is observed that the model never predicts 10 speakers. This is most likely due to the low percentage of 4-minute sequences containing 10 active speakers, which amount for 8.7\% of all sequences drawn from the training set in one epoch.

\subsubsection{Are the attractors' validity predictors necessary?}
\label{sec:predictors_necessary}

EEND-EDA is trained using two losses: one targets minimizing the per-speaker per-frame activity errors with BCE and the other is used to find the correct number of valid attractors. The former is calculated only for the valid attractors and the latter is calculated for the valid attractors and the first invalid one. DiaPer is trained with the same two losses but with a fixed number of attractors $A$, which is an upper bound on the expected number of speakers. In contrast to EEND-EDA, with DiaPer, both losses are calculated for all $A$ attractors, independently of which ones are valid or invalid. 

During inference, EEND-EDA decodes the attractors until one of them has existence probability below a certain threshold and the number of speakers in the output is directly determined by how many attractors are decoded. DiaPer always returns activities for all $A$ attractors and, since the per-speaker per-frame activities loss is calculated for all of them, the model should be able to determine the number of speakers directly from the posterior activities. While it is possible to use the attractor existence probabilities to decide for which speakers to return activity probabilities (as we did in all experiments), the frame-speaker activities are enforced to be 0 for invalid attractors. Thus, one could wonder if using the attractor existence probabilities is necessary for DiaPer.

We tested this hypothesis on CH1-2spk and CH1. When using DiaPer trained only on SC with 2 speakers and evaluating it on CH1-2spk, the DERs were exactly the same with both inference alternatives: using the attractor existence probabilities to determine the speakers or using the frame-speaker activities to determine the speakers naturally derived from the activities. This was expected since the model had always seen two speakers in each training sample, so it was biased to output that exact number of speakers at inference time. When adapting the model described above to the set consisting of SC with 2 to 7 speakers and evaluating it on CH1, the outputs were not exactly the same, but the errors differed only by 0.01\% DER and 0.05\% JER. This suggests that the attractors validity predictors might not be necessary for DiaPer, making the inference procedure much simpler than with EEND-EDA.

Nevertheless, having the attractor validity loss might affect the training procedure in DiaPer, perhaps helping it model valid and invalid attractors differently and, therefore, easing the activities prediction task. It might be possible that DiaPer can be trained without the attractors validity predictors (and without their corresponding loss), making the whole training procedure simpler. However, this idea has not been tried and it is left for future work.

\subsubsection{Wide-band scenarios}
\label{sec:wideband}
Most works on end-to-end models focus on the telephone scenario and use Callhome (which is a paid dataset) as benchmark. Even if Callhome is one of the standard baselines in diarization, we believe that it is popular for EEND training also because synthetic data (needed for training such models) match this condition quite well. However, there are many wide-band scenarios of interest when performing diarization and only a few authors have analyzed their systems on a wide variety of them~\cite{bredin23_interspeech,plaquet23_interspeech}. Following this direction, and pursuing a more democratic field, in this section we use DiaPer on a wide variety of corpora (most of which are of public and free access) and show the performance for the same model (before and after FT) across domains. 

We have observed that DiaPer performed better than EEND-EDA so the following results are only presented for DiaPer. An exhaustive comparison with state-of-the-art results is presented in Chapter~\ref{sec:conclusion}.

Since most of the scenarios include more than two speakers per conversation, all DiaPer models were adapted to the set of 1-10 speakers per recording using sequences of 4 minutes. The 8\,kHz model was trained on telephony SC and two 16\,kHz models were used. Both wide-band models were trained on LibriSpeech-based SC where one model had 10 attractors (like the 8\,kHz model) and another had 20 attractors to allow for more speakers. All models are evaluated without and with FT. For corpora where a multi-speaker training set is available, this set is used for FT until no more improvements are observed on the development set. If no training set is available, the dev set is used for FT until the performance on the test set does not improve further. Therefore, results on these latter corpora (denoted in gray in Table~\ref{tab:multi_wide-band_again}) should be taken with a grain of salt.

Looking at the results in Table~\ref{tab:multi_wide-band_again}, in some cases, there was overfitting when performing FT on the development set (since those corpora did not have a training set). In DipCo, this is most likely due to the limited amount of data. In VoxConverse, the distribution of the number of speakers per recording is skewed towards more speakers in the test set and FT on the dev set makes the model find fewer speakers than without FT. Moreover, recordings with more speakers are longer, making the overall error higher after FT on the test set. As for AliMeeting near, DiaPer (20 attractors) has slightly worse performance on the test set but the decision to stop the FT was made observing the performance on the dev set, for which there were improvements.

DiaPer, like any end-to-end system, is very sensitive to the type of training data. This is highly noticeable in the high error rates before fine-tuning for all far-field scenarios: AISHELL-4, AliMeeting far, AMI array, CHiME6 and DipCo;  and relatively lower errors for exclusively close-talk scenarios: AliMeeting near, AMI headset, Mixer6 and in the comparison between DIHARD2 and DIHARD3 full where the latter contains a large portion of telephone conversations. All SC (used to train the models) are generated with speech captured from short distances (telephone for the 8\,kHz system and LibriSpeech for the 16\,kHz ones). Use of reverberation as part of SC generation could improve the situation, but it has not been explored so far in this context.
Not having enough data matching the test scenario is a strong drawback for the fine-tuning of end-to-end models as observed with DipCo and VoxConverse. Conversely, Mixer6 and RAMC with large amounts of FT data and relatively simple setups are among the scenarios with the largest relative improvement given by the FT.

The main goal of this comparison was to present a unified framework evaluated across different corpora. More tailored models could be trained if we used SC with specific numbers of speakers per recording (matching the evaluation data). Likewise, the output post-processing (subsampling and median filter) could be adapted for each dataset. This should definitely result in better performance and is left for future work. 

Regarding the comparison between 8\,kHz and 16\,kHz DiaPers, in most cases, the latter reaches better performance both without and with FT. Even though the 8\,kHz model was trained with more conversational data, this does not provide advantages over the 16\,kHz model trained on LibriSpeech-based SC. However, the effect of FT is in most cases (for both sampling rates) considerably large, reducing the gap between 8\,kHz and 16\,kHz models. Creating synthetic training data that resembles real ones remains an open challenge for most scenarios.

With respect to the number of attractors in the model, we can observe that overall having more of them is beneficial. This is actually not a drawback for DiaPer since the quantity of attractors does not impact severely on the number of parameters or computations. It is left for future work to explore the effect of larger numbers of attractors (e.g. 40 or 80).

\begin{table}[H]
    \caption{DER (\%) comparison on a variety of test sets. All results are with DiaPer using models with 10 or 20 attractors. Results in gray were obtained running the FT for best performance on the test set so they could be overoptimistic.}
    \label{tab:multi_wide-band_again}
    \setlength{\tabcolsep}{6pt} 
    \centering
    \begin{tabular}{@{}
                  lll|llll
                  @{}}
    \toprule
    Sampling rate & \multicolumn{2}{c|}{8\,kHz} & \multicolumn{4}{c}{16\,kHz} \\
    \# attractors & \multicolumn{2}{c|}{10 attractors} & \multicolumn{2}{c}{10 attractors} & \multicolumn{2}{c}{20 attractors} \\
    Fine-tuning & \multicolumn{1}{c}{No} & \multicolumn{1}{c|}{Yes} & \multicolumn{1}{c}{No} & \multicolumn{1}{c}{Yes} & \multicolumn{1}{c}{No} & \multicolumn{1}{c}{Yes} \\
    \midrule
    AISHELL-4 & 49.29 & \textcolor{gray}{42.66} & 48.21 & \textcolor{gray}{41.43} & 47.86 & \textcolor{gray}{31.30} \\[0.2cm]
    AliMeeting far & 45.40 & 31.60 & 38.67 & 32.60 & 34.35 & 26.27 \\[0.2cm]
    AliMeeting near & 33.70 & 28.93 & 28.19 & 27.82 & 23.90 & 24.44 \\[0.2cm]
    AMI array & 54.69 & 50.49 & 57.07 & 49.75 & 52.29 & 50.97 \\[0.2cm]
    AMI headset & 41.31 & 36.41 & 36.36 & 32.94 & 35.08 & 30.49 \\[0.2cm]
    CHiME6 & 78.51 & 68.54 & 78.25 & 70.77 & 77.51 & 69.94 \\[0.2cm]
    DIHARD2 & 49.92 & \textcolor{gray}{34.10} & 43.75 & \textcolor{gray}{32.97} & 44.51 & \textcolor{gray}{31.23} \\[0.2cm]
    DIHARD3 full & 38.24 & \textcolor{gray}{24.08} & 34.21 & \textcolor{gray}{24.12} & 34.82 & \textcolor{gray}{22.77} \\[0.2cm]
    DipCo & 64.33 & \textcolor{gray}{45.09} & 48.26 & Overfit & 43.37 & Overfit \\[0.2cm]
    Mixer6 & 19.11 & 14.69 & 21.03 & 13.41 & 18.51 & 10.99 \\[0.2cm]
    MSDWild & 34.62 & 17.99 & 35.69 & 15.46 & 25.07 & 14.59 \\[0.2cm]
    RAMC & 32.61 & 20.90 & 38.05 & 21.11 & 32.08 & 18.69 \\[0.2cm]
    VoxConverse & 32.37 & \textcolor{gray}{31.63} & 23.20 & Overfit & 22.10 & Overfit \\[0.2cm]
    \bottomrule
    \end{tabular}
\end{table}

\subsection{Possible extensions}

Even though DiaPer presents some advantages over EEND-EDA, there are still a few aspects where it could be improved. One of them is that currently, the attractors are exclusively defined in terms of the frame embeddings. However, the frame embeddings could also be improved if they were adapted using the attractors. This is done up to a certain extent with the frame encoder conditioning and it has a significant effect on the results as shown in Table~\ref{tab:ablation_comparison}. However, more explicit configurations could be devised. Just like Perceiver blocks iteratively transform the latents attending to the frame embeddings, the frame embeddings could be iteratively transformed by another chain of Perceivers that attend to the attractors. This would result in a more symmetric architecture that could lead to better frame embeddings and, therefore, to better performance.

Another possible extension for DiaPer could focus on making it even more light-weight. We have shown that with a relatively low number of parameters, it is possible to achieve competitive performance. Another option would be to try parameter tying (i.e. utilizing the same parameters in different layers) in the Perceiver-based attractor decoder or the frame encoder which could lead to similar performance with substantially fewer parameters.

It was commented in Section~\ref{sec:predictors_necessary} that the attractors existence probabilities might not be necessary to train DiaPer. This idea could be explored further.

One aspect that was not investigated thoroughly in this work is how to better define the training sets. For example, training with specific numbers of speakers or acoustic augmentations matching the evaluation data could lead to better suited models. This remains a rather unexplored facet of training end-to-end models but could lead to better performance and/or faster and more efficient training procedures.

Related to the training sets is actually the training data per se. We analyzed the quality of simulated conversations which resemble real conversations better than previous approaches. However, the acoustic quality of the synthetic data is extremely relevant, especially for wide-band cases where the mismatch with real corpora is very large. Given the ample variety of acoustic scenarios, one option would be to create tailored setups that match the evaluation set. For example, better simulation of far-field recordings should help with meetings or dinner part scenarios where speakers are normally far from the microphones. Given that available data used to generate synthetic training data are usually collected with near-field microphones, they could be carefully reverberated and the volume adjusted to simulate the effects observed in real rooms.

\section{Final remarks}

In this chapter, we covered end-to-end diarization systems, starting with an overview of existing approaches and special emphasis on the EEND framework. 

Then, we focused on the generation of synthetic training data. End-to-end models require large amounts of diarization-annotated data but due to the scarcity of annotated real data, the compromise solution consists in generating synthetic recordings. We proposed an approach that allows such models to obtain better performance and reduces the need for fine-tuning on telephone conversations.

Finally, we presented DiaPer as an alternative to the most popular end-to-end model that can handle multiple speakers. We showed that it can achieve better performance even with fewer parameters and performed a thorough comparison between both approaches. Our approach - DiaPer - performs better than the popular EEND-EDA mainly in handling overlapped speech and it can find the number of speakers in a conversation more accurately.

\chapter{Conclusion}\label{sec:conclusion}

In this chapter, we present a final comparison between a cascaded system (based on VBx, as described in Chapter~\ref{sec:diarization_bhmm}) and an end-to-end system (DiaPer, as described in Chapter~\ref{sec:e2e_diaper}). The intention is to provide insights about the advantages of each approach across different scenarios. We also compare the results of these approaches with the best-published results at the time of writing this thesis. Then, we conclude the thesis with a brief summary of what are the challenges that we believe need to be addressed in the future of speaker diarization.

\section{Modular systems versus end-to-end systems}
\label{sec:comparison}

Table~\ref{tab:ch2two_other_methods} presents an exhaustive comparison with all competitive systems on the Callhome (2 speakers) dataset at the time of publication under the same conditions: all speech is evaluated and no oracle information is used. Data refers to the number of hours of data for supervision. For end-to-end models, it can be real or synthetic data and for the clustering-based method, it consists of all data used to train the x-vector extractor, VAD and OSD. Methods are divided into groups depending on if they are single or two-stage. Even though DiaPer does not present the best performance among all approaches, it reaches competitive results with fewer parameters and even without FT. On the other hand, the VBx-based method falls behind since its performance in handling overlapped speech is worse.

\begin{table}[H]
    \caption[capt]{DER (\%) comparison on CH2-2spk. For our results, we selected the model with the best performance on CH1 out of 5 runs. Type can be clustering (C), 1-stage (1-S) or 2-stage (2-S) system. (I) stands for iterative, meaning there is an iterative process at the inference time.}
    \label{tab:ch2two_other_methods}
    \setlength{\tabcolsep}{1pt} 
    \centering
    \begin{tabular}{@{}
                  l 
                  c
                  c
                  c
                  c
                  S[table-format=2.2] 
                  S[table-format=2.2]
                  @{}}
    \toprule
    \multirow{2}{*}{System} & \multirow{2}{*}{Type} & \multirow{2}{*}{Code} & \#Param. & Data & \multicolumn{1}{c}{No} & \multicolumn{1}{c}{With} \\
    &  &  & (Million) & (kHour) & \multicolumn{1}{c}{FT} & \multicolumn{1}{c}{FT} \\
    \midrule
    VAD + VBx + OSD & C & \checkmark & 17.9 & 9 & \multicolumn{1}{c}{N/A} & 9.92 \\
    \midrule
    EEND-EDA~\cite{horiguchi20_interspeech} & 1-S (I) & \checkmark & 6.4 & 2.4 & \multicolumn{1}{c}{--} & 8.07 \\
    EEND-EDA Conformer~\cite{yamashita22_odyssey} & 1-S (I) &  & 4 & 2.5 & 9.65 & 7.18\\
    CB-EEND~\cite{liu21j_interspeech} & 1-S &  & 4.2 & 4.7 & \multicolumn{1}{c}{--} & 6.82\\
    DIVE~\cite{zeghidour2021dive} & 1-S (I) &  & ?? & 2 & \multicolumn{1}{c}{--} & 6.7 \\
    RX-EEND~\cite{yu2022auxiliary} & 1-S &  & 12.8 & 2.4 & \multicolumn{1}{c}{--} & 7.37\\
    EDA-TS-VAD~\cite{wang2023target} & 1-S (I) &  & 16.1 & 16 & \multicolumn{1}{c}{--} & 7.04 \\
    EEND-OLA~\cite{wang2023told} & 1-S &  & $\approx$6.7 & 15.5 & \multicolumn{1}{c}{--} & 6.91 \\
    EEND-NA~\cite{fujita2023neural} & 1-S &  & 5.7 & 2.5 & 8.81 & 7.77\\
    EEND-NA-deep~\cite{fujita2023neural} & 1-S &  & 10.9 & 2.5 & 8.52 & 7.12\\
    EEND-IAAE~\cite{hao2023nn} (it=2) & 1-S (I) & \checkmark & 8.5 & 2.5 & 13.8 & 7.58 \\
    EEND-IAAE~\cite{hao2023nn} (it=5) & 1-S (I) & \checkmark & 8.5 & 2.5 & \multicolumn{1}{c}{--} & 7.36 \\
    AED-EEND~\cite{chen23n_interspeech} & 1-S (I) &  & 11.6 & 2.4 & \multicolumn{1}{c}{--} & 6.79 \\
    AED-EEND-EE~\cite{chen2023attention} & 1-S (I) &  & 11.6 & 24.7 & \multicolumn{1}{c}{--} & 5.69 \\
    \midrule
    EEND-VC~\cite{kinoshita21_interspeech}  & 2-S &  & $\approx$8 & 4.2 & \multicolumn{1}{c}{--} & 7.18 \\
    WavLM + EEND-VC~\cite{chen2022wavlm} & 2-S & \checkmark & $\approx$840 & 8 & \multicolumn{1}{c}{--} & 6.46 \\
    EEND-NAA~\cite{rybicka2022end} & 2-S (I) &  & 8 & 2.4 & \multicolumn{1}{c}{--} & 7.83\\
    Graph-PIT-EEND-VC~\cite{kinoshita22_interspeech} & 2-S &  & $\approx$5.5 & 5.5 & \multicolumn{1}{c}{--} & 7.1\\
    EEND-OLA + SOAP~\cite{wang2023told} &  2-S & \checkmark & 15.6 & 19.4 & \multicolumn{1}{c}{--} & 5.73 \\
    \midrule
    \midrule
    EEND-EDA & 1-S (I) & \checkmark & 6.4 & 2.5 & 8.77 & 7.96 \\
    DiaPer & 1-S & \checkmark & 4.6 & 2.5 & 8.05 & 7.51\tablefootnote{It is worth mentioning that out of the 5 runs, the best DER on Part 2 was 7.38 but that did not correspond to the lowest DER on Part 1. Analogously, for EEND-EDA it was 7.78.} \\
    \bottomrule
  \end{tabular}
\end{table}

Table~\ref{tab:ch2all_other_methods} shows the results when handling multiple speakers per conversation in the Callhome dataset. Results show that even if DiaPer has a competitive performance, many methods can reach considerably better results. The main advantage of DiaPer is its lightweight nature, having the least number of parameters in comparison with all other methods. Exploring larger versions of DiaPer (i.e. increasing the model dimension) which could lead to better performance in multi-speaker scenarios is left for future research.

Interestingly, in this scenario, the VBx-based method reaches similar and competitive results. In spite of its modular nature, the results are still on par with many end-to-end models showing the relevance of these types of systems even at the current time. 

\begin{table}[H]
    \caption[capt]{DER comparison on CH2. For our results, we selected the model with the best performance on CH1 out of 5 runs. Type can be clustering (C), 1-stage (1-S) or 2-stage (2-S) system. (I) stands for iterative, meaning there is an iterative process at the inference time.}
    \label{tab:ch2all_other_methods}
    \setlength{\tabcolsep}{1pt} 
    \centering
    \begin{tabular}{@{}
                  l 
                  c
                  c
                  c
                  c
                  c
                  S[table-format=2.2] 
                  S[table-format=2.2]
                  @{}}
    \toprule
    \multirow{2}{*}{System} & \multirow{2}{*}{Type} & \multirow{2}{*}{Code} & \#Param. & Data & \multicolumn{1}{c}{No} & \multicolumn{1}{c}{With} \\
    &  &  & (Million) & (kHour) & \multicolumn{1}{c}{FT} & \multicolumn{1}{c}{FT} \\
    \midrule
    VAD + VBx + OSD & C & \checkmark & 17.9 & 9 & \multicolumn{1}{c}{N/A} & 13.63 \\
    \midrule
    EEND-EDA~\cite{horiguchi20_interspeech} & 1-S (I) & \checkmark & 6.4 & 15.5 & \multicolumn{1}{c}{--} & 15.29 \\
    EDA-TS-VAD~\cite{wang2023target} & 1-S (I) &  & 16.1 & 16 & \multicolumn{1}{c}{--} & 11.18 \\
    EEND-OLA~\cite{wang2023told} & 1-S & \checkmark & 6.7 & 15.5 & \multicolumn{1}{c}{--} & 12.57 \\
    AED-EEND~\cite{chen23n_interspeech} & 1-S (I) &  & 11.6 & 15.5 & \multicolumn{1}{c}{--} & 14.22 \\
    AED-EEND-EE~\cite{chen2023attention} & 1-S (I) &  & 11.6 & 24.7 & \multicolumn{1}{c}{--} & 10.08 \\
    \midrule
    EEND-VC~\cite{kinoshita21_interspeech}  & 2-S &  & $\approx$8 & 4.2 & \multicolumn{1}{c}{--} & 12.49 \\
    EEND-GLA~\cite{horiguchi2021towards} & 2-S &  & 10.7 & 15.5 & \multicolumn{1}{c}{--} & 11.84 \\
    WavLM + EEND-VC~\cite{chen2022wavlm} & 2-S & \checkmark & $\approx$840 & 8 & \multicolumn{1}{c}{--} & 10.35 \\
    Graph-PIT-EEND-VC~\cite{kinoshita22_interspeech} & 2-S &  & $\approx$5.5 & 5.5 & \multicolumn{1}{c}{--} & 13.5\\
    EEND-OLA + SOAP~\cite{wang2023told} & 2-S & \checkmark & 15.6 & 19.4 & \multicolumn{1}{c}{--} & 10.14 \\
    EEND-VC MS-VBx~\cite{delcroix23_interspeech} & 2-S &  & $\approx$840 & 5.5 & \multicolumn{1}{c}{--} & 10.4 \\
    \midrule
    \midrule
    EEND-EDA & 1-S (I) & \checkmark & 6.4 & 15 & 16.70 & 15.29 \\
    DiaPer & 1-S & \checkmark & 4.6 & 15 & 14.86 & 13.60\tablefootnote{It is worth mentioning that out of the 5 runs, the best DER on Part 2 was 13.16 but that did not correspond to the lowest DER on Part 1.}
 \\
    \bottomrule
    \toprule
    \multicolumn{7}{c}{Scoring with collar 0\,s} \\
    \midrule
    VAD + VBx + OSD & C & \checkmark & 17.9 & 4 & \multicolumn{1}{c}{N/A} & 26.18 \\
    \midrule
    pyannote 2.1\cite{bredin23_interspeech} & 2-S & \checkmark & 23.6 & 2.9 & 32.4 & 29.3 \\
    \midrule
    \midrule
    EEND-EDA & 1-S (I) & \checkmark & 6.4 & 2.5 & 28.73 & 25.77 \\
    DiaPer & 1-S & \checkmark & 4.6 & 2.5 & 27.84 & 24.16\tablefootnote{It is worth mentioning that out of the 5 runs, the best DER on Part 2 was 23.81 but that did not correspond to the lowest DER on Part 1.} \\
    \bottomrule
    \end{tabular}
\end{table}

Table~\ref{tab:multi_wide-band} presents a comparison of a VBx-clustering-based system, DiaPer, and the best published results for each of the wide-band datasets described in Section~\ref{sec:vbx_evaluation_data}.
In comparison with the best results published at the time of writing the thesis, DiaPer performs considerably worse in most of the scenarios. However, it should be noted that in many cases, the best results correspond to systems submitted to challenges which usually consist of the fusion of a few carefully tuned models.
Even if in most cases the performance is not on par with other approaches, DiaPer's final performance is very competitive for MSDWild and RAMC.
We can also see that even a standard cascaded system can reach competitive results on a few datasets and perform better than DiaPer in approximately half of the scenarios: AISHELL-4, AliMeeting near, AMI array, AMI headset, DIHARD2, DIHARD3 full, RAMC and VoxConverse. Again, this shows the importance and relevance of these systems as baselines nowadays even when end-to-end solutions are the most studied in the community.

\begin{table}[H]
    \caption{DER (\%) comparison on a variety of test sets. Overlaps are evaluated and oracle VAD is NOT used. \underline{Underlined} results denote single systems and \textoverline{overlined} results correspond to fusions or more complex models. Results in gray were obtained running the FT for best performance on the test set so they could be overoptimistic.}
    \label{tab:multi_wide-band}
    \setlength{\tabcolsep}{5pt} 
    \centering
    \begin{tabular}{@{}
                  l|ccc|lll
                  @{}}
    \toprule
    \multirow{2}{*}{Corpus} & \multirow{2}{*}{VAD+VBx+OSD} & \multicolumn{2}{c|}{DiaPer (20 att.)} & \multicolumn{3}{c}{\multirow{2}{*}{Best published results}} \\
    & & No FT & With FT & & & \\
    \midrule
    AISHELL-4 & 15.84 & 47.86 & \textcolor{gray}{31.30} & \underline{16.76} [1] & \underline{14.0} [2] & \underline{13.2} [3] \\[0.2cm]
    AliMeeting far & 28.84 & 34.35 & 26.27 & \underline{23.8} [4] & \underline{23.3} [5] & \underline{23.5} [6] \\[0.2cm]
    AliMeeting near & 22.59 & 23.90 & 24.44 & --- & --- & --- \\[0.2cm]
    AMI array & 34.61 & 52.29 & 50.97 & \underline{22.2} [7] & \underline{22.0} [8] & \underline{19.53} [9] \\[0.2cm]
    AMI headset & 22.42 & 35.08 & 30.49 & \underline{18.0} [10] & \underline{16.95} [11] & \underline{13.00} [12] \\[0.2cm]
    CHiME6 & 70.42 & 77.51 & 69.94 & \textoverline{32.46} [13] & \textoverline{27.25} [14] & \textoverline{25.11} [15] \\[0.2cm]
    DIHARD2 & 26.67 & 44.51 & \textcolor{gray}{31.23} & \underline{26.4} [16] & \underline{26.88} [17] & \underline{24.64} [18] \\[0.2cm]
    DIHARD3 full & 20.28 & 34.82 & \textcolor{gray}{22.77} & \underline{17.32} [19] & \textoverline{16.94} [20] & \textoverline{16.76} [21] \\[0.2cm]
    DipCo & 49.22 & 43.37 & Overfit & \textoverline{22.36} [22] & \textoverline{22.04} [23] & \textoverline{16.36} [24] \\[0.2cm]
    Mixer6 & 35.60 & 18.51 & 10.99 & \textoverline{7.27} [25] & \textoverline{6.14} [26] & \textoverline{5.65} [27] \\[0.2cm]
    MSDWild & 16.86 & 25.07 & 14.59 & \underline{21.96} [28] & \underline{33.6} [29] & \underline{16.0} [30] \\[0.2cm]
    RAMC & 18.19 & 32.08 & 18.69 & \underline{22.2} [30] & \underline{19.90} [31] & \underline{14.37} [32] \\[0.2cm]
    VoxConverse & 6.12 & 22.10 & Overfit & \textoverline{4.0} [33] & \textoverline{4.39} [34] & \textoverline{4.35} [35] \\[0.2cm]
    \bottomrule
    \end{tabular}
\end{table}

Legend for Table~\ref{tab:multi_wide-band}:
\begin{multicols}{2}
\begin{itemize}\setlength\itemsep{-0.25em}
    \item $[1]$: \cite{chen22f_interspeech}
    \item $[2]$: \cite{bredin23_interspeech}
    \item $[3]$: \cite{plaquet23_interspeech}
    \item $[4]$: \cite{bredin23_interspeech}
    \item $[5]$: \cite{plaquet23_interspeech}
    \item $[6]$: \cite{raj23_interspeech}
    \item $[7]$: \cite{bredin23_interspeech}
    \item $[8]$: \cite{plaquet23_interspeech}
    \item $[9]$: \cite{he2023ansd}
    \item $[10]$: \cite{plaquet23_interspeech}
    \item $[11]$: \cite{he2023ansd}
    \item $[12]$: \cite{chen2023attention}
    \item $[13]$: \cite{kamontt}
    \item $[14]$: \cite{yeiacas}
    \item $[15]$: \cite{wan23_chime}
    \item $[16]$: \cite{delcroix23_interspeech}
    \item $[17]$: \cite{horiguchi2021end}
    \item $[18]$: \cite{chen2023attention}
    \item $[19]$: \cite{he2023ansd}
    \item $[20]$: \cite{horiguchi2021hitachi}
    \item $[21]$: \cite{he2023ansd}
    \item $[22]$: \cite{yeiacas}
    \item $[23]$: \cite{wan23_chime}
    \item $[24]$: \cite{wan23_chime}
    \item $[25]$: \cite{yeiacas}
    \item $[26]$: \cite{wan23_chime}
    \item $[27]$: \cite{kamontt}
    \item $[28]$: \cite{liu22t_interspeech}
    \item $[29]$: \cite{liu2022ber}
    \item $[30]$: \cite{plaquet23_interspeech}
    \item $[31]$: \cite{yang22h_interspeech}
    \item $[32]$: \cite{broughton23_interspeech}
    \item $[33]$: \cite{baroudipyannote}
    \item $[34]$: \cite{karamyan2023krisp}
    \item $[35]$: \cite{wang2023profile}
\end{itemize}
\end{multicols}

\begin{table}[H]
    \caption{Comparison on different sets in terms of DER and its three components, mean speaker counting error, JER, and precision and recall for VAD and OSD performances. DiaPer has 20 attractors and results are after FT (except for DipCo and VoxConverse). All results are obtained with 16\,kHz systems except for the results of Callhome where the 8\,kHz systems were used.}
    \label{tab:detailed_comparison}
    \setlength{\tabcolsep}{3pt} 
    \centering
    \begin{tabular}{@{}
                  ll |
                  S[table-format=2.2] |
                  S[table-format=2.2]
                  S[table-format=2.2] 
                  S[table-format=2.2] | 
                  S[table-format=1.2] |
                  S[table-format=2.2] |
                  S[table-format=2.1]
                  S[table-format=2.1] |
                  S[table-format=2.1]
                  S[table-format=2.1] 
                  @{}}
    \toprule
    \multirow{2}{*}{Corpus} & \multirow{2}{*}{System} & \multicolumn{1}{c|}{DER} & \multicolumn{1}{c}{Miss} & \multicolumn{1}{c}{FA} & \multicolumn{1}{c|}{Conf.} & \multicolumn{1}{c|}{MS} & \multicolumn{1}{c|}{JER} & \multicolumn{2}{c|}{VAD} & \multicolumn{2}{c}{OSD} \\
     &  & \multicolumn{1}{c|}{(\%)} & \multicolumn{1}{c}{(\%)} & \multicolumn{1}{c}{(\%)} & \multicolumn{1}{c|}{(\%)} & \multicolumn{1}{c|}{CE} & \multicolumn{1}{c|}{(\%)} & \multicolumn{1}{c}{P (\%)} & \multicolumn{1}{c|}{R (\%)} & \multicolumn{1}{c}{P (\%)} & \multicolumn{1}{c}{R (\%)} \\
    \midrule
    \multirow{2}{*}{AISHELL-4} & VBx & 15.84 & 9.98 & 1.8 & 4.06 & 0.45 & 23.46 & 98.4 & 94.0 & 66.4 & 15.5 \\
    & DiaPer & 31.30 & 15.33 & 2.65 & 13.32 & 1.75 & 52.91 & 98.4 & 88.3 & 35.2 & 15.0 \\
    \midrule
    AliMeeting & VBx & 28.84 & 20.64 & 3.4 & 4.8 & 0.15 & 33.56 & 96.6 & 94.2 & 85.1 & 27.6 \\
    far & DiaPer & 26.27 & 11.08 & 7.25 & 7.94 & 0.3 & 35.01 & 97.7 & 96.5 & 77.8 & 62.3 \\
    \midrule
    AliMeeting & VBx & 22.59 & 13.19 & 3.26 & 6.14 & 0.1 & 28.09 & 96.9 & 98.0 & 92.2 & 54.8 \\
    near & DiaPer & 24.44 & 9.83 & 5.91 & 8.70 & 0.25 & 32.12 & 98.2 & 97.5 & 82.4 & 65.7 \\
    \midrule
    AMI & VBx & 34.61 & 26.12 & 1.93 & 6.56 & 0.5 & 44.53 & 98.0 & 82.8 & 86.1 & 25.2 \\
    array & DiaPer & 50.97 & 30.96 & 2.95 & 17.06 & 1.63 & 67.43 & 97.1 & 77.1 & 78.5 & 24.1 \\
    \midrule
    AMI & VBx & 22.42 & 14.06 & 3.73 & 4.63 & 0.31 & 27.34 & 96.3 & 96.9 & 87.3 & 52.1 \\
    headset & DiaPer & 30.49 & 19.86 & 4.27 & 6.36 & 0.63 & 42.74 & 96.7 & 88.2 & 77.9 & 36.1 \\
    \midrule
    \multirow{2}{*}{Callhome} & VBx & 13.62 & 7.69 & 1.92 & 4.01 & 0.31 & 37.32 & 93.2 & 98.6 & 77.1 & 37.2 \\
    & DiaPer & 13.60 & 7.80 & 2.06 & 3.74 & 0.26 & 30.32 & 95.4 & 95.3 & 64.1 & 44.8 \\
    \midrule
    \multirow{2}{*}{CHiME6} & VBx & 70.42 & 47.72 & 0.91 & 21.79 & 3 & 75.38 & 98.3 & 67.7 & 89.1 & 7.4 \\
    & DiaPer & 69.94 & 35.3 & 12.77 & 21.87 & 1.75 & 79.16 & 88.7 & 85.2 & 47.5 & 21.1 \\
    \midrule
    \multirow{2}{*}{DIHARD2} & VBx & 26.59 & 14.14 & 4.5 & 7.95 & 1.01 & 48.41 & 95.9 & 93.1 & 66.1 & 23.4 \\
    & DiaPer & 31.23 & 14.98 & 6.98 & 9.27 & 0.54 & 57.00 & 95.5 & 90.9 & 50.2 & 33.7 \\
    \midrule
    DIHARD3 & VBx & 20.28 & 10.15 & 3.99 & 6.14 & 0.6 & 38.15 & 96.8 & 95.8 & 74.1 & 38.0 \\
    full & DiaPer & 22.77 & 11.3 & 5.91 & 5.56 & 0.72 & 44.58 & 96.7 & 93.5 & 58.7 & 45.9 \\
    \midrule
    \multirow{2}{*}{DipCo} & VBx & 49.22 & 40.66 & 1.05 & 7.51 & 2 & 61.93 & 98.6 & 68.1 & 71.0 & 6.0 \\
    & DiaPer & 43.37 & 15.46 & 9.72 & 18.19 & 0.8 & 63.05 & 95.8 & 94.0 & 57.4 & 37.5 \\
    \midrule
    \multirow{2}{*}{Mixer6} & VBx & 35.6 & 21.74 & 12.43 & 1.43 & 0.04 & 39.35 & 99.5 & 82.0 & 13.6 & 14.0 \\
    & DiaPer & 10.99 & 5.15 & 4.99 & 0.85 & 0 & 18.38 & 97.9 & 99.7 & 46.2 & 36.1 \\
    \midrule
    \multirow{2}{*}{MSDWild} & VBx & 16.86 & 5.81 & 2.75 & 8.3 & 1.05 & 56.94 & 94.2 & 97.0 & 61.9 & 33.4 \\
    & DiaPer & 14.59 & 5.25 & 3.61 & 5.73 & 0.44 & 43.14 & 94.6 & 95.5 & 56.9 & 45.1 \\
    \midrule
    \multirow{2}{*}{RAMC} & VBx & 18.19 & 0.39 & 13.15 & 4.65 & 0 & 19.37 & 89.9 & 99.6 & 0.0 & 14.4 \\
    & DiaPer & 18.69 & 4.41 & 5.34 & 8.94 & 0.12 & 24.95 & 94.9 & 95.6 & 0.0 & 0.0 \\
    \midrule
    \multirow{2}{*}{VoxConverse} & VBx & 6.12 & 1.6 & 2.02 & 2.5 & 1.12 & 29.44 & 96.5 & 98.5 & 62.5 & 52.9 \\
    & DiaPer & 22.1 & 6.99 & 4.93 & 10.18 & 2.52 & 57.65 & 94.4 & 93.1 & 32.7 & 44.0 \\
    \bottomrule
  \end{tabular}
\end{table}

Finally, Table~\ref{tab:detailed_comparison} compares VBx and DiaPer, the two approaches that have been developed in the context of this thesis, using different metrics across all datasets used in the thesis. The mean speaker counting error (MSCE) is reported, DER is presented with its three components and finally, the VAD and OSD performances are detailed for each model. For VBx, these are given by the VAD and OSD models used while for DiaPer those performances are derived from the outputs since there are no specific VAD or OSD modules.

We first discuss the results for each dataset separately and then comment on common patterns observed for all of them.

\begin{itemize}
    \item AISHELL-4: the clustering-based approach presents far better results than DiaPer. This is observed across all metrics (DER, JER and MSCE). The large percentage of missed speech for DiaPer is connected to its VAD capabilities, possibly due to the fact that this dataset was collected with far-field microphones. At the same time, the clustering-based approach has much higher precision for the same OSD recall, partly explaining the increased FA speech and confusion in DiaPer. Furthermore, this dataset presents 5.8 speakers per recording on average and while VBx finds 5.5, DiaPer finds 4.1 on average. This certainly explains the high confusion error in DiaPer and the more-than-twice increase in terms of JER.
    \item AliMeeting far: DiaPer presents almost 10\% relative improvement over the VBx-based approach and this can be traced mostly to overlap. While it is true that DiaPer performs slightly better in terms of VAD, the twice better recall regarding OSD with respect to the clustering-based approach gives substantial gains. This provides a clear advantage in AliMeeting where overlapped speech accounts for more than 18\% of the time in the test set. In this scenario, the ability of VBx to find the number of speakers more accurately does not compensate DiaPer's advantage in handling overlap.
    \item AliMeeting near: unlike the far-field scenario, when using near-field microphones, the clustering-based approach is more than 5\% relative better than DiaPer. The differences in terms of MSCE are similar to the far-field case but with near-field microphones, the OSD performance slightly improves for DiaPer while it improves substantially for the clustering-based system. The increased OSD recall enables the latter to considerably reduce the missed speech and the higher OSD precision also explains the lower FA.
    \item AMI array: both methods have very low recall percentages in terms of VAD, which contrasts with most other datasets. This is explained by the far-field characteristics of this dataset. The VAD system used as part of the clustering-based system is more robust to this situation and this explains the lower missed speech percentage with respect to DiaPer. This dataset presents 3.9 speakers on average in the test set and the clustering-based approach is relatively accurate (as seen in terms of MSCE) finding 3.4 while DiaPer only finds 2.3 on average. This implies that significant portions of speech are assigned to wrong speakers, explaining the high confusion error for DiaPer and the considerably worse JER.
    \item AMI headset: as expected, systems can perform better in the close-talk scenario than in the far-field and this is observed across all metrics (DER, JER, MSCE) for AMI headset vs. AMI array. However, DiaPer still struggles to find speech in the close-talk case as seen in the relatively low VAD recall. AMI presents recordings where different speakers have very different loudness levels, even if recorded with close-talk microphones. End-to-end models are considerably biased towards the training data and in this case, simulated conversations were generated using recordings from LibriSpeech for which recordings tend to have similar levels of loudness. We hypothesize that creating recordings with more variability in loudness could help in this particular scenario. Interestingly, the OSD performance for the clustering-based approach is considerably better than for DiaPer. However, it should be noted that AMI was utilized to train the OSD system used as part of the clustering-based approach and this could explain these differences. The combination of better VAD, OSD and more accurate speaker counting certainly explain the better performance of the clustering-based approach in this dataset.
    \item Callhome: both systems perform almost the same regarding DER. However, the slight advantage in terms of MSCE for DiaPer shows that it finds the numbers of speakers more accurately and this translates into a considerable difference in terms of JER. Even though this dataset can have up to 6 speakers in one conversation, most of the recordings have only 2 and on average there are only 2.6 speakers per conversation.
    \item CHiME6: both systems perform equally poorly. This is perhaps the most challenging scenario among all datasets considered in this work, with very long conversations happening in different rooms throughout a single session, with recordings captured with far-field devices and data presenting the highest percentage of overlap (about 29\% in the test set). However, the two systems fail due to different reasons. Even though all recordings have exactly 4 speakers, VBx finds on average 7 speakers while DiaPer finds 2.3 on average. This certainly explains the very high confusion errors and JER. VBx tends to separate the speech segments of one speaker into more than one label and eventually confuse them and DiaPer tends to assign segments of speech from different speakers to the same label. It should be noted that this dataset is very challenging and most of the time there are background noises that sound equally loud as the (usually quiet) speakers. Even though both systems are trained with data augmentation schemes including background noises, it seems insufficient. Using a speech enhancement module prior to the application of either system could help in both cases. This could also benefit the models regarding their OSD capabilities and especially regarding VAD where the performance is substandard.
    \item DIHARD2: the clustering-based approach presents around 15\% DER relative improvement over DiaPer; however, the latter finds the correct number of speakers more accurately. This set is composed of recordings from a variety of domains and while the number of speakers per recording can vary from 1 to 9, on average there are 3.4 speakers and 38\% of the recordings contain only 2 speakers. Even though VBx finds on average 2.5 speakers and is on average closer than DiaPer (with 2.2 on average) to the reference, it is almost twice less accurate according to MSCE. It seems that DiaPer is in general better able to handle recordings with fewer speakers and this is beneficial in this scenario with most recordings having very few speakers (78\% of recordings contain 4 or less speakers). Nevertheless, the clustering-based approach performs substantially better in terms of DER and this is mainly due to the lower FA and confusion errors. DIHARD2 is among the datasets with lowest percentage of overlap. Even though the OSD performance of both systems is far from ideal, the OSD for the clustering-based approach operates at a more ``conservative'' operating point with higher precision and lower recall than DiaPer. Given the low percentage of overlap in the test set (below 7\%), this means that finding less overlap is eventually more beneficial than finding more, which would potentially result in higher FA and more confusion errors as seen in this case with DiaPer.
    \item DIHARD3: Similar to DIHARD2, the clustering-based approach performs better than DiaPer. This dataset presents many similitudes with DIHARD2, where the main differences are the removal of day-long recordings with vocalizations from babies and their environments and the addition of conversational telephone speech which accounts for almost one quarter of the recordings in the whole set. The removal of very challenging recordings and addition of much easier ones explains the decrease in errors for both systems with respect to DIHARD2. The average number of speakers in this set is 2.9 and VBx finds 2.4 while DiaPer finds 2.2 on average. Having 61\% of the files with 2 speakers presents a simpler scenario possibly explaining why VBx is substantially more accurate than with DIHARD2. Regarding VAD and OSD, both systems perform better, most likely due to the high percentage of ``easy'' telephone conversations. However, once again, the conservative OSD in the clustering-based approach is most likely the reason for certain advantage of this method over DiaPer.
    \item DipCo: DiaPer presents more than 10\% DER relative improvement over the clustering-based approach; however, this is certainly due to the low VAD recall in the latter. This datasets presents a very similar scenario to CHiME6 with far-field microphones capturing dinner-party conversations where speech is very quiet and noises relatively loud. Once again, even though each conversation has exactly 4 speakers, VBx finds on average 5.6 speakers and DiaPer finds 3.2, therefore making DiaPer more accurate in terms of MSCE. Unlike for CHiME6, DiaPer reaches a reasonable VAD performance in DipCo and this explains the reduced DER. However, if a VAD more suited for this scenario was used for the clustering-based approach, this method would most likely reach even better performance since more than 40\% of the DER corresponds to missed speech.
    \item Mixer6: DiaPer presents far better results than the clustering-based approach. This dataset presents one of the easiest scenarios among all in this work: all conversations have 2 speakers and they consist of interviews in quiet environments. However, the conversations were recorded with a distant microphone so some speech segments are very quiet. We hypothesize that this caused the VAD in the clustering-based approach to miss a substantial amount of speech and this impacted the performance of the system quite significantly. DiaPer, alternatively, presents a remarkable performance in terms of VAD and most of the errors are due to overlapped speech since it finds the right number of speakers in all recordings.
    \item MSDWild: DiaPer has more than 10\% DER relative improvement with respect to the clustering-based approach while more than halving the MSCE. This dataset contains recordings with 2 to 4 speakers and 2.6 on average. VBx finds only 1.6 on average while DiaPer is more accurate with 2.3 on average. The recordings in this dataset correspond to videos from different sources and a high percentage of them have background music. At the same time, recordings are very short: 72\,s on average. With VBx, almost half of the recordings are labeled with only one speaker (in spite of all of them having at least 2). We hypothesize that this is due to very short recordings which imply few samples for each speaker are available. These combined factors possibly confuse VBx to believe that there is a single speaker.
    \item RAMC: both systems perform very similarly in terms of DER; however, VBx is slightly more accurate finding the correct number of speakers (no mistakes at all) and this probably explains the JER differences. Similarly to Mixer6, this dataset presents a very simple scenario of telephone conversations between two people. However, unlike Mixer6, close-talk microphones are utilized in this set, therefore imposing less challenging acoustic conditions which most likely explain why the VAD in the clustering-based approach performs substantially better than in Mixer6.
    \item VoxConverse: the clustering-based approach performs substantially better than DiaPer in all evaluation metrics. This dataset not only has the most variety in terms of numbers of speakers per recording but it also presents a higher average number of them in the test set with 6.5. DiaPer finds 4.6 speakers on average while VBx finds 5.7 and this explains to great extent the very high confusion error for DiaPer: finding fewer speakers means that, if detected, their speech is necessarily assigned to wrong speakers. This, combined with the low precision in terms of OSD, increases considerably the error when handling multiple speakers. It should be noted that in this dataset, some videos correspond to talk-shows or news shows where spectators whoop or laugh and these segments are sometimes detected as speech. Some recordings have background music with singers and these are not labeled in the reference annotations. It is debatable if those segments should be considered speech or not but, at any rate, more tailored systems should be trained for such situations, possibly by generating training data accordingly.
\end{itemize}

As observed in the discussions above, building a common system capable of dealing with different datasets presents several challenges for both clustering-based and end-to-end systems. It should be noted that both systems were fine-tuned using in-domain data for each dataset; however, VAD and OSD systems in the clustering-based approach were used with their standard configurations, i.e. without any adaptation. This implies that it might be possible to improve the performance for some datasets if, for example, we fine-tuned the hyperparameters of the VAD and OSD systems using their corresponding training sets. At the same time, these systems were trained on large amounts of data, probably leading to more robust models and that might be preferable in many cases. In this sense, end-to-end models present the advantage of being more easily adaptable to a particular scenario. However, as seen in these comparisons, clustering-based approaches can be more robust to certain challenging conditions and, therefore, lead to better performance.

If one divides the datasets into those with more than 4 speakers (AISHELL-4, Callhome, DIHARD2, DIHARD3 full, and VoxConverse), and those with is up to 4 speakers per conversation (AliMeeting, AMI, CHiME6, DipCo, Mixer6, MSDWild, and RAMC), we can observe some interesting patterns. 
The clustering approach obtains better DER in all sets in the first group (except for Callhome where the performance is about the same for both approaches). Regarding MSCE, the same pattern holds except for DIHARD2 where DiaPer has a lower error. In these datasets, the DER with DiaPer is higher mainly in terms of confusion error and most likely due to worse handling of overlapped speech (note the lower OSD precision and increased FA and confusion errors).
When observing the results for the second group of datasets with limited numbers of speakers, DiaPer shows better performance in some of them (DipCo, Mixer6 and MSDWild) and the clustering approach reaches better results for others (AliMeeting near, AMI array and headset) while both perform similarly in the remaining sets. 

Since JER is calculated without a collar, it can provide a common ground for comparing performances across datasets. We can observe that datasets with few and a fixed amount of speakers such as Mixer6 and RAMC allow for much lower JER (since they only have 2 speakers) than datasets with more variability: see VoxConverse, DIHARD2 and DIHARD3 full for which DERs are reasonably low but JERs are much higher. It should be noted that depending on the final application, one metric can be preferred over the other one. JER gives equal weight to all speakers so even those who speak very little in a conversation with more than ten participants will be very relevant for the metric. In contrast, DER does not penalize missing not-so-active speakers as long as most of the speech is correctly labeled. 

In general, differences between the two approaches are larger for JER than for DER (which could be partially influenced by the lack of forgiveness collar) and the better approach regarding DER is usually the better according to JER. 

In order to shed more light on the DER comparisons, Figure~\ref{fig:confidence_intervals} contains plots with confidence intervals based on bootstrap~\cite{efron1994introduction}\footnote{We computed the intervals based on \cite{Ferrer_Confidence_Intervals_for}.} for the DER values. Given $N$ recordings in a test set, the test set is sampled with replacement $N$ times to obtain a new test set of the same size as the original (where some recordings could be missing and others could be repeated) and the DER is calculated over this set for a given system.  This is repeated 10000 times to obtain an empirical distribution of the DER for the system. Computing the 2.5 and 97.5 percentiles provides an estimated range where the system's performance will fall for 95\% of the test sets of size $N$. 

Although differences of a few DER points are observed for many datasets, the confidence intervals do not allow us to claim that such differences are significant in most cases. For example, the more than 10\% relative improvement for the clustering approach on both DIHARD2 and DIHARD3 full is not supported by disjunct confidence intervals. While in challenges teams always pursue the lowest error in the test set, we should not forget that small differences are probably not enough to claim that one system is really better than another. Moreover, sets with very few files and/or very few hours are not very good proxies for such comparisons. One can observe how for DipCo the intervals are very long, most likely heavily influenced by the limited amount of files in the set.
Finally, this analysis also allows us to compare the variability one can expect with different methods if the test set is changed. For example, while the DER for both approaches on Callhome is virtually the same, the clustering-based method shows a shorter interval and it should be preferred for a real application on files with similar characteristics.

\begin{figure}[H]
    \centering
    
    \begin{subfigure}[b]{0.2\textwidth}
        \centering
        \includegraphics[height=6cm]{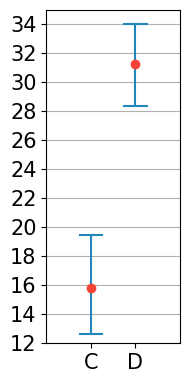}
        \caption{AISHELL-4}
    \end{subfigure}%
    \hfill
    \begin{subfigure}[b]{0.2\textwidth}
        \centering
        \includegraphics[height=6cm]{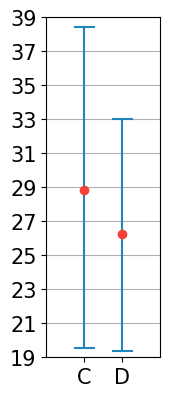}
        \caption{AliMeeting far}
    \end{subfigure}%
    \hfill
    \begin{subfigure}[b]{0.2\textwidth}
        \centering
        \includegraphics[height=6cm]{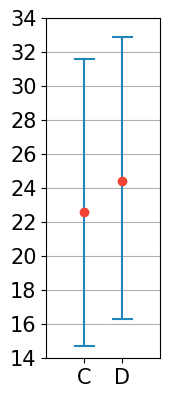}
        \caption{AliMeeting near}
    \end{subfigure}%
    \hfill
    \begin{subfigure}[b]{0.2\textwidth}
        \centering
        \includegraphics[height=6cm]{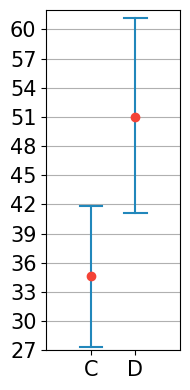}
        \caption{AMI array}
    \end{subfigure}%
    \hfill
    \begin{subfigure}[b]{0.2\textwidth}
        \centering
        \includegraphics[height=6cm]{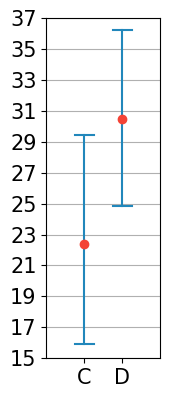}
        \caption{AMI headset}
    \end{subfigure}
    
    \begin{subfigure}[b]{0.2\textwidth}
        \centering
        \includegraphics[height=6cm]{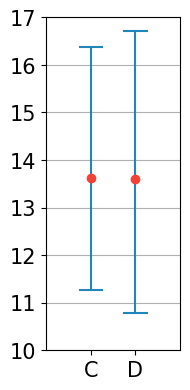}
        \caption{Callhome}
    \end{subfigure}%
    \hfill
    \begin{subfigure}[b]{0.2\textwidth}
        \centering
        \includegraphics[height=6cm]{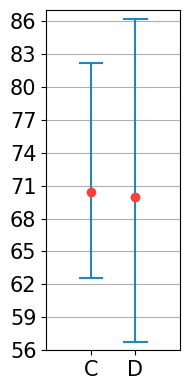}
        \caption{CHiME6}
    \end{subfigure}%
    \hfill
    \begin{subfigure}[b]{0.2\textwidth}
        \centering
        \includegraphics[height=6cm]{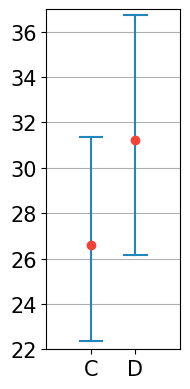}
        \caption{DIHARD2}
    \end{subfigure}%
    \hfill
    \begin{subfigure}[b]{0.2\textwidth}
        \centering
        \includegraphics[height=6cm]{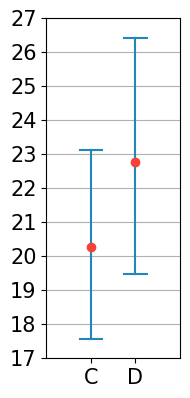}
        \caption{DIHARD3 full}
    \end{subfigure}%
    \hfill
    \begin{subfigure}[b]{0.2\textwidth}
        \centering
        \includegraphics[height=6cm]{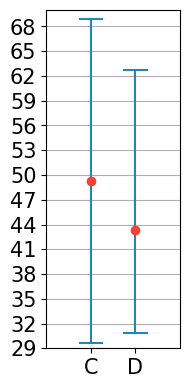}
        \caption{DipCo}
    \end{subfigure}
    
    \begin{subfigure}[b]{0.2\textwidth}
        \centering
        \includegraphics[height=6cm]{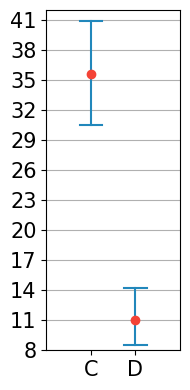}
        \caption{Mixer6}
    \end{subfigure}%
    \hfill
    \begin{subfigure}[b]{0.2\textwidth}
        \centering
        \includegraphics[height=6cm]{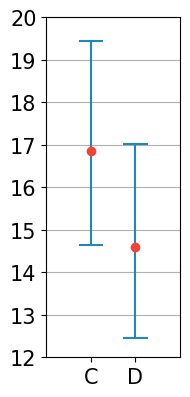}
        \caption{MSDWild}
    \end{subfigure}%
    \hfill
    \begin{subfigure}[b]{0.2\textwidth}
        \centering
        \includegraphics[height=6cm]{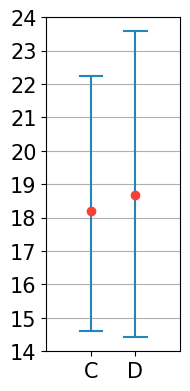}
        \caption{RAMC}
    \end{subfigure}%
    \hfill
    \begin{subfigure}[b]{0.2\textwidth}
        \centering
        \includegraphics[height=6cm]{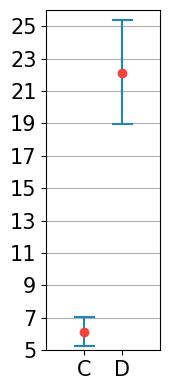}
        \caption{VoxConverse}
    \end{subfigure}
    
    \caption{Confidence intervals for $\alpha=0.05$ with bootstrapping using 10000 samples. The red dot denotes the DERs as shown in Table~\ref{tab:detailed_comparison} and the segments mark the intervals. ``C'' corresponds to the clustering-based method and ``D'' to DiaPer.}
    \label{fig:confidence_intervals}
\end{figure}

\newpage
\section{Moving forward}

As discussed in previous chapters, the last quinquennial has seen several changes in the field of speaker diarization. From adopting x-vectors in clustering-based approaches and abandoning speaker change detection in favor of uniform segmentation to the introduction of end-to-end models and single- or two-stage systems. 

While a few years ago it was common practice to evaluate systems (even in the context of challenges) assuming oracle voice activity detection, systems nowadays are expected to provide a full diarization solution. Obtaining good VAD performance is plausible in most scenarios and the quality of x-vector clustering has allowed to obtain remarkable performance in many situations. The main challenge in recordings captured with close-talk microphones and or quiet environments has become dealing with overlapped speech. However, this is not the case for more complex scenarios (i.e. far-field recordings, strong background noises).

End-to-end systems have emerged partly to address the challenges posed by overlapped speech. These methods provide much better performance in segments with overlap and provide good VAD performance without the need for external modules. However, end-to-end systems have considerably worse performance on recordings with many speakers and are outperformed by clustering-based systems in such scenarios. Furthermore, memory consumption can be a limiting factor in very long conversations for end-to-end systems.

Currently, the best compromise seems to be to use two-stage systems where an end-to-end model is run on relatively short segments, therefore handling VAD and overlap and a clustering-based second step reconciles the local decisions and allows for handling several speakers more accurately and longer recordings more efficiently. 

How to train these models is still an open problem. End-to-end models are ``data hungry'' by nature and training data have to be generated artificially. However, they are also very biased towards the data seen during training and the performance on some scenarios is far from useful. Generating training data appropriate for different scenarios is still a challenge. In contrast, clustering-based methods, nowadays based on speaker embeddings extraction, are very robust to different conditions and can leverage different sources of supervision without the need for synthetic data. How to combine these two aspects, namely to be able to train models in an end-to-end fashion while taking advantage of the supervision that embedding extractors are usually trained on, can be an avenue for better performing systems.

In this sense, both end-to-end and clustering-based approaches can complement each other in the following years even if not combined per se. On one hand, speaker embedding extractors tailored for diarization that can cope with silences and overlapped speech can help to bridge the gap and lead to clustering-based approaches that can deal with all aspects of diarization. On the other hand, end-to-end methods struggle to handle different scenarios so building speech encoders that are trained with a wide variety of data (as done for speaker embedding extractors) could help in this regard. Furthermore, the other current limitation of these models is handling several speakers. Different strategies to deal with such shortcomings need to be devised. Currently, TS-VAD-like models seem to be better than EEND-based ones when dealing with many speakers but that could change if the way speaker attractors are estimated improves.

Aside from models but equally important for the progress of the field, the availability of datasets representative of real applications needs to be considered. In recent years, many new corpora have been released to the public under free licenses. This is a major boost for democratization in the field allowing newcomers to delve into speaker diarization more easily. It is important, however, that new datasets have properly defined partitions and evaluation setups to avoid future problems when comparing results. Regarding results, many datasets present partitions into development/test sets but the development part is normally used for fine-tuning end-to-end systems; therefore, making decisions on the test part while developing the models. These results need to be taken with reservations since they might not hold on different data. This has become common practice in recent years and we need to be careful as a community in such regard. Moreover, comparisons are usually done on diarization error rate basis without showing any kind of statistical significance evidence which raises questions regarding the real improvements.

Finally, another aspect to encourage fair comparisons is the need for releasing models to the public. As shown in Tables~\ref{tab:ch2two_other_methods} and \ref{tab:ch2all_other_methods}, only a fraction of recent state-of-the-art works share their recipes making comparisons on different datasets very difficult. Sharing code for our published results should be the norm rather than an exception if we want a field that progresses openly and steadily.


  \else
    \input{projekt-01-kapitoly-chapters}
  \fi
  
  % ...

\ifslovak
  \makeatletter
  \def\@openbib@code{\addcontentsline{toc}{chapter}{Literatúra}}
  \makeatother
  \bibliographystyle{bib-styles/Pysny/skplain}
\else
  \ifczech
    \makeatletter
    \def\@openbib@code{\addcontentsline{toc}{chapter}{Literatura}}
    \makeatother
    \bibliographystyle{bib-styles/Pysny/czplain}
  \else 
    \makeatletter
    \def\@openbib@code{\addcontentsline{toc}{chapter}{Bibliography}}
    \makeatother
    \bibliographystyle{apalike}
  \fi
\fi
  \begin{flushleft}
  \bibliography{projekt}
  \end{flushleft}

  \iftwoside
    \cleardoublepage
  \fi

  \appendix
\ifczech
  \renewcommand{\appendixpagename}{Přílohy}
  \renewcommand{\appendixtocname}{Přílohy}
  \renewcommand{\appendixname}{Příloha}
\fi
\ifslovak
  \renewcommand{\appendixpagename}{Prílohy}
  \renewcommand{\appendixtocname}{Prílohy}
  \renewcommand{\appendixname}{Príloha}
\fi

  
\ifslovak
\else
  \ifczech
  \else
  \fi
\fi
  \startcontents[chapters]
  \setlength{\parskip}{0pt} 
  
  \ifODSAZ
    \setlength{\parskip}{0.5\bigskipamount}
  \else
    \setlength{\parskip}{0pt}
  \fi
  
  \iftwoside
    \cleardoublepage
  \fi
  
  \ifenglish









  \else
    \input{projekt-30-prilohy-appendices}
  \fi
  
  
\end{document}